\documentclass[a4paper,12pt]{article}
\usepackage{amssymb,latexsym,amsmath}

\usepackage{fullpage}
\usepackage{mathrsfs}
\usepackage{bbold}

\usepackage[pdftex]{graphicx}
\usepackage[table,dvipsnames]{xcolor}
\usepackage{multirow}
\usepackage{rotating}
\usepackage{graphicx}
\usepackage{ulem}
\usepackage[pdftex]{hyperref}

\makeatletter \@addtoreset{equation}{section}
\makeatother

\allowdisplaybreaks[1]

\begin{document}

\begin{titlepage}
	\thispagestyle{empty}
	
	\vspace{35pt}
	
	\begin{center}
	    { \LARGE{\bf Gauged $D=4$ ${\cal N}=4$ Supergravity }}
		
		\vspace{50pt}
		
		{G.~Dall'Agata$^{1,2}$, N.~Liatsos$^{3}$, R.~Noris$^{4}$ and M.~Trigiante$^{5,6}$}
		
		\vspace{25pt}

		{
		$^1${\it  Dipartimento di Fisica e Astronomia ``Galileo Galilei''\\
		Universit\`a di Padova, Via Marzolo 8, 35131 Padova, Italy}

		\vspace{15pt}

		$^2${\it   INFN, Sezione di Padova \\
		Via Marzolo 8, 35131 Padova, Italy}

		\vspace{15pt}

		$^3${\it Physics Division, National Technical University of Athens\\
		15780 Zografou Campus, Athens, Greece}

		\vspace{15pt}

		$^4${\it  CEICO, Institute of Physics of the Czech Academy of Sciences,\\
		Na Slovance 2, 182 21 Prague 8, Czech Republic}
		}
		
		\vspace{15pt}
		
		$^5${\it Dipartimento di Scienza Applicata e Tecnologia (DISAT), Politecnico di Torino, \\
		C.so Duca degli Abruzzi 24, I-10129 Torino, Italy}
		
		\vspace{15pt}

		$^6${\it   INFN, Sezione di Torino \\
		Via Pietro Giuria, 1, 10125 Torino, Italy}

		\vspace{40pt}
		
		{ABSTRACT}
	\end{center}
We present the full Lagrangian and supersymmetry transformation rules for the gauged $D=4$, ${\cal N}=4$ (half-maximal) supergravity coupled to an arbitrary number of vector multiplets. Using the embedding tensor formulation, the final results are universal and valid in arbitrary symplectic frames.
We also analyze the conditions for the critical points of the scalar potential and specify the full spectrum of the quadratic fluctuations about Minkowski vacua. 
This allows us also to exclude the appearance of quadratic divergences in the 1-loop corrections to the scalar potential for any Minkowski vacuum fully breaking supersymmetry.
We also provide some interesting byproducts of our analysis, like the field equations and the quadratic constraints for the fermion shifts characterizing the gauging (also known as T-tensor identities).

\vspace{10pt}

\end{titlepage}

\baselineskip 6 mm



\tableofcontents

\section{Introduction} 
\label{sec:introduction}

Half-maximal supergravities in four dimensions have played an important role in understanding several key aspects of string theory, like dualities \cite{Sen:1994fa}, the microscopic origin of black hole entropy \cite{Maldacena:1996gb,Dijkgraaf:1996it} and the existence of entire orbits of purely non-geometric string compactifications \cite{Dibitetto:2012rk}.
The main reason for the interest in these theories lies in the fact that they provide models with the maximum number of supersymmetries compatible with a consistent coupling of the gravity multiplet to matter multiplets.
This means that they enjoy the strong constraints deriving from supersymmetry, while keeping the freedom of adding an arbitrary number of matter vector multiplets.

While the first instances of four-dimensional pure ${\cal N}=4$ supergravities were constructed almost 50 years ago in \cite{Das:1977uy, Cremmer:1977tc, Cremmer:1977tt, Freedman:1978ra}, the coupling of ${\cal N}=4$ supergravity to vector multiplets, as well as some of its gaugings, were analyzed a few years later in \cite{deRoo:1984zyh, Bergshoeff:1985ms, deRoo:1985np, deRoo:1985jh, Perret:1987nk,Perret:1988jq}. 
More recently, sparked by the renewed interest in flux compactifications of string theory,  various gauged ${\cal N}=4$ supergravity models originating from type IIB or IIA orientifold compactifications \cite{Frey:2002hf, Kachru:2002he} were studied in detail \cite{DAuria:2002qje,DAuria:2003nhg,Berg:2003ri,Angelantonj:2003rq,Angelantonj:2003up,Villadoro:2004ci,Derendinger:2004jn,Villadoro:2005cu,DallAgata:2009wsi}, but always on a case by case basis.  

Currently, the most general analysis of the structure of the gauged theory is provided by \cite{Schon:2006kz}, where one can find a systematic discussion of the consistency conditions for the gauging procedure as well as various results concerning the bosonic Lagrangian, the supersymmetry transformations of the fermions and the relation of such models to flux compactifications.
However, as we will argue in the following, such analysis is incomplete and a proper general and unified framework for all possible gaugings of ${\cal N}=4$ supergravity is not readily available yet.

The contemporary understanding of four-dimensional gauged supergravities relies on the fact that any model is fully specified by the choice of symplectic frame and of embedding tensor. 
The first ingredient is related to the fact that one can formulate different equivalent classical ungauged supergravity models according to the different realizations of the rigid symmetry group of the Lagrangian $G_{\cal L}$, which is a subgroup of the duality group $G$ (for ${\cal N}=4$ supergravity coupled to $n$ vector multiplets, with a total of $n_v = 6+n$ vector fields, $G=\text{SL}(2,\mathbb{R})\times\text{SO}(6,$n$)$).
The group $G_{\cal L}$ is determined by the choice of which among the vector fields present in the theory, $A^{\Lambda}_\mu$, $\Lambda = 1,\ldots,n_v$, and their magnetic duals, $A_{\Lambda\mu}$, have a local description in the Lagrangian.
This choice in turn determines the embedding of $G$ inside the symplectic group $\text{Sp}(2n_v,\mathbb{R})$. 
Different choices of symplectic frames are indeed connected to one another by symplectic rotations and yield in general different Lagrangians that are not related to each other by local field redefinitions but are on-shell equivalent, as they lead to sets of Bianchi identities and equations of motion that can be mapped into each other by field redefinitions \cite{Gaillard:1981rj,Samtleben:2008pe,Trigiante:2016mnt,DallAgata:2021uvl}. 
The second ingredient, the embedding tensor $\Theta$, provides a duality covariant formulation of the gauging procedure, and specifies the decomposition of the gauge group generators in terms of the generators of $G$, of which the gauge group must be a subgroup.
The advantage of this description of the gauging is twofold. 
On the one hand, minimal couplings contain both electric and magnetic gauge fields in $G$-covariant combinations through the components of the embedding tensor, which ensures that the Bianchi identities and field equations of the gauged theory are formally invariant under global duality transformations, provided we treat the embedding tensor as a spurionic object that transforms under $G$. 
On the other hand, the gauge group is no longer required to be a subgroup of the rigid symmetry group of the original ungauged Lagrangian, which depends on the choice of the symplectic frame. This duality covariant method for gauging a supergravity theory was introduced in \cite{Cordaro:1998tx, Nicolai:2000sc, Nicolai:2001sv} and further developed in  \cite{deWit:2002vt,deWit:2004nw,deWit:2005ub,deWit:2007kvg} (see also \cite{Samtleben:2008pe,Trigiante:2016mnt,DallAgata:2021uvl} for reviews), while it was applied (with some limitations) to the cases of the gauged four- and five-dimensional ${\cal N}=4$ supergravities in \cite{Schon:2006kz}.  

In detail, \cite{Schon:2006kz} analyzed the consistency constraints on the embedding tensor, leading to the conclusion that all possible gaugings of ${\cal N}=4$ supergravity in four spacetime dimensions are parametrized by two real constant SL(2,${\mathbb R}$) $\times$ SO(6,$n$) tensors, $\xi_{\alpha M}$ and $f_{\alpha MNP}=f_{\alpha[MNP]}$, which are subject to a specific set of quadratic constraints that we will review in the following. 
However, only partial results for the Lagrangian and supersymmetry transformations were presented, also forcing a specific choice of symplectic frame, such that $G_{\cal L}=\text{SO}(1,1)\times\text{SO}(6,n)$. 
While this is a legitimate choice, it is so constraining that not even the maximally supersymmetric anti-de Sitter vacuum can be obtained by a pure electric gauging in this frame \cite{Louis:2014gxa}.

Our work overcomes these limitations by providing the full Lagrangian and supersymmetry rules for the gauged four-dimensional ${\cal N} = 4$ supergravity in an arbitrary symplectic frame.
This implies that any known (as well as yet unknown) vacuum of such a theory can be obtained from an electrically gauged theory, which will be incorporated in our general Lagrangian. 
Our general analysis allows us also to discuss the general structure of the vacua of any such theory and we therefore discuss both the conditions for the critical points of the scalar potential, as well the spectrum of the quadratic fluctuations about Minkowski vacua.
We then use this result to prove that the quadratic supertrace of the mass matrices is vanishing for any Minkowski vacuum that breaks all supersymmetries of any consistent ${\cal N} = 4$ gauged supergravity.
This is a rather non-trivial result, which extends what has already been  found in the case of the much more constrained maximal supergravity theory \cite{DallAgata:2012tne} and gives us a first insight into the quantum corrections of this class of theories.

All these results have been obtained by a careful reinterpretation of the quadratic consistency constraints in terms of the fermion shifts, which we also present in detail.
They will constitute the basis of possible further applications of this work, like the computation of the spectrum of fluctuations about anti-de Sitter vacua or the computation of higher-order supertrace relations.

\medskip

This paper is organized as follows: in section \ref{sec:the_ingredients_of_n_4_supergravity}, we give the field content of the four-dimensional ${\cal N}=4$ supergravity coupled to $n$ vector multiplets and describe the geometry of the coset space $\frac{\text{SL(2,${\mathbb R}$)}}{\text{SO(2)}} \times \frac{\text{SO(6,$n$)}}{\text{SO(6)} \times \text{SO($n$)}}$, parametrized by the scalar fields of the theory. 
In section \ref{sec:duality_and_symplectic_frames}, we briefly discuss the electric/magnetic duality in ${\cal N}=4$ supergravity, we introduce projectors, acting on symplectic vectors, which parametrize the choice of the symplectic frame and we give their explicit expressions for some of the symplectic frames in which the $D=4$, ${\cal N}=4$ supergravity has been formulated in the literature. 
In section \ref{sec:duality_covariant_gauging}, we describe the $\text{SL}(2,\mathbb{R}) \times \text{SO}(6,n)$-covariant formulation of the gauging procedure, which has also been discussed in detail in \cite{Schon:2006kz}, to keep our presentation self-contained. 
In section \ref{sec:lagrangian_and_supersymmetry_rules}, we give the complete Lagrangian in an arbitrary symplectic frame and the local supersymmetry transformation rules for the gauged $D=4$, ${\cal N}=4$ Poincar\'{e} supergravity coupled to $n$ vector multiplets, as well as some of the corresponding Bianchi identities and field equations and we compute the commutator of two consecutive local supersymmetry transformations. 
We end the section by discussing the relevant gauge fixings and by providing a constructive definition of the symplectic matrix which connects the chosen symplectic frame to the intrinsic electric frame of the embedding tensor. 
In section \ref{sec:vacua_masses_gradient_flow_and_supertrace_relations}, we derive the conditions satisfied by the critical points of the scalar potential, we specify the mass matrices of all the fields in the theory and we compute the supertrace of the squared mass eigenvalues for Minkowski vacua that completely break ${\cal N}=4$ supersymmetry. 
We summarize our conventions in appendix \ref{sec:conventions}, while in appendix \ref{sec:comparison_with_previous_articles}, we point out a discrepancy of our results with those of \cite{Schon:2006kz} and we compare our notation with that of \cite{Perret:1988jq}. 
In appendix \ref{sec:the_solution_of_the_bianchi_identities},
we provide the full derivation of the local supersymmetry transformations and of the Lagrangian for the ungauged and the gauged $D=4$, ${\cal N}=4$ matter-coupled Poincar\'{e} supergravities in an arbitrary symplectic frame, using the rheonomic approach. 
Finally, in appendix \ref{sec:t_tensor_identities}, we derive the quadratic constraints satisfied by the T-tensor by appropriately dressing the quadratric constraints on the embedding tensor with the coset representatives.


\section{The Ingredients of ${\cal N}=4$ Supergravity} 
\label{sec:the_ingredients_of_n_4_supergravity}
The ${\cal N}=4$ Poincar\'{e} supergravity in four dimensions is based on the  Poincar\'{e} superalgebra with four spinorial generators and ${\rm U}(4)$ R-symmetry group. 
We shall label the fundamental representation ${\bf 4}$ of the latter by the indices $i,j,\dots=1,\dots, 4$.
The theory allows for only two kinds of supermultiplets containing fields with spin not exceeding 2: the gravity and the vector ones.
The gravity multiplet contains the graviton $g_{\mu\nu}$, four gravitini $\psi_\mu^i$, six vectors $A_\mu^{ij}=-A_\mu^{ji}$, four spin-1/2 fermions $\chi_i$ (dilatini) and a complex scalar $\tau$, parameterizing the coset manifold $\frac{\text{SL(2,${\mathbb R}$)}}{\text{SO(2)}}$. 
This multiplet can be coupled to $n$ vector multiplets, which contain $n$ vector fields $A_\mu^{\underline{a}}$, $\underline{a} = 1, \ldots, n$, $4n$ gaugini $\lambda^{\underline{a}i}$, and $6n$ real scalar fields, parameterizing the scalar manifold $\frac{\text{SO(6,$n$)}}{\text{SO(6)} \times \text{SO($n$)}}$.
Overall, the scalar $\sigma$-model is described by the coset space \cite{deRoo:1984zyh,Bergshoeff:1985ms,deRoo:1985jh}
\begin{equation}
    \label{Mscalar}
    \mathcal{M} = \frac{G}{H} = \frac{\text{SL(2,${\mathbb R}$)}}{\text{SO(2)}} \times \frac{\text{SO(6,$n$)}}{\text{SO(6)} \times \text{SO($n$)}}\, .
\end{equation}
In the next two subsections, we shall focus on the scalar sector and describe the coset geometry of $ \mathcal{M}$. 
Subsequently, in subsection \ref{sub:fermions_as_sections_of_the_tangent_bundle}, we shall fix the relevant notations as far as the fermion fields are concerned.

\subsection{The scalar sector of the gravity multiplet} 
\label{sub:the_scalar_sector_of_the_gravity_multiplet}

As mentioned above, the two real scalar fields contained in the gravity multiplet are the coordinates of the SL(2,${\mathbb R}$)/SO(2) factor of the coset (\ref{Mscalar}).
As a homogeneous manifold, SL(2,${\mathbb R}$)/SO(2) can be described in terms of a coset representative  $S\in \text{SL}(2,\mathbb{R})$, which transforms under the isometry group SL(2,${\mathbb R}$) and the (local) isotropy group SO(2) as
\begin{equation}
    S \rightarrow g S h(x),
\end{equation} 
where global SL(2,${\mathbb R}$) transformations $g$ act on $S$ from the left, while local SO(2) transformations $h(x)$ act on $S$ from the right.
Following \cite{Schon:2006kz}, we will actually use the convenient representation in terms of a complex SL(2,${\mathbb R}$) vector 
\begin{equation}
\label{Valpha}
   \mathcal{V}_\alpha = {S_\alpha}^{\underline{\alpha}} v_{\underline{\alpha}} , 
\end{equation}
where $\alpha=+,-$ is an SL(2,${\mathbb R}$) index, ${\underline{\alpha}}=1,2$ is an SO(2) index and $v_{{\underline{\alpha}}}=(1,i)^T$. 
From the definition (\ref{Valpha}), one can immediately deduce that the ${\cal V}_{\alpha}$ vector satisfies
\begin{equation}
\label{VV*-V*V=e}
    \mathcal{V}_\alpha \mathcal{V}_\beta^* - \mathcal{V}_\alpha^*  \mathcal{V}_\beta = -2 i \epsilon_{\alpha \beta} \, , 
\end{equation}
where $\epsilon_{\alpha \beta} = - \epsilon_{\beta \alpha}$ and $\epsilon_{+-}=1$. 
Since conjugate 2-dimensional representations of SL(2,${\mathbb R}$) are equivalent, we can raise and lower SL(2,${\mathbb R}$) indices according to  the following convention
\begin{equation}
    \mathcal{V}^\alpha =  \mathcal{V}_\beta \epsilon^{\beta \alpha}, \qquad \mathcal{V}_\alpha =\epsilon_{\alpha \beta  } \mathcal{V}^\beta ,
\end{equation}
where $\epsilon^{\alpha \beta} = - \epsilon^{\beta \alpha}$, with $\epsilon^{+-}=1$ and $\epsilon^{\alpha \gamma} \epsilon_{\beta \gamma} = \delta^{\alpha}_\beta$.

The SO(2) $\cong$ U(1) action on $S$ implies that ${\cal V}_\alpha$ transforms as a charge +1 object
\begin{equation}
    {\cal V}_\alpha \rightarrow e^{i \theta(x)} {\cal V}_\alpha \, ,
\end{equation}
for a standard parameterization of 
\begin{equation}
		h(x) = \left(\begin{array}{cc}
		\cos \theta & \sin \theta \\[2mm]
		-\sin \theta & \cos \theta 
		\end{array}\right).
\end{equation}

In addition, it is useful to introduce the positive definite symmetric matrix 
\begin{equation}
    \label{Mab}
    M_{\alpha \beta} = {S_\alpha}^{\underline{\alpha}} {S_\beta}^{\underline{\beta}} \,\delta_{\underline{\alpha} \underline{\beta}} = \text{Re} ({\cal V}_{\alpha} {\cal V}_{\beta}^*) \, , 
\end{equation}
which satisfies 
\begin{equation}
    M^{\alpha \beta} M_{\beta \gamma} = \delta^{\alpha}_{\gamma}\,.
\end{equation}  

Using standard coset geometry, we can compute, for SL(2,${\mathbb R}$)/SO(2), the following complex vielbein 
\begin{equation}
\label{Pdef}
    P = \frac{i}{2} \epsilon^{\alpha \beta}  \,\mathcal{V}_\alpha d \mathcal{V}_\beta\,,
\end{equation}
in terms of which the metric on this manifold can be written as 
\begin{equation}
   ds^2=2\,P\, P^*, 
\end{equation}
and SO(2)-connection
\begin{equation}
    \label{Adef}
     \mathcal{A}  =  - \frac{1}{2} \epsilon^{\alpha \beta}  \,\mathcal{V}_\alpha d \mathcal{V}_\beta^*\, ,
\end{equation}
which follow from the usual decomposition of the left-invariant one-form $\Psi = S^{-1}dS$ along the basis $\{\sigma_1, i \sigma_2, \sigma_3\}$ of the Lie algebra $\mathfrak{sl}(2,\mathbb{R})$, where $i \sigma_2$ spans its compact $\mathfrak{so}(2)$ factor.
The corresponding Maurer--Cartan equation $d \Psi + \Psi \wedge \Psi = 0$ yields the relation
\begin{equation}
\label{scalarBianchi1}
    DP \equiv d P - 2i {\cal A} \wedge P = 0 
\end{equation}
and provides the SO(2)-curvature
\begin{equation}
\label{SO(2)curv}
 F \equiv d \mathcal{A} = i P^* \wedge P \, .
\end{equation}
With a little algebra, one can also derive the useful identity
\begin{equation}
\label{DV}
     D \mathcal{V}_\alpha \equiv d \mathcal{V}_\alpha - i \mathcal{A} \mathcal{V}_\alpha = P \mathcal{V}_\alpha^* \, ,
\end{equation}
which captures the full differential structure of the coset geometry.


\subsection{The scalar sector of the vector multiplets} 
\label{sub:the_vector_multiplet_scalar_sector}

The coset space parametrized by the scalars of the vector multiplets can be described by means of a coset representative ${L_M}^{\underline{M}}=({L_M}^{\underline{m}},{L_M}^{\underline{a}})$, where $M=1,\dots,n+6$ is a vector index of SO(6,$n$), $\underline{m}=1,\dots,6$ and $\underline{a}=1,\dots,n$ are indices of the fundamental representations of  SO(6) and SO($n$) respectively, while $\underline{M}$ is an index which, decomposed as $\underline{M}=(\underline{m},\underline{a})$, bears the local action of SO(6) $ \times$ SO($n$).

The matrix $L$ itself is an element of SO(6,\,$n$), meaning that 
\begin{equation}
\label{LSO6n}
   \eta_{MN} = \eta_{\underline{M} \underline{N}} {L_M}^{\underline{M}} {L_N}^{\underline{N}} = {L_M}^{\underline{M}} L_{N \underline{M}} ={L_M}^{\underline{m}} L_{N \underline{m}} + {L_M}^{\underline{a}} L_{N \underline{a}} ,
\end{equation}
where $\eta_{MN}=\eta_{\underline{M} \underline{N}} = \text{diag}(-1,-1,-1,-1,-1,-1,1,\dots,1)$.
The constant matrices $\eta_{MN}$ and $\eta_{\underline{M}\underline{N}}$ and their inverses $\eta^{MN}$ and $\eta^{\underline{M}\underline{N}}$ can be used as metrics to raise and lower the corresponding indices.

As for the scalar sector of the gravity multiplet, it is useful to introduce the positive definite symmetric matrix $M=LL^T$ with elements 
\begin{equation}
    \label{MMN}
     M_{MN} = -   {L_M}^{\underline{m}} L_{N \underline{m}} + {L_M}^{\underline{a}} L_{N \underline{a}} 
\end{equation}
and its inverse $M^{MN}$,
   \begin{equation}
\label{MM}
    M^{MN} M_{NP} = \delta^M_P.
\end{equation}

In this case, the $\sigma$-model geometry can be described in terms of a vielbein matrix $P_{\underline{a}}{}^{\underline{m}}$, together with SO(6) and SO($n$) connections $\omega_{\underline{m}}{}^{\underline{n}}$ and $\omega_{\underline{a}}{}^{\underline{b}}$ respectively, constructed from the left-invariant one-form
\begin{equation}
    \Omega = L^{-1} dL, 
\end{equation}
which, in the fundamental representation of SO(6,$n$), has the following matrix representation
\begin{equation}
    \label{MCf2}
    {\Omega_{\underline{M}}}^{\underline{N}} =  {L_{\underline{M}}}^{M} d {L_M}^{\underline{N}} = 
    \begin{pmatrix}
{\omega_{\underline{m}}}^{\underline{n}} & {P_{\underline{m}}}^{\underline{b}} \\
    {P_{\underline{a}}}^{\underline{n}} &  {\omega_{\underline{a}}}^{\underline{b}} 
    \end{pmatrix} . 
\end{equation}
In terms of the vielbein matrix, the metric on the coset manifold $\text{SO}(6,n)/(\text{SO}(6)\times\text{SO}(n))$ has the form $$ds^2=-P^{\underline{ma}}\,P_{\underline{ma}}\,.$$
Notice that $\Omega$ satisfies
\begin{equation}
    {\Omega_{\underline{M}}}^{\underline{N}}=- {\Omega^{\underline{N}}}_{\underline{M}} 
\end{equation}
and hence ${P_{\underline{m}}}^{\underline{a}} = - {P^{\underline{a}}}_{\underline{m}}$. 
The $\mathfrak{so}(6,n)$ Maurer--Cartan equations $d{\Omega_{\underline{M}}}^{\underline{N}}+{\Omega_{\underline{M}}}^{\underline{P}} \wedge {\Omega_{\underline{P}}}^{\underline{N}} = 0$ also imply the following relations 
    \begin{align}
        \label{DPam=0}
     D {P_{\underline{a}}}^{\underline{m}} &\equiv d {P_{\underline{a}}}^{\underline{m}} + {\omega_{\underline{a}}}^{\underline{b}}\wedge {P_{\underline{b}}}^{\underline{m}}  + {\omega^{\underline{m}}}_{\underline{n}} \wedge {P_{\underline{a}}}^{\underline{n}}=0 \,  , \\[2mm]
      \label{RSO(6)} 
     {R_{\underline{m}}}^{\underline{n}} &\equiv d  {\omega_{\underline{m}}}^{\underline{n}} + {\omega_{\underline{m}}}^{\underline{p}} \wedge {\omega_{\underline{p}}}^{\underline{n}} = -  {P_{\underline{m}}}^{\underline{a}} \wedge {P_{\underline{a}}}^{\underline{n}},\\[2mm]
     \label{RSO(n)}
     {R_{\underline{a}}}^{\underline{b}} & \equiv d  {\omega_{\underline{a}}}^{\underline{b}} + {\omega_{\underline{a}}}^{\underline{c}} \wedge {\omega_{\underline{c}}}^{\underline{b}} = -  {P_{\underline{a}}}^{\underline{m}} \wedge {P_{\underline{m}}}^{\underline{b}} \, , 
    \end{align}
which provide the definitions for the SO(6) and SO($n$) curvatures ${R_{\underline{m}}}^{\underline{n}}$ and ${R_{\underline{a}}}^{\underline{b}}$, respectively. 

The SO(6) factor in the coset has to be identified with the ${\mathbb Z}_2$ quotient of the SU(4) factor of the R-symmetry group.
It is therefore useful to note that an SO(6)-vector $v^{\underline{m}}$ can alternatively be described by an antisymmetric SU(4)-tensor $v^{ij}=-v^{ji}$, $i,j=1,\dots,4$, subject to the pseudo-reality constraint 
\begin{equation}
v_{ij}=(v^{ij})^*=\frac{1}{2} \epsilon_{ijkl} v^{kl}. 
\end{equation}
The map $v^{\underline{m}} \rightarrow v^{ij}$ can be constructed explicitly by using six antisymmetric 4$\times$4 matrices ${\Gamma}^{\underline{m} i j }$ interpolating between the two representations,
\begin{equation}
\label{map}
    v^{ij} = {\Gamma}^{\underline{m} i j } v_{\underline{m}} \,,
\end{equation}
normalized in such a way that 
\begin{equation}
     v^{\underline{m}} w_{\underline{m}} = - \frac{1}{2} \epsilon_{ijkl} v^{ij} w^{kl}  = - v^{ij} w_{ij} = - v_{ij} w^{ij}\,.
\end{equation}
Using this representation, equation \eqref{LSO6n} can be written as
\begin{equation}
    \label{LSO6SU4}
    \eta_{MN} = - {L_M}^{ij} L_{Nij} + {L_M}^{\underline{a}} L_{N \underline{a}} = - \frac{1}{2} \epsilon_{ijkl} {L_M}^{ij} {L_N}^{kl} +  {L_M}^{\underline{a}} L_{N \underline{a}},
\end{equation}
implying 
\begin{equation}
   {L_{(M}}^{ik} L_{N)jk} = - \frac{1}{4}\, \delta^i_j \,(\eta_{MN}-{L_M}^{\underline{a}} L_{N \underline{a}})  = \frac{1}{4} \,\delta^i_j\, {L_M}^{kl} L_{Nkl}\,,
\end{equation}
while the Bianchi identity for the vielbein 1-forms, now $P_{\underline{a}}{}^{ij}$, \eqref{DPam=0} may be written as 
 \begin{equation}
     \label{DPaij=0}
      D {P_{\underline{a}}}^{ij} \equiv d {P_{\underline{a}}}^{ij} + {\omega_{\underline{a}}}^{\underline{b}}\wedge {P_{\underline{b}}}^{ij} - {\omega^{ij}}_{kl} \wedge {P_{\underline{a}}}^{kl} =0 \, , 
 \end{equation}
where
\begin{equation}
    {\omega^{ij}}_{kl} = {{\Gamma}_{\underline{m}}}^{ij} {\Gamma}_{\underline{n} kl}\, \omega^{\underline{m} \underline{n}}\,.
\end{equation}
Since $\omega$ plays the role of an SU(4) connection, it can be shown that 
\begin{equation}
   {\omega^{ij}}_{kl} = 2 {\omega^{[i}}_{[k} \delta^{j]}_{l]} ,
\end{equation}
with $\omega^i{}_i = 0$ and ${\omega_i}^j = ( {\omega^i}_j )^*= -  {\omega^j}_i$, so that \eqref{DPaij=0} becomes 
\begin{equation}
    \label{DPaij=02}
     D {P_{\underline{a}}}^{ij} \equiv d {P_{\underline{a}}}^{ij} + {\omega_{\underline{a}}}^{\underline{b}}\wedge {P_{\underline{b}}}^{ij} - {\omega^i}_k \wedge {P_{\underline{a}}}^{kj}  - {\omega^j}_k \wedge {P_{\underline{a}}}^{ik} =0\,.
\end{equation}
In the same fashion, we can define the SU(4) curvature as
\begin{equation}
\label{RSU4}
    {R^{i}}_j = {R^{ik}}_{jk} = d {\omega^i}_j -  {\omega^i}_k \wedge  {\omega^k}_j\,  = P^{\underline{a}ik} \wedge P_{\underline{a} j k} \, ,  
\end{equation}
where $ {R^{ij}}_{kl} = {{\Gamma}_{\underline{m}}}^{ij} {\Gamma}_{\underline{n} kl} R^{\underline{m} \underline{n}}$, $R^i{}_i = 0$, ${R_i}^j = ( {R^i}_j )^*= -  {R^j}_i$, and the last equality in (\ref{RSU4}) follows from equation \eqref{RSO(6)}.
Also, the expression for the SO($n$) curvature in terms of the new vielbein 1-forms is
    \begin{equation}
    \label{RSOnPP}
   {R_{\underline{a}}}^{\underline{b}} = - P_{\underline{a}ij} \wedge P^{\underline{b}ij}\, . 
\end{equation}

We close this section by giving some useful relations following from the previous definitions.
These are the derivatives of the coset representatives, which satisfy
\begin{align}
\label{DLij}
    D {L_M}^{ij} & \equiv  d {L_M}^{ij} - {\omega^i}_k {L_M}^{kj}  - {\omega^j}_k {L_M}^{ik} =   {L_M}^{\underline{a}} {P_{\underline{a}}}^{ij} , \\[2mm]
    \label{DLa}
 D {L_M}^{\underline{a}} &\equiv  d {L_M}^{\underline{a}} + {\omega^{\underline{a}}}_{\underline{b}} {L_M}^{\underline{b}} = {L_M}^{ij} {P^{\underline{a}}}_{ij}\,.
\end{align}


\subsection{The fermion fields} 
\label{sub:fermions_as_sections_of_the_tangent_bundle}
As usual in supergravity theories, the fermion fields transform in representations of the holonomy group of the scalar manifold, which in our case, locally coincides with the isotropy group $H=\text{SO}(2) \times \text{SO}(6) \times \text{SO}(n)$. 
More precisely, the gravitini, the dilatini and the gaugini transform in the fundamental representation of SU(4), which is the universal cover of SO(6), while the gaugini alone transform in the fundamental representation of SO($n$) as well. 
Moreover, the SO(2) $\cong$ U(1) factor of $H$ acts on the fermions as a multiplication by a complex phase $e^{iq \Lambda(x)}$, where the charges $q$ of $\psi^i_\mu$, $\chi^i$ and $\lambda^{\underline{a}i}$ are 
\begin{equation}
    q(\psi^i_\mu) = - \frac{1}{2}, \quad q(\chi^i)=\frac{3}{2} \qquad \text{and} \qquad q(\lambda^{\underline{a}i})=\frac{1}{2}
\end{equation}
respectively. 
More details about fermions and their properties can be found in appendix~\ref{sec:conventions}.
We only remind here that $\psi^i_\mu$ and $\lambda^{\underline{a}i}$ are left-handed, while $\chi^i$ are right-handed, i.e.
\begin{equation}
    \label{LH}
    \gamma_5 \psi^i_\mu = \psi^i_\mu, \quad \gamma_5 \chi^i = - \chi^i, \quad \gamma_5 \lambda^{\underline{a}i} = \lambda^{\underline{a}i},
\end{equation}
and that their charge conjugates $\psi_{i \mu} = (\psi^i_\mu)^c$, $\chi_i=(\chi^i)^c$ and $\lambda_i^{\underline{a}} = (\lambda^{\underline{a}i})^c$ have opposite chiralities
\begin{equation}
    \gamma_5 \psi_{i \mu} = - \psi_{i \mu}, \quad \gamma_5 \chi_i = \chi_i, \quad \gamma_5 \lambda_i^{\underline{a}} = - \lambda_i^{\underline{a}}. 
\end{equation}



\section{Duality and Symplectic Frames} 
\label{sec:duality_and_symplectic_frames}

The sector of the ungauged Lagrangian specifying the vector field couplings at the 2-derivative level can be written as \cite{DallAgata:2021uvl}
\begin{equation}
\label{Lgen}
e^{-1} \mathcal{L} = \frac{1}{4} {\cal I}_{\Lambda \Sigma} F^{\Lambda}_{\mu \nu} F^{\Sigma \mu \nu} + \frac{1}{4} {\cal R}_{\Lambda \Sigma} F^{\Lambda}_{\mu \nu} (*F^\Sigma)^{\mu \nu} + \frac{1}{2} O_{\Lambda}^{\mu \nu} F^{\Lambda}_{\mu \nu} + e^{-1} \mathcal{L}_{\text{rest}}, 
\end{equation}
where $e=\text{det}(e_{\mu}^a)$, $A^\Lambda_\mu$, $\Lambda = 1, \ldots, n+6$, are the vector fields, $F^{\Lambda}_{\mu \nu} = 2 \partial_{[\mu} A^{\Lambda}_{\nu]}$ and $(*F^\Lambda)_{\mu \nu} = \frac{1}{2} \epsilon_{\mu \nu \rho \sigma} F^{\Lambda \rho \sigma}$ are the vector field strengths and their Hodge duals respectively. 
Furthermore, ${\cal I}_{\Lambda \Sigma}$ and ${\cal R}_{\Lambda \Sigma}$ are real symmetric matrices that depend on the scalar fields, with ${\cal I}_{\Lambda \Sigma}$ being negative definite,  $O_{\Lambda}^{\mu \nu}$ is an antisymmetric field dependent tensor that does not involve any of the vector fields and contains at most a single derivative and $\mathcal{L}_{\text{rest}}$ represents all the terms that do not depend on the vector fields.  

If we associate a magnetic dual $G_{\Lambda \mu \nu}$ to each field strength   $F^{\Lambda}_{\mu \nu}$ by defining
\begin{equation}
\label{GL}
    G_{\Lambda \mu \nu} \equiv - e^{-1} \epsilon_{\mu \nu \rho \sigma} \frac{\partial \mathcal{L}}{\partial F^{\Lambda}_{\rho \sigma}} = { \cal R}_{\Lambda \Sigma} F^{\Sigma}_{\mu \nu} - {\cal I}_{\Lambda \Sigma} (*F^\Sigma)_{\mu \nu} - (*O_{\Lambda})_{\mu \nu} \, ,
\end{equation}
the Bianchi identities and equations of motion of the vector fields can be condensed in the simple system
\begin{equation}
	\label{eomsib}
	\left\{
		\begin{array}{l}
		    \partial_{[\mu}F^{\Lambda}_{\nu \rho]} = 0 \, ,   \\[2mm]
		    \partial_{[\mu|}G_{\Lambda |\nu \rho]} = 0 \,,
		\end{array}\right.
\end{equation}
which also implies that for each vector field $A^\Lambda_\mu$ there is a dual magnetic vector $A_{ \Lambda \mu}$, local solution of the equations of motion, whose field strength is $G_{\Lambda \mu \nu}$. 
The vector fields $A^{\Lambda}_\mu$, which are those appearing in the ungauged Lagrangian, will be referred to as electric vectors. 

The set of equations (\ref{eomsib}) is invariant, in principle, under general GL(2($n+6$),${\mathbb R}$) transformations mixing $F^\Lambda$ and $G_{\Lambda}$
\begin{equation}
    \label{F',G'}
    \begin{pmatrix}
       F^{\Lambda}_{\mu \nu} \\
       G_{\Lambda \mu \nu}
    \end{pmatrix} \rightarrow \begin{pmatrix}
       F'^{\Lambda}_{\mu \nu} \\
       G'_{\Lambda \mu \nu} 
    \end{pmatrix} = \begin{pmatrix}
           {A^{\Lambda}}_\Sigma & B^{\Lambda \Sigma} \\
           C_{\Lambda \Sigma} & {D_{\Lambda}}^\Sigma 
       \end{pmatrix} \begin{pmatrix}
       F^{\Sigma}_{\mu \nu} \\
       G_{\Sigma \mu \nu}
    \end{pmatrix},
\end{equation}
which are restricted to the symplectic group Sp(2($n+6$),${\mathbb R}$) once we require that the $G'$ definition in terms of $F'$ is the same as (\ref{GL}), possibly for a modified lagrangian ${\cal L}'$ (see \cite{DallAgata:2021uvl} for a review and \cite{Gaillard:1981rj} for the original derivation).

A consistent choice of $n+6$ electric vector fields among the $2(n+6)$ vectors and dual vectors is called a choice of \textit{symplectic frame}.

Once one also takes into account the equations of motion of the scalar fields, one finds that, since ${\cal L}_{\rm rest}$ is only invariant under the symmetry group of the scalar $\sigma$-model, the U-duality group, which is the group of transformations that leave the full system of Bianchi identities and equations of motion of ${\cal N}=4$ supergravity invariant (up to possible suitable modifications of the Lagrangian), reduces to
\begin{equation}
		G = {\rm SL}(2,{\mathbb R}) \times {\rm SO}(6,n) \subset {\rm Sp}(2(n+6),{\mathbb R})\,.
\end{equation}
Clearly, SL(2,${\mathbb R}$) $\times$ SO(6,$n$) is a global symmetry group of the Bianchi identities and equations of motion but not of the Lagrangian, which is only invariant (up to a total derivative) under an electric subgroup $G_{\mathcal{L}}\subset$ SL(2,${\mathbb R}$) $\times$ SO(6,$n$).

Different choices of the symplectic frame give rise to different Lagrangians with different off-shell invariance groups $G_{\mathcal{L}}$, which are however on-shell equivalent in the sense that they lead to sets of Bianchi identities and equations of motion that can be mapped into each other by field redefinitions. 

In the theory at hand, the electric vector fields $A^{\Lambda}_\mu$ together with their magnetic duals $A_{\Lambda \mu}$ form an SL(2,${\mathbb R}$) $\times$ SO(6,$n$) vector $A^{M \alpha}_\mu=(A^{\Lambda}_\mu,A_{\Lambda \mu})$, which is also a symplectic vector of Sp($2(6+n),{\mathbb R}$). 
Following \cite{Schon:2006kz}, we can therefore introduce a composite SL(2,${\mathbb R}$) $\times$ SO(6,$n$) index $\mathcal{M}=M \alpha$ and an antisymmetric symplectic form ${\mathbb{C}}_{\mathcal{M}\mathcal{N}}$ defined by 
\begin{equation}
{\mathbb{C}}_{\mathcal{M}\mathcal{N}} = {\mathbb{C}}_{M \alpha N \beta} \equiv \eta_{MN} \epsilon_{\alpha \beta}   , \,    
\end{equation}
whose inverse is the opposite of 
\begin{equation}
    {\mathbb{C}}^{\mathcal{M}\mathcal{N}} = {\mathbb{C}}^{M \alpha N \beta} \equiv \eta^{MN} \epsilon^{\alpha \beta},
\end{equation}
so that
\begin{equation}
    {\mathbb{C}}^{\mathcal{M}\mathcal{N}} {\mathbb{C}}_{\mathcal{N}\mathcal{P}} =  {\mathbb{C}}^{M \alpha N \beta} {\mathbb{C}}_{N \beta P \gamma} = - \delta^M_P \, \delta^{\alpha}_{\gamma} \equiv - \delta^{\mathcal{M}}_{\mathcal{P}}. 
\end{equation}
Every electric/magnetic split $A^{\cal M}_\mu = A^{M \alpha}_\mu=(A^{\Lambda}_\mu,A_{\Lambda \mu})$, such that the $2(n+6) \times 2(n+6)$ matrix
${\mathbb{C}}^{\mathcal{M}\mathcal{N}}$ decomposes as 
\begin{equation}
    \label{CCinv}
    \mathbb{C}^{\mathcal{M} \mathcal{N}} = 
    \begin{pmatrix}
    \mathbb{C}^{\Lambda \Sigma} &   {\mathbb{C}^{\Lambda}}_\Sigma \\
    {\mathbb{C}_{\Lambda}}^\Sigma &  \mathbb{C}_{\Lambda \Sigma} \\ 
    \end{pmatrix}=
    \begin{pmatrix}
     0 & \delta^{\Lambda}_{\Sigma} \\
     - \delta_{\Lambda}^{\Sigma} & 0  
    \end{pmatrix} , 
\end{equation}
defines a symplectic frame and any two symplectic frames are related by a symplectic rotation. Note that composite SL(2,${\mathbb R}$) $\times$ SO(6,$n$) indices are lowered and raised according to
\begin{equation}
    V_{\mathcal{M}} = V_{M \alpha} = \eta_{MN} \epsilon_{\alpha \beta} V^{N \beta} = {\mathbb{C}}_{\mathcal{M} \mathcal{N}} V^\mathcal{N}, \, V^{\mathcal{M}}  = V^{M \alpha} =  V_{N \beta} \eta^{NM} \epsilon^{\beta \alpha} = V_{\cal{N}} {\mathbb{C}}^{{\cal N} {\cal M}} , 
\end{equation}
where $V^{\cal M}$ is an arbitrary SL(2,${\mathbb R}$) $\times$ SO(6,$n$) vector. 

It is convenient to parametrize the choice of the symplectic frame by means of projectors ${\Pi^{\Lambda}}_{\mathcal{M}}$ and $\Pi_{\Lambda \mathcal{M}}$ that extract the electric and magnetic components of a symplectic vector $V^{\mathcal{M}}=(V^{\Lambda}, V_{\Lambda})$ respectively, according to
\begin{equation}
    \label{projup}
    V^{\Lambda} = {\Pi^{\Lambda}}_{\mathcal{M}} V^{\mathcal{M}}, \qquad V_{\Lambda} = \Pi_{\Lambda \mathcal{M}} V^{\mathcal{M}}.
\end{equation}
In particular, we have that $A^{\Lambda}_\mu={\Pi^{\Lambda}}_{\mathcal{M}} A^{{\cal M} }_\mu$ and $A_{\Lambda \mu} = \Pi_{\Lambda \mathcal{M}} A^{{\cal M} }_\mu $. 
Since the symplectic form ${\mathbb{C}}^{{\cal M} {\cal N}}$ decomposes as in \eqref{CCinv} in any symplectic frame, these projectors must satisfy
\begin{align}
     \label{PelPel=0}
    {\Pi^{\Lambda}}_{\mathcal{M}} {\Pi^{\Sigma}}_{\mathcal{N}}\, {\mathbb{C}}^{\mathcal{M} \mathcal{N}}  & = 0\,, \\[2mm]
    \label{PelPmag=1}
    {\Pi^{\Lambda}}_{\mathcal{M}} \Pi_{\Sigma \mathcal{N}} \,{\mathbb{C}}^{\mathcal{M} \mathcal{N}} & =  \delta^{\Lambda}_{\Sigma}\,, \\[2mm] 
    \label{PmagPmag=0}
    \Pi_{\Lambda \mathcal{M}} \Pi_{\Sigma \mathcal{N}} \,{\mathbb{C}}^{\mathcal{M} \mathcal{N}} & = 0 \, .
\end{align}
On the other hand, for an object $W_{\cal M} = (W_{\Lambda}, W^{\Lambda})$ in the representation of SL(2,${\mathbb R}$) $\times$ SO(6,$n$) that is dual to the fundamental representation, we have
\begin{equation}
\label{projdown}
    W_\Lambda=   \Pi_{\Lambda \mathcal{M}} W^{\mathcal{M}}, \qquad W^\Lambda = - {\Pi^{\Lambda}}_{\mathcal{M}} W^{\mathcal{M}}.
\end{equation}
Furthermore, for any two symplectic vectors $Y^{\mathcal{M}}=(Y^{\Lambda}, Y_{\Lambda})$ and $Z^{\mathcal{M}}=(Z^{\Lambda}, Z_{\Lambda}) $ we have  
\begin{align*}
  Y^{\mathcal{M}} Z_{\mathcal{M}} & =  {\mathbb{C}}_{\mathcal{M} \mathcal{N}} Y^{\mathcal{M}} Z^{\mathcal{N}} \\
  & =  Y^{\Lambda} Z_{\Lambda} - Y_{\Lambda} Z^{\Lambda} \\
  & = ({\Pi^{\Lambda}}_{\mathcal{M}} \Pi_{\Lambda \mathcal{N}} - \Pi_{\Lambda \mathcal{M}} {\Pi^{\Lambda}}_{\mathcal{N}}  )Y^{\mathcal{M}} Z^{\mathcal{N}}  \, , 
\end{align*}
therefore 
\begin{equation}
     \label{PP=C}
    {\Pi^{\Lambda}}_{\mathcal{M}} \Pi_{\Lambda \mathcal{N}} - \Pi_{\Lambda \mathcal{M}} {\Pi^{\Lambda}}_{\mathcal{N}} =  {\mathbb{C}}_{\mathcal{M} \mathcal{N}}.
\end{equation}
Once the choice of frame has been made, the kinetic matrices for the electric vectors follow from decomposing the $2(6+n) \times 2(6+n)$ matrix
\begin{equation}
		{\cal M}_{\cal M \cal N} = M_{\alpha \beta} M_{MN}
\end{equation}
as
\begin{equation}
	\label{Mmatrix}
    {\cal M}_{\mathcal{M} \mathcal{N}} = 
    \begin{pmatrix}
        {\cal M}_{\Lambda \Sigma} & {{\cal M}_{\Lambda}}^{\Sigma} \\[2mm]
        {{\cal M}^{\Lambda}}_{\Sigma} &{\cal M}^{\Lambda \Sigma}
    \end{pmatrix}=
    \begin{pmatrix}
    -({\cal I}+{\cal R}{\cal I}^{-1}{\cal R})_{\Lambda \Sigma} & {({\cal R} {\cal I}^{-1})_\Lambda}^{\Sigma} \\[2mm]
    {({\cal I}^{-1}{\cal R})^\Lambda}_\Sigma & - ({\cal I}^{-1})^{\Lambda \Sigma} 
    \end{pmatrix}, 
\end{equation}
where the identifications are determined by
\begin{align}
    \label{Iinv}
   ( {\cal I}^{-1})^{\Lambda \Sigma} &= - {\Pi^{\Lambda}}_{\mathcal{M}} {\Pi^{\Sigma}}_{\mathcal{N}} {\cal M}^{\mathcal{M} \mathcal{N}}\,, \\[2mm]
   \label{RIinv}
    {({\cal R} {\cal I}^{-1})_\Lambda}^{\Sigma} & = - \Pi_{\Lambda \mathcal{M}} {\Pi^{\Sigma}}_{\mathcal{N}} {\cal M}^{\mathcal{M} \mathcal{N}}\,, \\[2mm]
    \label{IinvR}
    {({\cal I}^{-1}{\cal R})^\Lambda}_\Sigma &= - {\Pi^{\Lambda}}_{\mathcal{M}} \Pi_{\Sigma \mathcal{N}} {\cal M}^{\mathcal{M} \mathcal{N}}\,, \\[2mm]
    \label{I + RIinvR}
    ({\cal I}+{\cal R}{\cal I}^{-1}{\cal R})_{\Lambda \Sigma} & = -  \Pi_{\Lambda \mathcal{M}} \Pi_{\Sigma \mathcal{N}} {\cal M}^{\mathcal{M} \mathcal{N}}\, .
\end{align} 
This decomposition gives the most general form of a matrix ${\cal M}$ satisfying 
\begin{equation}
		{\cal M}_{\mathcal{M} \mathcal{P}} {\mathbb{C}}^{\mathcal{P} \mathcal{Q}}{\cal M}_{\mathcal{Q} \mathcal{N}} ={\mathbb{C}}_{\mathcal{M} \mathcal{N}},
\end{equation}
leading to the definition of the inverse as
\begin{equation}
		{\cal M}^{\mathcal{M} \mathcal{N}} ={\mathbb{C}}^{\mathcal{M} \mathcal{P}}{\mathbb{C}}^{\mathcal{N} \mathcal{Q}} {\cal M}_{\mathcal{P} \mathcal{Q}} \,.
\end{equation}

Moreover, the complex kinetic matrix of the vector fields
\begin{equation}
    \label{N}
    {\cal N}_{\Lambda \Sigma} \equiv {\cal R}_{\Lambda \Sigma} + i \,{\cal I}_{\Lambda \Sigma} 
\end{equation}
satisfies the following useful relations 
    \begin{align}
    {\cal N}_{\Lambda \Sigma} {{\Pi}^{\Sigma}}_{M \alpha} {\cal V}^{\alpha} L^{Mij} & = {\Pi}_{\Lambda M \alpha} {\cal V}^{\alpha} L^{Mij} ,\\[2mm]
   {\cal N}_{\Lambda \Sigma} {{\Pi}^{\Sigma}}_{M \alpha}  ({\cal V}^{\alpha})^* L^{M \underline{a}} &= {\Pi}_{\Lambda M \alpha} ({\cal V}^{\alpha})^* L^{M \underline{a}} \, , 
\end{align}
which are proven in appendix \ref{sec:the_solution_of_the_bianchi_identities}. 
\subsection{Examples of symplectic frames} 
\label{sub:examples_of_symplectic_frames}

Since the decomposition (\ref{Mmatrix}) can be obtained in several inequivalent ways, we discuss now the projectors ${{\Pi}^{\Lambda}}_{M\alpha}$, $\Pi_{\Lambda M \alpha}$ and the kinetic matrices of the electric vectors for some of the symplectic frames in which the $D=4$, ${\cal N}=4$ matter-coupled supergravity has been formulated in the literature. 
\paragraph{The standard frame.} 
The first such symplectic frame follows from requiring that the global symmetry group of the ungauged Lagrangian is $G_{\cal L}=$ SO(1,1) $\times$ SO(6,$n$)\, $\subset$ \,SL(2,${\mathbb R}$) $\times$ SO(6,$n$). 
This symplectic frame, which we shall refer to as \textit{standard frame} or ${\rm SO}(1,1)\times{\rm SO}(6,n)$-frame, corresponds to the electric/magnetic split $A^{M \alpha}_\mu= (A^{M+}_\mu,A_{M+ \mu})$, where the electric vector fields $A^{M+}_\mu$ form an SO(6,$n$) vector and carry SO(1,1) charge $+1$, while their dual magnetic vector fields $A_{M + \mu} = {{A_M}^-}_\mu$, which also form an SO(6,$n$) vector,  carry SO(1,1) charge $-1$.
The two factors in the on-shell global symmetry group are embedded in the symplectic one as follows:
\begin{align}
    \left(\begin{matrix}a & b\cr c & d\end{matrix}\right)&\in {\rm SL}(2,\mathbb{R})\,\rightarrow\,\,\left(\begin{matrix}a \, \mathbb{1}_{n+6} & b\,\eta\cr c\,\eta & d\, \mathbb{1}_{n+6} \end{matrix}\right)\,\in \,{\rm Sp}(2(6+n),\mathbb{R})\,,\,\,\,ad-bc=1\,,\nonumber\\
     g&\in {\rm SO}(6,n)\,\rightarrow\,\,\left(\begin{matrix}g  & {\bf 0}\cr  {\bf 0} & \eta\,g\,\eta\end{matrix}\right)\,\in \,{\rm Sp}(2(6+n),\mathbb{R})\,,
\end{align}
where $\mathbb{1}_{n+6}$ is the $(n+6)\times(n+6)$ identity matrix. 
It is apparent, from the above embeddings, that the off-shell global symmetry group is ${\rm SO}(1,1)\times {\rm SO}(6,\,n)$, as stated earlier.

It is in this symplectic frame that the ${\cal N}=4$ Poincar\'{e} supergravity has been described in \cite{deRoo:1984zyh,Bergshoeff:1985ms,deRoo:1985np,Perret:1988jq,Schon:2006kz} and in our notation with projectors we have
\begin{equation}
    A^{M+}_\mu = {\Pi^{M+}}_{N \alpha} A^{N \alpha}_\mu, \qquad A_{M + \mu} = {\Pi}_{M+ N \alpha} A^{N \alpha}_\mu, 
\end{equation}
where
\begin{equation}
\label{projSO6n}
{\Pi^{M+}}_{N \alpha} \equiv \delta^M_N \delta^+_\alpha , \qquad {\Pi}_{M+ N \alpha} \equiv \eta_{MN} \epsilon_{+ \alpha} \,.
\end{equation}
It is straightforward to show that these projectors satisfy conditions \eqref{PelPel=0}-\eqref{PmagPmag=0} and \eqref{PP=C}. 
Moreover, using equations \eqref{Iinv} and \eqref{RIinv}, we find that the kinetic matrices for the electric vectors $A^{M+}_\mu$ are given by 
\begin{equation}
    \label{kinmat}
    {\cal I}_{M+ N+} = - (\text{Im} \tau) M_{MN}, \, {\cal R}_{M+ N+} = - (\text{Re} \tau) \eta_{MN}, 
\end{equation}
where 
\begin{equation}
		\tau = \frac{1}{2} \left( \frac{{\cal V}_+}{{\cal V}_-} + \frac{{\cal V}_+^*}{{\cal V}_-^*} \right) + \frac{i}{|{\cal V}_-|^2} 
\end{equation}
is the complex scalar of the ${\cal N}=4$ supergravity multiplet. 
Therefore, the Lagrangian for the ungauged theory in this symplectic frame contains the following kinetic terms for the electric vector fields
\begin{equation}
    e^{-1} {\cal L} \supset - \frac{1}{4} (\text{Im} \tau) M_{MN} F^{M+}_{\mu \nu} F^{N+ \mu \nu} - \frac{1}{8} \epsilon^{\mu \nu \rho \sigma}  (\text{Re} \tau) \eta_{MN} F^{M+}_{\mu \nu} F^{N+}_{\rho \sigma} ,
    \end{equation}
where $F^{M+}_{\mu \nu}= 2 \partial_{[\mu} A^{M+}_{\nu]}$. 

While this simple choice allows for a clear distinction between electric and magnetic vectors and maintains SO(6,$n$) covariance, it has been shown \cite{Louis:2014gxa} that one cannot perform a simple electric gauging in this symplectic frame leading to a maximally supersymmetric $AdS$ vacuum.

Any consistent electric/magnetic split $A^{M \alpha}_\mu = (A^{\Lambda}_\mu, A_{\Lambda \mu})$ can be obtained from the standard frame by means of a symplectic rotation
\begin{equation}
\label{arbframe}
    \begin{pmatrix}
    A^{\Lambda}_\mu \\ A_{\Lambda \mu}
    \end{pmatrix} = \begin{pmatrix}
        {B^{\Lambda}}_M & C^{\Lambda M} \\ D_{\Lambda M} & {E_{\Lambda}}^M 
    \end{pmatrix}\begin{pmatrix}
            A^{M+}_\mu \\ A_{M+\mu}   
        \end{pmatrix}\,
\end{equation}
and the corresponding projectors are
\begin{equation}
    \label{genproj}
    {\Pi^{\Lambda}}_{M \alpha} = {B^\Lambda}_M \delta^+_\alpha + {C^\Lambda}_M \epsilon_{+ \alpha}, \,\quad \Pi_{\Lambda M \alpha} = D_{\Lambda M} \delta^+_\alpha + E_{\Lambda M} \epsilon_{+ \alpha}\,.
\end{equation}
The expressions for the matrices $\mathcal{I}_{\Lambda\Sigma},\,\mathcal{R}_{\Lambda\Sigma}$ in an arbitrary symplectic frame can be obtained from those in the SO(1,1) $\times$ SO(6,$n$)-frame, given by \eqref{kinmat}, by using the general transformation property of the complex kinetic matrix ${\cal N}_{\Lambda \Sigma}$ under the symplectic transformation relating the two frames (we suppress all indices):
\begin{equation}\label{Ntransform}
	\begin{split}
    \mathcal{N}&=({\bf E}\,\mathcal{N}_0+{\bf D})\,({\bf C}\,\mathcal{N}_0+{\bf B})^{-1}=  \\[2mm]
	&= \left[-{\bf E}\,\left(\text{Re}(\tau)\,\eta+i\,\text{Im}(\tau)\,M\right)+{\bf D}\right]\,
    \left[{\bf B}-{\bf C}\,\left(\text{Re}(\tau)\,\eta+i\,\text{Im}(\tau)\,M\right)\right]^{-1}\,,
	\end{split}
\end{equation}
where ${\bf E}\equiv ({E_{\Lambda}}^M),\,{\bf C}\equiv (C^{\Lambda M}),\,{\bf B}\equiv ( {B^{\Lambda}}_M)\,,{\bf D}\equiv ( D_{\Lambda M} )$ and 
\begin{equation}\mathcal{N}_0\equiv ( \mathcal{N}_{0\,M+\,N+})=-\left(\text{Re}(\tau)\,\eta+i\,\text{Im}(\tau)\,M\right)\,\label{N0}\end{equation}
is the complex kinetic matrix in the {standard} frame.

The standard frame naturally originates from compactifying heterotic superstring theory on a six-torus $T^6$. 
In this case, on a generic point in moduli space, the resulting $D=4$ supergravity is an $\mathcal{N}=4$ model with $22$ vector multiplets ($n=22$) which, at the classical level, features the global symmetry group  ${\rm SL}(2,\mathbb{R})\times {\rm SO}(6,22)$ \cite{Polchinski:1998rr}.
The vector fields, in this case, consist of the six Kaluza--Klein vectors $\mathcal{G}^m_\mu$, $m=1,\dots, 6$, six vectors $B_{m\mu}$ originating from the $D=10$ Kalb--Ramond field, and 16 vectors $A^\lambda_\mu$, $\lambda=1,\dots, 16$,  gauging the Cartan subalgebra of the ten-dimensional gauge group. 
The ${\rm SL}(2,\mathbb{R})/{\rm SO}(2)$ factor in the scalar manifold of the classical theory is spanned by the four-dimensional dilaton field $\phi_4$ and the axion dual to the 2-form $B_{\mu\nu}$, while the ${\rm SO}(6,22)/[{\rm SO}(6)\times {\rm SO}(22)]$  factor is parametrized by the internal metric moduli $G_{mn}$, the scalars $B_{mn}$ and $A^\lambda_m$, originating from the internal components of the Kalb--Ramond field and the internal components of the ten-dimensional gauge fields respectively. 

Below, we discuss various other instances of symplectic frames, besides the standard one, and their occurrence in superstring compactifications.
\paragraph{Frame in which  ${\rm SL}(2,\mathbb{R})$ is an off-shell symmetry.}  Another  interesting symplectic frame is the one in which the SL(2,${\mathbb R}$) factor of the U-duality group SL(2,${\mathbb R}$) $\times$ SO(6,$n$) is a global symmetry of the ungauged Lagrangian. 
This occurs when $n=6$ and the fundamental representation of ${\rm SO}(6,6)$ branches with respect to the ${\rm GL}(6,\mathbb{R})$ maximal subgroup as follows:
$${\bf 12}\,\rightarrow\,{\bf 6}'_{+\frac{1}{2}} + {\bf 6}_{-\frac{1}{2}}\,,$$
where the grading refers to the ${\rm O}(1,1)$ factor in  ${\rm GL}(6,\mathbb{R})$. 
Let us denote by $\tilde{\Lambda}=1,\dots, 6$ the index labeling the fundamental representation of ${\rm GL}(6,\mathbb{R})$ (and its conjugate).
The symplectic frame in which ${\rm SL}(2,\mathbb{R})$ is a global symmetry of the Lagrangian is the one in which this group has a  block-diagonal action and is obtained by rotating a vector $V^{\mathcal{M}}$ in the standard frame as follows:
$$(V^{\tilde{\Lambda}+},\,V_{\tilde{\Lambda}}{}^+,\,V_{\tilde{\Lambda} +},\,V^{\tilde{\Lambda}}{}_+)\,\rightarrow \,(V^{\tilde{\Lambda}+},\,V^{\tilde{\Lambda}}{}_+,\,V_{\tilde{\Lambda}+},\,-V_{\tilde{\Lambda}}{}^+)\,,$$
where
\begin{equation}
\label{SL(2,R)frame}
    V^{\tilde{\Lambda} \alpha}={\Pi^{\tilde{\Lambda} \alpha}}_{M \beta} V^{M \beta} , \, V_{\tilde{\Lambda} \alpha} = \Pi_{\tilde{\Lambda} \alpha M \beta} V^{M \beta} ,
\end{equation}
with the projectors ${\Pi^{\tilde{\Lambda} \alpha}}_{M \beta}$ and $\Pi_{\tilde{\Lambda} \alpha M \beta}$ that characterize this frame having the following forms: 
\begin{equation}
\label{PGL6R}
    {\Pi^{\Lambda}}_{M \alpha} = {\Pi^{\tilde{\Lambda} \beta}}_{M \alpha} = {\Pi^{\tilde{\Lambda} }}_M \delta^{\beta}_{\alpha}, \qquad \Pi_{\Lambda M \alpha}  = \Pi_{\tilde{\Lambda} \beta M \alpha} = \Pi_{\tilde{\Lambda} M} \epsilon_{\alpha \beta}.
\end{equation}
Thus, conditions \eqref{PelPel=0}-\eqref{PmagPmag=0} and \eqref{PP=C} are equivalent to 
\begin{equation}
     \label{Ptilde}   
    {\Pi^{\tilde{\Lambda} }}_M  {\Pi^{\tilde{\Sigma} }}_N \eta^{MN} = \Pi_{\tilde{\Lambda} M} \Pi_{\tilde{\Sigma} N} \eta^{MN}= 0, \qquad   {\Pi^{\tilde{\Lambda} }}_M \Pi_{\tilde{\Sigma} N} \eta^{MN} = - \delta^{\tilde{\Lambda}}_{\tilde{\Sigma}}
\end{equation}
and 
\begin{equation}
    \label{PtPt=eta}
    {\Pi^{\tilde{\Lambda} }}_M  \Pi_{\tilde{\Lambda} N} + \Pi_{\tilde{\Lambda} M}  {\Pi^{\tilde{\Lambda} }}_N = - \eta_{MN}.
\end{equation}
The $12\times 12$ matrix $\Pi^N{}_M\equiv ({\Pi^{\tilde{\Lambda} }}_M,\,\Pi_{\tilde{\Lambda} M} )$  satisfying the above constraints takes the following form
\begin{equation}
\label{Psol}
   {\Pi^{\tilde{\Lambda} }}_M  = \frac{1}{\sqrt{2}} \begin{pmatrix}
       \mathbb{1}_6 & \mathbb{1}_6
   \end{pmatrix}, \qquad  \Pi_{\tilde{\Lambda} M} =  \frac{1}{\sqrt{2}} \begin{pmatrix}
       \mathbb{1}_6 & -\mathbb{1}_6
   \end{pmatrix}, 
\end{equation}
$\mathbb{1}_6$ being the 6$\times$6 identity matrix. 
$\Pi^N{}_M$ is nothing but the matrix which transforms the original basis of the ${\bf 12}$ of SO(6,6) in which $\eta_{MN}$ is diagonal and an SO(6,6) vector has components $V^M=(V^{\underline{m}},\,V^{\underline{a}})$, into the one in which ${\rm GL}(6,\mathbb{R})$ has a block-diagonal action, $\eta$ is off-diagonal and an SO(6,6) vector has components $V^M=(V^{\tilde{\Lambda}},\,V_{\tilde{\Lambda}})$.

The kinetic matrices for the electric vector fields $A^{\tilde{\Lambda} \alpha}_\mu$ are given by 
\begin{equation}
    {\cal I}_{\tilde{\Lambda} 
 \alpha \tilde{\Sigma} \beta} = - (M^{-1})_{\tilde{\Lambda} \tilde{\Sigma}} M_{\alpha \beta}, \qquad {\cal R}_{\tilde{\Lambda} 
 \alpha \tilde{\Sigma} \beta}= - \epsilon_{\alpha \beta} \Pi_{\tilde{\Lambda} M} {\Pi^{\tilde{\Gamma} }}_N M^{MN} (M^{-1})_{\tilde{\Gamma} \tilde{\Sigma}}\, ,  
\end{equation}
where $(M^{-1})_{\tilde{\Lambda} \tilde{\Sigma}}$ is the inverse of $M^{\tilde{\Lambda} \tilde{\Sigma}} \equiv {\Pi^{\tilde{\Lambda} }}_M  {\Pi^{\tilde{\Sigma} }}_N M^{MN}$ and $\Pi_{\tilde{\Lambda} M} {\Pi^{\tilde{\Gamma} }}_N M^{MN} (M^{-1})_{\tilde{\Gamma} \tilde{\Sigma}}$ is antisymmetric in its indices. 
The ungauged Lagrangian for the $D=4$, ${\cal N}=4$ supergravity coupled to six vector multiplets in this symplectic frame has a global SL(2,${\mathbb R}$) $\times$ GL(6,$\mathbb R$) $\subset$ SL(2,${\mathbb R}$) $\times$ SO(6,6) symmetry and originates from compactification of type IIB supergravity on a $T^6/\mathbb{Z}_2$ orientifold \cite{Frey:2002hf,Kachru:2002he}. This corresponds to the $(T^0\times T^6)/\mathbb{Z}_2$ case reviewed, in more detail, at the end of this section.
The model and its electric gaugings have been studied in \cite{DAuria:2002qje,DAuria:2003nhg,Berg:2003ri}.

\paragraph{Electric gaugings with maximally supersymmetric ${\rm AdS}_4$ vacua.}
The most general gaugings of an $\mathcal{N}=4$ model which feature maximally supersymmetric anti-de Sitter vacua were studied in \cite{Louis:2014gxa} and their electric frame is different from the standard one. 
The simplest of these models involves no vector multiplets ($n=0$) and we shall characterize here its electric frame.
In this model the only components of the embedding tensor that need to be turned on are $f_{+123}$ and $f_{-456}$, where the indices run on the vector representation of the SO(6) R-symmetry group, which is broken to the SO(3)$_+ \times$ SO(3)$_-$ subgroup.
This gauging is purely electric in the symplectic frame where the electric vectors are $A^{\Lambda}_\mu=(A_\mu^{\hat{m}+}, A_{\mu}^{\tilde{m}-} )$  and their magnetic duals are $A_{\Lambda \mu}=(A_{\hat{m}+ \mu}, A_{\tilde{m}- \mu})$, where we have split the SO(6) index $M$ (recall $n=0$) as $M=(\hat{m},\tilde{m})$, where $\hat{m}=1,2,3$ and $\tilde{m}=4,5,6$ label the vector representations of two distinct SO(3) groups. 
The projectors defining this frame are
\begin{align}
    \Pi^\Lambda{}_{M \alpha}& = (\Pi^{\hat{m}+}{}_{M \alpha},\Pi^{\tilde{m}-}{}_{M \alpha} ) =(\delta^+_\alpha \delta^{\hat{m}}_{M},\delta^-_\alpha \delta^{\tilde{m}}_{M}) , \\ 
    \Pi_{\Lambda M \alpha} & = (\Pi_{\hat{m}+M\alpha},\Pi_{\tilde{m}-M\alpha})= ( \epsilon_{+ \alpha} \eta_{\hat{m} M},  \epsilon_{- \alpha} \eta_{\tilde{m} M})
\end{align}
and it is straightforward to show that they satisfy the properties \eqref{PelPel=0}-\eqref{PmagPmag=0} and \eqref{PP=C}. 
In this symplectic frame, the kinetic matrices for the electric vectors are
\begin{align}
    {\cal I}_{\Lambda \Sigma}= \begin{pmatrix}
        {\cal I}_{\hat{m}+ \hat{n}+} & {\cal I}_{\hat{m}+ \tilde{n}-} \\
        {\cal I}_{\tilde{m}- \hat{n}+} & {\cal I}_{\tilde{m}- \tilde{n}-}
    \end{pmatrix} 
    =  \text{Im} \tau
    \begin{pmatrix}
    \eta_{\hat{m} \hat{n}} & 0 \\
    0 & \frac{1}{|\tau|^2} \eta_{\tilde{m} \tilde{n}}
    \end{pmatrix} \,,
\end{align}
and 
\begin{align}
    {\cal R}_{\Lambda \Sigma} = \begin{pmatrix}
        {\cal R}_{\hat{m}+ \hat{n}+} & {\cal R}_{\hat{m}+ \tilde{n}-} \\
        {\cal R}_{\tilde{m}- \hat{n}+} & {\cal R}_{\tilde{m}- \tilde{n}-}
    \end{pmatrix} = \text{Re} \tau 
     \begin{pmatrix}
    -\eta_{\hat{m} \hat{n}} & 0 \\
    0 & \frac{1}{|\tau|^2} \eta_{\tilde{m} \tilde{n}}
    \end{pmatrix} \,.
\end{align}
This result can be written in a more compact form in terms of the complex kinetic matrix:
$$\mathcal{N}_{\hat{m}+\hat{n}+}= \bar{\tau}\delta_{\hat{m}\hat{n}}\,\,,\,\,\,\mathcal{N}_{\tilde{m}-\tilde{n}-}= -\frac{1}{\bar{\tau}}\,\delta_{\tilde{m}\tilde{n}}\,,$$
with all other entries being zero. 
The above expression for $\mathcal{N}$ is to be contrasted with the expression of the same matrix $\mathcal{N}_0$ in the original standard frame: $\mathcal{N}_{0\,M+\,N+}=\bar{\tau}\,\delta_{MN}$.
\paragraph{Symplectic frames from Type IIB compactified on $(T^{p-3}\times T^{9-p})/\mathbb{Z}_2$-orientifolds.} We now
consider the $D=4,\,\mathcal{N}=4$ supergravity models discussed in \cite{Angelantonj:2003rq}, which originate from  Type IIB supergravity compactified on $(T^{p-3}\times T^{9-p})/\mathbb{Z}_2$-orientifolds, in the presence of $Dp$-branes, whose worldvolume fills the whole non-compact $D=4$ spacetime (spacetime-filling branes) as well as $p-3$ directions (defining the sub-torus $T^{p-3}$) in the internal torus. 
We shall write the projection matrices defining the corresponding symplectic frames, while the kinetic matrices of the vector fields have been computed in this reference. 
As in \cite{Angelantonj:2003rq}, we shall restrict ourselves to the  bulk sector, which is described by a half-maximal theory with six vector multiplets ($n=6$). 
The $\mathbb{Z}_2$ is generated by the involution $I_{9-p}\,\Omega \, [(-1)^{F_L}]^{\left[\frac{9-p}{2}\right]}$, where $\Omega$ is the wordsheet parity, $I_{9-p}$ denotes the inversion on the directions of the transverse torus $T_{9-p}$ and $\left[\frac{9-p}{2}\right]$ the integer part of $(9-p)/2$. 
This quotient signals the presence of $\mathcal{O}{p}$-planes, parallel to the spacetime-filling $Dp$-branes. The directions of the internal six-torus split into $p-3$ Neumann (i.e. parallel to the $Dp$-branes), labeled by indices ${\tt i,j},\dots=1,\dots, p-3$, and $9-p$ Dirichlet directions (i.e. transverse to the $Dp$-branes), labeled by indices ${\tt a,b},\dots=p-3+1,\dots, 6$\footnote{Notice that we use a special font for the indices ${\tt i,j,\dots}$ and ${\tt a,b,\dots}$, not to confuse them with  ${ i,j,\dots}$ and ${ a,b,\dots}$, which, in the present paper, have a different meaning.}. 
Consequently, the ${\rm GL}(6,\mathbb{R})_g$ group acting transitively on the metric moduli $G_{{\tt ij}},\,G_{{\tt ia}},\,G_{{\tt ab}}$ of the torus in the un-orbifolded theory, is broken to ${\rm GL}(p-3,\mathbb{R})\times {\rm GL}(9-p,\mathbb{R})$ acting on  $G_{{\tt ij}},\,G_{{\tt ab}}$, which is contained in the global symmetry group of the four-dimensional Lagrangian.

It is useful to describe the fundamental representation of ${\rm SO}(6,6)$ in the basis in which the diagonal blocks describe the subgroup ${\rm GL}(6,\mathbb{R})$ and $\eta$ is off-diagonal. 
In this basis, the electric vector fields in the standard frame are $A^{\Lambda}_\mu = (A^{\tilde{\Lambda}+}_\mu, A_{\tilde{\Lambda}}{}^+{}_{\mu}) = (A^{\tilde{\Lambda}+}_\mu, - A_{\tilde{\Lambda}- \mu})$ and their magnetic duals are $A_{\Lambda\mu}=(A_{\tilde{\Lambda}+ \mu}, A^{\tilde{\Lambda}}{}_{+ \mu})= (A_{\tilde{\Lambda}+ \mu}, A^{\tilde{\Lambda}-}_\mu)$, where we recall that the index $\tilde{\Lambda}=1,\dots, 6$ labels the fundamental representation of ${\rm GL}(6,\mathbb{R})\subset {\rm SO}(6,6)$ and 
\begin{equation}
    A^{\tilde{\Lambda}\alpha}_\mu = \Pi^{\tilde{\Lambda} \alpha}{}_{M \beta} A^{M \beta}_\mu, \qquad A_{\tilde{\Lambda} \alpha \mu} = \Pi_{\tilde{\Lambda} \alpha M \beta} A^{M \beta}_\mu , 
\end{equation}
where $\Pi^{\tilde{\Lambda} \alpha}{}_{M \beta}$ and $\Pi_{\tilde{\Lambda} \alpha M \beta}$ are defined by equations 
\eqref{PGL6R} and \eqref{Psol}. 
A distinctive feature of these models is that this ${\rm GL}(6,\mathbb{R})$ does not coincide in general with ${\rm GL}(6,\mathbb{R})_g$, but intersects the latter in the subgroup ${\rm GL}(p-3,\mathbb{R})\times {\rm GL}(9-p,\mathbb{R})$ mentioned above. 
Indeed, ${\rm GL}(6,\mathbb{R})$ acts transitively on the moduli $G_{{\tt ij}},\,B_{{\tt ia}},\,G_{{\tt ab}}$. 
Finally, we notice that in its first $p-3$ values, the index $\tilde{\Lambda}$ coincides with ${\tt i}$ labeling the Neumann directions of $T^{p-3}$, while in the last $9-p$ values, it coincides with the index ${\tt a}$ of the dimensionally reduced fields, labeling the Dirichlet directions along $T^{9-p}$, though in the opposite position, due to the peculiar way ${\rm GL}(9-p,\mathbb{R})$ is embedded in ${\rm GL}(6,\mathbb{R})$.
Below we discuss the different cases.\\
\emph{Case $(T^6\times T^0)/\mathbb{Z}_2$:} This is a compactification in the presence of $D9$-branes and $\mathcal{O}9$-planes. 
The complex scalar in the ${\rm SL}(2,\mathbb{R})/{\rm SO}(2)$ factor is $\tau=c+i\,e^{\frac{\phi}{2}}\,V_6$, $c$ being the four-dimensional dual to the RR tensor $C_{\mu\nu}$, $\phi$ the ten-dimensional dilaton and $V_6$ the volume of $T^6$ in the Einstein frame. 
The scalars  $G_{{\tt ij}},\,C_{{\tt ij}}$, on the other hand, span the coset space ${\rm SO}(6,6)/[{\rm SO}(6)\times {\rm SO}(6)]$. 
In this case, the indices $\tilde{\Lambda}$ and ${\tt i}$ coincide and the symplectic frame is defined by the electric vectors $A^{{\tt i}+}_\mu=\mathcal{G}^{{\tt i}}_\mu,\,A_{\tt i}{}^+{}_\mu= - A_{{\tt i}- \mu} = C_{{\tt i}\mu}$, where $\mathcal{G}^{{\tt i}}_\mu$ are the Kaluza--Klein vectors. 
The projectors are given by \footnote{Here and in the following we always define the projectors as acting on the basis in which $\eta$ is diagonal.}
\begin{align}
    \Pi^{\Lambda}{}_{M \alpha}=& (\Pi^{{\tt i}+}{}_{M \alpha}, \Pi_{{\tt i}- M \alpha}) = (\Pi^{\tt i}{}_M \delta^+_\alpha, \Pi_{{\tt i} M} \epsilon_{- \alpha}), \\
    \Pi_{\Lambda M \alpha} = & (\Pi_{{\tt i}+ M \alpha}, \Pi^{{\tt i}-}{}_{M \alpha}) = (-\Pi_{{\tt i} M} \epsilon_{+ \alpha}, \Pi^{\tt i}{}_M \delta^-_\alpha ), 
    \end{align}
where $\Pi^{\tt i}{}_M = \Pi^{\tilde{\Lambda}}{}_M$ and $\Pi_{{\tt i} M} = \Pi_{\tilde{\Lambda} M}$ are given by \eqref{Psol}. 
This symplectic frame is equivalent to the standard electric/magnetic split $A^{M \alpha}_\mu = (A^{M+}_\mu, A_{M + \mu})$, since it is related to the latter by a symplectic rotation of the form \eqref{arbframe} that is block-diagonal, i.e. $C^{\Lambda M} = D_{\Lambda M} = 0$. 
\\
\emph{Case $(T^0\times T^6)/\mathbb{Z}_2$:} This is a compactification in the presence of $D3$-branes and $\mathcal{O}3$-planes. 
The scalars consist of $\tau=C_{(0)}+i\,e^{-\phi}$ parametrizing ${\rm SL}(2,\mathbb{R})/{\rm SO}(2)$, $C_{(0)}$ being the ten-dimensional RR axion, and $G_{{\tt ab}},\,C^{{\tt ab}}=\epsilon^{{\tt ab a_1\dots a_4}}\,C_{{\tt a_1\dots a_4}}$ spanning the $\frac{{\rm SO}(6,6)}{{\rm SO}(6)\times {\rm SO}(6)}$ submanifold ($C_{{\tt a_1\dots a_4}}$ are the internal components of the RR 4-form field). 
In this case, the index $\tilde{\Lambda}$ of the ${\rm GL}(6,\mathbb{R})$ and the index ${\tt a}$ of the dimensionally reduced string modes coincide, aside from their upper/lower positions, as commented above. 
The symplectic frame is defined by the electric vectors $A^{{\tt a}+}_\mu=B_{{\tt a}\mu},\,A^{{\tt a}}{}_{+\,\mu}= A^{\tt{a}-}_\mu=C_{{\tt a}\mu}$. 
The projection matrices are:
\begin{equation}
    \Pi^{\Lambda}{}_{M \alpha} = \Pi^{\tt{a} \beta}{}_{M \alpha} = \Pi^{\tt{a}}{}_M \delta^\beta_\alpha, \qquad \Pi_{\Lambda M \alpha} = \Pi_{{\tt{a}} \beta M \alpha} = \Pi_{{\tt{a}} M} \epsilon_{\alpha \beta}\, , 
\end{equation}
where $\Pi^{\tt{a}}{}_M = \Pi^{\tilde{\Lambda}}{}_M$ and $\Pi_{{\tt a} M} = \Pi_{\tilde{\Lambda} M}$ are given by \eqref{Psol}. 
This is the model constructed in \cite{Frey:2002hf,Kachru:2002he} and studied, in its gauged version, in \cite{DAuria:2002qje,DAuria:2003nhg,Berg:2003ri}, as mentioned above.\\
\emph{Case $(T^2\times T^4)/\mathbb{Z}_2$:} This is a compactification in the presence of $D5$-branes and $\mathcal{O}5$-planes. 
The scalars consist of $\tau=C_{{\tt ij}}+i\,e^{-\frac{\phi}{2}}\,V_2$ parametrizing ${\rm SL}(2,\mathbb{R})/{\rm SO}(2)$ and $G_{{\tt ij}},\,G_{{\tt ab}},\,C_{{\tt ab}},\,B_{{\tt ia}},\,C_{{\tt iabc}},\,c$ spanning ${\rm SO}(6,6)/[{\rm SO}(6)\times {\rm SO}(6)]$. 
The symplectic frame is defined by the electric vectors $A^{{\tt i}+}_\mu=\mathcal{G}^{{\tt i}}_\mu,\,A^{{\tt a}+}_\mu=B_{{\tt a}\mu},\,{A}^{{\tt i}}{}_{+\,\mu}= A^{{\tt i}-}_\mu=\epsilon^{{\tt ij}}C_{{\tt j}\mu},\,A_{\tt a}{}^+{}_{\mu} = - A_{{\tt a}- \mu}=\epsilon^{{\tt abcd}}\,C_{{\tt bcd}\mu}$. 
The projection matrices are:
\begin{align}
    \Pi^{\Lambda}{}_{M \alpha} & = ( \Pi^{\tilde{\Lambda}+}{}_{M \alpha}, \Pi^{{\tt i}-}{}_{M \alpha}, \Pi_{{\tt a}- M \alpha} ) = ( \Pi^{\tilde{\Lambda}}{}_M \delta^+_\alpha, \Pi^{\tt i}{}_M \delta^{-}_\alpha, \Pi_{{\tt a} M} \epsilon_{- \alpha} ), \\
    \Pi_{\Lambda M \alpha} & = (\Pi_{\tilde{\Lambda}+ M \alpha}, \Pi_{{\tt i} - M \alpha}, \Pi^{{\tt a}-}{}_{M \alpha}) = (-\Pi_{\tilde{\Lambda} M} \epsilon_{+ \alpha}, - \Pi_{{\tt i} M} \epsilon_{- \alpha}, \Pi^{\tt a}{}_M \delta^-_{\alpha}),
\end{align}
where the GL(6,$\mathbb{R}$) index $\tilde{\Lambda}$ is decomposed as $\tilde{\Lambda}=({\tt i},{\tt a})$, $\Pi^{\tt i}{}_M$ and $\Pi_{{\tt i}M}$ are the $2\times12$ matrices that consist of the first two rows of the matrices $\Pi^{\tilde{\Lambda}}{}_M$ and $\Pi_{\tilde{\Lambda}M}$ of \eqref{Psol} respectively, while $\Pi^{\tt a}{}_M$ and $\Pi_{{\tt a}M}$ are the $4\times12$ matrices consisting of the last four rows of $\Pi^{\tilde{\Lambda}}{}_M$ and $\Pi_{\tilde{\Lambda}M}$ respectively.\\
\emph{Case $(T^4\times T^2)/\mathbb{Z}_2$:} The compactification is perfomed in the presence of $D7$-branes and $\mathcal{O}7$-planes. 
The scalars consist of $\tau=C_{{\tt ijkl}}+i\,V_4$ parametrizing ${\rm SL}(2,\mathbb{R})/{\rm SO}(2)$ and $G_{{\tt ij}},\,G_{{\tt ab}},\,C_{{\tt ia}},\,B_{{\tt ia}},\,C_{(0)},\,C_{{\tt ijab}}$ spanning the coset manifold ${\rm SO}(6,6)/[{\rm SO}(6)\times {\rm SO}(6)]$. 
The symplectic frame is defined by the electric vectors $A^{{\tt i}+}_\mu=\mathcal{G}^{{\tt i}}_\mu,\,A^{{\tt a}+}_\mu=B_{{\tt a}\mu},\,A^{{\tt i}}{}_{+\mu} = A^{{\tt i} -}_\mu=\epsilon^{{\tt ijkl}}C_{{\tt jkl}\mu},\,A_{\tt a}{}^+{}_\mu = - A_{{\tt a} - \mu}=\epsilon^{{\tt ab}}\,C_{{\tt b}\mu}$. 
The projection matrices are:
\begin{align}
    \Pi^{\Lambda}{}_{M \alpha} & = ( \Pi^{\tilde{\Lambda}+}{}_{M \alpha}, \Pi^{{\tt i}-}{}_{M \alpha}, \Pi_{{\tt a}- M \alpha} ) = ( \Pi^{\tilde{\Lambda}}{}_M \delta^+_\alpha, \Pi^{\tt i}{}_M \delta^{-}_\alpha, \Pi_{{\tt a} M} \epsilon_{- \alpha} ), \\
    \Pi_{\Lambda M \alpha} & = (\Pi_{\tilde{\Lambda}+ M \alpha}, \Pi_{{\tt i} - M \alpha}, \Pi^{{\tt a}-}{}_{M \alpha}) = (-\Pi_{\tilde{\Lambda} M} \epsilon_{+ \alpha}, - \Pi_{{\tt i} M} \epsilon_{- \alpha}, \Pi^{\tt a}{}_M \delta^-_{\alpha}),
\end{align}
where again $\tilde{\Lambda}=({\tt i},{\tt a})$, $\Pi^{\tt i}{}_M$ and $\Pi_{{\tt i}M}$ are the $4\times12$ matrices that consist of the first four rows of the matrices $\Pi^{\tilde{\Lambda}}{}_M$ and $\Pi_{\tilde{\Lambda}M}$ of \eqref{Psol} respectively, while $\Pi^{\tt a}{}_M$ and $\Pi_{{\tt a}M}$ are the $2\times12$ matrices consisting of the last two rows of $\Pi^{\tilde{\Lambda}}{}_M$ and $\Pi_{\tilde{\Lambda}M}$ respectively.
Gaugings of these models, originating from internal fluxes, were studied in \cite{Angelantonj:2003rq,DallAgata:2005fjb}.



\section{Duality Covariant Gauging} 
\label{sec:duality_covariant_gauging}

The gauging procedure consists in promoting a suitable subgroup $G_g$ of the global symmetry group $G_{\cal L}$ of the Lagrangian to a local symmetry group gauged by a subset of the electric vector fields $A^{\Lambda}_\mu$ of the theory. 
Gauging a group $G_g$ requires the introduction of minimal couplings of the gauge fields to the other fields and the modification of the Lagrangian and the local supersymmetry transformation rules in such a way that the resulting theory features the same amount of supersymmetry (${\cal N}=4$) as the original ungauged one. 

The choice of the symplectic frame is not physically relevant in the ungauged theory, as it affects the Lagrangian description, but not the set of equations of motion and Bianchi identities. 
However, the introduction of minimal couplings explicitly breaks the original on-shell global SL(2,${\mathbb R}$) $\times$ SO(6,$n$) invariance of the ungauged model, and the initial choice of the symplectic frame has physical implications on the resulting gauged theory because different frames correspond to different Lagrangians with different global symmetry groups $G_{\cal L}\subset$ SL(2,${\mathbb R}$) $\times$ SO(6,$n$) and thus different choices of possible gauge groups $G_g$. 

Nevertheless, there exists an SL(2,${\mathbb R}$) $\times$ SO(6,$n$)-covariant formulation of the gauging procedure that does not depend on the symplectic frame in which the ungauged theory is written.
 This formulation involves the introduction of gauge fields $A^{\cal M}_\mu$ that decompose into electric gauge fields $A^{\Lambda}_\mu$ and magnetic gauge fields $A_{\Lambda \mu}$ and gauge group generators $X_{\cal M}=(X_{\Lambda},X^{\Lambda})$. 
 Since the gauge group $G_g$ is a subgroup of the duality group SL(2,${\mathbb R}$) $\times$ SO(6,$n$), these generators can be expressed as linear combinations of the generators $t_A$ of SL(2,${\mathbb R}$)$\times$SO(6,$n$), where $A$ is an index labeling the adjoint representation of SL(2,${\mathbb R}$) $\times$ SO(6,$n$), according to 
\begin{equation}
    \label{X_M}
    X_{\mathcal{M}} = {{\Theta}_{\cal M}}^A t_A \, , 
\end{equation}
where ${{\Theta}_{\cal M}}^A = ({{\Theta}_{\Lambda}}^A,{\Theta}^{\Lambda A})$ is a constant tensor, called the \textit{embedding tensor}, which encodes all the information about the embedding of $G_g$ in SL(2,${\mathbb R}$) $\times$ SO(6,$n$). 
The index $A$ decomposes as $A=([MN],(\alpha \beta)$), where $[MN]$ labels the adjoint representation of SO(6,$n$) and $(\alpha \beta)$ labels the adjoint representation of SL(2,${\mathbb R}$), so equation \eqref{X_M} can be written as 
\begin{equation}
    X_{\mathcal{M}} = {{\Theta}_{\cal M}}^{NP} t_{NP} +  {{\Theta}_{\cal M}}^{\beta \gamma} t_{\beta \gamma} ,
\end{equation}
where $t_{NP}=t_{[NP]}$ and $t_{\beta \gamma}=t_{(\beta \gamma)}$ are the generators of SO(6,$n$) and SL(2,${\mathbb R}$) respectively and ${{\Theta}_{\cal M}}^{NP}={{\Theta}_{\cal M}}^{[NP]}$, while ${{\Theta}_{\cal M}}^{\beta \gamma} ={{\Theta}_{\cal M}}^{(\beta \gamma)} $. 
Furthermore, the gauge connection is defined by
\begin{equation}
    \label{Og}
    \Omega_{g \mu} \equiv g A^{\cal M}_{\mu} X_{\cal M}  , 
\end{equation}
where $g$ is the gauge coupling constant. 

The main advantage of this description of the gauging is that the Bianchi identities and equations of motion of the gauged theory are formally invariant under global SL(2,${\mathbb R}$) $\times$ SO(6,$n$) transformations, as is the case in the ungauged theory, provided we treat the embedding tensor ${\Theta_{\cal M}}^A$ as a spurionic object that transforms under SL(2,${\mathbb R}$) $\times$ SO(6,$n$). 
When freezing ${\Theta_{\cal M}}^A$ to a constant, this formal on-shell SL(2,${\mathbb R}$) $\times$ SO(6,$n$)-invariance is broken. 

This procedure of gauging a supergravity theory has been introduced in \cite{Cordaro:1998tx, Nicolai:2000sc, Nicolai:2001sv} and developed, in the form presented here, in  \cite{deWit:2002vt,deWit:2004nw,deWit:2005ub,deWit:2007kvg} (see also \cite{Samtleben:2008pe,Trigiante:2016mnt,DallAgata:2021uvl} for reviews). 
We should note that a quite detailed discussion of this procedure for ${\cal N}=4$ supergravity has been given in \cite{Schon:2006kz}, though with some clear limitations, as discussed in the introduction.
In any case, our presentation aims at being self-contained.

Consistency of the gauging procedure, namely the possibility of constructing a locally $G_g$-invariant and ${\cal N}=4$ supersymmetric action, requires the embedding tensor $({{\Theta}_{\cal M}}^{NP},{{\Theta}_{\cal M}}^{\beta \gamma} )$ to satisfy a set of linear and quadratic SL(2,${\mathbb R}$) $\times$ SO(6,$n$)-covariant constraints. 
The linear constraint is
\begin{equation}
    \label{linear}
    X_{({\cal M} {\cal N} {\cal P})} = {X_{({\cal M} {\cal N}}}^{\cal Q} \mathbb{C}_{{\cal P}) {\cal Q}}=0 \, , 
\end{equation}
where ${X_{{\cal M} {\cal N}}}^{\cal P} \equiv{{\Theta}_{\cal M}}^{QR} {{(t_{QR})}_{\cal N}}^{\cal P} + {{\Theta}_{\cal M}}^{\delta \epsilon} {{(t_{\delta \epsilon})}_{\cal N}}^{\cal P}$ are the matrix elements of the gauge generators $X_{\cal M}$ in the fundamental representation of SL(2,${\mathbb R}$) $\times$ SO(6,$n$). 
The linear constraint restricts the embedding tensor to a particular representation of SL(2,${\mathbb R}$) $\times$ SO(6,$n$). 
More precisely, the embedding tensor $({{\Theta}_{\alpha M}}^{NP},{{\Theta}_{\alpha M}}^{\beta \gamma} )$ formally transforms in the tensor product of the fundamental $(\mathbf{2},\mathbf{n+6})$ and the adjoint $(\mathbf{3},\mathbf{1})+\left(\mathbf{1},\mathbf{\frac{1}{2}}\mathbf{(n+6)(n+5)}\right)$ representations of SL(2,${\mathbb R}$) $\times$ SO(6,$n$), which decomposes according to 
\begin{align}
\label{decomp}
    &(\mathbf{2},\mathbf{n+6}) \times \left[ (\mathbf{3},\mathbf{1})+\left(\mathbf{1},\mathbf{\frac{1}{2}}\mathbf{(n+6) (n+5)}\right)\right] \nonumber \\ &=  \,2 \cdot (\mathbf{2},\mathbf{n+6}) + (\mathbf{4},\mathbf{n+6}) + \left( \mathbf{2}, \mathbf{\binom{n+6}{3}} \right) + \left(\mathbf{2}, \mathbf{\frac{1}{3} (n+6) ((n+6)^2-4)} \right) . 
\end{align}
The linear constraint \eqref{linear} removes all the representations in the above decomposition that are contained in the 3-fold symmetric product of the $(\mathbf{2},\mathbf{n+6})$ representation
\begin{align}
     X_{({\cal M} {\cal N} {\cal P})} \in &((\mathbf{2},\mathbf{n+6}) \times (\mathbf{2},\mathbf{n+6}) \times (\mathbf{2},\mathbf{n+6}))_{\text{sym.}} \nonumber \\
     & =  (\mathbf{2},\mathbf{n+6}) + (\mathbf{4},\mathbf{n+6}) + \left(\mathbf{2}, \mathbf{\frac{1}{3} (n+6) ((n+6)^2-4)} \right) \\
      & \hspace{0.5cm}+ \left( \textbf{4}, \mathbf{\frac{1}{6}(n+6)(n+10)(n+5)}\right). \nonumber
\end{align}
Hence, the linear constraint restricts the embedding tensor to the $(\mathbf{2},\mathbf{n+6}) + \left( \mathbf{2}, \mathbf{\binom{n+6}{3}} \right)$ representation of SL(2,${\mathbb R}$) $\times$ SO(6,$n$), and the possible gaugings of the four-dimensional ${\cal N}=4$ supergravity are therefore parametrized by two real constant SL(2,${\mathbb R}$) $\times$ SO(6,$n$) tensors, $\xi_{\alpha M}$ and $f_{\alpha MNP}=f_{\alpha[MNP]}$, corresponding to these representations \cite{Schon:2006kz}.
Once we make explicit this constraint, the components of the embedding tensor are expressed in terms of the $\xi$ and $f$ tensors as
\begin{equation}
    \label{Theta}
 {{\Theta}_{\alpha M}}^{NP} = {f_{\alpha M}}^{NP}  - \xi^{[N}_{\alpha} \delta^{P]}_M, \qquad {{\Theta}_{\alpha M}}^{\beta \gamma} = \delta^{(\beta}_{\alpha} \xi^{\gamma)}_M \, . 
    \end{equation}
Thus, the quantities ${X_{{\cal M} {\cal N}}}^{\cal P}$ are given by
  \begin{equation}
      \label{XMNP}
      {X_{{\cal M} {\cal N}}}^{\cal P} =     {X_{M \alpha N \beta}}^{P \gamma} = - \delta^\gamma_\beta {f_{\alpha M N}}^P + \frac{1}{2} \left( \delta^P_M \delta^\gamma_\beta \xi_{\alpha N} - \delta^P_N \delta^\gamma_\alpha \xi_{\beta M} - \eta_{MN} \delta^\gamma_\beta \xi^P_\alpha  + \delta^P_N \epsilon_{\alpha \beta} \xi^\gamma_M\right)
  \end{equation}
and satisfy the constraint \eqref{linear} by construction \cite{Schon:2006kz}. 

Gauge invariance requires the embedding tensor to be invariant under the action of the gauge group $G_g$ that it defines. 
This implies the quadratic constraint
\begin{equation}
      \label{quad}
      0 = {{\Theta}_{\cal M}}^{B} t_{B} {{\Theta}_{\cal N}}^A = {{\Theta}_{\cal M}}^B {(t_B)_{\cal N}}^{\cal P} {{\Theta}_{\cal P}}^A + {{\Theta}_{\cal M}}^B 
      {{\Theta}_{\cal N}}^C 
      {f_{BC}}^A  , 
\end{equation}
where we used the fact that the generators of SL(2,${\mathbb R}$) $\times$ SO(6,$n$) in the adjoint representation are given by ${(t_B)_C}^A = -  {f_{BC}}^A$, where ${f_{BC}}^{A}$ are the structure constants of the Lie algebra of SL(2,${\mathbb R}$) $\times$ SO(6,$n$) defined by $[t_A,t_B]={f_{AB}}^C t_C $. 
By contracting the last equation with the generators $t_A$, we obtain
\begin{equation}
\label{closure}
[X_{\cal M}, X_{\cal N}] = - {X_{\mathcal{M} \mathcal{N}}}^{\cal P} X_{\cal P} , 
\end{equation}
which amounts to the closure of the gauge algebra. 
It was found in \cite{Schon:2006kz} that the above constraint is equivalent to the following quadratic constraints on the tensors $\xi_{\alpha M}$ and $f_{\alpha MNP}$
\begin{align}
    \label{xixi}
    &\xi^M_{\alpha} \xi_{\beta M}  = 0\,, \\[2mm]
    \label{xif}
    &\xi^P_{(\alpha} f_{\beta) PMN}  = 0\,, \\[2mm]
    \label{ff + xif}
    &3 f_{\alpha R [MN|} {f_{\beta |PQ]}}^R + 2 \xi_{(\alpha|[M|} f_{|\beta) |NPQ]}  = 0\,, \\[2mm]
    \label{xif + xixi}
    &\epsilon^{\alpha \beta} (\xi^P_{\alpha} f_{\beta PMN} + \xi_{\alpha M} \xi_{\beta N})  = 0\,, \\[2mm]
    \label{ff + 3 xi f}
    &\epsilon^{\alpha \beta} (f_{\alpha MNR} {f_{\beta P Q}}^R - \xi^R_{\alpha} f_{\beta R[M[P} \eta_{Q]N]} - \xi_{\alpha[M|} f_{\beta|N]PQ} + \xi_{\alpha [P|} f_{\beta|Q]MN})  = 0\,.
\end{align}
These quadratic constraints also solve
 \begin{equation}
     \label{locality}
     \mathbb{C}^{{\cal M} {\cal N}} {\Theta_{\cal M}}^A {\Theta_{\cal N}}^B = 0 \, , 
 \end{equation}
which implies the existence of a symplectic frame in which the magnetic components $\Theta^{\Lambda A}$ of the embedding tensor vanish (electric frame). 
Equation \eqref{locality} is known as the locality constraint on the embedding tensor and guarantees that the dimension of the gauge group $G_g$ does not exceed the number $n+6$ of the vector fields that are present in the ungauged Lagrangian and are available for the gauging.  

In the gauged theory, the ordinary exterior derivative $d$ is replaced by a gauge-covariant one which acts on objects ($p$-forms) in an arbitrary representation of SL(2,${\mathbb R}$) $\times$ SO(6,$n$) as
\begin{equation}
    \label{dhat}
    \hat{d} = d - g A^{\cal M} X_{\cal M} = d - g A^{M \alpha} {{\Theta}_{\alpha M}}^{NP} t_{NP} + g A^{M(\alpha} \epsilon^{\beta)\gamma} \xi_{\gamma M} t_{\alpha \beta} \, , 
\end{equation}
where we have introduced the connection one-forms $A^{\cal M} = A^{M \alpha} =A^{M \alpha}_\mu d x^\mu$, which we assume to transform under a gauge transformation with infinitesimal parameters $\zeta^{\cal M}(x) = \zeta^{M \alpha} (x)$ as
\begin{equation}
    \delta_{\zeta} A^{\cal M} = \hat{d} \zeta^{\cal M} = d \zeta^{\cal M} + g {X_{{\cal N} {\cal P}}}^{\cal M} A^{\cal N} \zeta^{\cal P}.
\end{equation}
Using the relation for the closure of the gauge algebra for the generators in (\ref{closure})
\begin{equation}
    {X_{{\cal M} {\cal Q}}}^{\cal S} {X_{{\cal N} {\cal S}}}^{\cal R} - {X_{{\cal N}{\cal Q}}}^{\cal S} {X_{{\cal M} {\cal S}}}^{\cal R} = - {X_{{\cal M} {\cal N}}}^{\cal P} {X_{{\cal P} {\cal Q}}}^{\cal R}, 
\end{equation}
we find that
\begin{equation}
    {\hat d}^2 = - g {\hat F}^{\cal M} X_{\cal M}  , 
\end{equation} 
where 
\begin{equation}
    \label{hat F}
     {\hat F}^{\cal M}  = \frac{1}{2} {\hat F}^{\cal M}_{\mu \nu} d x^\mu \wedge d x^\nu \equiv d A^{\cal M} + \frac{g}{2} {X_{{\cal N} {\cal P}}}^{\cal M} A^{\cal N} \wedge A^{\cal P}
\end{equation}
are the usual non-abelian field strengths of the vector fields (in form notation). 
This can also be rewritten as 
\begin{equation}
    \label{hatFMa}
    {\hat F}^{M \alpha} = d A^{M \alpha} - \frac{g}{2} {{\hat{f}}_{\beta N P}}^{\hspace{0.65cm} M} A^{N \beta} \wedge A^{P \alpha}, 
\end{equation}
where we have defined 
\begin{equation}
    \label{hatf}
   \hat{{f}}_{\alpha MNP}  = f_{\alpha MNP} - \xi_{\alpha[M} \eta_{P]N} - \frac{3}{2} \xi_{\alpha N} \eta_{MP},
\end{equation}
following \cite{Schon:2006kz}. 

It is important to stress that the field strengths \eqref{hat F} do not transform covariantly under gauge transformations, because
\begin{equation}
    \label{noncov}
     \delta_{\zeta} {\hat F}^{\cal M} = - g {X_{{\cal N} {\cal P}}}^{\cal M} {\zeta}^{\cal N} {\hat F}^{\cal P} +  g  {X_{({\cal N} {\cal P})}}^{\cal M} \left( 2  \zeta^{\cal N} {\hat F}^{\cal P} - A^{\cal N} \wedge \delta_{\zeta} A^{\cal P} \right) \ne - g {X_{{\cal N} {\cal P}}}^{\cal M} {\zeta}^{\cal N} {\hat F}^{\cal P}.
\end{equation}
In order to construct gauge covariant quantities describing the vector fields, we introduce the two-form gauge fields $B^{MN} = B^{[MN]}= \frac{1}{2} B^{MN}_{\mu \nu} dx^\mu \wedge dx^\nu $ and $B^{\alpha \beta} = B^{(\alpha \beta)} = \frac{1}{2} B^{\alpha \beta}_{\mu \nu} dx^\mu \wedge dx^\nu $, transforming in the adjoint representations of SO(6,$n$) and SL(2,${\mathbb R}$) respectively and we modify the field strengths as follows\footnote{While it is clear that in four dimensions one can always dualize a massless tensor field to a scalar and a massive tensor field to a massive vector, very often the natural low-energy Lagrangians of supergravity theories that come from string compactifications contain tensor fields from the beginning \cite{Louis:2002ny}. 
This sparked the necessity to be able to clearly identify the gauged supergravity theories containing tensor fields as physical degrees of freedom \cite{DallAgata:2003sjo} and for a better analysis of the corresponding gauge structure, which takes the form of a free differential algebra \cite{DallAgata:2005xvr}.
 As we will see later, the embedding tensor formulation we present here allows for an elegant and general solution to these issues.} \cite{deWit:2005hv, deWit:2004nw,deWit:2005ub,Schon:2006kz}
\begin{equation}
    \label{HMa}
    H^{M \alpha}  = \frac{1}{2} H^{M \alpha}_{\mu \nu} dx^\mu \wedge dx^\nu \equiv {\hat F}^{M \alpha} - \frac{g}{2} {\Theta^{\alpha M}}_{NP} B^{NP} + \frac{g}{2} \xi^M_\beta B^{\alpha \beta}.
\end{equation}
These modified field strengths transform covariantly under gauge transformations 
\begin{equation}
    \label{gaugecov}
    \delta_{\zeta} H^{\cal M} = - g {X_{{\cal N} {\cal P}}}^{\cal M} {\zeta}^{\cal N} H^{\cal P},
\end{equation}
provided the two-form gauge fields transform as (see for example \cite{Trigiante:2016mnt})
\begin{align}
      \label{dzBMN}
    \delta_{\zeta} B^{MN} &= \epsilon_{\alpha \beta} \left(-2 \zeta^{[M| \alpha} H^{|N] \beta} + A^{[M| \alpha} \wedge \delta_{\zeta} A^{|N] \beta}\right), \\[2mm]
    \label{dzBab}
    \delta_{\zeta} B^{\alpha \beta} & = \eta_{MN} \left( 
2 \zeta^{M (\alpha|} H^{N |\beta)} - A^{M (\alpha|} \wedge \delta_{\zeta} A^{N |\beta)} \right).
\end{align}
A consistent definition of the two-form gauge fields $B^{MN}$ and $B^{\alpha \beta}$ requires the theory to also be invariant under tensor gauge transformations parametrized by one-forms $\Xi^{MN} = \Xi^{[MN]} = \Xi^{MN}_\mu dx^\mu$ and $\Xi^{\alpha \beta}=\Xi^{(\alpha \beta)} = \Xi^{\alpha \beta}_\mu dx^\mu $ acting on the vector and two-form gauge fields as \cite{Schon:2006kz}
\begin{align}
    \label{dXiAMa}
    \delta_{\Xi} A^{M \alpha} & = \frac{g}{2} \, {\Theta^{\alpha M}}_{NP} \Xi^{NP}  - \frac{g}{2} \xi^M_\beta \Xi^{\alpha \beta}, \\
    \label{dXiBMN}
    \delta_{\Xi} B^{MN} & = \hat{d}  {\Xi}^{MN} + \epsilon_{\alpha \beta} A^{[M|\alpha} \wedge \delta_{\Xi} A^{|N]\beta}, \\[2mm]
    \label{dXiBab}
    \delta_{\Xi} B^{\alpha \beta} & = \hat{d} \Xi^{\alpha \beta} - \eta_{MN} A^{M(\alpha|} \wedge \delta_{\Xi} A^{N |\beta)} , 
\end{align}
where
\begin{equation}
    \hat{d}  {\Xi}^{MN} \equiv d  {\Xi}^{MN} + 2 g {\Theta_{\alpha PQ}}^{[M|} A^{P \alpha} \wedge \Xi^{|N]Q}
\end{equation}
and
\begin{equation}
    \hat{d} {\Xi}^{\alpha \beta} \equiv d {\Xi}^{\alpha \beta} - g \xi^{(\alpha|M} A_{M \gamma} \wedge \Xi^{|\beta) \gamma} - g \xi_{\gamma M} A^{M (\alpha} \wedge \Xi^{\beta) \gamma}. 
\end{equation}
The transformation rules \eqref{dXiAMa}-\eqref{dXiBab} ensure that $\delta_{\Xi} H^{M \alpha} = 0 $. 

In the scalar sector, gauging a subgroup of the duality group means gauging the isometries of the scalar $\sigma$-model. 
This can be accounted for by constructing gauged Maurer--Cartan forms from which we recover the gauged vielbeins and connections of the scalar manifold.
For the coset space SL(2,${\mathbb R}$)/SO(2), the gauged Maurer--Cartan left-invariant one-form is given by
\begin{equation}
    \label{gaugedMC1}   {\hat{\Psi}}_{\underline{\alpha}}{}^{\underline{\beta}} = {(S^{-1})_{\underline{\alpha}}}^\alpha \hat{d} {S_{\alpha}}^{\underline{\beta}} =  {(S^{-1})_{\underline{\alpha}}}^\alpha d {S_{\alpha}}^{\underline{\beta}} + \frac{1}{2} g {(S^{-1})_{\underline{\alpha}}}^\alpha \xi_{\alpha M} A^{M \beta} {S_{\beta}}^{\underline{\beta}} + \frac{1}{2} g {(S^{-1})_{\underline{\alpha}}}^\alpha \xi^{ \beta M} A_{M \alpha}  {S_{\beta}}^{\underline{\beta}} , 
\end{equation}
and, in our conventions, has the following expansion 
\begin{equation}
    \hat{\Psi} = (\text{Re}\hat{P}) \sigma_3 \, + (\text{Im} \hat{P}) \sigma_1 + i \hat{{\cal A}} \, \sigma_2 \, ,
\end{equation}
where  we have suppressed the SO(2) indices.
We then see that 
\begin{equation}\label{hatPdef}
	\hat{P}=\frac{i}{2} \epsilon^{\alpha \beta} {\cal V}_{\alpha} \hat{d} {\cal V}_\beta
\end{equation} 
is the gauged SL(2,${\mathbb R}$)/SO(2) zweibein and 
\begin{equation}\label{hatcalA}
	\hat{\cal A} = - \frac{1}{2} \epsilon^{\alpha \beta} {\cal V}_{\alpha}  \hat{d} {\cal V}_\beta^* 
\end{equation} 
is the gauged SO(2) connection, where  
\begin{equation}
    \hat{d} {\cal V}_\alpha \equiv d {\cal V}_\alpha + \frac{1}{2} g   \xi_{\alpha M} A^{M \beta} {\cal V}_{\beta} + \frac{1}{2} g \xi^{M \beta} A_{M \alpha} {\cal V}_{\beta} \, .
\end{equation}
The one-form \eqref{gaugedMC1} satisfies the gauged Maurer--Cartan equation
\begin{align}
    d \hat{\Psi} + \hat{ \Psi} \wedge \hat{\Psi} = & \, \frac{i}{4} g \xi_{\alpha M} \left[  {\cal V}^\alpha {\cal V}_\beta - ({\cal V}^\alpha)^* {\cal V}_\beta^* \right]  {H}^{M \beta} \sigma_3 \nonumber \\
    & +  \frac{g}{4}  \xi_{\alpha M} \left[  {\cal V}^\alpha {\cal V}_\beta + ({\cal V}^\alpha)^* {\cal V}_\beta^* \right]  {H}^{M \beta} 
    \sigma_1  \\ 
      & +  \frac{i}{4} g \xi^\alpha_M \left(  {\cal V}_\alpha {\cal V}_\beta^* + {\cal V}_\alpha^* {\cal V}_\beta \right)  {H}^{M \beta} \sigma_2 \, ,\nonumber 
\end{align}
which implies the relation
\begin{equation}
    \label{DhatPhat}
    \hat{D} \hat{P} \equiv d \hat{P} - 2 i \hat{\cal A} \wedge \hat{P} = \frac{i}{2} g \xi_{\alpha M} {\cal V}^\alpha  {\cal V}_\beta {H}^{M \beta}  
    \end{equation}
and gives the following expression for the gauged SO(2) curvature
\begin{equation}
\label{gSO(2)curv}
     \hat{F} \equiv d \hat{\cal A} = \, i \hat{P}^* \wedge \hat{P} + \frac{g}{4}  \xi^\alpha_M  \left(  {\cal V}_\alpha {\cal V}_\beta^* + {\cal V}_\alpha^* {\cal V}_\beta \right)H^{M \beta}  .
\end{equation}
Once again, with some algebra, one can also derive the useful identity
\begin{equation}
\label{hatDV=hatPV}
    \hat{D} {\cal V}_{\alpha} \equiv \hat{d}  {\cal V}_{\alpha} - i \hat{\cal A} {\cal V}_{\alpha} = \hat{P} {\cal V}_{\alpha}^* \, .
\end{equation}

On the other hand, the gauged Maurer--Cartan left-invariant one-form for the coset space SO(6,$n$)/SO(6) $ \times$ SO($n$) is given by
\begin{equation}
     \label{gaugedMC2}    {\hat{\Omega}}_{\underline{M}}{}^{\underline{N}} = {L_{\underline{M}}}^M \hat{d} {L_M}^{\underline{N}} =  {L_{\underline{M}}}^M d {L_M}^{\underline{N}} + g A^{M \alpha} {\Theta_{\alpha M N }}^P  {L_{\underline{M}}}^N {L_P}^{\underline{N}} , 
    \end{equation}
which satisfies $\hat{\Omega}_{\underline{M}}{}^{\underline{N}}=- \hat{\Omega}^{\underline{N}}{}_{\underline{M}}$ and has the following matrix form in the fundamental representation of SO(6,$n$)
\begin{equation}
    \label{MC2matrix}    {\hat{\Omega}}_{\underline{M}}{}^{\underline{N}} = \begin{pmatrix}    \hat{\omega}_{\underline{m}}{}^{\underline{n}} & {\hat{P}}_{\underline{m}}{}^{\underline{b}} \\    {\hat{P}}_{\underline{a}}{}^{\underline{n}} &  {\hat{\omega}}_{\underline{a}}{}^{\underline{b}} 
    \end{pmatrix} , 
\end{equation}
 where $  \hat{\omega}_{\underline{m}}{}^{\underline{n}}$ is the gauged SO(6) connection,  $\hat{\omega}_{\underline{a}}{}^{
\underline{b}}$ is the gauged SO($n$) connection and  $\hat{P}_{\underline{a}}{}^{\underline{n}}$
is the gauged SO(6,$n$)/SO(6) $ \times$ SO($n$) vielbein. 
The one-form \eqref{gaugedMC2} satisfies the gauged Maurer--Cartan equations 
\begin{equation}
    d {{\hat{\Omega}}_{\underline{M}}}{}^{ \underline{N}} + {{\hat{\Omega}}_{\underline{M}}}{}^{\underline{P}} \wedge {{\hat{\Omega}}_{\underline{P}}}{}^{ \underline{N}} = g {{\Theta}_{\alpha M N}}^P {L_{\underline{M}}}^N {L_P}^{\underline{N}} H^{M \alpha} , 
\end{equation}
which, using the gauged SU(4) connection
\begin{equation}
    {\hat{\omega}}^i{}_{ j} = {\hat{\omega}}^{ik}{}_{ jk} \, 
\end{equation}
and the SU(4) covariant expressions for the vielbeins
\begin{equation}
\label{hPaijdef}
{\hat P}_{\underline{a}}{}^{ij} = L_{\underline{a}}{}^M {\hat d} L_M{}^{ij},
\end{equation}
 imply that 
\begin{align}
    \label{scalarBianchi3}
    \hat{D} {{\hat{P}}_{\underline{a}}}{}^{ij}  \equiv \, & d {{\hat{P}}_{\underline{a}}}{}^{ij} + {{\hat{\omega}}_{\underline{a}}}{}^{\underline{b}} \wedge {{\hat{P}}_{\underline{b}}}{}^{ij} - {\hat{\omega}^i}{}_{k} \wedge {{\hat{P}}_{\underline{a}}}{}^{kj} - {\hat{\omega}^j}{}_{k} \wedge {{\hat{P}}_{\underline{a}}}{}^{ik} 
     = g {{\Theta}_{\alpha M }}^{N P}   L_{N \underline{a}} {L_P}^{ij}  H^{M \alpha}, \\[2mm]
	  \label{gSU(4)curv}
     {{\hat{R}}^i}{}_{j}  \equiv \, &  d {\hat{\omega}^i}{}_{j} - {\hat{\omega}^i}{}_{k} \wedge {\hat{\omega}^k}{}_{j}  
      =  {\hat{P}}^{\underline{a} ik } \wedge {\hat{P}}_{\underline{a} jk}  + g {{\Theta}_{\alpha M }}^{N P}   {L_N}^{ik} L_{Pjk} H^{M \alpha}, \\[2mm]
      \label{gSO(n)curv}{{\hat{R}}_{\underline{a}}}{}^{ \underline{b}} \equiv \, &  d {{\hat{\omega}}_{\underline{a}}}{}^{\underline{b}} + {{\hat{\omega}}_{\underline{a}}}{}^{ \underline{c}} \wedge {{\hat{\omega}}_{\underline{c}}}{}^{\underline{b}}  = 
 - {\hat{P}}_{\underline{a}ij} \wedge {\hat P}^{\underline{b} i j} + g {{\Theta}_{\alpha M }}^{N P}   L_{N \underline{a}} {L_P}^{\underline{b}} H^{M \alpha} ,
\end{align}
where ${{\hat{R}}^i}{}_{j}$ and $ {{\hat{R}}_{\underline{a}}}{}^{ \underline{b}}$ are the gauged SU(4) and SO($n$) curvatures respectively. 
Again, one can also derive the following useful relations 
\begin{align}\label{hatDLij}
  \hat{D} {L_M}^{ij} & \equiv  \hat{d} {L_M}^{ij} - {\hat{\omega}^i}{}_{k} {L_M}^{kj}  - {{\hat{\omega}}^j}{}_{k} {L_M}^{ik} =   {L_M}^{\underline{a}} {\hat{P}_{\underline{a}}}{}^{ij} , \\[2mm]
\label{hatDLa}
 \hat{D} {L_M}^{\underline{a}} &\equiv  \hat{d} {L_M}^{\underline{a}} + {\hat{\omega}^{\underline{a}}}{}_{\underline{b}} {L_M}^{\underline{b}} = {L_M}^{ij} {\hat{P}^{\underline{a}}}{}_{ij} \, .  
\end{align}


\section{The Lagrangian
 {and Supersymmetry Transformation Rules}} 
\label{sec:lagrangian_and_supersymmetry_rules}

The full procedure to build the supersymmetric Lagrangian and derive the supersymmetry transformation rules of the gauged $D=4$, ${\cal N}=4$ matter-coupled Poincar\'{e} supergravity in an arbitrary symplectic frame using the geometric approach can be found in appendix \ref{sec:the_solution_of_the_bianchi_identities}.
Here we provide the results, namely the Lagrangian and the local supersymmetry transformations of the fields, and comment on both the equations of motion and the closure of the supersymmetry algebra.

\subsection{{The Lagrangian}} 
\label{sub:the_lagrangian}
The ${\cal N}=4$ supergravity Lagrangian can be split in 6 terms as follows
\begin{align}
\label{L4formg}
    {\cal L}= & \, {\cal L}_{\text{kin}} + {\cal L}_{\text{Pauli}} +  {\cal L}_{\stackrel{\text{fermion}}{\text{mass}}} +  {\cal L}_{\text{pot}}  +  {\cal L}_{\text{top}} + {\cal L}_{\text{4fermi}} \, , 
\end{align}
where ${\cal L}_{\text{kin}} $ contains the kinetic terms of the various fields, ${\cal L}_{\text{Pauli}}$ the Pauli-like couplings of the scalar and vector field strengths to the fermions, ${\cal L}_{\text{fermion mass}}$ is the self-explanatory fermion mass part, ${\cal L}_{\text{pot}}$ the scalar potential, ${\cal L}_{\text{top}}$ the necessary couplings of the 2-form fields that, according to the embedding tensor choice, lead to non-dynamical field equations that ensure that we did not add new degrees of freedom by changing the explicit Lagrangian and, finally, ${\cal L}_{\text{4fermi}}$ are the remaining 4-fermion couplings.

We now list all the terms and the corresponding relevant definitions.
\begin{align}
  \label{Lkinspg}
  e^{-1} {\cal L}_{\text{kin}} = \, &  \frac{1}{2} R + \frac{i}{2} \epsilon^{\mu \nu \rho \sigma} \left({\bar\psi}^i_\mu \gamma_{\nu} {\hat\rho}_{i \rho \sigma} - {\bar\psi}_{i \mu} \gamma_{\nu} {\hat\rho}^i_{\rho\sigma} \right) \nonumber \\[2mm]
  & - \frac{1}{2} \left( {\bar\chi}^i \gamma^\mu {\hat D}_{\mu} \chi_i + {\bar\chi}_i \gamma^\mu {\hat D}_{\mu} \chi^i \right) - \left( {\bar\lambda}^{\underline{a}}_i \gamma^\mu {\hat D}_{\mu} \lambda^i_{\underline{a}} +  {\bar\lambda}_{\underline{a}}^i \gamma^\mu {\hat D}_{\mu} \lambda_i^{\underline{a}} \right) \\[2mm]
  & - {\hat P}_\mu^* {\hat P}^\mu - \frac{1}{2} {\hat P}_{\underline{a} i j \mu} {\hat P}^{\underline{a} i j \mu} + \frac{1}{4} {\cal I}_{\Lambda \Sigma} H^{\Lambda}_{\mu \nu} H^{\Sigma \mu \nu} + \frac{1}{8} \epsilon^{\mu \nu \rho \sigma} {\cal R}_{\Lambda \Sigma} H^{\Lambda}_{\mu \nu} {H}^{\Sigma}_{\rho \sigma}\,, \nonumber \\[2mm]
  \label{LPaulispg}
  e^{-1} {\cal L}_{\text{Pauli}} = \, & {\hat P}_\mu^* \left( {\bar\chi}^i \psi^{\mu}_i - {\bar\chi}^i \gamma^{\mu \nu} \psi_{i\nu} \right) + {\hat P}_\mu \left( {\bar\chi}_i \psi^{i \mu} - {\bar\chi}_i \gamma^{\mu \nu} \psi^i_\nu \right) \nonumber \\[2mm] 
  & - 2 {\hat P}_{\underline{a} i j \mu} \left( {\bar\lambda}^{\underline{a}i} \psi^{j \mu} - {\bar\lambda}^{\underline{a}i} \gamma^{\mu \nu} \psi^j_\nu \right) - 2 {\hat P}^{\underline{a} i j \mu} \left( {\bar\lambda}_{\underline{a}i} \psi_{j \mu} - {\bar\lambda}_{\underline{a}i} \gamma_{\mu \nu} \psi_j^\nu\right) \\[2mm] 
  & + \frac{1}{2} H^{\Lambda}_{\mu \nu} O_{\Lambda}^{\mu \nu}\,, \nonumber \\[2mm]
  \label{Lfmasssp}
  e^{-1} {\cal L}_{\stackrel{\text{fermion}}{\text{mass}}} = \, & - 2 g {\bar A}_2{}^{\underline{a} j}{}_i {\bar\chi}^i \lambda_{\underline{a}j} + 2 g {\bar A}_2{}^{\underline{a} i}{}_i {\bar\chi}^j \lambda_{\underline{a}j}+ 2 g A_{\underline{a} \underline{b}}{}^{ij} {\bar\lambda}^{\underline{a}}_i \lambda^{\underline{b}}_j + \frac{2}{3} g A_2^{ij} {\bar\lambda}^{\underline{a}}_i \lambda_{\underline{a} j} \nonumber \\[2mm] 
  & + \frac{2}{3} g {\bar A}_{2ij} {\bar\chi}^i \gamma^\mu \psi^j_\mu + 2 g A_{2 \underline{a} j}{}^i  {\bar\lambda}^{\underline{a}}_i \gamma^\mu \psi^j_\mu   - \frac{2}{3} g {\bar A}_{1ij} {\bar\psi}^i_\mu \gamma^{\mu \nu} \psi^j_\nu + c.c.\,,  \\[2mm] 
  \label{Lpotsp}
  e^{-1} {\cal L}_{\text{pot}} = & \, g^2 \left(\frac{1}{3} A_1^{ij} {\bar A}_{1ij} - \frac{1}{9} A_2^{ij} {\bar A}_{2 ij} - \frac{1}{2} A_{2 \underline{a} i}{}^j {\bar A}_2{}^{\underline{a}i}{}_j \right)\,, \\[2mm] 
  e^{-1} {\cal L}_{\text{top}} = & \, \frac{1}{8} g \epsilon^{\mu \nu \rho \sigma} {{\Pi}^{\Lambda}}_{M \alpha} \Pi_{\Lambda N \beta} \left( {{\Theta}^{\alpha M}}_{PQ} B^{PQ}_{\mu \nu} - \xi^M_\gamma B^{\alpha\gamma}_{\mu \nu} \right) \times \nonumber \\[2mm]
  & \left( 2 \partial_\rho A^{N \beta}_\sigma -g {\hat f}_{\delta R S}^{\hspace{0.55cm}N} A^{R \delta}_\rho A^{S \beta}_{\sigma}  - \frac{1}{4} g {{\Theta}^{\beta N}}_{RS} B^{RS}_{\rho \sigma} + \frac{1}{4} g \xi^N_{\delta} B^{\beta \delta}_{\rho \sigma} \right) \nonumber \\[2mm] 
  & - \frac{1}{6} g \epsilon^{\mu \nu \rho \sigma} \left( {{\Pi}^\Lambda}_{R \epsilon}  {\Pi}_{\Lambda S \zeta}  + 2 {\Pi}_{\Lambda R \epsilon} {{\Pi}^\Lambda}_{S \zeta}  \right) {X_{ M \alpha N \beta}}^{R \epsilon} A^{M \alpha}_\mu A^{N\beta}_\nu  \times \\[2mm]
  &\left( \partial_\rho A^{S \zeta}_\sigma + \frac{1}{4} g {X_{P \gamma Q \delta}}^{S \zeta} A^{P \gamma}_\rho A^{Q \delta}_\sigma \right)\,, \nonumber  \\[2mm]
    \label{L4fsp}
    e^{-1} {\cal L}_{\text{4fermi}} =  &  - {\bar\chi}_i \psi^i_\mu {\bar\chi}^j \psi^\mu_j - 4 {\bar\lambda}^{\underline{a}}_i \psi_{j \mu} {\bar\lambda}^{[i}_{
  \underline{a}} \psi^{j] \mu} - \epsilon^{ijkl} {\bar\lambda}_{\underline{a}i} \psi_{j \mu} {\bar\lambda}^{\underline{a}}_k \psi^\mu_l - \epsilon_{ijkl} {\bar\lambda}^i_{\underline{a}} \psi^j_\mu {\bar\lambda}^{\underline{a}k} \psi^{l \mu} \nonumber \\
  & + \frac{3}{8} {\bar\chi}^i \chi^j {\bar\chi}_i \chi_j - \frac{1}{2}  {\bar\chi}^i \lambda^{\underline{a}}_j {\bar\chi}_i \lambda_{\underline{a}}^j - {\bar\chi}^i \lambda^{\underline{a}}_i {\bar\chi}_j \lambda_{\underline{a}}^j  - \frac{1}{2} {\bar\lambda}^{\underline{a}}_i {\lambda}^{\underline{b}}_j {\bar\lambda}_{\underline{a}}^i {\lambda}_{\underline{b}}^j\nonumber \\ 
  & - {\bar\lambda}^{\underline{a}}_i \lambda_{\underline{a}j} {\bar\lambda}_{\underline{b}}^i \lambda^{\underline{b}j} + 2 {\bar\lambda}^{\underline{a}}_i {\lambda}^{\underline{b}}_j {\bar\lambda}_{\underline{b}}^i {\lambda}_{\underline{a}}^j  - {\bar\lambda}^{\underline{a}}_i \lambda_{\underline{a}j}  {\bar\chi}^i \gamma^\mu \psi^j_\mu - {\bar\lambda}_{\underline{a}}^i \lambda^{\underline{a}j} {\bar\chi}_i \gamma^\mu \psi_{j \mu} \nonumber \\
  & + i \epsilon^{\mu \nu \rho \sigma} \left( \frac{1}{2} {\bar\chi}_i \gamma_\mu \chi^j {\bar\psi}^i_\nu \gamma_\rho \psi_{j \sigma} + {\bar\lambda}^{\underline{a}}_i \gamma_\mu \lambda^j_{\underline{a}} {\bar\psi}^i_\nu \gamma_\rho \psi_{j \sigma} -{\bar\lambda}^{\underline{a}}_i \gamma_\mu \lambda^i_{\underline{a}} {\bar\psi}^j_\nu \gamma_\rho \psi_{j \sigma} \right) \nonumber \\
  & + \epsilon^{ijkl} \left( {\bar\chi}_i \gamma^\mu \psi^\nu_j {\bar\psi}_{k \mu} \psi_{l \nu}  - \frac{i}{2}  \epsilon^{\mu \nu \rho \sigma} {\bar\chi}_i \gamma_\mu \psi_{j \nu} {\bar\psi}_{k \rho} \psi_{l \sigma}\right)  \\ 
  & + \epsilon_{ijkl} \left( {\bar\chi}^i \gamma^\mu \psi^{j \nu} {\bar\psi}^k_\mu \psi^l_\nu  + \frac{i}{2}  \epsilon^{\mu \nu \rho \sigma} {\bar\chi}^i \gamma_\mu \psi^j_\nu 
  {\bar\psi}^k_\rho \psi^l_\sigma\right) \nonumber \\ & + {\bar\chi}_i \gamma_{\mu\nu} \lambda^{\underline{a}i} {\bar\lambda}^j_{\underline{a}} \gamma^\mu \psi^\nu_j + {\bar\chi}^i \gamma_{\mu\nu} \lambda_{\underline{a}i} {\bar\lambda}^{\underline{a}}_j \gamma^\mu \psi^{j \nu} \nonumber \\
  & - 2 {\bar\chi}^{[i} \gamma_\mu \psi^{j]}_\nu {\bar\chi}_{i} \gamma^{[\mu} \psi^{\nu]}_{j} - 2 {\bar\lambda}^i_{\underline{a}} \gamma^\mu \psi^\nu_i {\bar\lambda}^{\underline{a}}_j \gamma_{[\mu} \psi^j_{\nu]} \nonumber \\ 
  & - 2 {\bar\psi}^i_\mu \psi^j_\nu {\bar\psi}^{\mu}_{[i} \psi^\nu_{j]} + \frac{1}{8} ({\cal I}^{-1})^{\Lambda \Sigma} O_{\Lambda \mu \nu} O_{\Sigma}^{\mu \nu},  \nonumber
\end{align} 
where
\begin{align}
    \label{OLsp}
    O_{\Lambda \mu \nu} = & \, {\cal I}_{\Lambda \Sigma} {\Pi^{\Sigma}}_{M \alpha} \big{(} - 2 ({\cal V}^\alpha)^* L^{Mij} {\bar{\psi}}_{i \mu} \psi_{j \nu} - i \epsilon_{\mu \nu \rho \sigma} ({\cal V}^\alpha)^* L^{Mij}  {\bar{\psi}}^{\rho}_i {\psi}_j^{\sigma} \nonumber \\[2mm]
    & + {\mathcal{V}}^{\alpha}  L^{Mij} {\bar{\lambda}}_{\underline{a} i} \gamma_{\mu \nu}  \lambda^{\underline{a}}_j  - {\mathcal{V}}^{\alpha} L^{M \underline{a}} {\bar\chi}_i \gamma_{\mu \nu} \lambda^i_{\underline{a}}  + 2 ({\cal V}^\alpha)^* {L^M}_{ij} {\bar\chi}^i \gamma_{[\mu} \psi^j_{\nu]} \nonumber \\[2mm]  
    & + i  \epsilon_{\mu \nu \rho \sigma} ({\cal V}^\alpha)^* {L^M}_{ij} {\bar\chi}^i \gamma^{\rho} \psi^{j \sigma} + 2 {\cal V}^\alpha L^{M \underline{a}} {\bar\lambda}_{\underline{a}i} \gamma_{[\mu} \psi_{\nu]}^i \\[2mm] 
    & + i \epsilon_{\mu \nu \rho \sigma} {\cal V}^\alpha  L^{M \underline{a}} {\bar\lambda}_{\underline{a}i} \gamma^{\rho} \psi^{i \sigma} + \text{c.c.} \big{)}, \nonumber
\end{align}
${\cal I}_{\Lambda \Sigma}$ and ${\cal R}_{\Lambda \Sigma}$ follow from the solution of (\ref{Iinv}) and (\ref{RIinv}) in the chosen symplectic frame specified by the projectors $\Pi^\Lambda{}_{M \alpha}$ and $\Pi_{\Lambda M \alpha}$.
Moreover, ${\hat P}_\mu$ and ${\hat P}_{\underline{a} ij \mu}$ are the components of the spacetime one-forms ${\hat P}$ and ${\hat P}_{\underline{a}ij}$ defined in (\ref{hatPdef}) and (\ref{hPaijdef}) respectively, i.e.  ${\hat P} = {\hat P}_\mu d x^\mu$ and ${\hat P}_{\underline{a}ij} = {\hat P}_{\underline{a} ij \mu} d x^\mu$. 
In addition, we have defined  $H^{\Lambda}_{\mu \nu} \equiv {{\Pi}^{\Lambda}}_{M \alpha} H^{M \alpha}_{\mu \nu} $, where the field strengths $ H^{M \alpha}_{\mu \nu}$ were introduced in \eqref{HMa}.

The field strengths of the fermionic fields have the following expressions
\begin{align}
{\hat\rho}_{i \mu \nu} \equiv & \, 2 \partial_{[\mu|} \psi_{i|\nu]} + \frac{1}{2} {\omega_{[\mu|}}^{ab}(e,\psi) \gamma_{ab} \psi_{i|\nu]} - i \hat{\cal A}_{[\mu|} \psi_{i|\nu]} - 2 {\hat\omega}_i{}^j{}_{[\mu|} \psi_{j|\nu]}, \\[2mm]
{\hat D}_{\mu} \chi_i \equiv & \, \partial_\mu \chi_i + \frac{1}{4} {\omega_{\mu}}^{ab}(e,\psi) \gamma_{ab} \chi_i + \frac{3i}{2} \hat{\cal A}_{\mu} \chi_i -  {\hat\omega}_i{}^j{}_\mu \chi_j \, , \\[2mm]
{\hat D}_{\mu} \lambda_{\underline{a}i} \equiv & \, \partial_\mu \lambda_{\underline{a}i} + \frac{1}{4} {\omega_{\mu}}^{ab}(e,\psi) \gamma_{ab} \lambda_{\underline{a}i} + \frac{i}{2} \hat{\cal A}_{\mu} \lambda_{\underline{a}i} -  {\hat\omega}_i{}^j{}_\mu \lambda_{\underline{a}j} + {\hat\omega}_{\underline{a}}{}^{\underline{b}}{}_\mu \lambda_{\underline{b}i} \, , 
\end{align}
where ${\hat{\cal A}}_\mu$, ${\hat \omega}_i{}^j{}_{\mu}$ and ${\hat\omega}_{\underline{a}}{}^{\underline{b}}{}_\mu$ are the components of the spacetime one-forms ${\hat{\cal A}}$, ${\hat \omega}_i{}^{j}$ and ${\hat\omega}_{\underline{a} }{}^{\underline{b}}$ respectively, i.e. ${\hat{\cal A}}={\hat{\cal A}}_\mu dx^\mu$, ${\hat \omega}_i{}^{j} ={\hat \omega}_i{}^j{}_{\mu} dx^\mu $, and $ {\hat\omega}_{\underline{a} }{}^{\underline{b}} = {\hat\omega}_{\underline{a}}{}^{\underline{b}}{}_\mu dx^\mu $ and $\omega_{\mu a b} (e,\psi)$ is the solution of the supertorsion constraint  (\ref{Ra=0}), $T^a = 0$, projected on spacetime for the spin connection as a function of the vielbein and gravitini.

Finally, the fermion mass matrices, which also appear in the scalar potential,  are
\begin{align}     
    \label{gravshift}
   A_1^{ij} & =   f_{\alpha MNP} ({\cal V}^\alpha)^* L^M{}_{kl} L^{Nik} L^{Pjl}, \\[2mm]
    \label{gaushift}
    A_{2 \underline{a} i}{}^j & =  f_{\alpha MNP} {\cal V}^\alpha {L_{\underline{a}}}^M {L^N}_{ik} L^{Pjk} - \frac{1}{4} \delta^j_i \xi_{\alpha M}  {\cal V}^\alpha {L_{\underline{a}}}^M, \\[2mm] 
     \label{dilshift}
   A_2^{ij} & =  f_{\alpha MNP} {\cal V}^\alpha {L^M}_{kl} L^{Nik} L^{P jl} + \frac{3}{2} \xi_{\alpha M} {\cal V}^\alpha L^{M ij}, \\[2mm]
   \label{Aabij}
   A_{\underline{a} \underline{b}}{}^{ij} &= f_{\alpha MNP} {\cal V}^{\alpha} {L^M}{}_{\underline{a}} {L^N}{}_{\underline{b}} L^{Pij}.
\end{align}
Using the quadratic constraints \eqref{xixi}-\eqref{ff + 3 xi f} one can show that  
\begin{equation}
      \label{Wardmain}
     \frac{2}{3} A_1^{jk} {\bar A}_{1 ik} - \frac{2}{9} A_2^{kj} {\bar A}_{2 ki} - A_{2 \underline{a} i}{}^k {\bar A}_2{}^{\underline{a}j}{}_k = \frac{1}{4} \delta^j_i \left(  \frac{2}{3} A_1^{kl} {\bar A}_{1 kl} - \frac{2}{9} A_2^{kl} {\bar A}_{2 kl} - A_{2 \underline{a} k}{}^l {\bar A}_2{}^{\underline{a}k}{}_l   \right).
\end{equation}
Note that we explicitly introduced factors of $g$ for the terms arising from the gauging procedure.


\subsection{The supersymmetry transformation rules} 
\label{sub:the_susy_rules}

Using the geometric approach presented in appendix \ref{sec:the_solution_of_the_bianchi_identities}, one can also deduce, from the spacetime projections of the Lie derivatives of the various superfields, the local supersymmetry transformations of the corresponding spacetime fields.
For the fermionic fields we find
\begin{align}
     \label{dpsig}
     \delta_{\epsilon} \psi_{i \mu} = \, & {\hat D}_{\mu} \epsilon_i + \frac{1}{4} {\cal I}_{\Lambda \Sigma} {{\Pi}^{\Lambda}}_{M \alpha} {\cal V}^\alpha {L^M}_{ij}  {\hat{ \cal H}}^{\Sigma}_{\nu \rho} \gamma^{\nu \rho} \gamma_\mu \epsilon^j - \frac{1}{4} \epsilon_{ijkl} ({\bar\lambda}^j_{\underline{a}} \gamma_{\mu \nu} \lambda^{\underline{a}k})\gamma^\nu \epsilon^l \nonumber \\[2mm]
     & + \frac{1}{4} (\bar\chi_i \gamma_\mu \chi^j)\epsilon_j  -  \frac{1}{4} (\bar\chi_j \gamma_\mu \chi^j)\epsilon_i - \frac{1}{4} (\bar\chi_i \gamma^\nu \chi^j)\gamma_{\mu \nu} \epsilon_j \nonumber \\[2mm]
     & + \frac{1}{8} (\bar\chi_j \gamma^\nu \chi^j)\gamma_{\mu \nu} \epsilon_i + \frac{1}{2} ({\bar\lambda}^{\underline{a}}_i \gamma_\mu \lambda^j_{\underline{a}}) \epsilon_j - \frac{1}{2} ({\bar\lambda}^{\underline{a}}_i \gamma^\nu \lambda^j_{\underline{a}}) \gamma_{\mu\nu} \epsilon_j \\[2mm]
     & + \frac{1}{4}  ({\bar\lambda}^{\underline{a}}_j \gamma^\nu \lambda^j_{\underline{a}}) \gamma_{\mu\nu} \epsilon_i - \epsilon_{ijkl} \chi^j {\bar\epsilon}^k \psi^l_\mu - \frac{1}{3} g {\bar A}_{1ij} \gamma_\mu \epsilon^j, \nonumber \\[2mm]
      \label{dlambdag}
      \delta_{\epsilon} \lambda_{\underline{a}i} = & - \frac{1}{4} {\cal I}_{\Lambda \Sigma} {{\Pi}^{\Lambda}}_{M \alpha} ({\cal V}^\alpha)^* {L^M}_{\underline{a}} {\hat{ \cal H}}^{\Sigma}_{\mu \nu} \gamma^{\mu \nu} \epsilon_i \nonumber \\[2mm]
      & - \gamma^\mu \epsilon^j ( {\hat P}_{\underline{a}ij\mu} + 2 {\bar\lambda}_{\underline{a}[i} \psi_{j]\mu} 
+ \epsilon_{ijkl} {\bar\lambda}^k_{\underline{a}} \psi^l_\mu ) \\[2mm]
 & + (\bar{\chi}_i \lambda^j_{\underline{a}})\epsilon_j - \frac{1}{2} (\bar{\chi}_j \lambda^j_{\underline{a}}) \epsilon_i + g {\bar A}_{2 \underline{a}}{}^j{}_i \epsilon_j,\nonumber \\[2mm]
     \label{dchig}
      \delta_{\epsilon} \chi_i =  & - \frac{1}{2} {\cal I}_{\Lambda \Sigma} {{\Pi}^{\Lambda}}_{M \alpha} ({\cal V}^\alpha)^* {L^M}_{ij} {\hat{ \cal H}}^{\Sigma}_{\mu \nu} \gamma^{\mu \nu} \epsilon^j \nonumber \\
      & + \gamma^\mu \epsilon_i ( {\hat P}_{\mu}^* - {\bar\chi}_j \psi^j_\mu  ) - ({\bar\lambda}_{\underline{a}i} \lambda^{\underline{a}}_j) \epsilon^j + \frac{2}{3} g {\bar A}_{2ij} \epsilon^j ,
\end{align}
while for the bosonic fields we have
\begin{align}
    \label{deag}
    \delta_{\epsilon} e^a_\mu = \, & {\bar\epsilon}^i \gamma^a \psi_{i \mu} + {\bar\epsilon}_i \gamma^a \psi^i_\mu, \\[2mm]
       \label{dVg}  
       \delta_{\epsilon} {\cal V}_\alpha = \, & {\cal V}_{\alpha}^* {\bar\epsilon}_i \chi^i, \\[2mm]
       \label{dLMijg}
        \delta_{\epsilon} L_{Mij} = \, & L_{M \underline{a}} ( 2 {\bar\epsilon}_{[i} \lambda^{\underline{a}}_{j]} + \epsilon_{ijkl} {\bar\epsilon}^k \lambda^{\underline{a}l} ), \\[2mm]
        \label{dLMag}
        \delta_{\epsilon} {L_M}^{\underline{a}} = \, & 2 {L_M}^{ij} {\bar\epsilon}_i \lambda^{\underline{a}}_j + c.c. ,\\[2mm]
    \label{dAMag}
     \delta_{\epsilon} A^{M \alpha}_\mu = \, & ({\cal V}^\alpha)^* {L^M}_{ij} {\bar\epsilon}^i \gamma_\mu \chi^j - {\cal V}^\alpha L^{M \underline{a}} {\bar\epsilon}^i \gamma_\mu \lambda_{\underline{a}i} + 2 {\cal V}^\alpha {L^M}_{ij} {\bar\epsilon}^i \psi^j_\mu + c.c. ,\\[2mm]
    \label{dB}
        \delta_{\epsilon} B^{M\alpha}_{\mu \nu}  = \, & 2i {\Theta}^{\alpha MNP} {L_N}^{\underline{a}} {L_P}^{ij} {\bar\epsilon}_i \gamma_{\mu \nu} \lambda_{\underline{a}j} + \frac{1}{2} \xi^M_\beta ({\cal V}^\alpha)^* ({\cal V}^\beta)^* {\bar\epsilon}_i \gamma_{\mu \nu} \chi^i \nonumber \\[2mm]
        & - 2 i  {\Theta}^{\alpha MNP} {L_N}^{\underline{a}} L_{Pij} {\bar\epsilon}^i \gamma_{\mu\nu} \lambda^j_{\underline{a}} + \frac{1}{2} \xi^M_\beta {\cal V}^\alpha  {\cal V}^\beta
        {\bar\epsilon}^i \gamma_{\mu\nu} \chi_i \nonumber \\[2mm]
        & - 4i {\Theta}^{\alpha MNP} {L_N}^{ik} L_{Pjk} \left({\bar\epsilon}^j \gamma_{[\mu|} \psi_{i|\nu]} +{\bar\epsilon}_i \gamma_{[\mu} \psi^j_{\nu]}\right) \\[2mm]
        & +  \xi^M_\beta M^{\alpha \beta} \left({\bar\epsilon}^i \gamma_{[\mu|} \psi_{i|\nu]} +{\bar\epsilon}_i \gamma_{[\mu} \psi^i_{\nu]}\right) \nonumber \\[2mm] 
        & - {{\Theta}^{\alpha M}}_{NP}  
       \epsilon_{\beta \gamma} A^{N \beta}_{[\mu|} \delta_{\epsilon} A^{P \gamma}_{|\nu]} \nonumber - \xi^M_\beta \eta_{NP} A^{N(\alpha|}_{[\mu|}\delta_{\epsilon} A^{P|\beta)}_{|\nu]} ,
\end{align}
where $B^{M \alpha}_{\mu \nu} \equiv - \frac{1}{2} {{\Theta}^{\alpha M}}_{NP} B^{NP}_{\mu \nu} + \frac{1}{2} \xi^M_\beta B^{\alpha \beta}_{\mu \nu}$, 
\begin{equation}
    \label{Dhatmepsilon}
    {\hat D}_\mu \epsilon_i \equiv \partial_\mu \epsilon_i + \frac{1}{4} \omega_{\mu a b} (e,\psi) \gamma^{ab} \epsilon_i - \frac{i}{2} {\hat{\cal A}}_\mu \epsilon_i - {\hat \omega}_i{}^j{}_{\mu} \epsilon_j,
\end{equation}
 and ${\hat{\cal H}}^{\Lambda}_{\mu \nu} = \Pi^{\Lambda}{}_{M \alpha}  {\hat{\cal H}}^{M \alpha}_{\mu \nu}$, where 
\begin{align}
   {\hat{\cal H}}^{M \alpha}_{\mu \nu} \equiv & \, H^{M \alpha}_{\mu \nu} +  \bigg{[ } - 2  ({\cal V}^\alpha)^* L^{M ij} {\bar\psi}_{i \mu} \psi_{j \nu} + \frac{1}{2} {\cal V}^\alpha L^{Mij} {\bar\lambda}_{\underline{a}i} \gamma_{\mu \nu} \lambda^{\underline{a}}_j \nonumber \\[2mm]
   & - \frac{1}{2} ({\cal V}^\alpha)^* L^{M \underline{a}} {\bar\chi}^i \gamma_{\mu \nu} \lambda_{\underline{a}i} + 2 ({\cal V}^\alpha)^* {L^M}_{ij} {\bar\chi}^i \gamma_{[\mu} \psi^j_{\nu]} \\[2mm] 
   & + 2 {\cal V}^\alpha L^{M \underline{a}} {\bar\lambda}_{\underline{a}i} \gamma_{[\mu} \psi^i_{\nu]} + c.c. \nonumber \bigg{]}.
\end{align}

Introducing the symplectic vector ${\cal G}^{M \alpha}_{\mu \nu} = (H^{\Lambda}_{\mu \nu}, {\cal G}_{\Lambda \mu \nu})$, where 
\begin{equation}
    \label{GLg}
    {\cal G}_{\Lambda \mu \nu} \equiv - e^{-1} \epsilon_{\mu \nu \rho \sigma} \frac{\partial {\cal L}}{\partial H^{\Lambda}_{\rho \sigma}} =  {\cal R}_{\Lambda \Sigma} H^{\Sigma}_{\mu \nu} - {\cal I}_{\Lambda \Sigma} (*H^\Sigma)_{\mu \nu} - (*O_{\Lambda})_{\mu \nu} \, ,
\end{equation}
we can write the terms in the local supersymmetry transformations of the fermions that involve ${\hat{\cal H}}^{\Lambda}_{\mu \nu}$ in a manifestly SL(2,$\mathbb{R}$) $\times$ SO(6,$n$)-covariant form as 
\begin{align}
    \delta_\epsilon \chi_i \supset &  - \frac{1}{2} {\cal I}_{\Lambda \Sigma} {{\Pi}^{\Lambda}}_{M \alpha} ({\cal V}^\alpha)^* {L^M}_{ij} {\hat{ \cal H}}^{\Sigma}_{\mu \nu} \gamma^{\mu \nu} \epsilon^j \nonumber \\[2mm]
    = & - \frac{i}{4} {\cal V}_{\alpha}^* L_{M ij} {\cal G}^{M \alpha}_{\mu \nu} \gamma^{\mu \nu} \epsilon^j + \gamma_{\mu \nu} \epsilon^j {\bar\chi}_{[i} \gamma^\mu \psi^\nu_{j]} - \frac{1}{2} \epsilon_{ijkl} \gamma^{\mu \nu} \epsilon^j {\bar\psi}^k_\mu \psi^l_\nu ,\\[2mm]
    \delta_\epsilon \lambda_{\underline{a} i} \supset &  - \frac{1}{4} {\cal I}_{\Lambda \Sigma} {{\Pi}^{\Lambda}}_{M \alpha} ({\cal V}^\alpha)^* {L^M}_{\underline{a}} {\hat{ \cal H}}^{\Sigma}_{\mu \nu} \gamma^{\mu \nu} \epsilon_i \nonumber \\[2mm]
    = & \,\frac{i}{8} {\cal V}_{\alpha}^* L_{M \underline{a}} {\cal G}^{M \alpha}_{\mu \nu} \gamma^{\mu \nu} \epsilon_i + \frac{1}{2} \gamma^{\mu \nu} \epsilon_i {\bar\lambda}_{\underline{a}j} \gamma_\mu \psi^j_\nu, \\[2mm]
     \delta_\epsilon \psi_{i \mu} \supset & \,  \frac{1}{4} {\cal I}_{\Lambda \Sigma} {{\Pi}^{\Lambda}}_{M \alpha} {\cal V}^\alpha {L^M}_{ij}  {\hat{ \cal H}}^{\Sigma}_{\nu \rho} \gamma^{\nu \rho} \gamma_\mu \epsilon^j \nonumber \\[2mm]
     = & - \frac{i}{8} {\cal V}_\alpha L_{M ij} {\cal G}^{M \alpha}_{\nu \rho} \gamma^{\nu \rho} \gamma_\mu  \epsilon^j + \frac{1}{2} \gamma^{\nu \rho} \gamma_\mu  \epsilon^j {\bar\psi}_{i \nu} \psi_{j \rho} - \frac{1}{4} \epsilon_{ijkl} \gamma^{\nu \rho} \gamma_\mu  \epsilon^j {\bar\chi}^k \gamma_\nu \psi^l_\rho \, .
\end{align}
We note that ${\cal G}^{M \alpha}_{\mu \nu}$ satisfies the twisted self-duality condition 
\begin{align}
    \label{twist}
    \epsilon_{\mu \nu \rho \sigma} {\cal G}^{M \alpha \rho \sigma} = & \, 2 \eta^{M N} \epsilon^{\alpha \beta} M_{NP} M_{\beta \gamma} {\cal G}^{P \gamma}_{\mu \nu} + 2 \big{(} -2 i ({\cal V}^\alpha)^* L^{Mij} {\bar{\psi}}_{i \mu} \psi_{j \nu}  \nonumber \\ & + \epsilon_{\mu \nu \rho \sigma} ({\cal V}^\alpha)^* L^{Mij}  {\bar{\psi}}^{\rho}_i {\psi}_j^{\sigma} 
     - i {\mathcal{V}}^{\alpha}  L^{Mij} {\bar{\lambda}}_{\underline{a} i} \gamma_{\mu \nu} \lambda^{\underline{a}}_j  - i {\mathcal{V}}^{\alpha} L^{M \underline{a}} {\bar\chi}_i \gamma_{\mu \nu} \lambda^i_{\underline{a}} \nonumber\\ & + 2 i ({\cal V}^\alpha)^* {L^M}_{ij} {\bar\chi}^i \gamma_{[\mu} \psi^j_{\nu]}  - \epsilon_{\mu \nu \rho \sigma} ({\cal V}^\alpha)^* {L^M}_{ij} {\bar\chi}^i \gamma^{\rho} \psi^{j \sigma} \\ & + 2 i {\cal V}^\alpha L^{M \underline{a}} {\bar\lambda}_{\underline{a} i } \gamma_{[\mu} \psi^i_{\nu]}- \epsilon_{\mu \nu \rho \sigma} {\cal V}^\alpha  L^{M \underline{a}} {\bar\lambda}_{\underline{a}i} \gamma^{\rho} \psi^{ i \sigma} + \text{c.c.} \big{)}\,. \nonumber
\end{align}

The Lagrangian \eqref{L4formg} is invariant, up to a total derivative, under the local supersymmetry transformations 
\eqref{dpsig}-\eqref{dB} 
and under vector-gauge transformations, provided the transformation rules \eqref{dzBMN} and \eqref{dzBab} for the two-form gauge fields are modified as \cite{deWit:2005ub,Trigiante:2016mnt}
\begin{align}
    \label{dzBMNmod}
    \delta_{\zeta} B^{MN}_{\mu \nu} & = -2 \epsilon_{\alpha \beta} \left(\zeta^{[M|\alpha} {\cal G}^{|N] \beta}_{\mu \nu} - A^{[M|\alpha}_{[\mu|} \delta_{\zeta} A^{|N] \beta}_{|\nu]} \right)\,, \\[2mm]
    \label{dzBabmod}
     \delta_{\zeta} B^{\alpha \beta}_{\mu \nu} & = 2 \eta_{MN} \left( \zeta^{M(\alpha|} {\cal G}^{N|\beta)}_{\mu \nu} - A^{M(\alpha|}_{[\mu|} \delta_{\zeta} A^{N|\beta)}_{|\nu]} \right) .
\end{align}
It is also invariant under the tensor-gauge transformations \eqref{dXiAMa}-\eqref{dXiBab}. 
Furthermore, there is an additional gauge invariance parametrized by rank-2 tensors $\Delta^{MN \Sigma}_{\mu \nu}=\Delta^{[MN] \Sigma}_{\mu \nu}$ and $\Delta^{\alpha \beta \Sigma}_{\mu \nu}=\Delta^{(\alpha \beta) \Sigma}_{\mu \nu}$ which acts only on the the antisymmetric tensor fields ${{\Pi}^\Lambda}_{M \alpha} B^{M \alpha}_{\mu \nu}$ as \cite{deWit:2007kvg,deVroome:2007unr} 
\begin{equation}
    \label{Deltasym}
    \delta_\Delta({{\Pi}^\Lambda}_{M \alpha} B^{M \alpha}_{\mu \nu}) = {\Delta^{\Lambda \Sigma \rho}}_{\rho} \left({\cal G}_{\Sigma \mu \nu} - H_{\Sigma \mu \nu} \right) - 6 {\Delta^{(\Lambda \Sigma) \rho}}_{[\rho|} \left({\cal G}_{\Sigma |\mu \nu]} - H_{\Sigma |\mu \nu]} \right) \, , 
\end{equation}
where 
\begin{equation}
    \Delta^{\Lambda \Sigma}_{\mu \nu} \equiv - {{\Pi}^\Lambda}_{M \alpha} {\Theta^{\alpha M}}_{NP} \Delta^{NP \Sigma}_{\mu \nu} + {{\Pi}^\Lambda}_{M \alpha} \xi^M_\beta \Delta^{\alpha \beta \Sigma}_{\mu \nu}.  
\end{equation}

\subsection{ Bianchi identities and field equations} 
\label{sub:Field_Equations_and_Bianchi_Identities}
The field strengths of the two-form gauge fields are defined by \cite{deWit:2005hv}
\begin{align}
    {\cal H}^{(3)MN}_{\mu \nu \rho} \equiv & \, 3 \partial_{[\mu} B^{MN}_{\nu \rho]} + 6 g {\Theta_{\alpha P Q}}^{[M|} A^{P \alpha}_{[\mu} B^{|N]Q}_{\nu \rho]} \nonumber \\[2mm]
    & + 6 \epsilon_{\alpha \beta} A^{[M|\alpha}_{[\mu|} \left( \partial_{|\nu} A^{|N]\beta}_{\rho]} + \frac{g}{3} {X_{P \gamma Q \delta}}^{|N] \beta} A^{P \gamma}_{|\nu} A^{Q \delta}_{\rho]} 
    \right), \\[2mm]
    {\cal H}^{(3) \alpha \beta}_{\mu \nu \rho} \equiv & \, 3 \partial_{[\mu} B^{\alpha \beta}_{\nu \rho]} - 3 g \xi^{(\alpha|M} A_{M \gamma [\mu} B^{\beta) \gamma}_{\nu \rho]} - 3 g \xi_{\gamma M} A^{M (\alpha}_{[\mu} B^{\beta) \gamma}_{\nu \rho]} \nonumber  \\[2mm] 
    & - 6 \eta_{MN} A^{M(\alpha|}_{[\mu|} \left(  \partial_{|\nu} A^{N|\beta)}_{\rho]} + \frac{g}{3} {X_{P \gamma Q \delta}}^{N |\beta)} A^{P \gamma}_{|\nu} A^{Q \delta}_{\rho]}\right).   \end{align}
The field strengths of the vector and the two-form gauge fields satisfy the Bianchi identities 
\begin{align}
\label{DH=gH3}
     {\hat D}_{[\mu} H^{M \alpha}_{\nu \rho]} & = - \frac{g}{6} \left( \Theta^{\alpha M}{}_{NP} {\cal H}^{(3)NP}_{\mu \nu \rho} - \xi^M_\beta   {\cal H}^{(3) \alpha \beta}_{\mu \nu \rho} \right) , \\
    \label{DH3=XHH}
    - \Theta^{\alpha M}{}_{NP} {\hat D}_{[\mu} {\cal H}^{(3)NP}_{\nu \rho \sigma]} + \xi^M_{\beta} {\hat D}_{[\mu} {\cal H}^{(3)\alpha \beta}_{\nu \rho \sigma]} & = 3 X_{N \beta P \gamma}{}^{M \alpha} H^{N \beta}_{[\mu \nu} H^{P \gamma}_{\rho \sigma]} \, , 
\end{align}
where the covariant derivatives of the field strengths appearing in the above equations are defined as follows
\begin{align}
    {\hat D}_{\mu} H^{M \alpha}_{\nu \rho} &\equiv \partial_\mu H^{M \alpha}_{\nu \rho} + g X_{N \beta P \gamma}{}^{M \alpha} A^{N \beta}_\mu H^{P \gamma}_{\nu \rho},\\[2mm]
    {\hat D }_\mu {\cal H}^{(3)MN}_{\nu \rho \sigma} &\equiv \partial_\mu {\cal H}^{(3)MN}_{\nu \rho \sigma} 
 + 2 g \Theta_{\alpha P Q}{}^{[M|} A^{P \alpha}_\mu {\cal H}^{(3)|N]Q}_{\nu \rho \sigma},\\[2mm]
 {\hat D }_\mu {\cal H}^{(3)\alpha \beta}_{\nu \rho \sigma} &\equiv \partial_\mu  {\cal H}^{(3)\alpha \beta}_{\nu \rho \sigma} - g \xi^{(\alpha|M} A_{M \gamma \mu} {\cal H}^{(3)|\beta)\gamma}_{\nu \rho \sigma} - g \xi_{\gamma M} A^{M (\alpha|}_\mu {\cal H}^{(3)|\beta)\gamma}_{\nu \rho \sigma}.
\end{align}

The equations of motion for the two-form gauge fields $B^{MN}_{\mu \nu}$ and $B^{\alpha \beta}_{\mu \nu}$, which do not have kinetic terms, take the following form
\begin{equation}
    \label{BMNeom}
    {{\Pi}^{\Lambda}}_{M \alpha} {\Theta^{\alpha M}}_{NP} \left( H_{\Lambda \mu \nu} - {\cal G}_{\Lambda \mu \nu} \right)  = 0,
\end{equation}
\begin{equation}   
    \label{Babeom}
     {{\Pi}^{\Lambda}}_{M (\alpha} \xi^M_{\beta)} \left( H_{\Lambda \mu \nu} - {\cal G}_{\Lambda \mu \nu} \right)  = 0 \, ,
\end{equation}
where $H_{\Lambda \mu \nu} = \Pi_{\Lambda M \alpha} H^{M \alpha}_{\mu \nu}$.\\
  
The field equations for the vector gauge fields $A^{M\alpha}_\mu$ are
\begin{equation}
    \label{veceom}
    \frac{1}{2} \epsilon^{\mu \nu \rho \sigma} \,\hat{D}_{\nu}\mathcal{G}^{M \alpha}_{\rho\sigma}   =  g J^{M \alpha \mu}\,,  
\end{equation}
where we have used the property 
\begin{equation}
\label{X(H-G)=0}
    X_{P \gamma N \beta}{}^{M \alpha} \left( 
      H^{P \gamma}_{\rho \sigma} - {\cal G}^{P \gamma}_{\rho \sigma}\right) =0\, ,
\end{equation}
  which holds on-shell by virtue of \eqref{BMNeom} and \eqref{Babeom}. 
  The current on the right-hand side of \eqref{veceom} is defined as
\begin{align}
      \label{JMa}
     J^{M \alpha \mu} \equiv & \, {\Theta^{\alpha M}}_{NP} \Big{[} {L^N}_{\underline{a}} {L^P}_{ij} {\hat P}^{\underline{a} i j \mu} + {L^N}_{ik} L^{P jk} \left( {\bar\chi}_j \gamma^\mu \chi^i + 2  {\bar\lambda}^{\underline{a}}_j \gamma^\mu \lambda^i_{\underline{a}} + 2i \epsilon^{\mu \nu \rho \sigma}   {\bar\psi}^i_\nu
      \gamma_\rho  \psi_{j \sigma}  \right) \nonumber \\
      &  + 2 L^{N \underline{a}} L^{P \underline{b}} {\bar\lambda}_{\underline{a}i} \gamma^\mu \lambda^i_{\underline{b}} + 2 {L^N}_{\underline{a}} {L^P}_{ij} \left({\bar\lambda}^{\underline{a}i} \psi^{j \mu} - {\bar\lambda}^{\underline{a}i} \gamma^{\mu \nu} \psi^j_\nu \right) \nonumber \\ 
      & + 2 {L^N}_{\underline{a}} L^{Pij} \left( {\bar\lambda}^{\underline{a}}_i \psi^{\mu}_j - {\bar\lambda}^{\underline{a}}_i \gamma^{\mu \nu} \psi_{j \nu}  \right) \Big{]} + \xi^M_{\beta} \bigg{[} \frac{i}{2} {\cal V}^\alpha {\cal V}^\beta ({\hat P}^\mu)^* - \frac{i}{2} ( {\cal V}^\alpha)^* ({\cal V}^\beta)^* {\hat P}^\mu \nonumber \\ & + M^{\alpha \beta} \left( \frac{3i}{4} {\bar\chi}_i \gamma^\mu \chi^i + \frac{i}{2} {\bar\lambda}^{\underline{a}}_i \gamma^\mu {\lambda}_{\underline{a}}^i      + \frac{1}{2} \epsilon^{\mu \nu \rho \sigma} {\bar\psi}^i_\nu
      \gamma_\rho  \psi_{i \sigma} \right) \\ & - \frac{i}{2} {\cal V}^{\alpha} {\cal V}^{\beta} \left( {\bar\chi}_i \psi^{i \mu} - {\bar\chi}_i \gamma^{\mu \nu} \psi^i_\nu \right) + \frac{i}{2} ( {\cal V}^\alpha)^* ({\cal V}^\beta)^* \left( {\bar\chi}^i \psi_i^\mu - {\bar\chi}^i \gamma^{\mu \nu} \psi_{i \nu} \right) \bigg{]}. \nonumber
\end{align}
Multiplying \eqref{veceom} by the projectors $\Pi^{\Lambda}{}_{M \alpha}$, we obtain the equations of motion for the magnetic vector fields $A_{\Lambda \mu}$. 
Using the Bianchi identity \eqref{DH=gH3}, the linear constraint \eqref{linear} on the embedding tensor and the on-shell condition \eqref{X(H-G)=0}, we can write the latter as 
\begin{align}
    \label{mageom}
    & - \frac{1}{12} \epsilon^{\mu \nu \rho \sigma} {{\Pi}^{\Lambda}}_{M \alpha} \Big{[} 
        {\Theta^{\alpha M}}_{NP} {\cal H}^{(3) N P}_{\nu \rho \sigma } - \xi^M_{\beta} {\cal H}^{(3) \alpha \beta}_{\nu \rho \sigma }  \nonumber \\[2mm] 
      &\qquad\qquad + 6 \Pi^{\Sigma}{}_{P \gamma} X^{M \alpha}{}_{N \beta}{}^{P \gamma} A^{N \beta}_\nu  \left( H_{\Sigma \rho \sigma} - {\cal G}_{\Sigma \rho \sigma} \right) \Big{]}   =  \Pi^{\Lambda}{}_{M \alpha} J^{M \alpha \mu }\,.
\end{align}

Furthermore, the equations of motion for the fermionic fields are
\begin{align}
    \label{dileom}
    \gamma^\mu {\hat D}_\mu \chi_i = & \, \gamma^\mu \gamma^\nu \psi_{i \mu} \left( {\hat P}_\nu^* - {\bar\chi}_j \psi^j_\nu \right) + 2 {\cal I}_{\Lambda \Sigma} \Pi^{\Lambda}{}_{M \alpha} ({\cal V}^{\alpha})^* L^M{}_{ij} {\hat{\cal H}}^{\Sigma -}_{\mu \nu} \gamma^\mu \psi^{j \nu } \nonumber \\
    & - \frac{1}{2} {\cal I}_{\Lambda \Sigma} \Pi^{\Lambda}{}_{M \alpha} ({\cal V}^{\alpha})^* L^M{}_{\underline{a}} {\hat{\cal H}}^{\Sigma }_{\mu \nu} \gamma^{\mu \nu} \lambda^{\underline{a}}_i - \gamma^\mu \psi^j_\mu {\bar\lambda}_{\underline{a}i} \lambda^{\underline{a}}_j+ \frac{3}{4} \chi^j {\bar\chi}_i \chi_j   \\
    &  - \frac{1}{2} \lambda^{\underline{a}}_j {\bar\lambda}^j_{\underline{a}} \chi_i -  \lambda^{\underline{a}}_i {\bar\lambda}^j_{\underline{a}} \chi_j    + \frac{2}{3} g {\bar A}_{2 ij} \gamma^\mu \psi^j_\mu - 2 g {\bar A}_2{}^{\underline{a}j}{}_i \lambda_{\underline{a}j} + 2 g {\bar A}_2{}^{\underline{a}j}{}_j \lambda_{\underline{a}i} \, ,  \nonumber \\
    \label{gaugeom}
    \gamma^\mu {\hat D}_\mu \lambda_{\underline{a} i} = & - \gamma^\mu \gamma^\nu \psi^j_\mu \left( {\hat P}_{\underline{a}ij \nu} + 2 {\bar\lambda}_{\underline{a}[i} \psi_{j] \nu} + \epsilon_{ijkl} {\bar\lambda}^k_{\underline{a}} \psi^l_\nu \right) \nonumber  \\
    & + {\cal I}_{\Lambda \Sigma} \Pi^{\Lambda}{}_{M \alpha} ({\cal V}^\alpha)^* L^M{}_{\underline{a}} {\hat{\cal H}}^{\Sigma+}_{\mu \nu} \gamma^\mu \psi_i^\nu + \frac{1}{2} {\cal I}_{\Lambda \Sigma} \Pi^{\Lambda}{}_{M \alpha} ({\cal V}^\alpha)^* L^M{}_{ij} \hat{\cal H}^{\Sigma}_{\mu \nu} \gamma^{\mu \nu} \lambda^j_{\underline{a}} \nonumber \\ 
    & + \frac{1}{4} {\cal I}_{\Lambda \Sigma} \Pi^{\Lambda}{}_{M \alpha} {\cal V}^{\alpha} L^M{}_{\underline{a}} \hat{\cal H}^{\Sigma}_{\mu \nu} \gamma^{\mu \nu}  \chi_i + \gamma^\mu \psi_{j \mu} {\bar\chi}_i \lambda^j_{\underline{a}} - \frac{1}{2} \gamma^\mu \psi_{i \mu} {\bar\chi}_j \lambda^j_{\underline{a}} \\
    & - \frac{1}{2} \lambda^j_{\underline{b}} {\bar\lambda}^{\underline{b}}_j \lambda_{\underline{a} i } - \lambda^j_{\underline{a}} {\bar\lambda}_{\underline{b}i} \lambda^{\underline{b}}_j + 2 \lambda^j_{\underline{b}} {\bar\lambda}^{\underline{b}}_i \lambda_{\underline{a}j} - \frac{1}{4} \chi_j {\bar\chi}^j \lambda_{\underline{a}i} -  \frac{1}{2} \chi_i {\bar\chi}^j \lambda_{\underline{a}j}   \nonumber \\
    &  + g {\bar A}_{2 \underline{a}}{}^j{}_i \gamma^\mu \psi_{j \mu} - g A_{2 \underline{a} i}{}^j \chi_j + g A_{2 \underline{a} j}{}^j \chi_i +  2 g {\bar A}_{\underline{a} \underline{b} ij} \lambda^{\underline{b} j} + \frac{2}{3} g {\bar A}_{2(ij)} \lambda^j_{\underline{a}}\,, \nonumber \\
    \label{graveomsp}
    \gamma^\mu {\hat\rho}_{i \mu \nu} = & \, \chi_i \left( {\hat P}_\nu - {\bar\chi}^j \psi_{j \nu} \right)  + 2 \lambda^{\underline{a} j} \left( {\hat P}_{\underline{a} ij \nu} + 2 {\bar\lambda}_{\underline{a}[i} \psi_{j] \nu} + \epsilon_{ijkl} {\bar\lambda}_{\underline{a}}^k \psi^l_\nu \right) \nonumber \\
    &  + {\cal I}_{\Lambda \Sigma} \Pi^{\Lambda}{}_{M \alpha} {\cal V}^\alpha L^M{}_{ij} \hat{\cal H}^{\Sigma+}_{\mu \nu} \left( \psi^{j \mu} - \gamma^{\mu \rho} \psi^j_\rho \right) \nonumber \\
    & - {\cal I}_{\Lambda \Sigma} \Pi^{\Lambda}{}_{M \alpha} {\cal V}^\alpha L^M{}_{\underline{a}} \hat{\cal H}^{\Sigma-}_{\mu \nu} \gamma^\mu \lambda^{\underline{a}}_i + {\cal I}_{\Lambda \Sigma} \Pi^{\Lambda}{}_{M \alpha} ({\cal V}^\alpha)^* L^M{}_{ij} \hat{\cal H}^{\Sigma-}_{\mu \nu} \gamma^\mu \chi^j  \nonumber \\ 
    & - \frac{1}{2} \epsilon_{ijkl} \gamma^{\mu \rho} \psi^j_{[\mu|} {\bar\lambda}^k_{\underline{a}} \gamma_{|\nu]\rho} \lambda^{\underline{a} l} + \frac{1}{4} \epsilon_{ijkl} \psi^{j \mu} {\bar\lambda}^k_{\underline{a}} \gamma_{\mu \nu} \lambda^{\underline{a} l} \nonumber \\
    & + \frac{1}{2} \gamma_{(\mu|} \psi_{j |\nu)} {\bar\chi}_i \gamma^\mu \chi^j + \frac{1}{4} \gamma^\mu \psi_{j \mu} {\bar\chi}_i \gamma_\nu \chi^j - \frac{1}{4} \gamma_{\mu \nu \rho} \psi^\mu_j {\bar\chi}_i \gamma^\rho \chi^j \nonumber \\
    & - \frac{1}{8} \gamma_\nu \psi_{i \mu} {\bar\chi}_j \gamma^\mu \chi^j - \frac{1}{4} \gamma^\mu \psi_{i \mu} {\bar\chi}_j \gamma_\nu \chi^j + \frac{1}{8} \gamma_{\mu \nu \rho} \psi^\mu_i {\bar\chi}_j \gamma^\rho \chi^j \nonumber \\
    & + \gamma_{(\mu|} \psi_{j |\nu)} {\bar\lambda}^{\underline{a}}_i \gamma^\mu \lambda^j_{\underline{a}} + \frac{1}{2} \gamma^\mu \psi_{j \mu } {\bar\lambda}^{\underline{a}}_i \gamma_\nu \lambda^j_{\underline{a}} - \frac{1}{2} \gamma_{\mu \nu \rho} \psi^\mu_j {\bar\lambda}^{\underline{a}}_i \gamma^\rho \lambda^j_{\underline{a}} \nonumber \\
    & - \frac{1}{2} \left( \gamma_\mu \psi_{i \nu} + \frac{1}{2} \gamma_\nu \psi_{i \mu} \right) {\bar\lambda}^{\underline{a}}_j \gamma^\mu \lambda^j_{\underline{a}} + \frac{1}{4} \gamma_{\mu \nu \rho} \psi^\mu_i {\bar\lambda}^{\underline{a}}_j \gamma^\rho \lambda^j_{\underline{a}} \nonumber \\
    & - \epsilon_{ijkl} \gamma^\mu \chi^j {\bar\psi}^k_\mu \psi^l_\nu - \frac{1}{2} \gamma_\nu \lambda_{\underline{a} j } {\bar\lambda}^{\underline{a}}_i \chi^j + g {\bar A}_{1ij} \left(\psi^j_\nu - \frac{1}{3} \gamma_{\mu \nu} \psi^{j \mu} \right) \nonumber \\
    & + \frac{1}{3} {\bar A}_{2ji} \gamma_\nu \chi^j + g A_{2 \underline{a}i}{}^j \gamma_\nu \lambda^{\underline{a}}_j \,.
\end{align}
The terms on the right-hand sides of equations \eqref{dileom}-\eqref{graveomsp} that contain ${\hat{\cal H}}^{\Lambda}_{\mu \nu} $ can be written in a manifestly SL(2,$\mathbb{R}$) $\times$ SO(6,$n$)-covariant form in terms of ${\cal G}^{M \alpha}_{\mu \nu}$ as 
\begin{align}
    \gamma^{\mu} {\hat D}_\mu \chi_i \supset  &\, 2 {\cal I}_{\Lambda \Sigma} \Pi^{\Lambda}{}_{M \alpha} ({\cal V}^{\alpha})^* L^M{}_{ij} {\hat{\cal H}}^{\Sigma -}_{\mu \nu} \gamma^\mu \psi^{j \nu } \nonumber - \frac{1}{2} {\cal I}_{\Lambda \Sigma} \Pi^{\Lambda}{}_{M \alpha} ({\cal V}^{\alpha})^* L^M{}_{\underline{a}} {\hat{\cal H}}^{\Sigma }_{\mu \nu} \gamma^{\mu \nu} \lambda^{\underline{a}}_i \nonumber \\ 
    =& - \frac{i}{4} {\cal V}_\alpha^* L_{M ij} { {\cal G}}^{M \alpha}_{\nu \rho} \gamma^\mu \gamma^{\nu \rho} \psi^j_\mu  - \frac{1}{2} \epsilon_{ijkl} \gamma^\mu \gamma^{\nu \rho } \psi^j_\mu {\bar\psi}^k_\nu \psi^l_\rho + \gamma^\mu \gamma_{\nu \rho } \psi^j_\mu {\bar\chi}_{[i} \gamma^\nu \psi^\rho_{j]}  \\
    &+ \frac{i}{4} {\cal V}_\alpha^* L_{M \underline{a}} {\cal G}^{M \alpha}_{\mu \nu} \gamma^{\mu \nu} \lambda^{\underline{a}}_i + \gamma^{\mu \nu} \lambda^{\underline{a}}_i {\bar\lambda}_{\underline{a} j} \gamma_\mu \psi^j_\nu\,, \nonumber \\
    \gamma^{\mu} {\hat D}_\mu \lambda_{\underline{a} i } \supset & \, {\cal I}_{\Lambda \Sigma} \Pi^{\Lambda}{}_{M \alpha} ({\cal V}^\alpha)^* L^M{}_{\underline{a}} {\hat{\cal H}}^{\Sigma+}_{\mu \nu} \gamma^\mu \psi_i^\nu + \frac{1}{2} {\cal I}_{\Lambda \Sigma} \Pi^{\Lambda}{}_{M \alpha} ({\cal V}^\alpha)^* L^M{}_{ij} \hat{\cal H}^{\Sigma}_{\mu \nu} \gamma^{\mu \nu} \lambda^j_{\underline{a}} \nonumber \\ 
    & + \frac{1}{4} {\cal I}_{\Lambda \Sigma} \Pi^{\Lambda}{}_{M \alpha} {\cal V}^{\alpha} L^M{}_{\underline{a}} \hat{\cal H}^{\Sigma}_{\mu \nu} \gamma^{\mu \nu}  \chi_i \nonumber\\
    =& \, \frac{i}{8} {\cal V}_\alpha^* L_{M \underline{a}} {\cal G}^{M \alpha}_{\nu \rho} \gamma^\mu \gamma^{\nu \rho} \psi_{i \mu} + \frac{1}{2} \gamma^\mu \gamma^{\nu \rho} \psi_{i \mu} {\bar\lambda}_{\underline{a} j} \gamma_\nu \psi^j_\rho \\ 
    & + \frac{i}{4} {\cal V}_\alpha^* L_{Mij} {\cal G}^{M \alpha}_{\mu \nu} \gamma^{\mu \nu} \lambda^j_{\underline{a}} + \frac{1}{2} \epsilon_{ijkl} \gamma^{\mu \nu} \lambda^j_{\underline{a}} {\bar\psi}^k_\mu \psi^l_\nu - \gamma^{\mu \nu} \lambda^j_{\underline{a}} {\bar\chi}_{[i|} \gamma_\mu \psi_{|j]\nu} \nonumber \\
    & + \frac{i}{8} {\cal V}_\alpha L_{M \underline{a}} {\cal G}^{M \alpha}_{\mu \nu} \gamma^{\mu \nu} \chi_i -\frac{1}{2} \gamma^{\mu \nu} \chi_i {\bar\lambda}^j_{\underline{a}} \gamma_\mu \psi_{j \nu}\,, \nonumber \\
    \gamma^\mu {\hat\rho}_{i \mu \nu} \supset  &  \,{\cal I}_{\Lambda \Sigma} \Pi^{\Lambda}{}_{M \alpha} {\cal V}^\alpha L^M{}_{ij} \hat{\cal H}^{\Sigma+}_{\mu \nu} \left( \psi^{j \mu} - \gamma^{\mu \rho} \psi^j_\rho \right) \nonumber \\
    & - {\cal I}_{\Lambda \Sigma} \Pi^{\Lambda}{}_{M \alpha} {\cal V}^\alpha L^M{}_{\underline{a}} \hat{\cal H}^{\Sigma-}_{\mu \nu} \gamma^\mu \lambda^{\underline{a}}_i + {\cal I}_{\Lambda \Sigma} \Pi^{\Lambda}{}_{M \alpha} ({\cal V}^\alpha)^* L^M{}_{ij} \hat{\cal H}^{\Sigma-}_{\mu \nu} \gamma^\mu \chi^j  \nonumber \\
    = & - \frac{i}{8} {\cal V}_\alpha L_{Mij} {\cal G}^{M \alpha}_{\rho \sigma} \gamma^\mu \gamma^{\rho \sigma} \gamma_\nu \psi^j_\mu + \frac{1}{2} \gamma^\mu \gamma^{\rho \sigma} \gamma_\nu \psi^j_\mu \left( {\bar\psi}_{i \rho} \psi_{j \sigma} - \frac{1}{2} \epsilon_{ijkl} {\bar\chi}^k \gamma_\rho \psi^l_\sigma \right) \\
    & - \frac{i}{8} {\cal V}_\alpha L_{M \underline{a}} {\cal G}^{M \alpha}_{\mu \rho} \gamma^{\mu \rho} \gamma_\nu \lambda^{\underline{a}}_i + \frac{1}{2} \gamma^{\mu \rho} \gamma_\nu \lambda^{\underline{a}}_i {\bar\lambda}^j_{\underline{a}} \gamma_\mu \psi_{j \rho} \nonumber \\
    & + \frac{i}{8} {\cal V}_\alpha^* L_{Mij} {\cal G}^{M \alpha}_{\mu \rho} \gamma^{\mu \rho} \gamma_\nu \chi^j + \frac{1}{4} \gamma_{\mu \rho} \gamma_\nu \chi^j \left( \epsilon_{ijkl} {\bar\psi}^{k \mu} \psi^{l \rho} -2 {\bar\chi}_{[i} \gamma^\mu \psi_{j]}^\rho \right)\,. \nonumber
\end{align}
 \subsection{Closure of the supersymmetry algebra}\label{sub:closure_susy_rules}
Let us now discuss the closure of the supersymmetry transformation rules of section \ref{sub:the_susy_rules}. 
The commutator of two consecutive local supersymmetry transformations, $\delta_Q (\epsilon_1)$ and $\delta_Q (\epsilon_2)$, parametrized by left-handed Weyl spinors $\epsilon_1^i$ and $\epsilon_2^i$ respectively and their charge conjugates, has the following expression:
\begin{align}
  \label{QQcomg}
  [\delta_Q (\epsilon_1), \delta_Q (\epsilon_2)] = & \, \delta_{\text{cgct}}(\xi^\mu) \,  + \delta_{\text{Lorentz}} (\lambda_{ab}) \, + \delta_Q (\epsilon_3) + \delta_{\text{SO}(2)} (\Lambda) \nonumber \\ 
  &   + \delta_{\text{SU}(4)} ({\Lambda_i}^j) + \delta_{\text{SO}(n)} ({\Lambda_{\underline{a}}}^{\underline{b}}) + \delta_{\text{gauge}}(\zeta^{M \alpha}) + \delta_{\text{tensor}}(\Xi^{MN}_\mu, \Xi^{\alpha \beta}_\mu) \, , 
  \end{align}
where the first term denotes a covariant general coordinate transformation with parameters
 \begin{equation}
     \label{xim}
     \xi^\mu = \bar{\epsilon}_{2 i} \gamma^\mu \epsilon^i_1 + \bar{\epsilon}^i_2 \gamma^\mu \epsilon_{1i} \, , 
 \end{equation}
which is defined by \cite{deWit:1975veh, Jackiw:1978ar} (see \cite{Freedman:2012zz} for a review)
\begin{align}
\label{cgctg}
\delta_{\text{cgct}} (\xi^\mu) \equiv & \,  \delta_{\text{gct}} (\xi^\mu) - \delta_{\text{Lorentz}} (\xi^\mu \omega_{\mu a b}) - \delta_Q (\xi^\mu \psi^i_\mu) - \delta_{\text{SO}(2)} (\xi^\mu {\cal A}_\mu) \nonumber \\[2mm]
& - \delta_{\text{SU}(4)} (\xi^\mu {\omega}_{i \hspace{0.1cm} \mu}^{\hspace{0.1cm}j}) - \delta_{\text{SO}(n)} (\xi^\mu {\omega}_{\underline{a} \hspace{0.15cm}\mu}^{\hspace{0.15cm} \underline{b}}) - \delta_{\text{gauge}} (\xi^\mu A^{M \alpha}_\mu) \\[2mm]
& - \delta_{\text{tensor}} \left( \xi^\nu B^{M N}_{\nu \mu} + \epsilon_{\alpha \beta} \xi^\nu A^{[M| \alpha}_\nu A^{|N] \beta}_{\mu},  \xi^\nu B^{\alpha \beta}_{\nu \mu} - \eta_{MN} \xi^\nu A^{M(\alpha|}_\nu A^{N|\beta)}_\mu \right) ,  \nonumber 
\end{align}
where $\delta_{\text{gct}}(\xi^\mu)$ is a general coordinate transformation and ${\cal A}_\mu$, $\omega_i{}^j{}_\mu$ and $\omega_{\underline{a}}{}^{\underline{b}}{}_\mu$ are the components of the ungauged SO(2), SU(4) and SO($n$) one-form connections ${\cal A}$, $\omega_i{}^j$ and $\omega_{\underline{a}}{}^{\underline{b}}$ respectively, which have been defined in section \ref{sec:the_ingredients_of_n_4_supergravity}. 
The parameters of the remaining transformations that appear on the right-hand side of \eqref{QQcomg} are given by 
\begin{align}
    \label{params}
     \lambda_{ab} = & \left( \frac{1}{2} \epsilon_{ijkl} \bar{\epsilon}^i_1 \epsilon_2^j \bar{\lambda}^k_{\underline{a}} \gamma_{ab} \lambda^{\underline{a}l} + 2 {\cal I}_{\Lambda \Sigma} {\Pi^{\Lambda}}_{M \alpha} {\cal V}^{\alpha} {L^M}_{ij} {\bar\epsilon}^i_1 \epsilon^j_2 e_a^\mu e_b^\nu {\hat{\cal H}}^{\Sigma +}_{\mu \nu}+ c.c. \right) \nonumber 
    \\
     & + \frac{1}{2} \left( {\bar\epsilon}_{1i} \gamma_{abc} \epsilon_2^j - {\bar\epsilon}_{2i} \gamma_{abc} \epsilon_1^j \right) {\bar\chi}_j \gamma^c \chi^i - \frac{1}{4} \left( {\bar\epsilon}_{1i} \gamma_{abc} \epsilon_2^i - {\bar\epsilon}_{2i} \gamma_{abc} \epsilon_1^i \right) {\bar\chi}_j \gamma^c \chi^j \nonumber  \\
    & + \left( {\bar\epsilon}_{1i} \gamma_{abc} \epsilon_2^j - {\bar\epsilon}_{2i} \gamma_{abc} \epsilon_1^j \right) {\bar\lambda}^{\underline{a}}_j \gamma^c\lambda^i_{\underline{a}}  - \frac{1}{2} \left( {\bar\epsilon}_{1i} \gamma_{abc} \epsilon_2^i - {\bar\epsilon}_{2i} \gamma_{abc} \epsilon_1^i \right) {\bar\lambda}^{\underline{a}}_j \gamma^c\lambda^j_{\underline{a}}  \\
    & + \left( - \frac{2}{3} g {\bar A}_{1ij} {\bar\epsilon}_1^i \gamma_{ab} \epsilon_2^j + c.c.\right)\,, \nonumber \\
    \epsilon_{3i} = & \, \epsilon_{ijkl} \chi^j {\bar\epsilon}^k_1 \epsilon^l_2\,, \\ 
    \Lambda = & - \frac{i}{2} \left( {\bar\epsilon}_{1i} \gamma_\mu \epsilon^j_2 - {\bar\epsilon}_{2i} \gamma_\mu \epsilon_1^j \right) {\bar\chi}_j \gamma^\mu \chi^i\,, \\ 
    {\Lambda_i}^j = & \bigg{(} {\bar\epsilon}_{2 i} \gamma_\mu \epsilon_1^j {\bar\lambda}^{\underline{a}}_k \gamma^\mu \lambda^k_{\underline{a}} + {\bar\epsilon}_{2 k} \gamma_\mu \epsilon_1^k {\bar\lambda}^{\underline{a}}_i \gamma^\mu \lambda^j_{\underline{a}} - \frac{1}{2} \delta^j_i {\bar\epsilon}_{2 k} \gamma_\mu \epsilon_1^k {\bar\lambda}^{\underline{a}}_l \gamma^\mu \lambda^l_{\underline{a}}  \nonumber \\
    &  - {\bar\epsilon}_{2 k} \gamma_\mu \epsilon_1^j {\bar\lambda}^{\underline{a}}_i \gamma^\mu \lambda^k_{\underline{a}} - {\bar\epsilon}_{2 i} \gamma_\mu \epsilon_1^k {\bar\lambda}^{\underline{a}}_k \gamma^\mu \lambda^j_{\underline{a}} + \frac{1}{2} \delta_i^j {\bar\epsilon}_{2 k} \gamma_\mu \epsilon_1^l {\bar\lambda}^{\underline{a}}_l \gamma^\mu \lambda^k_{\underline{a}} - (1 \leftrightarrow 2) \bigg{)} \nonumber \\
    & + \epsilon_{iklm} {\bar\epsilon}^k_1 \epsilon^l_2 {\bar\lambda}^j_{\underline{a}} \lambda^{\underline{a} m}  + \frac{1}{4} \epsilon_{iklm} {\bar\epsilon}^{(j}_1 \gamma_{\mu \nu} \epsilon_2^{k)} {\bar\lambda}^l_{\underline{a}} \gamma^{\mu \nu} \lambda^{\underline{a} m} \\ 
    & - \epsilon^{jklm} {\bar\epsilon}_{1 k} \epsilon_{2 l} {\bar\lambda}_{\underline{a}i} \lambda^{\underline{a}}_m - \frac{1}{4} \epsilon^{jklm} {\bar\epsilon}_{1(i|} \gamma_{\mu \nu} \epsilon_{2 |k)} {\bar\lambda}^{\underline{a}}_l \gamma^{\mu \nu} \lambda_{\underline{a} m }\,, \nonumber \\
    {\Lambda_{\underline{a}}}^{\underline{b}} = & \, 2 {\bar\epsilon}^i_1 \lambda^j_{\underline{a}} \left( 
      2 {\bar\epsilon}_{2[i} \lambda^{\underline{b}}_{j]} + \epsilon_{ijkl} {\bar\epsilon}_2^k \lambda^{\underline{b} l } \right) - (1 \leftrightarrow 2) + \text{c.c.}\,, \\
        \zeta^{M \alpha} = & -2 ({\cal V}^\alpha)^* L^{Mij} \bar{\epsilon}_{1i} \epsilon_{2j} + \text{c.c.}\,, \\
        \Xi_{MN \mu} = & \, 4 i {L_{[M}}^{ik} L_{N] jk} \left( {\bar\epsilon}_{1 i} \gamma_\mu \epsilon_2^j - {\bar\epsilon}_{2 i} \gamma_\mu \epsilon_1^j\right)\,, \\
    \Xi_{\alpha \beta \mu} = & M_{\alpha \beta} \left( 
{\bar\epsilon}_{1 i} \gamma_\mu \epsilon_2^i - {\bar\epsilon}_{2 i} \gamma_\mu \epsilon_1^i \right) \, .
\end{align}  
   
In particular, for the vector gauge fields $A^{M \alpha}_\mu$ we have 
\begin{align}
  [\delta_Q(\epsilon_1), \delta_Q (\epsilon_2)] A^{M \alpha}_\mu = & - \xi^\nu {\cal G}^{M \alpha}_{\mu \nu} - \delta_Q (\xi^\nu \psi^i_\nu) A^{M \alpha}_\mu + \delta_Q (\epsilon_3) A^{M \alpha}_\mu + \delta_{\text{gauge}} (\zeta^{N \beta})  A^{M \alpha}_\mu \nonumber \\[2MM]
  & + \delta_{\text{tensor}}(\Xi^{NP}_\nu, \Xi^{\beta \gamma}_\nu) A^{M \alpha}_\mu 
\end{align}
and, since 
\begin{align}
      - \xi^\nu H^{M \alpha}_{\mu \nu} = & \, \delta_{\text{gct}} (\xi^\nu) A^{M \alpha}_\mu - \delta_{\text{gauge}} ( \xi^\nu A^{N \beta}_\nu) A^{M \alpha}_\mu \nonumber \\[2mm]
      & -  \delta_{\text{tensor}} \left( \xi^\rho B^{N P}_{\rho \nu} + \epsilon_{\beta \gamma} \xi^\rho A^{[N| \beta}_\rho A^{|P] \gamma}_{\nu},  \xi^\rho B^{\beta \gamma}_{\rho \nu} - \eta_{NP} \xi^\rho A^{N(\beta|}_\rho A^{P|\gamma)}_\nu \right) A^{M \alpha}_\mu
  \end{align}
and ${\cal G}^{\Lambda}_{\mu \nu} \equiv H^{\Lambda}_{\mu \nu}$, the commutator of two supersymmetry transformations closes on the electric vectors $A^{\Lambda}_{\mu}$. It also closes on the linear combinations ${{\Pi}^{\Lambda}}_{M \alpha} {\Theta^{\alpha M}}_{NP} A_{\Lambda \mu}$ and ${{\Pi}^{\Lambda}}_{M (\alpha} \xi^M_{\beta)} A_{\Lambda \mu}$ of the magnetic vector fields, if the equations of motion \eqref{BMNeom} and \eqref{Babeom} respectively hold.

  Furthermore, for the two-form gauge fields $B^{M \alpha}_{\mu \nu}$ we find 
  \begin{align}
      [\delta_Q(\epsilon_1), \delta_Q (\epsilon_2)] B^{M \alpha}_{\mu \nu} =& \,  \delta_Q (\epsilon_3) B^{M \alpha}_{\mu \nu} + \delta_{\text{gauge}} (\zeta^{N \beta}) B^{M \alpha}_{\mu \nu} +  \delta_{\text{tensor}}(\Xi^{NP}_\rho, \Xi^{\beta \gamma}_\rho) B^{M \alpha}_{\mu \nu} \nonumber \\
      & + \epsilon_{\mu \nu \rho \sigma} \xi^\rho \bigg{[} {\Theta^{\alpha M}}_{NP} \Big{(} {L^N}_{\underline{a}} {L^P}_{ij} {\hat P}^{\underline{a} i j \sigma} + 2 L^{N \underline{a}} L^{P \underline{b}} {\bar\lambda}_{\underline{a}i} \gamma^\sigma \lambda^i_{\underline{b}}  \nonumber \\
      & + {L^N}_{ik} L^{P jk} {\bar\chi}_j \gamma^\sigma \chi^i + 2 {L^N}_{ik} L^{P jk} {\bar\lambda}^{\underline{a}}_j \gamma^\sigma \lambda^i_{\underline{a}} \Big{)} \nonumber \\
      & + i \xi^M_\beta \bigg{(} \frac{1}{2} {\cal V}^\alpha {\cal V}^\beta ({\hat P}^\sigma)^*  - \frac{1}{2} ({\cal V}^\alpha)^* ({\cal V}^\beta)^* {\hat P}^\sigma  \\
      & + \frac{3}{4} M^{\alpha \beta} {\bar\chi}_i \gamma^\sigma \chi^i + \frac{1}{2} M^{\alpha \beta} {\bar\lambda}^{\underline{a}}_i \gamma^\sigma \lambda^i_{\underline{a}} \bigg{)} \bigg{]} \nonumber \\ & + {\Theta^{\alpha M}}_{NP} \epsilon_{\beta \gamma} \xi^\rho A^{N \beta}_{[\mu} {\cal G}^{P \gamma}_{\nu] \rho} + \delta^{\alpha}_{(\beta} \xi^M_{\gamma)} \eta_{NP} \xi^\rho A^{N \beta}_{[\mu} {\cal G}^{P \gamma}_{\nu] \rho}\,, \nonumber
\end{align}
up to terms that contain the gravitini. 
If the equations of motion \eqref{mageom} hold, the action of the commutator $[\delta_Q(\epsilon_1), \delta_Q (\epsilon_2)]$ on the antisymmetric tensor fields ${{\Pi}^{\Lambda}}_{M \alpha} B^{M \alpha}_{\mu \nu}$ is given by \eqref{QQcomg} with an additional term that corresponds to a transformation of the form \eqref{Deltasym} with
\begin{equation}
    \label{DLS}
    \Delta^{\Lambda 
\Sigma}_{\mu \nu} = - \frac{1}{2} {{\Pi}^{\Lambda}}_{M \alpha} {\Pi^{\Sigma N}}_\beta {\Theta^{\alpha M}}_{NP} \xi_\mu A^{P \beta}_\nu + \frac{1}{2} {{\Pi}^{\Lambda}}_{M \alpha} \Pi^{\Sigma \hspace{0.15cm}  (\alpha|}_{\hspace{0.15cm}N} \xi^M_\beta \xi_\mu A^{N |\beta)}_\nu . 
\end{equation}  

In addition, the commutator $[\delta_Q(\epsilon_1), \delta_Q (\epsilon_2)]$ closes on the fermionic fields, provided the equations of motion for the fermions hold.
\subsection{Comments} 
\label{sub:comments}

Equations \eqref{BMNeom} and \eqref{Babeom} relate the field strengths of the magnetic vector fields  $H_{\Lambda \mu \nu}$ to the dual field strengths ${\cal G}_{\Lambda \mu \nu}$, at least as far as those components projected by the embedding tensor are concerned, allowing to express the former in terms of $H^{\Lambda}_{\mu \nu}$ and the matter fields via \eqref{GLg}.
On the other hand, equation \eqref{mageom} is a duality equation between the two-form gauge fields and the scalars that relates the field strengths of the former to the gauge and the matter fields. 
Therefore, equations \eqref{BMNeom}, \eqref{Babeom} and \eqref{mageom} determine the field strengths of the magnetic vectors and the two-form gauge fields in terms of the other fields.
As pointed out in \cite{deWit:2005ub}, altogether these equations are not dynamical, but, together with the vector and tensor gauge invariances, they ensure that the number of propagating degrees of freedom has not changed upon the introduction of magnetic vector and two-form gauge fields in the gauged theory. 
In fact, this gauge fixing can be implemented in various ways, thus determining different descriptions of the propagating degrees of freedom in terms of the fields of the theory. 
For instance, one can always dispose of the antisymmetric tensor fields by fixing the tensor-gauge transformations and solving equations \eqref{BMNeom}, \eqref{Babeom} in the tensor fields as functions of the other fields. 
The result is a theory in the electric frame of the embedding tensor, with no tensor fields and magnetic vectors \cite{deWit:2005ub}. 
Alternatively, in certain cases, the gauge invariance associated with the magnetic vector fields $A_{\Lambda\mu}$ can be fixed in order to eliminate a number of scalar fields. 
Then equation \eqref{mageom} is solved in $A_{\Lambda\mu}$ as functions of the remaining fields including the tensor ones. 
Upon inserting these expressions for $A_{\Lambda\mu}$ in the Lagrangian, the net result is a gauged supergravity, in the original symplectic frame, in which a number of scalar fields have been dualized to tensor ones, which now encode propagating degrees of freedom.

As is often the case in string/M-theory  compactifications, the low-energy degrees of freedom in the resulting four-dimensional consistent truncation are represented by dynamical tensor fields rather than the corresponding dual scalars. 
Half-maximal gauged models of this kind are obtained, within the general setting described here, by partly fixing the gauge freedom and solving equation (\ref{mageom}) along the lines explained above.

Let us end this section by expanding on the notion of the electric frame of the embedding tensor. 
The general formulation of the gauging procedure discussed here, along the lines of \cite{deWit:2005ub}, features a characteristic redundancy in the description of the propagating degrees of freedom, due to the presence of antisymmetric tensor fields and magnetic vector potentials. 
Yhese extra ingredients are needed since the gauging is performed starting from an ungauged model which is formulated in a generic symplectic frame that does not necessarily coincide with the electric frame of the embedding tensor. 
The latter is defined as  the frame in which the gauging only involves electric vector fields and thus the embedding tensor has only electric components. 
As a characteristic feature of the embedding tensor, this frame can be defined in a $G$-invariant fashion as follows. 
The embedding tensor is described by the rectangular matrix $\Theta_\mathcal{M}{}^A $, where $A =1,\dots, {\rm dim}(G)$ is the index of the adjoint representation of $G$: $A =((\alpha\beta),[MN])$. 
If $r$ is its rank, this matrix can be rewritten using the rank-factorization, in the following form \cite{Trigiante:2016mnt}:
\begin{align}
 \Theta_\mathcal{M}{}^A =\sum_{I=1}^r \vartheta_\mathcal{M}{}^I\,\mathcal{W}_I{}^A \,,\label{rankf}
\end{align}
where $\vartheta^I\equiv (\vartheta_\mathcal{M}{}^I)$ are $r$ independent vectors in the $2(6+n)$-dimensional symplectic vector space $\mathbb{V}_v$ of the electric and magnetic vector fields, while $\mathcal{W}_I{}\equiv  (\mathcal{W}_I{}^A )$ are $r$ independent vectors in the vector space of the Lie algebra of $G$. 
The locality constraint \eqref{locality} then implies:
\begin{equation}
    \mathbb{C}^{\mathcal{M}\mathcal{N}}\,\vartheta_\mathcal{M}{}^I\vartheta_\mathcal{N}{}^J=0\,\,,\,\,\,\,\forall I,J=1,\dots, r\,,
\end{equation}
that is $\vartheta^I$ generate an isotropic subspace of the symplectic vector space $\mathbb{V}_v$ and thus $r\le 6+n$. 
We can complete ${\rm Span}(\vartheta^I)$ to a Lagrangian (i.e. maximal isotropic) subspace of  $\mathbb{V}_v$ by adding $6+n-r$ vectors $\vartheta^i$, $i=1,\dots,6+n-r$, to define a system of $6+n$ vectors $\vartheta^{\hat{\Lambda}}\equiv\{\vartheta^I,\,\vartheta^i\}$ satisfying the property:
\begin{equation}
    \mathbb{C}^{\mathcal{M}\mathcal{N}}\,\vartheta_\mathcal{M}{}^{\hat{\Lambda}}\vartheta_\mathcal{N}{}^{\hat{\Sigma}}=0\,\,,\,\,\,\,\forall \hat{\Lambda},\,\hat{\Sigma}=1,\dots, 6+n\,.
\end{equation}
The choice of $\vartheta^i$ is not unique and we will choose them such that $\vartheta^{\Lambda\,i}=0$.
Given the Lagrangian subspace ${\rm Span}(\vartheta^{\hat{\Lambda}})$ of $\mathbb{V}_v$ we can find another Lagrangian subspace ${\rm Span}(\vartheta_{\hat{\Lambda}})$, disjoint from the former, and choose their bases such that the following condition is satisfied:
\begin{equation}
    \mathbb{C}^{\mathcal{M}\mathcal{N}}\,\vartheta_\mathcal{M}{}^{\hat{\Lambda}}\vartheta_{\mathcal{N}\,\hat{\Sigma}}=\delta^{\hat{\Lambda}}_{\hat{\Sigma}}\,\,,\,\,\,\,\forall \hat{\Lambda},\,\hat{\Sigma}\,.
\end{equation}
The matrix
\begin{equation}
    E_{\mathcal{M}}{}^{\hat{\mathcal{N}}}\equiv (\vartheta_\mathcal{M}{}^{\hat{\Lambda}},\,\vartheta_{\mathcal{M}\,\hat{\Lambda}})\,
\end{equation}
is then symplectic and maps the original frame to the new one labeled by the index $\hat{\mathcal{M}}$: $V^{\hat{\mathcal{M}}}=(V^{\hat{\Lambda}},\,V_{\hat{\Lambda}})$. 
The latter is the electric frame of the embedding tensor. 
To see this we first write the inverse matrix $(E^{-1})_{\hat{\mathcal{M}}}{}^{\mathcal{M}}$:
\begin{equation}
  (E^{-1})_{\hat{\Lambda}}{}^\mathcal{M}=\mathbb{C}^{\mathcal{M}\mathcal{N}}\,\vartheta_{\mathcal{N}\,\hat{\Lambda}}\,\,,\,\,\,(E^{-1})^{\hat{\Lambda}\mathcal{M}}=-\mathbb{C}^{\mathcal{M}\mathcal{N}}\,\vartheta_{\mathcal{N}}{}^{\hat{\Lambda}}\,,
\end{equation}
and then the embedding tensor in the new frame:
\begin{equation}
    \Theta_{\hat{\mathcal{M}}}{}^A =(E^{-1})_{\hat{\mathcal{M}}}{}^{\mathcal{M}}\,\Theta_{{\mathcal{M}}}{}^A \,.
\end{equation}
We find
\begin{equation}
\Theta_{I}{}^A =\mathcal{W}_I{}^A \,,\,\,\Theta_{i}{}^A =\Theta^{I\,A }=\Theta^{i\,A }=0\,.
\end{equation}
Since the electric frame is a characteristic feature of $\Theta$, its definition is  $G$-invariant, being based on the factorization \eqref{rankf} 
in which the index $I$ is $G$-invariant.

Of the tensor fields $B_{A \,\mu\nu}$, 
only the combinations $$\Theta^{\Lambda\,A }\,B_{A \,\mu\nu}=\vartheta^{\Lambda\, I}\,\mathcal{W}_I{}^{A }\,B_{A \,\mu\nu}\,,$$
namely the $r$ independent tensor fields
$$B_{I\,\mu\nu}\equiv \mathcal{W}_I{}^{A }\,B_{A \,\mu\nu}\,,$$
enter the Lagrangian. 
This formulation allows us to intrinsically distinguish those vector fields $A^I_\mu$ which enter the gauge connection (and whose field strengths are covariantly closed) from those $A_{I\mu}$ which are St\"uckelberg-coupled to the tensor fields.
This is done by writing the vector potentials in the electric frame:
$$A^{\hat{\mathcal{M}}}_\mu=E_\mathcal{M}{}^{\hat{\mathcal{M}}}\,A^\mathcal{M}_\mu=(A^I_\mu,\,A^i_\mu,\,A_{I\mu},\,A_{i\mu})\,,$$
so that the symplectic-invariant gauge connection takes the form
$$g\,A^{{\mathcal{M}}}_\mu\,X_{{\mathcal{M}}}=g\,A^{\hat{\mathcal{M}}}_\mu\,X_{\hat{\mathcal{M}}}=g\,A^I_\mu \,X_I\,,$$
where $X_I\equiv \mathcal{W}_I{}^A \,t_A $ are the independent gauge generators.  
The components of the modified field strengths $H^{\hat{\mathcal{M}}}_{\mu\nu}$, defined in \eqref{HMa}, in the electric frame are (in form-notation)
\begin{equation}
    H^I={\hat F}^I\,,\,\,H^i={\hat F}^i\,,\,\,H_I={\hat F}_I-\frac{g}{2}\,B_I\,,\,\,H_i={\hat F}_i\,.
\end{equation}
From \eqref{DH=gH3} it follows that $\hat{D}_{[\mu } {\hat F}^I_{\nu\rho]}=\hat{D}_{[\mu } {\hat F}^i_{\nu\rho]}=\hat{D}_{[\mu } {\hat F}_{i\,\nu\rho]}=0$, while ${\hat F}_{I\,\mu\nu}$ are the only components of the field strengths for which $\hat{D}_{[\mu } {\hat F}_{I\,\nu\rho]}\neq 0$. 
We also see that only the vectors $A_{I\mu}$, which are magnetic in the electric frame, are St\"uckelberg-coupled to the tensor fields and transform, under a tensor-gauge transformation \eqref{dXiAMa}, as
\begin{equation}
   \delta_{\Xi} A_{I\mu}=\frac{g}{2}\,\Xi_{I\,\mu}\,, 
\end{equation}
where $\Xi_{I\,\mu}\equiv \mathcal{W}_I{}^A\,\Xi_{A\,\mu}$.
All other components of $A^{\hat{\mathcal{M}}}$ are inert under the  transformations \eqref{dXiAMa}. 
Choosing $g\,\Xi_{I\,\mu}=-2\,A_{I\,\mu}$  we can dispose of $A_{I\,\mu}$. 
As explained above, equations \eqref{BMNeom}, \eqref{Babeom} can then be solved in the transformed tensor fields $B'_I$ as functions of the other fields. 
Replacing then the resulting expressions for $B'_I$ in the Lagrangian amounts to effectively performing the rotation to the electric frame.

The rotation to the electric frame can also be done directly at the level of the field equations and Bianchi identities, which are formally symplectic covariant, by means of the matrix $E$. 
This amounts to replacing everywhere the index $\mathcal{M}$ by $\hat{\mathcal{M}}$. 
In particular, the twisted self-duality condition implies that ${\cal G}_{\hat{\Lambda}}$ can be expressed as the variation, with respect to ${\cal G}^{\hat{\Lambda}}$, of a new Lagrangian, in which the kinetic terms of the vector fields are written in terms of ${\cal I}_{\hat{\Lambda}\hat{\Sigma}},\,{\cal R}_{\hat{\Lambda}\hat{\Sigma}}$, ${\cal G}^{\hat{\Lambda}}$ and  ${}^*{\cal G}^{\hat{\Lambda}}$.
The fact that that ${\cal G}^{\hat{\Lambda}}=H^{\hat{\Lambda}}$ follows directly from equations \eqref{BMNeom}, \eqref{Babeom} and from having chosen $\vartheta^{\Lambda\,i}=0$\footnote{Indeed $H^I-\mathcal{G}^I=\vartheta_{\mathcal{M}}{}^I(H^{\mathcal{M}}-\mathcal{G}^{\mathcal{M}})=0$, by virtue of \eqref{BMNeom}, \eqref{Babeom}, while $H^i-\mathcal{G}^i=\vartheta_{\mathcal{M}}{}^i(H^{\mathcal{M}}-\mathcal{G}^{\mathcal{M}})=\vartheta_{\Lambda}{}^i(H^{\Lambda}-\mathcal{G}^{\Lambda})=0$, since $\vartheta^{\Lambda\,i}=0$ and $\mathcal{G}^{\Lambda}=H^{\Lambda}$ in the original frame. }.

\section{Vacua, Masses, Gradient Flow and Supertrace Relations} 
\label{sec:vacua_masses_gradient_flow_and_supertrace_relations}

\subsection{Gradient flow relations} 
\label{sub:gradient_flow_relations}

It is known that in gauged supergravities the scalar potential is related to the fermion shifts of the supersymmetry transformations \cite{DallAgata:2021uvl}.
As noted in \cite{DAuria:2001rlt} and reviewed in \cite{DallAgata:2021uvl}, supergravity actually provides a structure of gradient flow relations between the fermion shifts and the fermion mass matrices that are needed in establishing supersymmetry invariance, though they are largely due to the properties and structure of the scalar $\sigma$-model.
Since this type of relations played a rather important role in establishing and understanding properties of various vacua, black hole and domain-wall solutions, we give here the relevant expressions:
\begin{align}
    \label{Drhoij}
    {\hat D} A_1^{ij}  = & \,A_2^{(ij)} {\hat P}^* +3 
    {\bar A}_2{}^{\underline{a}(i}{}_k {\hat P}_{\underline{a}}{}^{j)k}, \\[2mm]
    \label{DVij}
    {\hat D} A_2^{ij}  = & - 3 A_2{}^{\underline{a}}{}_k{}^{(i} {\hat P}_{\underline{a}}{}^{j)k} - \frac{3}{2} A_2{}^{\underline{a}}{}_k{}^{k}  {\hat P}_{\underline{a}}{}^{ij} + \frac{1}{2} \epsilon^{ijkl} {\bar A}_{2kl} {\hat P} + A_1^{ij}{\hat P}, \\[2mm]
    \label{DLaij}
    {\hat D} A_2{}^{\underline{a}}{}_j{}^i  = & - {\bar A}_2{}^{\underline{a}i}{}_j \hat{P} + \frac{1}{2} \delta^i_j {\bar A}_2{}^{\underline{a}k}{}_k \hat{P}  + 2 A^{\underline{a} \underline{b} ik} {\hat P}_{\underline{b} jk} - \frac{1}{2}  \delta^i_j A^{\underline{a} \underline{b} kl} {\hat P}_{\underline{b} kl} \nonumber \\[2mm]
    & - \frac{1}{6} \delta^i_j A_2^{kl} {\hat P}^{\underline{a}}{}_{kl} - \frac{2}{3} {\bar A}_{1jk} {\hat P}^{\underline{a} ik} + \frac{2}{3} A_2^{(ik)} {\hat P}^{\underline{a}}{}_{jk}\,, \\[2mm]
	\label{DMabij}
	{\hat D} A_{\underline{a} \underline{b}}{}^{ij} =& \frac{1}{2} \epsilon^{ijkl} {\bar A}_{\underline{a} \underline{b} kl} {\hat P} - 4 A_{2[\underline{a}|k}{}^{[i} {\hat P}_{|\underline{b}]}{}^{j]k}  - A_{2[\underline{a}|k}{}^k {\hat P}_{|\underline{b}]}{}^{ij} + A_{\underline{a} \underline{b} \underline{c}} {\hat P}^{\underline{c} i j} ,
\end{align}
where
\begin{equation}
	    \label{Sabc}
	    A_{\underline{a} \underline{b} \underline{c}} \equiv f_{\alpha MNP} {\cal V}^{\alpha} L^M{}_{\underline{a}} L^N{}_{\underline{b}} L^P{}_{\underline{c}} .
\end{equation}
The derivation follows straightforwardly from (\ref{hatDV=hatPV}), (\ref{hatDLij}), (\ref{hatDLa}) and the definition of the various $A$ tensors.


\subsection{Vacua} 
\label{sub:vacua}

The same relations can be used as a guide to compute derivatives of the scalar potential
\begin{equation}
\label{scalpot}
    V = -  e^{-1} {\cal L}_{\text{pot}} =   g^2 \left(-\frac{1}{3} A_1^{ij} {\bar A}_{1ij} + \frac{1}{9} A_2^{ij} {\bar A}_{2 ij} + \frac{1}{2} A_{2 \underline{a} i}{}^j {\bar A}_2{}^{\underline{a}i}{}_j \right).
\end{equation}
In particular, the critical points of (\ref{scalpot}) will provide us with the vacua of the gauged ${\cal N}=4$ supergravity models.
In order to derive the conditions satisfied by these vacua, we follow \cite{deWit:1983gs} and compute the variation of the scalar potential that is induced by the action of an infinitesimal rigid SL(2,$\mathbb{R}$) $\times $ SO(6,$n$) transformation that is orthogonal to the isotropy group SO(2) $\times$ SU(4) $\times$ SO($n$) on the coset representatives ${\cal V}_{\alpha}$ and $L_M{}^{\underline{M}}$. 
Such a transformation can be written as
\begin{equation}
\label{scfl}
    \delta {\cal V}_{\alpha} = \Sigma {\cal V}_{\alpha}^*, \qquad   \delta L_M{}^{ij} = \Sigma_{\underline{a}}{}^{ij} L_M{}^{\underline{a}} , \qquad \delta L_M{}^{\underline{a}} = \Sigma^{\underline{a}}{}_{ij} L_M{}^{ij} , 
\end{equation}
where $\Sigma$ denotes the complex SL(2,${\mathbb R}$)/SO(2) scalar fluctuation and $\Sigma_{\underline{a}ij} = (\Sigma_{\underline{a}}{}^{ij})^* = \frac{1}{2} \epsilon_{ijkl} \Sigma_{\underline{a}}{}^{kl}$ are the SO(6,$n$)/[SO(6) $\times$ SO($n$)] scalar fluctuations. The variations of the $A$ tensors \eqref{gravshift}-\eqref{Aabij} under \eqref{scfl} are given by the gradient flow relations \eqref{Drhoij}-\eqref{DMabij} with the replacements ${\hat D}\rightarrow \delta$, ${\hat P} \rightarrow \Sigma$ and ${\hat P}_{\underline{a}ij} \rightarrow \Sigma_{\underline{a}ij}$.
Then, it follows that the variation of the scalar potential is given by
\begin{equation}
\label{deltaV}
    \delta V = g^2\, \left(X \Sigma + X^* \Sigma^* + X^{\underline{a}ij} \Sigma_{\underline{a}ij} \right) , 
\end{equation}
where 
\begin{align}
    \label{X} 
    X = & \,-\frac{2}{9} A_1^{ij} {\bar A}_{2ij} + \frac{1}{18} \epsilon^{ijkl} {\bar A}_{2ij} {\bar A}_{2kl} - \frac{1}{2} {\bar A}_{2 \underline{a}}{}^i{}_j {\bar A}_2{}^{\underline{a}j}{}_i  + \frac{1}{4} {\bar A}_{2 \underline{a}}{}^i{}_i {\bar A}_2{}^{\underline{a}j}{}_j,  \\
    \label{Xaij}
    X^{\underline{a}ij} = & \, - \frac{2}{3} A_1^{[i|k} A_2{}^{\underline{a}}{}_k{}^{|j]} - \frac{1}{3} A_2^{[i|k} {\bar A}_2{}^{\underline{a} |j]}{}_k - \frac{1}{3} A_2^{k[i|} {\bar A}_2{}^{\underline{a} |j]}{}_k 
    - \frac{1}{4} A_2^{[ij]} {\bar A}_{2}{}^{\underline{a}k}{}_k \nonumber \\[2mm]
    & - A^{\underline{a} \underline{b} [i| k} {\bar A}_{2 \underline{b}}{}^{|j]}{}_k + \frac{1}{4} A^{\underline{a} \underline{b} ij} {\bar A}_{2 \underline{b}}{}^{k}{}_k + \epsilon^{ijlm} \bigg{(} - \frac{1}{3} {\bar A}_{1kl} {\bar A}_2{}^{\underline{a} k}{}_m - \frac{1}{3} {\bar A}_{2 (kl)} A_2{}^{\underline{a}}{}_m{}^k \\[2mm]
    & - \frac{1}{8} {\bar A}_{2lm} A_2{}^{\underline{a}}{}_k{}^k + \frac{1}{2}  {\bar A}^{\underline{a} \underline{b}}{}_{kl} A_{2 \underline{b} m}{}^k + \frac{1}{8} {\bar A}^{\underline{a} \underline{b}}{}_{lm}  A_{2 \underline{b} k}{}^k \bigg{)}. \nonumber
\end{align}
Note that, by construction, $X^{\underline{a}}{}_{ij} \equiv (X^{\underline{a} i j})^* = \frac{1}{2} \epsilon_{ijkl} X^{\underline{a} kl}$. 
The stationary points of the scalar potential correspond to solutions of the following system of $6n+2$ real equations
\begin{equation}
    \label{crit}
    X=0\, , \qquad X^{\underline{a}ij}=0 \, .  
\end{equation}


\subsection{Masses} 
\label{sub:masses}

When analyzing supergravity vacua, one important element is the resulting spectrum of the fluctuations.
We therefore focus now on the computation of the mass matrices of all the fields in our theory, assuming a Minkowski vacuum.
While most of the formulae for the mass matrices do not depend on the value of the cosmological constant, the supersymmetry breaking pattern depends heavily on the vacuum energy, because of the super-Higgs mechanism by which some or all gravitini acquire a mass, which eventually affects the correct definition of the spin-1/2 mass matrix.

\subsubsection{Scalar masses}

We can compute the mass spectrum of the scalar fields by taking the second variation of the scalar potential under \eqref{scfl}. 
Using \eqref{deltaV}-\eqref{Xaij} and the gradient flow equations \eqref{Drhoij}-\eqref{DMabij}.
The result however does not describe proper masses unless the scalar fluctuations are canonically normalized.
For this reason we introduce the real scalar fluctuations 
\begin{equation}
	\Sigma_1 = \sqrt{2} \,\text{Re} \Sigma, \quad  \Sigma_2 = \sqrt{2} \,\text{Im} \Sigma, \quad  \Sigma_{\underline{a} \underline{m}} = - \Gamma_{\underline{m} i j} \Sigma_{\underline{a}}{}^{ij},
\end{equation}
and substitute the expansions of the coset representatives around their vacuum expectation values  $\langle {\cal V}_\alpha \rangle$, $\langle L_M{}^{ij} \rangle$ and $\langle L_M{}^{\underline{a}} \rangle$, namely
\begin{align}
    {\cal V}_{\alpha} & = \langle {\cal V}_\alpha \rangle + \langle {\cal V}_{\alpha}^* \rangle \Sigma + \mathcal{O}(\Sigma^2) \, , \\[2mm]
    L_M{}^{ij} & = \langle L_M{}^{ij} \rangle + \langle L_M{}^{\underline{a}} \rangle \Sigma_{\underline{a}}{}^{ij} + \mathcal{O}(\Sigma_{\underline{a}ij}^2) \, , \\[2mm]
    L_M{}^{\underline{a}} & = \langle L_M{}^{\underline{a}} \rangle + \langle L_M{}^{ij} \rangle \Sigma^{\underline{a}}{}_{ij} + \mathcal{O}(\Sigma_{\underline{a}ij}^2) \, , 
\end{align}
into the kinetic terms for the scalars, 
\begin{align}
\label{kinsc}
    e^{-1} {\cal L}_{\text{scalar kin}} =&  - {\hat P}_\mu^* {\hat P}^\mu - \frac{1}{2} {\hat P}_{\underline{a} ij \mu} {\hat P}^{\underline{a} ij \mu} \nonumber \\ 
    \supset & -  \frac{1}{4} ({\cal V}^\alpha)^* {\cal V}^\beta \partial_\mu {\cal V}_\alpha^* \partial^\mu {\cal V}_\beta - \frac{1}{2} L^{M \underline{a}} L_{N \underline{a}} \partial_\mu L_{M ij} \partial^\mu L^{N ij} 
\end{align}
so that the kinetic and mass terms for the scalar fluctuations take the following form:
\begin{align}
\label{quadsc}
    e^{-1} {\cal L}  \supset & - \frac{1}{2} (\partial_\mu \Sigma_1) (\partial^\mu \Sigma_1) - \frac{1}{2} (\partial_\mu \Sigma_2) (\partial^\mu \Sigma_2) - \frac{1}{2} \delta^{\underline{a} \underline{b}} \delta^{\underline{m} \underline{n}} (\partial_\mu \Sigma_{\underline{a} \underline{m}}) (\partial^\mu \Sigma_{\underline{b} \underline{n}}) \nonumber \\[2mm]
    & - \frac{1}{2} ({\cal M}_0^2)^{1,1} \Sigma_1^2 - \frac{1}{2} ({\cal M}_0^2)^{2,2} \Sigma_2^2 - ({\cal M}_0^2)^{1, \underline{a} \underline{m}} \Sigma_1 \Sigma_{\underline{a} \underline{m}} - ({\cal M}_0^2)^{2, \underline{a} \underline{m}} \Sigma_2 \Sigma_{\underline{a} \underline{m}} \\[2mm]
    & - \frac{1}{2} ({\cal M}_0^2)^{ \underline{a} \underline{m}, \underline{b} \underline{n}} \Sigma_{\underline{a} \underline{m}} \Sigma_{\underline{b} \underline{n}} \, , \nonumber
\end{align}
which is the appropriate one for canonically normalized fluctuations.
The explicit form of the squared mass matrix for the scalars ${\cal M}_0^2$ is then given by 
\begin{align}
({\cal M}_0^2)^{1,1} =& ({\cal M}_0^2)^{2,2}=  \,  g^2 \left( - \frac{2}{9} A_1^{ij} {\bar A}_{1ij} - \frac{2}{9} A_2^{(ij)} {\bar A}_{2ij} +  \frac{2}{9} A_2^{[ij]} {\bar A}_{2ij}  + A_{2 \underline{a} i}{}^j {\bar A}_2{}^{\underline{a}i}{}_j  \right)\,, \\
({\cal M}_0^2)^{1, \underline{a} \underline{m}} =& ({\cal M}_0^2)^{ \underline{a} \underline{m}, 1} = \, \frac{\sqrt{2}}{4} g^2 \left(  - {\bar A}_{2ij} {\bar A}_{2 }{}^{\underline{a} k}{}_k + 4 {\bar A}^{\underline{a} \underline{b}}{}_{ik} {\bar A}_{2 \underline{b}}{}^{ k}{}_j - {\bar A}^{\underline{a} \underline{b}}{}_{ij} {\bar A}_{2 \underline{b}}{}^{ k}{}_k \right) \Gamma^{\underline{m} i j } + c.c.\,, \\
({\cal M}_0^2)^{2, \underline{a} \underline{m}} =& ({\cal M}_0^2)^{ \underline{a} \underline{m}, 2} =  \, \frac{i \sqrt{2}}{4} g^2 \left(  - {\bar A}_{2ij} {\bar A}_{2 }{}^{\underline{a} k}{}_k + 4 {\bar A}^{\underline{a} \underline{b}}{}_{ik} {\bar A}_{2 \underline{b}}{}^{ k}{}_j - {\bar A}^{\underline{a} \underline{b}}{}_{ij} {\bar A}_{2 \underline{b}}{}^{ k}{}_k \right) \Gamma^{\underline{m} i j } + c.c.\,, \\
({\cal M}_0^2)^{ \underline{a} \underline{m}, \underline{b} \underline{n}} = & \, \frac{1}{2} g^2 \left(  2 {\bar A}_2{}^{\underline{a}j}{}_k A_2{}^{\underline{b}}{}_l{}^i - A^{\underline{a} \underline{c} ij} {\bar A}^{\underline{b}}{}_{\underline{c}kl} \right) \Gamma^{\underline{m}}{}_{ij} \Gamma^{\underline{n}kl} \nonumber\\
& + \frac{1}{2} g^2 \Big{(} - 2 A_2{}^{\underline{a}}{}_k{}^j {\bar A}_2{}^{\underline{b} k}{}_l + 2 {\bar A}_2{}^{\underline{a} j}{}_k A_2{}^{\underline{b}}{}_l{}^k - 2 A_2{}^{\underline{a}}{}_l{}^k {\bar A}_2{}^{\underline{b} j}{}_k + A_2{}^{\underline{a}}{}_k{}^k {\bar A}_2{}^{\underline{b} j}{}_l \nonumber \\ &  + A_2{}^{\underline{a}}{}_l{}^j {\bar A}_2{}^{\underline{b} k}{}_k - \frac{1}{3} \epsilon_{klmn} A_1^{jk} A^{\underline{a} \underline{b} mn} - \frac{1}{3} \epsilon^{jkmn} {\bar A}_{1kl} {\bar A}^{\underline{a} \underline{b}}{}_{mn}  + 2 A_2^{(jk)} {\bar A}^{\underline{a} \underline{b}}{}_{kl}   \nonumber \\
  & + 2 {\bar A}_{2 (kl)} A^{\underline{a} \underline{b} jk}   + A^{\underline{a} \underline{b} \underline{c}} {\bar A}_{2 \underline{c}}{}^j{}_l - {\bar A}^{\underline{a} \underline{b} \underline{c}} A_{2 \underline{c} l}{}^j -  4 A^{\underline{a} \underline{c} jk} {\bar A}^{\underline{b}}{}_{\underline{c}kl}  \Big{)} \Gamma^{\underline{m}}{}_{ij} \Gamma^{\underline{n} i l} \nonumber \\
  & + \frac{1}{4} g^2 A_2{}^{\underline{b}}{}_k{}^k {\bar A}_2{}^{\underline{a}l}{}_l \Gamma^{\underline{m}}{}_{ij}  \Gamma^{\underline{n} i j}  \\
  & + \frac{1}{2} g^2 \left(  \frac{1}{3} A_2^{ij} {\bar A}_{2kl}  - 2 A_{2 \underline{c} l}{}^i {\bar A}_2{}^{\underline{c}j}{}_k \right)  \delta^{\underline{a} \underline{b}} \Gamma^{\underline{m}}{}_{ij}  \Gamma^{\underline{n} k l} \nonumber \\
  & + \frac{1}{2} g^2 \Big{(} - \frac{8}{9} A_1^{jk} {\bar A}_{1kl} + 2 A_{2 \underline{c} l}{}^k {\bar A}_2{}^{\underline{c} j}{}_k - A_{2 \underline{c} k}{}^k {\bar A}_2{}^{\underline{c} j}{}_l -  A_{2 \underline{c} l}{}^j {\bar A}_2{}^{\underline{c} k}{}_k  \nonumber \\ 
  & + \frac{8}{9} A_2^{(jk)} {\bar A}_{2(kl)} \Big{)} \delta^{\underline{a} \underline{b}}  \Gamma^{\underline{m}}{}_{ij} \Gamma^{\underline{n} i l} + \frac{1}{8} g^2 A_{2 \underline{c} k}{}^k {\bar A}_2{}^{\underline{c}l}{}_l \delta^{\underline{a} \underline{b}}  \Gamma^{\underline{m}}{}_{ij} \Gamma^{\underline{n} i j}  \nonumber \\
  & + (\underline{a} \leftrightarrow \underline{b}, \underline{m} \leftrightarrow \underline{n} ). \nonumber
\end{align}

\subsubsection{Vector masses}

In order to identify the squared mass matrix for the vector gauge fields $A^{M \alpha }_\mu$, we recall from subsection \eqref{sub:Field_Equations_and_Bianchi_Identities} that 
the equations of motion for the electric and the magnetic vectors are given by 
\begin{equation}
    \label{vectoreom}
    \epsilon^{\mu \nu \rho \sigma} \partial_\nu {\mathcal G}^{M \alpha}_{\rho \sigma} = i g\, \xi^M_\beta \left( {\cal V}^{\alpha} {\cal V}^{\beta} ({\hat P}^{\mu})^* - ({\cal V}^{\alpha})^* ({\cal V}^{\beta})^* {\hat P}^{\mu} \right) + 2 g\, \Theta^{\alpha M}{}_{NP} L^N{}_{\underline{a}} L^P{}_{ij} {\hat P}^{\underline{a} ij \mu} + \dots \, , 
\end{equation}
where the ellipses represent terms of higher order in the fields that are not relevant for the present analysis. 
Using the duality relation \eqref{twist}
 and that ${\cal G}^{M \alpha}_{\mu \nu}$ is on-shell identified with $H^{M \alpha}_{\mu \nu}$, we can write \eqref{vectoreom} as 
\begin{equation}
    e^{-1} \partial_\nu (e H^{M \alpha \nu \mu}) = ({\cal M}_1^2)^{M \alpha}{}_{N \beta} A^{N \beta \mu} + \dots \, , 
\end{equation}
where 
\begin{align}
    \label{M1}
     ({\cal M}_1^2)^{M \alpha}{}_{N \beta}= & \, \frac{i}{4} g^2 M^{MP} \xi_{\gamma P} \xi^{\delta}_N \left( ({\cal V}^\alpha)^*  ({\cal V}^\gamma)^* {\cal V}_\beta {\cal V}_\delta - {\cal V}^\alpha  {\cal V}^\gamma  {\cal V}_\beta^* {\cal V}_\delta^* \right) \nonumber \\
     & + g^2 \Theta_{\gamma PQR} \Theta_{\beta NST} M^{MP} M^{\alpha \gamma} L^Q{}_{\underline{a}} L^{S \underline{a}} L^R{}_{ij} L^{T ij}
\end{align}
is the squared mass matrix of the vector fields.

The matrix \eqref{M1} is a $(12+2n)\times(12+2n)$ matrix. 
However, the locality constraint on the embedding tensor implies that $6+n$ vector fields are not physical. 
Therefore, at least half of the eigenvalues of this matrix are zero at any vacuum. 

\subsubsection{Fermion masses}

For the computation of the fermion mass matrices one has to focus on the subsector of the Lagrangian reported here
\begin{align}
\label{L1/2}
    e^{-1} {\cal L} \supset & \, \frac{1}{2} R(e) + \bigg{(} i \epsilon^{\mu \nu \rho \sigma} {\bar\psi}^i_\mu \gamma_\nu {\cal D}_\rho \psi_{i \sigma} - \frac{1}{2} {\bar\chi}^i \gamma^\mu {\cal D}_\mu \chi_i - {\bar\lambda}^i_{\underline{a}} \gamma^\mu {\cal D}_\mu \lambda^{\underline{a}}_i \nonumber \\
    & - 2 g {\bar A}_2{}^{\underline{a} j}{}_i {\bar\chi}^i \lambda_{\underline{a}j} + 2g 
     {\bar A}_2{}^{\underline{a} i}{}_i {\bar\chi}^j \lambda_{\underline{a}j} + 2 g A^{\underline{a} \underline{b} ij} {\bar\lambda}_{\underline{a} i} \lambda_{\underline{b} j} + \frac{2}{3} g A_2^{ij} {\bar\lambda}_{\underline{a}i} \lambda^{\underline{a}}_j \\
    & - g {\bar\psi}^i_\mu \gamma^\mu G_i - \frac{2}{3} g {\bar A}_{1ij} {\bar\psi}^i_\mu \gamma^{
    \mu \nu} \psi^j_\nu + c.c. \bigg{)} \, ,  \nonumber
\end{align}
where $R(e)$ is the Ricci scalar associated with the the torsion-free spin connection $\omega_{\mu a b}(e)$,
\begin{equation}
{\cal D}_\mu \psi_{i \nu} \equiv \partial_\mu \psi_{i \nu} + \frac{1}{4} \omega_{\mu a b}(e) \gamma^{ab} \psi_{i \nu}
\end{equation}
and similarly for the spin-1/2 fermions and the mixing terms between the gravitini and the spin-1/2 fields single out the combination 
\begin{equation}
    \label{goldstini}
    G_i \equiv \frac{2}{3} {\bar A}_{2ji} \chi^j + 2 A_{2 \underline{a} i}{}^j \lambda^{\underline{a}}_j,
\end{equation}
which provides the goldstini of the broken supersymmetries, and the coset representatives are understood to be replaced by their vacuum expectation values. 

In order to disentangle the spin-3/2 and the spin-1/2 fields we need to fix the vacuum and describe the super-Higgs mechanism.
From now on, we therefore assume that we are at a critical point of the scalar potential where
\begin{equation}
    \label{Minkvacua}
    V=0 \quad \iff \quad \frac{2}{3} A_1^{ij} {\bar A}_{1ij} - \frac{2}{9}  A_2^{ij} {\bar A}_{2ij}  - A_{2 \underline{a} i}{}^j {\bar A}_2{}^{\underline{a}i}{}_j = 0 \, .
\end{equation}
At such points, the goldstini transform linearly under supersymmetry as
\begin{equation}
    \delta_{\epsilon} G_i = \frac{4}{3} g {\bar A}_{1ij} { A}_1^{jk} \epsilon_k \, , 
\end{equation}
where we have used that the Ward identity \eqref{Wardmain} and  the vanishing cosmological constant implies
\begin{equation}
    \label{WardV=0}
    \frac{2}{3}  A_1^{jk} {\bar A}_{1ik} = \frac{2}{9}  A_2^{kj} {\bar A}_{2ki} +  A_{2 \underline{a} i}{}^k {\bar A}_2{}^{\underline{a}j}{}_k \, .
\end{equation}
The number of unbroken supersymmetries is equal to the number of linearly independent SU(4) vectors $\epsilon_i$ that are solutions of the equation $\delta_\epsilon G_i = 0$, which is the number of zero eigenvalues of the matrix in SU(4) space ${\bar A}_{1ij} { A}_1^{jk}$. 
For computational simplicity, we consider Minkowski vacua that completely break ${\cal N}=4$ supersymmetry, which means that the matrix ${\bar A}_{1ij} { A}_1^{jk}$ has no zero eigenvalue and thus is invertible, but the final results can be easily applied to vacua with partially broken supersymmetry with the appropriate modifications. 
In any case, from now on we assume that the symmetric matrix in SU(4) space $A_1^{ij}$ is invertible and we denote its inverse by $(A_1^{-1})_{ij}$. 

In order to eliminate the mass mixing terms between the gravitini and the spin-1/2 fermions, 
\begin{equation}
    \label{mix}
    e^{-1} {\cal L}_{\text{mix}} = - g {\bar\psi}^i_\mu \gamma^\mu G_i + c.c. \, , 
\end{equation}
we follow \cite{Ferrara:2016ntj} and we perform the following redefinition of the gravitini
\begin{equation}
    \label{gravredef}
    \psi_{i \mu} \rightarrow \psi_{i \mu} + \frac{3}{4 g} (A_1^{-1})_{ij} ({\bar A}_1^{-1})^{jk} {\cal D}_\mu G_k - \frac{1}{4} (A_1^{-1})_{ij} \gamma_\mu G^j,
\end{equation}
followed by a shift of the vielbein
\begin{equation}
\label{vielredef}
    e_\mu^a \rightarrow e_\mu^a + \frac{3}{4 g} \left( ({\bar A}_1^{-1})^{ij} (A_1^{-1})_{jk} {\bar G}^k \gamma^a \psi_{i \mu} + c.c. \right),
\end{equation}
and a further redefinition of the vielbein as 
\begin{equation}
    \label{viel2ndredef}
     e_\mu^a \rightarrow e_\mu^a + \left[ \frac{3}{32 g^2} ({\bar A}_1^{-1})^{ij} (A_1^{-1})_{jk} (A_1^{-1})_{il} {\bar G}^k \left( 3( {\bar A}_1^{-1})^{lm} \gamma^a {\cal D}_\mu G_m - g e^a_\mu G^l \right)  + c.c. \right].
\end{equation}
After all these steps \eqref{L1/2} becomes (up to terms at least quartic in the fermions)
\begin{align}
    \label{L1/2redef}
    e^{-1} {\cal L} \supset & \, \frac{1}{2} R(e) + \bigg{(} i \epsilon^{\mu \nu \rho \sigma} {\bar\psi}^i_\mu \gamma_\nu {\cal D}_\rho \psi_{i \sigma} - \frac{1}{2} {\bar\chi}^i \gamma^\mu {\cal D}_\mu \chi_i - {\bar\lambda}^i_{\underline{a}} \gamma^\mu {\cal D}_\mu \lambda^{\underline{a}}_i  \nonumber \\[2mm]
    & + \frac{3}{8}  ({\bar A}_1^{-1})^{ij} (A_1^{-1})_{ik} {\bar G}_j \gamma^\mu {\cal D}_\mu G^k - 2 g {\bar A}_2{}^{\underline{a} j}{}_i {\bar\chi}^i \lambda_{\underline{a} j} + 2g  {\bar A}_2{}^{\underline{a} i}{}_i {\bar\chi}^j \lambda_{\underline{a}j} \\[2mm]
    & + 2 g A^{\underline{a} \underline{b} ij} {\bar\lambda}_{\underline{a} i} \lambda_{\underline{b} j} + \frac{2}{3} g A_2^{ij} {\bar\lambda}_{\underline{a}i} \lambda^{\underline{a}}_j - \frac{1}{2} g ({\bar A}_1^{-1})^{ij} {\bar G}_i G_j - \frac{2}{3} g {\bar A}_{1ij} {\bar\psi}^i_\mu \gamma^{
    \mu \nu} \psi^j_\nu + c.c. \bigg{)} \, . \nonumber
\end{align}
In particular, the kinetic terms for the spin-1/2 fermions are
\begin{align}
    \label{Lkin1/2}
    e^{-1} {\cal L}_{\frac{1}{2}, \text{kin}} = & - \frac{1}{2} {\bar\chi}_i \gamma^\mu {\cal D}_\mu \chi^i - {\bar\lambda}^i_{\underline{a}} \gamma^\mu {\cal D}_\mu \lambda^{\underline{a}}_i   + \frac{3}{8}  (A_1^{-1})_{ij} ({\bar A}_1^{-1})^{ik} {\bar G}^j \gamma^\mu {\cal D}_\mu G_k + c.c. \nonumber \\[2mm]
    = & - \frac{1}{2} \begin{pmatrix}
        {\bar\chi}_i & \sqrt{2} {\bar\lambda}^{\underline{a}i} 
    \end{pmatrix} {\cal K}_{\frac{1}{2}}  \gamma^\mu 
  {\cal D}_\mu
  \begin{pmatrix}
   \chi^j \\ \sqrt{2} \lambda^{\underline{b}}_j  
    \end{pmatrix} + c.c.   \, , 
\end{align}
where 
\begin{align}
    \label{K1/2}
    {\cal K}_{\frac{1}{2}} & = \begin{pmatrix}
         ({\cal K}_{\frac{1}{2}})^i{}_j & ({\cal K}_{\frac{1}{2}})^{i,}{}_{\underline{b}}{}^j \\[2mm]  ({\cal K}_{\frac{1}{2}})_{\underline{a}i,j} & ({\cal K}_{\frac{1}{2}})_{\underline{a}i,\underline{b}}{}^j
    \end{pmatrix} \nonumber \\[2mm]
    & \equiv \begin{pmatrix}
        \delta^i_j - \frac{1}{3} ({\bar A}_1^{-1})^{kl} (A_1^{-1})_{lm} A_2^{im} {\bar A}_{2jk} & - \frac{\sqrt{2}}{2} ({\bar A}_1^{-1})^{kl} (A_1^{-1})_{lm}   A_2^{im}  A_{2 \underline{b} k}{}^j \\[2mm]
        - \frac{\sqrt{2}}{2} ({\bar A}_1^{-1})^{kl} (A_1^{-1})_{lm}   {\bar A}_{2 \underline{a}}{}^m{}_i {\bar A}_{2jk} & \delta_{\underline{a} \underline{b}} \delta^j_i - \frac{3}{2} ({\bar A}_1^{-1})^{kl} (A_1^{-1})_{lm} {\bar A}_{2 \underline{a}}{}^m{}_i A_{2 \underline{b} k}{}^j
    \end{pmatrix}
\end{align}
is the kinetic matrix of the spin-1/2 fermions, while the mass terms for these fermions are given by 
\begin{align}
     \label{Lmass1/2}
     e^{-1} {\cal L}_{\frac{1}{2}, \text{mass}} = &  - 2 g {\bar A}_2{}^{\underline{a} j}{}_i {\bar\chi}^i \lambda_{\underline{a} j} + 2g  {\bar A}_2{}^{\underline{a} i}{}_i {\bar\chi}^j \lambda_{\underline{a}j} + 2 g A^{\underline{a} \underline{b} ij} {\bar\lambda}_{\underline{a} i} \lambda_{\underline{b} j} \\[2mm]
    &  + \frac{2}{3} g A_2^{ij} {\bar\lambda}_{\underline{a}i} \lambda^{\underline{a}}_j - \frac{1}{2} g ({\bar A}_1^{-1})^{ij} {\bar G}_i G_j + c.c. \nonumber \\[2mm]
    = & \, \frac{1}{2} \begin{pmatrix}
        {\bar\chi}^i  & \sqrt{2}{\bar\lambda}_{\underline{a}i} 
    \end{pmatrix} {\cal M}_{\frac{1}{2}} \begin{pmatrix}
        \chi^j \\ \sqrt{2} \lambda_{\underline{b}j}
    \end{pmatrix} + c.c. \, , 
\end{align}
where 
\begin{align}
    \label{M1/2}
    {\cal M}_{\frac{1}{2}}   = & \, \begin{pmatrix}
        ({\cal M}_{\frac{1}{2}})_{ij} & ({\cal M}_{\frac{1}{2}})_i{}^{\underline{b}j} \\[2mm]
        ({\cal M}_{\frac{1}{2}})^{\underline{a}i}{}_j & ({\cal M}_{\frac{1}{2}})^{\underline{a}i, \underline{b}j}  \end{pmatrix} \nonumber \\[2mm] 
         \equiv & \,  g \begin{pmatrix}
            0  & - \sqrt{2} {\bar A}_2{}^{\underline{b} j}{}_i + \sqrt{2} \delta^j_i {\bar A}_2{}^{\underline{b} k}{}_k  \\[2mm] - \sqrt{2} {\bar A}_2{}^{\underline{a} i}{}_j + \sqrt{2} \delta^i_j {\bar A}_2{}^{\underline{a} k}{}_k
             & 2 A^{\underline{a} \underline{b} ij} + \frac{2}{3} \delta^{\underline{a} \underline{b}} A_2^{(ij)} 
              \end{pmatrix} \\[2mm]
              & + g \begin{pmatrix}
            - \frac{4}{9} ({\bar A}_1^{-1})^{kl} {\bar A}_{2ik} {\bar A}_{2jl} &  -\frac{2\sqrt{2}}{3} ({\bar A}_1^{-1})^{kl} {\bar A}_{2ik} A_2{}^{\underline{b}}{}_l{}^j \\[2mm] -\frac{2\sqrt{2}}{3} ({\bar A}_1^{-1})^{kl} {\bar A}_{2jk} A_2{}^{\underline{a}}{}_l{}^i
             & -2 ({\bar A}_1^{-1})^{kl} 
             A_2{}^{\underline{a}}{}_k{}^i   A_2{}^{\underline{b}}{}_l{}^j 
              \end{pmatrix} \nonumber
\end{align}
is the mass matrix for the spin-1/2 fermions. 
In the $(\chi_j, \sqrt{2} \lambda^{\underline{a}j})$ basis, the goldstini $G^i = \frac{2}{3} A_2^{ji} \chi_j  + 2 {\bar A}_{2 \underline{a}}{}^i{}_j \lambda^{\underline{a}j}$ are represented by the column vectors
\begin{equation}
    G^i = \begin{pmatrix}
        G^{ij} \\[2mm] G^i{}_{\underline{a}j}
    \end{pmatrix} \equiv \begin{pmatrix}
       \frac{2}{3} A_2^{ji} \\[2mm] \sqrt{2} {\bar A}_{2 \underline{a}}{}^i{}_j 
    \end{pmatrix}
\end{equation}
and they are null eigenvectors of the kinetic matrix (\ref{K1/2}).
This can be verified using \eqref{WardV=0}, which implies 
\begin{align}
({\cal K}_{\frac{1}{2}})^i{}_j G^{kj} + ({\cal K}_{\frac{1}{2}})^{i, \underline{a}j} G^k{}_{\underline{a}j} &= 0,   \\[2mm]
({\cal K}_{\frac{1}{2}})_{\underline{a}i,j} G^{kj} + ({\cal K}_{\frac{1}{2}})_{\underline{a}i}{}^{\underline{b}j} G^k{}_{\underline{b}j} &= 0 \, .
\end{align}
Therefore, the goldstini have disappeared from the kinetic Lagrangian. 
Furthermore, using \eqref{WardV=0}, the quadratic constraints on the embedding tensor expressed in terms of the $A$ tensors \eqref{T1}, \eqref{T3} and \eqref{T10} and the critical point condition $X=0$, we obtain
 \begin{equation}
     \label{M1/2G=01}
     ({\cal M}_{\frac{1}{2}})_{ij} G^{kj} + ({\cal M}_{\frac{1}{2}})_i{}^{\underline{a}j} G^k{}_{\underline{a}j} = 0 \, . 
 \end{equation}
On the other hand, by making use of \eqref{WardV=0}, the  constraints \eqref{T7}, \eqref{T14}, \eqref{T23}, \eqref{T2610} and \eqref{T266}, as well as the vacuum conditions $X^{\underline{a} ij}=0$, one finds
\begin{equation}
\label{M1/2G=02}
    ({\cal M}_{\frac{1}{2}})^{\underline{a}i}{}_j G^{kj} + ({\cal M}_{\frac{1}{2}})^{\underline{a}i, \underline{b}j} G^k{}_{\underline{b}j} = 0 \, . 
\end{equation}
These equations show that the goldstini are also null eigenvectors of the mass matrix ${\cal M}_{\frac{1}{2}}$. 
Thus, the goldstini have been removed from the fermionic mass terms as well. 
This is the super-Higgs mechanism, in which the goldstini are ``eaten'' by the gravitini, 
 which become massive. 

The same redefinitions \eqref{gravredef}-\eqref{viel2ndredef} also diagonalize the equations of motion for the gravitini, which now become
\begin{equation}
    \label{gravredeom}
    \gamma^{\mu \nu \rho} {\cal D}_\nu \psi_{i \rho} = - \frac{2}{3} g {\bar A}_{1ij} \gamma^{\mu \nu} \psi^j_\nu + \dots \, ,
\end{equation}
so the mass matrix of the gravitini is given by
\begin{equation}
     ({\cal M}_{\frac{3}{2}})_{ij} = - \frac{2}{3} g {\bar A}_{1ij}.
\end{equation}


\subsection{Supertrace relations} 
\label{sub:supertrace_relations}

Having computed the mass matrices for all the fields of the theory at any supersymmetry breaking Minkowski vacuum, it is natural to ask ourselves what is the expression of the supertrace of the squared mass matrices
\begin{align}
    \label{Strdef}
    \text{STr}({\cal M}^2) \equiv & \displaystyle\sum_{\text{spins} \, J}  (-1)^{2J} (2 J + 1 ) \text{Tr} ({\cal M}_J^2) \nonumber \\[2mm] 
     = & \, \text{Tr} \left({\cal M}_0^2\right) - 2 \text{Tr}\left({\cal M}_{\frac{1}{2}}^{\dagger} {\cal M}_{\frac{1}{2}} \right) + 3 \text{Tr} \left({\cal M}_1^2\right) - 4  \text{Tr}\left({\cal M}_{\frac{3}{2}}^{\dagger}{\cal M}_{\frac{3}{2}}\right) \, . 
\end{align}
This supertrace (and the analogous ones $\text{STr}({\cal M}^{2k})$ for $k>1$) can be used as a phenomenological guide on the possible mass splittings of the vacuum, but it also gives us some interesting information on the ultraviolet behaviour of the theory.
For instance, it is known \cite{Coleman:1973jx,Weinberg:1973ua} that $\text{STr}({\cal M}^2)$ controls the quadratic divergences of the one-loop potential and in ${\cal N}=1$ supergravity it is in general non vanishing, while the quartic supertrace $\text{STr}({\cal M}^4)$ controls the logarithmic divergences of the one-loop effective potential. 
Very little is known on the properties of the quadratic and higher supertraces in gauged extended supergravities.
In the case of maximal (${\cal N}=8$) supergravity in four spacetime dimensions, it has been recently shown \cite{DallAgata:2012tne}, by using the vacuum conditions and the quadratic constraints on the embedding tensor, that $\text{STr}({\cal M}^2) = \text{STr}({\cal M}^4) = 0$ for all Minkowski vacua that completely break ${\cal N}=8$ supersymmetry in general and even $\text{STr}({\cal M}^6) = 0$ at such vacua for special classes of gaugings.
Here we make the first step in half-maximal supergravity, proving that $\text{STr}({\cal M}^2) = 0$ at any Minkowski vacuum with completely broken supersymmetry.

The first step is to compute the traces of the squared mass matrices for all the fields and then simplify them by using the constraints on the $A$ tensors following from the quadratic constraints on the embedding tensor (see appendix \ref{sec:t_tensor_identities}), the critical point conditions as well as the vanishing of the vacuum energy.

For the gravitini we have a very simple expression:
\begin{equation}
\label{TrM3/2}
    \text{Tr}\left({\cal M}_{\frac{3}{2}}^{\dagger} {\cal M}_{\frac{3}{2}}\right) = \left(\bar{{\cal M}}_{\frac{3}{2}}\right)^{ij} \left({\cal M}_{\frac{3}{2}}\right)_{ij} = \frac{4}{9} g^2 A_1^{ij} {\bar A}_{1ij} \, . 
\end{equation}

For the vector fields we find 
\begin{equation}
    \label{TrM1}
    \text{Tr}({\cal M}_1^2) = ({\cal M}_1^2)^{M \alpha}{}_{M \alpha} = \left( \frac{4}{3} + \frac{1}{9} n \right) g^2 A_2^{[ij]} {\bar A }_{2ij} + 2 g^2 A_{2 \underline{a} i}{}^j {\bar A}_2{}^{\underline{a}i}{}_j + g^2 A^{\underline{a} \underline{b} ij} {\bar A}_{\underline{a} \underline{b} ij} \, , 
\end{equation}
where we have used the definition of $M^{MN}$ and the quadratic constraint \eqref{T2}.

For the spin-1/2 fields we have
 \begin{align}
     \label{TrM1/2}
     \text{Tr}\left({\cal M}_{\frac{1}{2}}^{\dagger} {\cal M}_{\frac{1}{2}} \right) = & \, \left({\bar{\cal M}}_{\frac{1}{2}}\right)^{ij} \left({\cal M}_{\frac{1}{2}}\right)_{ij} + 2   \left({\bar{\cal M}}_{\frac{1}{2}}\right)_{\underline{a}i}{}^j  \left({\cal M}_{\frac{1}{2}}\right)_{j}{}^{\underline{a}i} +  \left({\bar{\cal M}}_{\frac{1}{2}}\right)_{\underline{a}i, \underline{b} j} \left({\cal M}_{\frac{1}{2}}\right)^{\underline{a}i, \underline{b} j} \nonumber \\[2mm]
     = &  -\frac{16}{9} g^2 A_1^{ij} {\bar A}_{1ij} + 4 g^2 A_{2 \underline{a} i}{}^j {\bar A}_2{}^{\underline{a} i}{}_j   + \frac{4}{9} n g^2 A_2^{(ij)} {\bar A}_{2ij} \\[2mm] 
	 &+ 4 g^2 A^{\underline{a} \underline{b} ij} {\bar A}_{\underline{a} \underline{b} ij} + \frac{32}{9} g^2 A_2^{[ij]} {\bar A}_{2ij}, \nonumber
\end{align}
which can be shown by using the conditions \eqref{crit}, \eqref{Minkvacua} and \eqref{WardV=0} satisfied by Minkowski vacua and the quadratic constraints \eqref{T1}, \eqref{T2}, \eqref{T3}, \eqref{T7}, \eqref{T10}, \eqref{T14}, \eqref{T23}, \eqref{T2610} and \eqref{T266} on the $A$ tensors.

Finally, for the scalar fields we find
\begin{align}
\label{TrM0}
    \text{Tr}({\cal M}_0^2) = & \, ({\cal M}_0^2)^{1,1} + ({\cal M}_0^2)^{2,2} + \delta_{\underline{a} \underline{b}} \delta_{\underline{m} \underline{n}} ({\cal M}_0^2)^{ \underline{a} \underline{m}, \underline{b} \underline{n}} \nonumber \\
    = & - \frac{4}{9} ( 3n + 1 ) g^2  A_1^{ij}  {\bar A}_{1ij} + \frac{4}{9}(3n-1)g^2 A_2^{(ij)} {\bar A}_{2ij} + \frac{1}{9} \left( n + 24 \right) g^2 A_2^{[ij]} {\bar A}_{2ij}  \\
    & + 2 n g^2 A_{2 \underline{a} i}{}^j {\bar A}_2{}^{\underline{a} i}{}_j + 5 g^2 A^{\underline{a} \underline{b} ij} {\bar A}_{\underline{a} \underline{b} ij} \nonumber \, , 
\end{align}
where we have used the quadratic constraint \eqref{T2}.

Altogether, we have that the supertrace of the squared mass eigenvalues is
\begin{equation}
     \label{Str = 0}
     \text{STr}({\cal M}^2) = 4 (n-1) V = 0 
\end{equation}
for any number of vector multiplets and for any gauging. 




\section{Conclusions and Discussion}
We have constructed the complete Lagrangian that incorporates all gauged ${\cal N}=4$ matter-coupled supergravities in four spacetime dimensions. 
The choice of the symplectic frame has been conveniently parametrized by means of projectors ${\Pi^{\Lambda}}_{\mathcal{M}}$ and $\Pi_{\Lambda \mathcal{M}}$ that extract the electric and magnetic components of a symplectic vector $V^{\mathcal{M}}=(V^{\Lambda}, V_{\Lambda})$. 
These projectors must satisfy certain properties following from the decomposition of the symplectic form $\mathbb{C}^{{\cal M} {\cal N}}$ in any symplectic frame. 
We have also proven that the supertrace of the squared mass eigenvalues vanishes for Minkowski vacua that completely break ${\cal N}=4$ supersymmetry irrespective of the number of vector multiplets and the choice of the gauge group. 
This implies that the one-loop effective potential at such vacua has no quadratic divergence. 

An interesting but quite involved computation would be that of the quartic supertrace of the mass matrices for the same class of vacua of ${\cal N}=4$ supergravity. 
As mentioned in the previous section, it has been shown in \cite{DallAgata:2012tne} that $\text{STr}({\cal M}^4)=0$ for all Minkowski vacua of any gauged four-dimensional ${\cal N}=8$ supergravity that completely break supersymmetry. 
Therefore, this should also hold for the gauged $D=4$, ${\cal N}=4$ supergravities with six vector multiplets that can be obtained by a truncation of a gauged $D=4$, ${\cal N}=8$ supergravity, and, combined with $\text{STr}({\cal M}^0)=\text{STr}({\cal M}^2)=0$, it implies that the one-loop effective potential is finite at all classical Minkowski vacua with completely broken supersymmetry of this particular class of ${\cal N}=4$ supergravities.  
It has been proven in \cite{Dibitetto:2011eu} that the irreducible components $f_{\alpha MNP}$ of the embedding tensor that parametrizes this class of ${\cal N}=4$ gaugings satisfy two additional quadratic constraints:
\begin{equation}
    \label{trunc}
    f_{\alpha MNP} f_{\beta}{}^{MNP} = 0  ~~~~\text{and}~~~~ \epsilon^{\alpha \beta} f_{\alpha[MNP|} f_{\beta|QRS]} \big{|}_{\text{SD}}  = 0 \, , 
\end{equation}
where the second condition picks out the self-dual part of the SO(6,6) six-form $\epsilon^{\alpha \beta} f_{\alpha[MNP|}$ $f_{\beta|QRS]}$. 
However, we have no reason to expect the quartic supertrace to vanish for all Minkowski vacua of any gauged $D=4$, ${\cal N}=4$ supergravity that completely break supersymmetry, unless an explicit calculation like the one presented in this work shows it.   

\section*{Acknowledgments}

G.D.~would like to thank S.~De Angelis and D.~Partipilo for discussions and preliminary calculations that led to revisit gauged ${\cal N}=4$ supergravity.
This work is supported in part by MIUR-PRIN contract 2017CC72MK00, the MUR Excellent Departments Project 2023-2027. The research work was also supported by the Hellenic Foundation for Research and Innovation (HFRI) under the 3rd Call for HFRI PhD Fellowships (Fellowship Number: 6554). N. L. would like to thank the University of Padova for hospitality during the early stages of the project.
R. N. acknowledges GAČR grant EXPRO 20-25775X for financial support.

\includegraphics[scale=0.4]{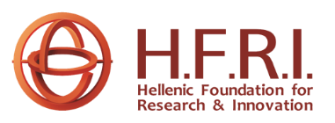}
             

\appendix

\section{Conventions} 
\label{sec:conventions}

Index conventions:
\begin{align*}
   \mu,\nu,\dots&=0,\dots,3 &&:\text{spacetime indices}\\
   a,b,\dots&=0,\dots,3 &&:\text{Lorentz indices}  \\
\alpha, \beta, \dots&=+,-  &&:\text{SL(2,${\mathbb R}$) indices} \\ 
M,N,\dots&=1,\dots,n+6 &&:\text{SO(6,$n$) indices} \\
\underline{\alpha}, \underline{\beta},\dots&=1,2 &&:\text{SO(2) indices}\\
\underline{m},\underline{n},\dots&=1,\dots,6 &&:\text{SO(6) indices}\\
i,j,\dots&=1,\dots,4 &&:\text{SU(4) indices}\\
\underline{a},\underline{b},\dots&=1,\dots,n &&:\text{SO($n$) indices}
\end{align*}
We also use underlined capital Latin letters $\underline{M}$, $\underline{N}$, \dots \, for SO(6) $\times$ SO($n$) indices, which we decompose as $\underline{M}=(\underline{m},\underline{a})$.

We use the gamma matrix, spinor and duality conventions of  \cite{DallAgata:2021uvl}. 
The Minkowski metric is given by 
\begin{equation}
\label{Mink}
    \eta_{ab} = \text{diag}(-1,1,1,1).
\end{equation}
The gamma matrices, $\gamma_a$, obey the following basic relations 
\begin{eqnarray}
	\label{Clifford}
	&&\left\{\gamma_a, \gamma_b\right\} = 2 \,\eta_{ab} \,{\mathbb 1}_{4}, \\ 
 \label{gammadagger}
 	&&\gamma_{0}^{\dagger}=-\gamma_{0},\quad  
	\gamma_{i}^{\dagger}=+\gamma_{i},\quad
	\gamma_{a}^{T}=\pm\gamma_{a}, \\
	&&\gamma_{a_{1}\ldots a_{p}}\equiv\gamma_{[a_{1}}\gamma_{a_{2}}\ldots \gamma_{a_{p}]} , \\
	&&	\label{gamm5} 	\gamma_{5}\equiv \gamma^5 \equiv -i \gamma^0 \gamma^1 \gamma^2 \gamma^3 = +i \gamma_0 \gamma_1 \gamma_2 \gamma_3, \\
	&& 	(\gamma_{5})^{2} = {\mathbb 1}_{4}, \quad 
	\{\gamma_{5},\gamma_{a}\}=0,
\end{eqnarray}
where the last of equations \eqref{gammadagger} means that each gamma matrix is either symmetric or antisymmetric, as well as the duality relations
\begin{equation}
   		\begin{array}{l}
		\displaystyle \gamma^{abc} = i\, \epsilon^{abcd} \gamma_d \gamma_5, \qquad i \,\gamma_a \gamma_5 = \frac{1}{3!} \epsilon_{abcd}\gamma^{bcd},\\[3mm]
		\displaystyle \gamma^{abcd} = - i\, \epsilon^{abcd} \gamma_5, \qquad i \,\gamma_5 =  \frac{1}{4!}\epsilon_{abcd}\gamma^{abcd}, \\[3mm]
		\displaystyle \gamma^{ab} = \frac{i}{2} \epsilon^{abcd} \gamma_{cd} \gamma_5,
	\end{array}
   \end{equation}
where $\epsilon_{abcd}$ is the totally antisymmetric epsilon tensor with 
\begin{equation}
\epsilon_{0123} = 1.
\end{equation}
We define $\epsilon_{\mu \nu \rho \sigma}$ as a totally antisymmetric tensor rather than a tensor density, 
\begin{equation}
\epsilon_{\mu \nu \rho \sigma} \equiv \epsilon_{abcd} e_\mu^a e_\nu^b e_\rho^c e_\sigma^d\,. 
\end{equation} 
We also introduce the charge conjugation matrix $C$, satisfying
\begin{eqnarray}
   	&& 	C^T = - C = C^{-1} =C^{\dagger}, \\
   	&& \gamma_a^T = - C \gamma_a C^{-1},
\end{eqnarray}
which imply the following symmetry properties 
\begin{equation}
	\begin{array}{lclcl}
		C^T = -C, \qquad  (C \gamma^{a})^T = (C \gamma^{a}), \qquad (C \gamma^{ab})^T = (C \gamma^{ab}), \\[3mm]
		(C \gamma^{abc})^T = - (C \gamma^{abc}),  \qquad  (C \gamma^{abcd})^T = -(C \gamma^{abcd}).
	\end{array}
\end{equation}

In terms of $C$, the charge conjugate spinor of a four-component spinor $\psi$ is defined as  
\begin{equation}
	\psi^c = C \overline{\psi}^T = i C \gamma^{0 T} \psi^* , \label{ccspinor} 
\end{equation}
where 
\begin{equation}
\overline{\psi} \equiv i \psi^{\dagger} \gamma^0
\end{equation}
is the Dirac conjugate of $\psi$. A {\textit{Majorana spinor} is then a spinor that equals its own charge conjugate, 
\index{Majorana spinor} 
\begin{equation}
	\psi^c = \psi \, ,  \qquad \textrm{(Majorana condition)}
\label{Majodef} 
\end{equation}
and for such a spinor the Dirac conjugate can also be written as 
\begin{equation}
    \overline{\psi} = \psi^T C \, .
\end{equation}
We therefore find that, for anti-commuting Majorana spinors, the following symmetry properties hold
\begin{equation}    \label{Majflips}
	{\overline{\psi}}_{1}M\psi_{2} = \left\{ 
	\begin{array}{l}
		+{\overline{\psi}}_{2}M\psi_{1} \qquad \textrm{for } M={\mathbb 1}_{4}, \gamma_{abc}, \gamma_{abcd}\,,\\[4mm]
		-{\overline{\psi}}_{2}M\psi_{1} \qquad \textrm{for } M=\gamma_{a}, \gamma_{ab} \,.
	\end{array}
	\right.
\end{equation}

We also introduce chirality projectors
\begin{equation}
    P_L \equiv \frac{1}{2} \left( \mathbb{1}_4 + \gamma_5\right) , \qquad  P_R \equiv \frac{1}{2} \left( \mathbb{1}_4 - \gamma_5\right).
\end{equation}
Left- and right-handed Weyl spinors $\psi_{L,R}$ satisfy the conditions
\begin{equation}
P_{L,R} \psi_{L,R} = \psi_{L,R} \quad \iff \quad \gamma_5 \psi_{L,R} = \pm \psi_{L,R},  
\end{equation}
where the upper sign is for left-handed spinors and the lower for right-handed spinors.

We will often use chirality projections also for Majorana spinors $\psi$, in which case one has the relations
\begin{equation}
	(\psi_L)^c = \psi_R, \qquad (\psi_R)^c = \psi_L,
\label{dependencepsi2} 
\end{equation}
where $\psi_{L,R} \equiv P_{L,R} \psi$, which make manifest the Majorana nature of the field. We also define 
\begin{equation}
{\overline \psi}_L \equiv \overline{\psi_R} = {\overline\psi} P_L = (\psi_L)^T C,  \qquad {\overline \psi}_R \equiv \overline{\psi_L} = {\overline\psi} P_R = (\psi_R)^T C\,.
\end{equation}

We will often need to rewrite 3 or 4-fermion terms and hence Fierz identities will be extremely useful.
We list here the main ones for two spinors: 
\begin{eqnarray}
	\psi_R \overline{\chi}_R &=& -\frac12 \overline\chi_R \psi_R\, \, P_R + \frac18 \overline{\chi}_R \gamma_{ab}\psi_R \; \gamma^{ab}\, P_R, \label{FierzRR} \\[2mm]
	\psi_R \overline{\chi}_L &=& -\frac12 \overline{\chi}_L \gamma^a \psi_R \; \gamma_a\, P_L, \label{FierzRL} 
\end{eqnarray}
where, for the sake of clarity, we explicitly left the projectors on the right-hand side.

The components of a spacetime $p$-form $\omega^{(p)}$ are normalized as 
\begin{equation}
    \label{pform}
    \omega^{(p)} = \frac{1}{p!} \omega_{\mu_1 \dots \mu_p} dx^{\mu_1} \wedge \dots \wedge dx^{\mu_p}
\end{equation}
and we assume that the exterior derivative $d$ acts from the left as
\begin{equation}
    d \omega^{(p)} = \frac{1}{p!} \partial_\mu \omega_{\mu_1 \dots \mu_p} dx^{\mu} \wedge dx^{\mu_1} \wedge \dots \wedge dx^{\mu_p}\,.
\end{equation}

SU(4) indices are raised and lowered by complex or charge conjugation. For an SU(4) vector $v^i$ that is a scalar in spinor space, we have 
\begin{equation}
    (v^i)^* = {\bar v}_i \, .
\end{equation}
On the other hand, for a chirally projected spinor $\phi^i$ in the fundamental representation of SU(4), we have
\begin{equation}
    (\phi^i)^c = i C \gamma^{0T} (\phi^i)^* = \phi_i \, 
\end{equation}
and we define
\begin{equation}
    {\bar\phi}^i \equiv \overline{\phi_i} \equiv i (\phi_i)^{\dagger} \gamma^0 =  (\phi^i)^T C  , \qquad {\bar\phi}_i \equiv \overline{\phi^i} \equiv i (\phi^i)^{\dagger} \gamma^0 =  (\phi_i)^T C\,,
\end{equation}
so that 
\begin{eqnarray}
\begin{aligned}
  {\bar\phi}^i \phi^j & =    {\bar\phi}^j \phi^i, \, {\bar\phi}^i \gamma^a \phi_j = - {\bar\phi}_j \gamma^a \phi^i , \, {\bar\phi}^i \gamma^{ab} \phi^j =  -  {\bar\phi}^j \gamma^{ab} \phi^i , \\
  {\bar\phi}^i \gamma^{abc} \phi_j & = {\bar\phi}_j \gamma^{abc} \phi^i , \, {\bar\phi}^i \gamma^{abcd} \phi^j = {\bar\phi}^j   \gamma^{abcd} \phi^i 
\end{aligned}  
\end{eqnarray}
and for example 
\begin{equation}
({\bar\phi}^i \phi^j )^* = {\bar\phi}_i \phi_j ,  \qquad ({\bar\phi}^i \gamma^a \phi_j)^* =  {\bar\phi}_i \gamma^a \phi^j. 
\end{equation}

SO(6,$n$) and SO(6) $\times$ SO($n$) indices are raised and lowered with the $\eta$ metrics
\begin{equation}
    v^M=\eta^{MN} v_N, \quad  v_M=\eta_{MN} v^N, \quad  v^{\underline{M}}=\eta^{\underline{M}\underline{N}} v_{\underline{N}}, \quad  v_{\underline{M}}=\eta_{\underline{M}\underline{N}} v^{\underline{N}},
\end{equation}
where $\eta^{MN}=\eta_{MN}=\eta^{\underline{M}\underline{N}}=\eta_{\underline{M}\underline{N}}=\text{diag}(-1,-1,-1,-1,-1,-1,1,\dots,1)$. 

SL(2,${\mathbb R}$) indices are raised and lowered as 
\begin{equation}
    \mathcal{V}^\alpha =  \mathcal{V}_\beta \epsilon^{\beta \alpha}, \qquad \mathcal{V}_\alpha =\epsilon_{\alpha \beta  } \mathcal{V}^\beta  \, , 
\end{equation}
where $\epsilon^{\alpha \beta} = - \epsilon^{\beta \alpha}$, $\epsilon_{\alpha \beta} = - \epsilon_{\beta \alpha}$ and $\epsilon^{+-}=\epsilon_{+-} = 1$. 

A real SO(6) vector $v^{\underline{m}}$ can alternatively be described by an antisymmetric SU(4) tensor $v^{ij}=-v^{ji}$ subject to the pseudo-reality constraint 
\begin{equation}
v_{ij}=(v^{ij})^*=\frac{1}{2} \epsilon_{ijkl} v^{kl},
\end{equation}
by introducing the map $v^{\underline{m}} \rightarrow v^{ij}$ defined by
\begin{equation}
\label{SO6toSU4}
    v^{ij} = {\Gamma}^{\underline{m} i j } v_{\underline{m}} \,,
\end{equation}
where ${\Gamma}^{\underline{m} i j }$ are intertwiners between the two representations, which satisfy
\begin{align}
 {\Gamma^{\underline{m}}}_{ij} = ({\Gamma}^{\underline{m} i j})^* &= \frac{1}{2} \epsilon_{ijkl} {\Gamma}^{\underline{m} kl}\,, \\[2mm]
 \label{Cliffalg} 
 {\Gamma}^{(\underline{m}|ik} {{\Gamma}^{|\underline{n})}}_{jk} &= -\frac{1}{4} \eta^{\underline{m} \underline{n}} \delta^i_j\,, \\[2mm]
 \label{complete}
 {\Gamma}^{\underline{m} ij} {\Gamma}_{\underline{m} kl} &= - \delta^{[i}_{k} \delta^{j]}_l\,.
 \end{align}
A possible explicit choice is given by the following antisymmetric 4$\times$4 matrices:
\begin{align*}
 {\Gamma}^{1ij}&=\frac{1}{2}
 \begin{pmatrix}
0 & 1 & 0 & 0 \\
 -1 & 0 & 0 & 0 \\
 0 & 0 & 0 & 1 \\
 0 & 0 & -1 & 0
 \end{pmatrix}, \hspace{0.5cm}
 {\Gamma}^{2ij}=\frac{1}{2}
 \begin{pmatrix}
 0&0&1&0 \\
 0&0&0&-1 \\
 -1&0&0&0 \\
 0&1&0&0
 \end{pmatrix}, \\
 {\Gamma}^{3ij} &= \frac{1}{2} 
 \begin{pmatrix}
 0&0&0&1 \\
 0&0&1&0 \\
 0&-1&0&0 \\
-1&0&0&0
 \end{pmatrix}, \hspace{0.5cm}
 {\Gamma}^{4ij}=\frac{1}{2}
 \begin{pmatrix}
 0&i&0&0 \\
 -i&0&0&0 \\
 0&0&0&-i \\
 0&0&i&0
 \end{pmatrix}, \\
 {\Gamma}^{5ij} &= \frac{1}{2}
 \begin{pmatrix}
 0&0&i&0 \\
 0&0&0&i \\
 -i&0&0&0 \\
 0&-i&0&0
 \end{pmatrix}, \hspace{0.5cm}
 {\Gamma}^{6ij} = \frac{1}{2}
 \begin{pmatrix}
 0&0&0&i \\
 0&0&-i&0 \\
 0&i&0&0 \\
 -i&0&0&0 \\
 \end{pmatrix}.
 \end{align*}
{}From the definition \eqref{SO6toSU4} and equation \eqref{Cliffalg}, it also follows that 
 \begin{equation}
v^{\underline{m}} = - {{\Gamma}^{\underline{m}}}_{ij} v^{ij} = - {\Gamma}^{\underline{m} i j } v_{ij},      
\end{equation}
and using the completeness relation \eqref{complete} we find 
\begin{equation}
    v^{\underline{m}} w_{\underline{m}} = - \frac{1}{2} \epsilon_{ijkl} v^{ij} w^{kl}  = - v^{ij} w_{ij} = - v_{ij} w^{ij} .
\end{equation}

The exterior derivative $D$ is covariant with respect to local Lorentz, SO(2), SU(4) and SO($n$) transformations, while the exterior derivative $\hat{D}$ is covariant with respect to local Lorentz, SO(2), SU(4), SO($n$) and gauge transformations. 

The Lie derivative of a p-form $A_p$ along the flow of a vector field $V$ is defined as 
\begin{equation}
		\ell_V A_p = \lim_{t\to 0} \frac1t \left(\sigma_{t}^* A_p(\sigma_{t}(x)) - A_p(x)\right),
\end{equation}
where $\sigma_{t}^*$ is the pull back of the differential form along the flow generated by the vector field $V$.
When applied to a scalar valued $p$-form this reduces to
\begin{equation}		
		\ell_V A_p = (\imath_V d + d \imath_V)A_p.
\end{equation}

For an antisymmetric tensor $T_{\mu \nu}$ we define the self-dual combination $T^{+}_{\mu\nu}$  and the anti-self-dual combination $T^{-}_{\mu \nu}$ by 
\begin{equation}
    T^{\pm}_{\mu \nu} \equiv \frac{1}{2} \left( T_{\mu \nu} \mp \frac{i}{2} \epsilon_{\mu \nu \rho \sigma} T^{\rho \sigma} \right), 
\end{equation}
which satisfy 
\begin{equation}
  \frac{1}{2} \epsilon_{\mu \nu \rho \sigma} T^{\pm \rho \sigma} = \pm i T^{\pm}_{\mu \nu} \,.
\end{equation}

The generators of SO(6,$n$) and SL(2,${\mathbb R}$) in the fundamental representation can be chosen as ${(t_{MN})_P}^Q = \delta^Q_{[M} \eta_{N]P}$ and ${(t_{\alpha \beta})_\gamma}^\delta=\delta^\delta_{(\alpha} \epsilon_{\beta) \gamma}$ respectively and there exists a $2(n+6)$-dimensional symplectic representation of SL(2,${\mathbb R}$) $\times$ SO(6,$n$) with generators
\begin{equation}
\label{sympgen}
    {(t_{MN})_{\cal P}}^{\cal Q} = {(t_{MN})_{P \gamma}}^{Q \delta} = \delta^Q_{[M} \eta_{N]P} \delta^{\delta}_{\gamma}\, , \qquad {(t_{\alpha \beta})_{\cal P}}^{\cal Q} = {(t_{\alpha \beta})_{P \gamma}}^{Q \delta} = \delta^\delta_{(\alpha} \epsilon_{\beta) \gamma} {\delta}^Q_P\,, 
\end{equation}
which satisfy
\begin{equation}
 {(t_{MN})_{\cal P}}^{\cal R} {\mathbb{C}}_{\mathcal{Q} \mathcal{R}} = {(t_{MN})_{\cal Q}}^{\cal R} {\mathbb{C}}_{\mathcal{P} \mathcal{R}}, \qquad{(t_{\alpha \beta})_{\cal P}}^{\cal R} {\mathbb{C}}_{\mathcal{Q} \mathcal{R}} = {(t_{\alpha \beta})_{\cal Q}}^{\cal R} {\mathbb{C}}_{\mathcal{P} \mathcal{R}}. 
\end{equation}
This representation is identified with the fundamental representation of SL(2,${\mathbb R}$) $\times$ SO(6,$n$). An infinitesimal global SL(2,${\mathbb R}$) $\times$ SO(6,$n$) transformation acts on a symplectic vector $V_{M \alpha}$ as 
\begin{equation}
    \delta_{\Lambda} V_{M \alpha} = {\Lambda}^{NP} {(t_{NP})_{M \alpha}}^{Q \delta} V_{Q \delta} + \Lambda^{\beta \gamma} {(t_{\beta \gamma})_{M \alpha}}^{Q \delta} V_{Q \delta} = -{\Lambda_M}^N V_{N \alpha} - {{\Lambda}_\alpha}^\beta V_{M \beta} , 
\end{equation}
where $\Lambda^{MN}=\Lambda^{[MN]}$ and $\Lambda^{\alpha\beta} = \Lambda^{(\alpha\beta)}$ are constant parameters. 


\section{Comparison with Previous Articles} 
\label{sec:comparison_with_previous_articles}

When comparing our results concerning the supersymmetry transformation rules with the ones in \cite{Schon:2006kz}, we find a crucial difference regarding the fermion shifts of the dilatini. 
More precisely, in \cite{Schon:2006kz} the shifts of the dilatini supersymmetry transformations are
\begin{equation}
  \label{dchiSW}
  \delta_{\epsilon,g} \chi^i = - \frac{4i}{3} \,g\, A_2^{ji} \epsilon_j, 
\end{equation}
while in the present work
\begin{equation}
    \label{dchihere}
  \delta_{\epsilon,g}\chi^i = \frac{2}{3} \,g\, A_2^{ij} \epsilon_j.
\end{equation}
Furthermore, equation (2.41) of \cite{Schon:2006kz}, which expresses the scalar potential $V$ in terms of the fermion shifts, takes the form
\begin{equation}
\label{potSW}
   \frac{1}{3} A_1^{ik} {\bar A}_{1jk} - \frac{1}{9} A_2^{ik} {\bar A}_{2jk} - \frac{1}{2} A_{2 \underline{a} j}{}^k {\bar A}_2{}^{\underline{a} i}{}_k = - \frac{1}{4 g^2} \delta^i_j V  , 
   \end{equation}
where we have rescaled $g^2 V \rightarrow V$, while our expression for the supersymmetric Ward-identity is
\begin{equation}
    \label{pothere}
   \frac{1}{3} A_1^{ik} {\bar A}_{1jk} - \frac{1}{9} A_2^{ki} {\bar A}_{2kj} - \frac{1}{2} A_{2 \underline{a} j}{}^k {\bar A}_2{}^{\underline{a} i}{}_k = - \frac{1}{4 g^2} \delta^i_j V.
\end{equation}
It is therefore clear that in the expansion of the second term we find a crucial sign difference with respect to \cite{Schon:2006kz}, which however disappears when tracing the expression, because of the symmetry properties of the various terms.

For the ungauged theory, it is also useful to list a dictionary between the conventions used in the paper by Perret \cite{Perret:1988jq} and ours.

\setlength{\tabcolsep}{20pt}
\renewcommand{\arraystretch}{1.5}

\begin{center}
\begin{tabular}{|c|c|}
\hline
Perret                              &  Our conventions                                             \\ \hline
$A,B=1,\ldots,4$                    &  $i,j=1,\dots,4$                                             \\ \hline
$\gamma^a$                          &  $-i\gamma^a$                                                \\ \hline
$\gamma_a$                          &  $+i\gamma_a$                                                \\ \hline
$\Sigma_{mn}=\frac{\gamma_{mn}}{2}$ &  $-\frac{\gamma_{ab}}{2}$ \\ \hline
$\psi_A$                            &  $\psi_i$                                                    \\ \hline
$P$                                 &  $-P^*$                                                   \\ \hline
$\Phi$                              &  $-\mathcal V_{-}$                                             \\ \hline
$P_{iAB}$                           &  $\sqrt 2 P_{\underline a ij}$                               \\ \hline
$V^{[A}W^{B]}=V^AW^B-V^BW^A$        &  $2V^{[A}W^{B]}$                                             \\ \hline
$\mathcal A$                        &  $i\frac{\mathcal V_+}{\mathcal V_-}$                        \\ \hline
$\chi_{iA}$                         &  $\frac{1}{\sqrt 2}\lambda_{\underline a i}$                 \\ \hline
$\lambda_A$                         &  $-\frac12 \chi_i$                                           \\ \hline 
\end{tabular}
\end{center}


\section{The Solution of the Bianchi Identities and the Construction of the Superspace Lagrangian} 
\label{sec:the_solution_of_the_bianchi_identities}

In this appendix, we provide the full derivation of the local supersymmetry transformations and of the Lagrangian for the ungauged and the gauged $D=4$, ${\cal N}=4$ matter-coupled Poincar\'{e} supergravities in an arbitrary symplectic frame, using the geometric or rheonomic approach (for a review see \cite{Castellani:1991eu}). 

The first step is to extend the spacetime fields of the ungauged theory to superfields in ${\cal N}=4$ superspace: this means that the spacetime one-forms $e^a=e^a_{\mu} dx^\mu$, $\psi^i=\psi^i_\mu dx^\mu$, $\psi_i=\psi_{i\mu} dx^\mu$, $A^{M \alpha} = A^{M \alpha}_\mu dx^\mu$ and $\omega_{ab} = \omega_{\mu a b} dx^\mu$, where $\omega_{\mu a b}$ is the spin connection, and the spacetime zero-forms $\chi^i$, $\chi_i$, $\lambda^{\underline{a}i}$, $\lambda^{\underline{a}}_i$, ${\cal V}_\alpha$, ${\cal V}_{\alpha}^*$, ${L_M}^{ij}$ and ${L_M}^{\underline{a}}$ are promoted to super-one-forms and super-zero-forms in  ${\cal N}=4$ superspace respectively. 
These superforms depend on the superspace coordinates $(x^\mu, \theta^i_\alpha, \theta_{i \alpha})$ (where $\theta^i_\alpha$ and $\theta_{i \alpha}$, $i,\alpha=1,2,3,4$, are anticommuting fermionic coordinates and are the components of left-handed Weyl spinors $\theta^i$ and their charge conjugates $\theta_i$ respectively) in such a way that their projections on the spacetime submanifold, i.e. the $\theta^i=d \theta^i = 0 $ hypersurface, are equal to the corresponding spacetime quantities. 

A basis of one-forms in ${\cal N}=4$ superspace is given by the supervielbein $\{ e^a, \psi^i_\alpha, \psi_{i \alpha} \}$, where $e^a$ is the bosonic vielbein, while $\psi^i_\alpha$ and $\psi_{i \alpha}$, which are the spinor components of the left-handed gravitino super-one-forms $\psi^i$ and their charge conjugates $\psi_i$ respectively, constitute the fermionic vielbein. 

We start by defining the supercurvatures of the various super-$p$-forms in ${\cal N}=4$ superspace as follows
\begin{align}
    R^{ab}&= d \omega^{ab} + \omega^{ac} \wedge {\omega_c}^b,  \\[2mm]
T^a &= d e^a + {\omega^a}_b \wedge e^b - {\bar\psi}^i \wedge \gamma^a \psi_i = D e^a  - {\bar\psi}^i \wedge \gamma^a \psi_i,  \\[2mm]
\rho_i  = D \psi_i &= d \psi_i + \frac{1}{4} \omega^{ab} \wedge \gamma_{ab} \psi_i - \frac{i}{2} \mathcal{A} \wedge \psi_i - {\omega_i}^j \wedge \psi_j ,\\[2mm]
V_i = D \chi_i & = d \chi_i + \frac{1}{4} \omega^{ab} \gamma_{ab} \chi_i + \frac{3 i}{2} \mathcal{A} \chi_i - {\omega_i}^j \chi_j,  \\[2mm]
 \Lambda_{\underline{a}i} =D \lambda_{\underline{a}i}&= d \lambda_{\underline{a}i} + \frac{1}{4} \omega^{ab}  \gamma_{ab}  \lambda_{\underline{a}i}  + \frac{i}{2} \mathcal{A}\lambda_{\underline{a}i} -  {\omega_i}^j  \lambda_{\underline{a}j} + {\omega_{\underline{a}}}^{\underline{b}} \lambda_{\underline{b}i},  \\[2mm]
 \label{Fdef}
{\cal F}^{M \alpha} & = d A^{M \alpha} - (\mathcal{V}^\alpha)^* L^{Mij} {\bar\psi}_i \wedge \psi_j - \mathcal{V}^\alpha  {L^M}_{ij} {\bar \psi}^i \wedge \psi^j ,\\[2mm]
 P & = \frac{i}{2} \epsilon^{\alpha \beta}  \mathcal{V}_\alpha d \mathcal{V}_\beta, \\[2mm]
 P_{\underline{a} ij} & = {L_{\underline{a}}}^M d L_{Mij}\,,
 \end{align}
where ${\cal A}$, ${\omega_i}^j$ and ${\omega_{\underline{a}}}^{\underline{b}}$ are super-one-forms, whose projections on spacetime are the spacetime SO(2), SU(4) and SO($n$) connections respectively, which have been defined in the description of the scalar manifold in section 2 and $D$ is the exterior derivative that is covariant with respect to local Lorentz, SO(2), SU(4) and SO($n$) transformations. The supercurvatures $R^{ab}$, $T^a$ and $\rho_i$ have been defined in such a way that by setting them to zero and deleting the composite connections ${\cal A}$ and ${\omega_i}^j$ we obtain the Maurer--Cartan equations of the ${\cal N}=4$ super-Poincar\'{e} algebra 
\begin{align}
[M_{ab},M_{cd}]&= - \eta_{ac} M_{bd} + \eta_{ad} M_{bc} + \eta_{bc} M_{ad} - \eta_{bd} M_{ac} \, ,   \\[2mm] 
 [P_a, M_{bc}] & = \eta_{ab} P_c - \eta_{ac} P_b \, , \\[2mm]
 [M_{ab},Q^i_\alpha] & = -\frac{1}{2} {{(\gamma_{ab})}_\alpha}^\beta Q^i_\beta \, , \\[2mm] 
 [M_{ab},Q_{i\alpha}] & = -\frac{1}{2} {{(\gamma_{ab})}_\alpha}^\beta Q_{i\beta} \, , \\[2mm]
 \{ Q^i_\alpha, {\bar{Q}}_j^\beta\}&= -  \delta^i_j {(P_R \gamma^a)_\alpha}^\beta P_a \, , \\[2mm] 
 \{ Q_{i \alpha}, {\bar Q}^{j \beta} \} & = -  \delta^j_i {(P_L \gamma^a)_\alpha}^\beta P_a \, ,
\end{align}
where $\alpha, \beta= 1,2,3,4$ are spinor indices, $\gamma_5 Q_i = Q_i, \, \gamma_5 Q^i = - Q^i$ and the one-forms $\omega^{ab}$, $e^a$, $\psi^i$ and $\psi_i$ are dual to the generators $M^{ab}$, $P^a$, $Q^i$ and $Q_i$ respectively. 

By acting on the supercurvatures with the exterior derivative $d$ and using the fact that $d^2=0$, we obtain the following Bianchi identities 
\begin{align}
\label{DRab}
D R^{ab} =& \,  0, \\[2mm]
\label{DRa}
D T^a  = & \,  {R^a}_b \wedge e^b +  {\bar\psi}_i \wedge \gamma^a \rho^i + {\bar\psi}^i \wedge \gamma^a \rho_i, \\[2mm]
\label{Drho}
D \rho_i   = & \, \frac{1}{4} R^{ab} \wedge \gamma_{ab} \psi_i - \frac{i}{2} F \wedge \psi_i - {R_i}^j \wedge \psi_j ,\\[2mm]
\label{DX}
D V_i    =& \, \frac{1}{4} R^{ab}  \gamma_{ab} \chi_i + \frac{3i}{2} F \chi_i -  {R_i}^j \chi_j, \\[2mm]
\label{DLam}
D  \Lambda_{\underline{a}i}   = & \, \frac{1}{4} R^{ab}  \gamma_{ab} \lambda_{\underline{a}i} + \frac{i}{2} F  \lambda_{\underline{a}i} -  {R_i}^j \lambda_{\underline{a}j} + {R_{\underline{a}}}^{\underline{b}} \lambda_{\underline{b}i}, \\[2mm]
\label{DF}
D {\cal F}^{M \alpha}  = & \, - \mathcal{V}^\alpha L^{Mij}  P^* \wedge \bar\psi_i \wedge \psi_j - (\mathcal{V}^\alpha )^* L^{M \underline{a}} {P_{\underline{a}}}^{ij} \wedge \bar\psi_i \wedge \psi_j + 2 (\mathcal{V}^\alpha)^* L^{Mij}  \bar\psi_i \wedge \rho_j \nonumber  \\[2mm] 
 & - (\mathcal{V}^\alpha)^* {L^M}_{ij}  P \wedge \bar\psi^i \wedge \psi^j - \mathcal{V}^\alpha L^{M \underline{a}} P_{\underline{a}ij} \wedge \bar\psi^i \wedge \psi^j + 2 \mathcal{V}^\alpha  {L^M}_{ij}  \bar\psi^i \wedge \rho^j, \\[2mm]
 \label{DP}
 DP=  & \,  0, \\[2mm]
 \label{DPaij}
D P_{\underline{a}ij} =& \, 0 \, ,
\end{align}
where $F$, ${R_i}^j$ and ${R_{\underline{a}}}^{\underline{b}}$ are the superspace SO(2), SU(4) and SO($n$) curvatures given by equations \eqref{SO(2)curv}, \eqref{RSU4} and \eqref{RSOnPP} respectively, which are now to be viewed as superspace equations. 

The solution of the Bianchi identities can be obtained as follows: first, one notes that the one-form supercurvatures can be expanded along the supervielbein basis $\{ e^a, \psi^i_\alpha, \psi_{i \alpha} \}$, while the two-form supercurvatures can be expanded along the intrinsic basis of two-forms $\{ e^a \wedge e^b, \psi^i_\alpha \wedge e^a, \psi_{i \alpha} \wedge e^a, \psi^i_\alpha \wedge \psi^j_\beta, \psi^i_\alpha \wedge \psi_{j \beta}, \psi_{i \alpha} \wedge \psi_{j \beta} \}$ in ${\cal N}=4$  superspace. 
Then, one requires that all the components of the supercurvatures along the basis elements that involve at least one of $\psi^i_\alpha$, $\psi_{i \alpha}$ (outer components) be expressed in terms of the supercurvature components along the basis elements $e^a$ and $e^a \wedge e^b$ (inner components) and the physical superfields. This requirement is known as the {rheonomy principle} and ensures that no new degrees of freedom are introduced in the theory. 
Furthermore, the expansions of the supercurvatures along the intrinsic bases of one- and two-forms in superspace are referred to as the rheonomic parametrizations of the supercurvatures. 

The next step is to write down these expansions in a form that is compatible with all the symmetries of the theory, that is: covariance under local SO(2), SU(4), SO($n$) and Lorentz transformations. 
It is also very useful to take into account the invariance of the scalar $\sigma$-model equations \eqref{DV}, \eqref{DLij} and \eqref{DLa} extended to ${\cal N}=4$ superspace and the Bianchi identities \eqref{DRab}-\eqref{DPaij} under the following rigid rescalings of the various super-$p$-forms (and the corresponding supercurvatures)
\begin{align}
    (\omega^{ab}, {\cal V}_\alpha, {L_M}^{ij}, {L_M}^{\underline{a}}) & \rightarrow  (\omega^{ab}, {\cal V}_\alpha, {L_M}^{ij}, {L_M}^{\underline{a}}), \\[2mm]
    (e^a, A^{M \alpha}) & \rightarrow \lambda (e^a, A^{M \alpha}), \\[2mm]
    \psi^i & \rightarrow \lambda^{\frac{1}{2}} \psi^i \, . 
\end{align}
Furthermore, the spin-1/2 fermions scale as
\begin{equation}
    (\chi^i, \lambda^{\underline{a}i}) \rightarrow \lambda^{-\frac{1}{2}} (\chi^i, \lambda^{\underline{a}i})\, , 
\end{equation}
because they must appear contracted with the gravitino super-one-forms in the rheonomic parametrizations of the supercurvatures $P$ and $P_{\underline{a} ij}$, which are taken to be 
\begin{align}
\label{P}
    P  = & P_a e^a + {\bar\psi}_i \chi^i, \\[2mm]
    \label{Paij}
P_{\underline{a}ij} = & P_{\underline{a}ij a} e^a + 2 {\bar{\psi}}_{[i|} \lambda_{\underline{a} |j]} + \epsilon_{ijkl} {\bar{\psi}}^k \lambda^l_{\underline{a}}  . 
\end{align}

The most general rheonomic parametrizations of the other supercurvatures that are compatible with the symmetries of the theory and have the correct scaling behaviours are
\begin{align}
\label{Xi}
V_i  = & V_{ia} e^a  + b_1 L_{Mij} \mathcal{V}_\alpha^* {\cal F}^{M \alpha}_{ab}\gamma^{ab} \psi^j + b_2 ({\bar\lambda}_{\underline{a} i} \lambda^{\underline{a}}_j) \psi^j + b_3 \gamma^a P_a^* \psi_i,
\\[2mm]
\label{Lai}
 \Lambda_{\underline{a}i} = & \Lambda_{\underline{a}i a}  e^a + c_1 P_{\underline{a}ij a} \gamma^a \psi^j + c_2 L_{M \underline{a}} \mathcal{V}_\alpha^* {\cal F}^{M \alpha}_{ab}\gamma^{ab} \psi_i  + c_3 ({\bar\chi}_i \lambda^j_{\underline{a}}) \psi_j + c_4 ({\bar\chi}_j \lambda^j_{\underline{a}}) \psi_i, \\[2mm]
 \label{FMa}
{\cal F}^{M \alpha}  =&  \frac{1}{2} {\cal F}^{M \alpha}_{ab} e^a \wedge e^b + \big{(} d_3 \mathcal{V}^\alpha L^{Mij} {\bar\lambda}_{\underline{a}i} \gamma_{ab} \lambda^{\underline{a}}_j \, e^a \wedge e^b + d_4 \mathcal{V}^\alpha L^{M \underline{a}} {\bar\chi}_i \gamma_{ab} \lambda^i_{\underline{a}} \, e^a \wedge e^b \nonumber  \\[2mm]
&+ d_1 (\mathcal{V}^\alpha)^* {L^M}_{ij} {\bar \chi}^i \gamma_a \psi^j \wedge e^a + d_2 (\mathcal{V}^\alpha)^* L^{M \underline{a}} {\bar\lambda}^i_{\underline{a}} \gamma_a \psi_i \wedge e^a  + c.c. \big{)},\\[2mm]
\label{rhoi}
\rho_i =& \frac{1}{2} \rho_{iab} e^a \wedge e^b + f_1 L_{Mij} \mathcal{V}_\alpha {\cal F}^{M \alpha}_{ab} \gamma^b \psi^j \wedge e^a  + f_2 L_{Mij} \mathcal{V}_\alpha \epsilon_{abcd} {\cal F}^{M \alpha cd} \gamma^b \psi^j \wedge e^a \nonumber \\[2mm]
&+ f_3 \epsilon_{ijkl} ({\bar{\lambda}}^j_{\underline{a}} \gamma_{ab} \lambda^{\underline{a} k }) \gamma^a \psi^l \wedge e^b
 + f_4 (\bar{\chi}_i \gamma_a \chi^j) \psi_j \wedge e^a + f_5  (\bar{\chi}_j \gamma_a \chi^j) \psi_i \wedge e^a  \nonumber \\[2mm]
& + f_6 (\bar{\chi}_i \gamma^a \chi^j) \gamma_{ab} \psi_j \wedge e^b + f_7 (\bar{\chi}_j \gamma^a \chi^j) \gamma_{ab} \psi_i \wedge e^b \\[2mm]
& + g_1 ({\bar\lambda}^{\underline{a}}_i \gamma_a \lambda^j_{\underline{a}} ) \psi_j \wedge e^a + g_2 ({\bar\lambda}^{\underline{a}}_j \gamma_a \lambda^j_{\underline{a}} ) \psi_i \wedge e^a + g_3  ({\bar\lambda}^{\underline{a}}_i \gamma^a \lambda^j_{\underline{a}} ) \gamma_{ab} \psi_j \wedge e^b
\nonumber \\[2mm]
& + g_4 ({\bar\lambda}^{\underline{a}}_j \gamma^a \lambda^j_{\underline{a}} ) \gamma_{ab} \psi_i \wedge e^b + g_5 \epsilon_{ijkl} \chi^j ({\bar\psi}^k \wedge \psi^l) , \nonumber 
\end{align}
where $b_1,b_2,b_3,c_1,c_2,c_3,c_4,d_1,d_2,d_3,d_4,f_1,f_2,f_3,f_4,f_5,f_6,f_7,g_1,g_2,g_3,g_4$ and $g_5$ are constant coefficients. 
We also impose the kinematic constraint
\begin{equation}
    \label{Ra=0}
    T^a = 0 \, , 
\end{equation}
which amounts to the vanishing of the supertorsion and relates the spin connection to the vielbein and the gravitini, reducing the gravitational degrees of freedom to the correct ones. By substituting the parametrizations \eqref{P}-\eqref{rhoi} and the constraint \eqref{Ra=0} into the Bianchi identities, one can determine the values of the coefficients, which are
\begin{align}
b_1&=-\frac{i}{4}, \, b_2=-1, \, b_3=1, \nonumber \\     
c_1&=-1, \, c_2=\frac{i}{8}, \, c_3=1, \, c_4=-\frac{1}{2}, \nonumber \\ 
d_1&=1, \, d_2=1, \, d_3=-\frac{1}{4}, \, d_4=\frac{1}{4}, \nonumber \\ 
f_1&=\frac{i}{4}, \, f_2=\frac{1}{8}, \, f_3=\frac{1}{4}, \nonumber\, f_4=\frac{1}{4}, \, f_5=-\frac{1}{4}, \, f_6 = \frac{1}{4} , \, f_7 = - \frac{1}{8} , \nonumber \\
g_1&=\frac{1}{2}, \, g_2=0, \, g_3=\frac{1}{2}, \, g_4=-\frac{1}{4}, \, g_5=-\frac{1}{2} \,, \nonumber
\end{align}
and find that ${\cal F}^{M \alpha}_{ab}$ must satisfy
\begin{equation}
\label{moddual}
\epsilon_{abcd} {\cal F}^{M \alpha cd}= - 2\,{M^M}_N {M^{\alpha}}_{\beta} { \cal F}^{N \beta}_{ab} ,
\end{equation}
which is a twisted self-duality constraint implying that only $6+n$ vectors are physical.
Furthermore, from the Bianchi identity \eqref{DRa} one obtains the rheonomic parametrization of the supercurvature $R_{ab}$:
\begin{align}
    \label{Rab}
    R_{ab} = & \frac{1}{2} R_{cdab} e^c \wedge e^d + {\bar \theta}^i_{abc} \psi_i \wedge e^c + {\bar\theta}_{iabc} \psi^i \wedge e^c \nonumber \\[2mm]
& + \frac{i}{4} \mathcal{V}_\alpha L_{Mij} {\cal F}^{M \alpha}_{ab} {\bar\psi}^i \wedge \psi^j + \frac{1}{8} \mathcal{V}_\alpha L_{Mij} \epsilon_{abcd} {\cal F}^{M \alpha cd} {\bar\psi}^i \wedge \psi^j \nonumber \\[2mm]
& - \frac{i}{4} \mathcal{V}_\alpha^* {L_M}^{ij} {\cal F}^{M \alpha}_{ab} {\bar\psi}_i \wedge \psi_j + \frac{1}{8}  \mathcal{V}_\alpha^* {L_M}^{ij}  \epsilon_{abcd} {\cal F}^{M \alpha cd} {\bar\psi}_i \wedge \psi_j \nonumber \\[2mm] 
& - \frac{1}{4} \epsilon_{ijkl} ({\bar{\lambda}}^i_{\underline{a}} \gamma_{ab} \lambda^{\underline{a} j }) {\bar\psi}^k \wedge \psi^l - \frac{1}{4} \epsilon^{ijkl} ({\bar{\lambda}}_i^{\underline{a}} \gamma_{ab} \lambda_{\underline{a} j }) {\bar\psi}_k \wedge \psi_l 
\\[2mm]
& + \frac{1}{2} ({\bar\chi}_i \gamma^c \chi^j) {\bar\psi}^i \wedge \gamma_{abc} \psi_j - \frac{1}{4} ({\bar\chi}_j \gamma^c \chi^j) {\bar\psi}^i \wedge \gamma_{abc} \psi_i \nonumber \\[2mm]  \nonumber
& + ({\bar\lambda}^{\underline{a}}_i \gamma^c \lambda^j_{\underline{a}}) {\bar\psi}^i \wedge \gamma_{abc} \psi_j - \frac{1}{2} ({\bar\lambda}^{\underline{a}}_j \gamma^c \lambda^j_{\underline{a}}) {\bar\psi}^i \wedge \gamma_{abc} \psi_i \, , 
\end{align}
where 
\begin{equation}
    \theta^i_{abc} =  \gamma_{[a} \rho^i_{b]c} - \frac{1}{2} \gamma_c \rho^i_{ab}\,.
\end{equation}

In addition, the Bianchi identities impose differential constraints on the inner components of the supercurvatures, whose projections on spacetime are identified with the equations of motion of the theory. Indeed, the closure of the Bianchi identities is equivalent to the closure of the ${\cal N}=4$ supersymmetry algebra on the spacetime fields modulo local symmetry transformations, which happens only when the equations of motion are satisfied. In particular, the ${\bar{\psi}}^i \wedge \gamma^a \psi_i $ sector of the Bianchi identity \eqref{DX} implies the following superspace equations of motion for the dilatini 
\begin{equation}
    \label{choeom}
    \gamma^a V_{i a} =  \frac{i}{4} \mathcal{V}_{\alpha}^* L_{M \underline{a}} {\cal F}^{M \alpha}_{ab} \gamma^{ab} \lambda^{\underline{a}}_i + \frac{3}{4} \chi^j\, \bar{\chi}_i \chi_j - \frac{1}{2} \lambda^{\underline{a}}_j\, {\bar\lambda}^j_{\underline{a}} \chi_i - \lambda^{\underline{a}}_i\, {\bar\lambda}^j_{\underline{a}} \chi_j , 
\end{equation}
while the corresponding sector of the Bianchi identity \eqref{DLam} gives the following superspace equations of motion for the gaugini  
\begin{align}
 \label{lameom}
    \gamma^a \Lambda_{\underline{a} i a} = & \, \frac{i}{4} \mathcal{V}_{\alpha}^* L_{Mij}  {\cal F}^{M \alpha}_{ab} \gamma^{ab} \lambda^j_{\underline{a}} + \frac{i}{8} \mathcal{V}_{\alpha} L_{M \underline{a}} {\cal F}^{M \alpha}_{ab} \gamma^{ab} \chi_i  \nonumber \\[2mm]
    & - \frac{1}{2} \lambda_{\underline{b}}^j\, {\bar\lambda}_j^{\underline{b}} \lambda_{\underline{a}i} - \lambda_{\underline{a}}^j\,  {\bar\lambda}_{\underline{b} i } \lambda^{\underline{b}}_j + 2 \lambda_{\underline{b}}^j\, {\bar\lambda}_i^{\underline{b}} \lambda_{\underline{a}j} - \frac{1}{4} \chi_j \,\bar{\chi}^j \lambda_{\underline{a}i} - \frac{1}{2}\, \chi_i\, {\bar\chi}^j \lambda_{\underline{a}j} .
\end{align}
Furthermore, by considering the ${\bar{\psi}}^i \wedge \gamma^a \psi_i \wedge e^b$ sector of the Bianchi identity \eqref{Drho}, one can specify the superspace equations of motion for the gravitini
\begin{align}
    \label{graveom}
    \gamma^b \rho_{iba} = & \, \frac{i}{2} \, \mathcal{V}_\alpha L_{M \underline{a}} {\cal F}^{M \alpha}_{ab} \gamma^b \lambda^{\underline{a}}_i - \frac{i}{2}\,  \mathcal{V}_{\alpha}^*  L_{Mij}  {\cal F}^{M \alpha}_{ab} \gamma^b \chi^j\nonumber\\[2mm]
    &- \frac{1}{2} \gamma_a \lambda_{\underline{a} j}\, {\bar\lambda}^{\underline{a}}_i \chi^j + P_a \chi_i + 2 {P^{\underline{a}}}_{ija} \lambda^j_{\underline{a}} \,.
\end{align}

Let us now study the implications of the constraint \eqref{moddual}. We first define the symmetric $2(n+6)\times2(n+6)$ matrix 
\begin{equation}
    {\cal M}_{\mathcal{M} \mathcal{N}} = {\cal M}_{M \alpha N \beta} =  M_{MN} M_{\alpha \beta} \, , 
\end{equation}
which satisfies 
\begin{equation}
   {\cal M}_{\mathcal{M} \mathcal{N}} {\mathbb{C}}^{\mathcal{N}\mathcal{P}}  {\cal M}_{\mathcal{P} \mathcal{Q}} = {\mathbb{C}}_{\mathcal{M}\mathcal{Q}}.
\end{equation}
By equating the right-hand sides of \eqref{Fdef}, which gives the definition of the supercurvature ${\cal F}^{M \alpha}$, and \eqref{FMa}, which gives its rheonomic parametrization,  and considering the $\theta^i = d \theta^i = 0 $ projection of the resulting relation we obtain 
\begin{align}
    e^a_\mu e^b_\nu {\cal F}^{M \alpha}_{ab}|_{\theta^i=0} = &F^{M \alpha}_{\mu \nu} + \bigg{[ } - 2  ({\cal V}^\alpha)^* L^{M ij} {\bar\psi}_{i \mu} \psi_{j \nu} + \frac{1}{2} {\cal V}^\alpha L^{Mij} {\bar\lambda}_{\underline{a}i} \gamma_{\mu \nu} \lambda^{\underline{a}}_j \nonumber \\[2mm]
   & - \frac{1}{2} ({\cal V}^\alpha)^* L^{M \underline{a}} {\bar\chi}^i \gamma_{\mu \nu} \lambda_{\underline{a}i} + 2 ({\cal V}^\alpha)^* {L^M}_{ij} {\bar\chi}^i \gamma_{[\mu} \psi^j_{\nu]} \\[2mm] 
   & + 2 {\cal V}^\alpha L^{M \underline{a}} {\bar\lambda}_{\underline{a}i} \gamma_{[\mu} \psi^i_{\nu]} + c.c. \nonumber \bigg{]} \equiv {\hat{\cal F}}^{M \alpha}_{\mu \nu} ,
\end{align}
where $F^{M \alpha}_{\mu \nu} = 2 \partial_{[\mu} A^{M \alpha}_{\nu]}$, which decomposes in an arbitrary symplectic frame as 
\begin{equation}
\label{FLambda}
    F^{M \alpha}_{\mu \nu} = (F^{\Lambda}_{\mu \nu}, F_{\Lambda \mu \nu}) = 2 (\partial_{[\mu} A^{\Lambda}_{\nu]}, \partial_{[\mu|} A_{\Lambda|\nu]}).
\end{equation}
The quantities ${\hat{\cal F}}^{M \alpha}_{\mu \nu}$ are referred to as the supercovariant field strengths of the vector fields $A^{M \alpha}_\mu$. 
Then, restricting the superspace equation \eqref{moddual} to spacetime, by setting $\theta^i = 0 $, we find 
\begin{align}
    (*F^{M \alpha})_{\mu \nu} = \, & {\mathbb{C}}^{M \alpha \mathcal{N}} {\cal M}_{\mathcal{N} \mathcal{P}} F^{\mathcal{P}}_{\mu \nu} + \big{(} -2 i ({\cal V}^\alpha)^* L^{Mij} {\bar{\psi}}_{i \mu} \psi_{j \nu} + \epsilon_{\mu \nu \rho \sigma} ({\cal V}^\alpha)^* L^{Mij}  {\bar{\psi}}^{\rho}_i {\psi}_j^{\sigma}  \nonumber \\[2mm] \nonumber
    & - i {\mathcal{V}}^{\alpha}  L^{Mij} {\bar{\lambda}}_{\underline{a} i} \gamma_{\mu \nu} \lambda^{\underline{a}}_j  - i {\mathcal{V}}^{\alpha} L^{M \underline{a}} {\bar\chi}_i \gamma_{\mu \nu} \lambda^i_{\underline{a}} + 2 i ({\cal V}^\alpha)^* {L^M}_{ij} {\bar\chi}^i \gamma_{[\mu} \psi^j_{\nu]} \\[2mm] \label{dualeqfull}
    & - \epsilon_{\mu \nu \rho \sigma} ({\cal V}^\alpha)^* {L^M}_{ij} {\bar\chi}^i \gamma^{\rho} \psi^{j \sigma} + 2 i {\cal V}^\alpha L^{M \underline{a}} {\bar\lambda}_{\underline{a} i } \gamma_{[\mu} \psi^i_{\nu]} \\[2mm] \nonumber
  &- \epsilon_{\mu \nu \rho \sigma} {\cal V}^\alpha  L^{M \underline{a}} {\bar\lambda}_{\underline{a}i} \gamma^{\rho} \psi^{ i \sigma} + \text{c.c.} \big{)} .
\end{align}
The Hodge duals of the electric field strengths can be obtained by multiplying the above equation by the projectors ${\Pi^{\Lambda}}_{M \alpha}$,
\begin{align}
    (*F^{\Lambda})_{\mu \nu} = \, & {{\cal M}^{\Lambda}}_{\Sigma} F^{\Sigma}_{\mu \nu} +{\cal M}^{\Lambda \Sigma} F_{\Sigma \mu \nu} + {\Pi^{\Lambda}}_{M \alpha} \big{(} -2 i ({\cal V}^\alpha)^* L^{Mij} {\bar{\psi}}_{i \mu} \psi_{j \nu} \nonumber \\[2mm] \nonumber & + \epsilon_{\mu \nu \rho \sigma} ({\cal V}^\alpha)^* L^{Mij}  {\bar{\psi}}^{\rho}_i {\psi}_j^{\sigma}  \nonumber
     - i {\mathcal{V}}^{\alpha}  L^{Mij} {\bar{\lambda}}_{\underline{a} i} \gamma_{\mu \nu} \lambda^{\underline{a}}_j  - i {\mathcal{V}}^{\alpha} L^{M \underline{a}} {\bar\chi}_i \gamma_{\mu \nu} \lambda^i_{\underline{a}} \nonumber \\[2mm]
     & + 2 i ({\cal V}^\alpha)^* {L^M}_{ij} {\bar\chi}^i \gamma_{[\mu} \psi^j_{\nu]} \label{dualeqel}
    - \epsilon_{\mu \nu \rho \sigma} ({\cal V}^\alpha)^* {L^M}_{ij} {\bar\chi}^i \gamma^{\rho} \psi^{j \sigma} \\[2mm] & + 2 i {\cal V}^\alpha L^{M \underline{a}} {\bar\lambda}_{\underline{a} i } \gamma_{[\mu} \psi^i_{\nu]}  - \epsilon_{\mu \nu \rho \sigma} {\cal V}^\alpha  L^{M \underline{a}} {\bar\lambda}_{\underline{a}i} \gamma^{\rho} \psi^{ i \sigma} + \text{c.c.}  \big{)}, \nonumber
\end{align}
while multiplying \eqref{dualeqfull} by $\Pi_{\Lambda M \alpha}$ we get the Hodge duals of the magnetic field strengths
\begin{align}
    (*F_{\Lambda})_{\mu \nu} = & - {\cal M}_{\Lambda \Sigma} F^{\Sigma}_{\mu \nu } - {{\cal M}_{\Lambda}}^{\Sigma} F_{\Sigma \mu \nu} + {\Pi}_{\Lambda M \alpha}   \big{(} -2 i ({\cal V}^\alpha)^* L^{Mij} {\bar{\psi}}_{i \mu} \psi_{j \nu} \nonumber \\[2mm] \nonumber & + \epsilon_{\mu \nu \rho \sigma} ({\cal V}^\alpha)^* L^{Mij}  {\bar{\psi}}^{\rho}_i {\psi}_j^{\sigma}  \nonumber
     - i {\mathcal{V}}^{\alpha}  L^{Mij} {\bar{\lambda}}_{\underline{a} i} \gamma_{\mu \nu} \lambda^{\underline{a}}_j  - i {\mathcal{V}}^{\alpha} L^{M \underline{a}} {\bar\chi}_i \gamma_{\mu \nu} \lambda^i_{\underline{a}} \nonumber \\[2mm]
     & + 2 i ({\cal V}^\alpha)^* {L^M}_{ij} {\bar\chi}^i \gamma_{[\mu} \psi^j_{\nu]} \label{dualeqmag}
    - \epsilon_{\mu \nu \rho \sigma} ({\cal V}^\alpha)^* {L^M}_{ij} {\bar\chi}^i \gamma^{\rho} \psi^{j \sigma} \\[2mm] &+ 2 i {\cal V}^\alpha L^{M \underline{a}} {\bar\lambda}_{\underline{a} i } \gamma_{[\mu} \psi^i_{\nu]}  - \epsilon_{\mu \nu \rho \sigma} {\cal V}^\alpha  L^{M \underline{a}} {\bar\lambda}_{\underline{a}i} \gamma^{\rho} \psi^{ i \sigma} + \text{c.c.} \big{)}  . \nonumber
\end{align}
From equations \eqref{dualeqel} and \eqref{dualeqmag} one can determine the symmetric matrices ${\cal I}_{\Lambda \Sigma}$, ${\cal R}_{\Lambda \Sigma}$ and the antisymmetric tensor $O_{\Lambda \mu \nu}$ that appear in the parametrization \eqref{Lgen} of the ungauged Lagrangian. 
Indeed, from the expression \eqref{GL} for the magnetic duals $G_{\Lambda \mu \nu}$ of the field strengths $F^{\Lambda}_{\mu \nu}$ of the electric vectors it follows that 
\begin{eqnarray}
 \begin{aligned}
 \label{dual}
 \begin{pmatrix}
   (*F^\Lambda)_{\mu \nu} \\[2mm]
   (*G_\Lambda)_{\mu \nu} 
 \end{pmatrix} = &
 \begin{pmatrix}
 {({\cal I}^{-1}{\cal R})^\Lambda}_\Sigma & - ({\cal I}^{-1})^{\Lambda \Sigma} \\[2mm]
 ({\cal I} + {\cal R} {\cal I}^{-1} {\cal R})_{\Lambda \Sigma} & - {( {\cal R} {\cal I}^{-1})_\Lambda}^{\Sigma} 
 \end{pmatrix} 
 \begin{pmatrix}
  F^\Sigma_{\mu \nu} \\[2mm]
  G_{\Sigma \mu \nu }
 \end{pmatrix} 
  \\[2mm]
 & +\begin{pmatrix}
  - ({\cal I}^{-1})^{\Lambda \Sigma}  (*O_{\Sigma})_{\mu \nu} \\[2mm]
  O_{\Lambda \mu \nu} -  {({\cal R} {\cal I}^{-1})_\Lambda}^{\Sigma}  (*O_{\Sigma})_{\mu \nu} 
 \end{pmatrix}.
        \end{aligned}
 \end{eqnarray}
On-shell, $G_{\Lambda \mu \nu}$ are identified with the field strengths $F_{\Lambda \mu \nu}$ of the magnetic vector fields $A_{\Lambda \mu}$. 
Therefore, by comparing the above matrix equation with the relations \eqref{dualeqel} and \eqref{dualeqmag}, we find that the matrix ${\cal M}_{{\cal M}{\cal N}}$ decomposes as 
\begin{equation}
\label{Mmatrixapp}
    {\cal M}_{\mathcal{M} \mathcal{N}} = 
    \begin{pmatrix}
        {\cal M}_{\Lambda \Sigma} & {{\cal M}_{\Lambda}}^{\Sigma} \\ 
        {{\cal M}^{\Lambda}}_{\Sigma} &{\cal M}^{\Lambda \Sigma}
    \end{pmatrix}=
    \begin{pmatrix}
    -({\cal I}+{\cal R}{\cal I}^{-1}{\cal R})_{\Lambda \Sigma} & {({\cal R} {\cal I}^{-1})_\Lambda}^{\Sigma} \\
    {({\cal I}^{-1}{\cal R})^\Lambda}_\Sigma & - ({\cal I}^{-1})^{\Lambda \Sigma} 
    \end{pmatrix}, 
\end{equation}
implying
\begin{align}
    \label{Iinvapp}
   ( {\cal I}^{-1})^{\Lambda \Sigma} &= - {\Pi^{\Lambda}}_{\mathcal{M}} {\Pi^{\Sigma}}_{\mathcal{N}} {\cal M}^{\mathcal{M} \mathcal{N}}\,, \\[2mm]
   \label{RIinvapp}
    {({\cal R} {\cal I}^{-1})_\Lambda}^{\Sigma} & = - \Pi_{\Lambda \mathcal{M}} {\Pi^{\Sigma}}_{\mathcal{N}} {\cal M}^{\mathcal{M} \mathcal{N}}\,, \\[2mm]
    \label{IinvRapp}
    {({\cal I}^{-1}{\cal R})^\Lambda}_\Sigma &= - {\Pi^{\Lambda}}_{\mathcal{M}} \Pi_{\Sigma \mathcal{N}} {\cal M}^{\mathcal{M} \mathcal{N}}\,, \\[2mm]
    \label{I + RIinvRapp}
    ({\cal I}+{\cal R}{\cal I}^{-1}{\cal R})_{\Lambda \Sigma} & = -  \Pi_{\Lambda \mathcal{M}} \Pi_{\Sigma \mathcal{N}} {\cal M}^{\mathcal{M} \mathcal{N}} .
    \end{align}
Furthermore, we have that  
\begin{align}
    \label{OL}
    O_{\Lambda \mu \nu} = & \, {\cal I}_{\Lambda \Sigma} {\Pi^{\Sigma}}_{M \alpha} \big{(} - 2 ({\cal V}^\alpha)^* L^{Mij} {\bar{\psi}}_{i \mu} \psi_{j \nu} - i \epsilon_{\mu \nu \rho \sigma} ({\cal V}^\alpha)^* L^{Mij}  {\bar{\psi}}^{\rho}_i {\psi}_j^{\sigma} \nonumber \\[2mm]
    & + {\mathcal{V}}^{\alpha}  L^{Mij} {\bar{\lambda}}_{\underline{a} i} \gamma_{\mu \nu}  \lambda^{\underline{a}}_j  - {\mathcal{V}}^{\alpha} L^{M \underline{a}} {\bar\chi}_i \gamma_{\mu \nu} \lambda^i_{\underline{a}}  + 2 ({\cal V}^\alpha)^* {L^M}_{ij} {\bar\chi}^i \gamma_{[\mu} \psi^j_{\nu]} \nonumber \\[2mm]  
    & + i  \epsilon_{\mu \nu \rho \sigma} ({\cal V}^\alpha)^* {L^M}_{ij} {\bar\chi}^i \gamma^{\rho} \psi^{j \sigma} + 2 {\cal V}^\alpha L^{M \underline{a}} {\bar\lambda}_{\underline{a}i} \gamma_{[\mu} \psi_{\nu]}^i \\[2mm] 
    & + i \epsilon_{\mu \nu \rho \sigma} {\cal V}^\alpha  L^{M \underline{a}} {\bar\lambda}_{\underline{a}i} \gamma^{\rho} \psi^{i \sigma} + \text{c.c.} \big{)} \nonumber
\end{align}
and 
\begin{align}
\label{OL-*OL}
O_{\Lambda \mu \nu} - {{({\cal R} {\cal I}^{-1})}_{\Lambda}}^{\Sigma} (*O_{\Sigma})_{\mu \nu} = & \, {\Pi}_{\Lambda M \alpha} \big{(} - 2 i ({\cal V}^\alpha)^* L^{Mij} {\bar{\psi}}_{i \mu} \psi_{j \nu} + \epsilon_{\mu \nu \rho \sigma} ({\cal V}^\alpha)^* L^{Mij}  {\bar{\psi}}^{\rho}_i {\psi}_j^{\sigma} \nonumber \\[2mm]
    & - i  {\mathcal{V}}^{\alpha}  L^{Mij} {\bar{\lambda}}_{\underline{a} i} \gamma_{\mu \nu}  \lambda^{\underline{a}}_j  - i {\mathcal{V}}^{\alpha} L^{M \underline{a}} {\bar\chi}_i \gamma_{\mu \nu} \lambda^i_{\underline{a}}  + 2 i ({\cal V}^\alpha)^* {L^M}_{ij} {\bar\chi}^i \gamma_{[\mu} \psi^j_{\nu]} \nonumber \\[2mm]
    & -  \epsilon_{\mu \nu \rho \sigma} ({\cal V}^\alpha)^* {L^M}_{ij} {\bar\chi}^i \gamma^{\rho} \psi^{j \sigma} + 2 i {\cal V}^\alpha L^{M \underline{a}} {\bar\lambda}_{\underline{a}i} \gamma_{[\mu} \psi_{\nu]}^i \\[2mm] 
    & - \epsilon_{\mu \nu \rho \sigma} {\cal V}^\alpha  L^{M \underline{a}} {\bar\lambda}_{\underline{a}i} \gamma^{\rho} \psi^{i \sigma} + \text{c.c.} \big{)}  .\nonumber
\end{align}
Consistency of \eqref{OL-*OL} with \eqref{OL} requires the complex kinetic matrix ${\cal N}_{\Lambda \Sigma}$ to satisfy
\begin{align}
    {\cal N}_{\Lambda \Sigma} {{\Pi}^{\Sigma}}_{M \alpha} {\cal V}^{\alpha} L^{Mij} & = {\Pi}_{\Lambda M \alpha} {\cal V}^{\alpha} L^{Mij} ,\\[2mm]
   {\cal N}_{\Lambda \Sigma} {{\Pi}^{\Sigma}}_{M \alpha}  ({\cal V}^{\alpha})^* L^{M \underline{a}} &= {\Pi}_{\Lambda M \alpha} ({\cal V}^{\alpha})^* L^{M \underline{a}}.
\end{align}

In addition, by multiplying equation \eqref{moddual} by ${{\Pi}^{\Lambda}}_{M \alpha}$ and using \eqref{Mmatrixapp}, we can express the inner components ${\cal F}_{\Lambda ab} = \Pi_{\Lambda M \alpha} {\cal F}^{M \alpha}_{ab}$ of the supercurvatures ${\cal F}_{\Lambda} = \Pi_{\Lambda M \alpha}{\cal F}^{M \alpha}$ of the magnetic super-one-forms $A_{\Lambda} = \Pi_{\Lambda M \alpha} A^{M \alpha}$ in terms of the inner components ${\cal F}^{\Lambda}_{ab} = {{\Pi}^{\Lambda}}_{M \alpha} {\cal F}^{M \alpha}_{ab} $ of the supercurvatures ${\cal F}^{\Lambda} = {{\Pi}^{\Lambda}}_{M \alpha} {\cal F}^{M \alpha}$ of the electric super-one-forms $A^{\Lambda} = {{\Pi}^{\Lambda}}_{M \alpha} A^{M \alpha}$. 
The result is
\begin{equation}
    \label{MagEl}
    {\cal F}_{\Lambda ab} = - \frac{1}{2} \epsilon_{abcd} {\cal I}_{\Lambda \Sigma} {\cal F}^{\Sigma cd} + {\cal R}_{\Lambda \Sigma} {\cal F}^{\Sigma}_{ab} \, .
\end{equation}
Using the above equation and \eqref{PP=C}, we can express all the terms in the rheonomic parametrizations of the fermionic supercurvatures and the superspace equations of motion for the fermions that contain 
${\cal F}^{M \alpha}_{ab}$ solely in terms of ${\cal F}^{\Lambda}_{ab}$.  
We find that those terms can be written as 
\begin{align}
        \label{Vel}
         V_i \supset & - \frac{i}{4} L_{Mij} \mathcal{V}_\alpha^* {\cal F}^{M \alpha}_{ab}\gamma^{ab} \psi^j \nonumber \\[2mm]
    = & - \frac{i}{4} {\Pi}_{\Lambda M \alpha} {L^M}_{ij} ({\cal V}^\alpha)^* {\mathcal{F}}^{\Lambda}_{ab} \gamma^{ab} \psi^j  + \frac{i}{4} {\cal N}_{\Lambda \Sigma} {{\Pi}^{\Lambda}}_{M \alpha} {L^M}_{ij} ({\cal V}^\alpha)^* {\mathcal{F}}^{\Sigma}_{ab} \gamma^{ab} \psi^j, \\[2mm]
    \label{Lel}
    \Lambda_{\underline{a} i} \supset & \, \frac{i}{8} L_{M \underline{a}} \mathcal{V}_\alpha^*  {\cal F}^{M \alpha}_{ab}\gamma^{ab} \psi_i \nonumber \\[2mm]
    =  & \, \frac{i}{8} {\Pi}_{\Lambda M \alpha} {L^M}_{\underline{a}} ({\cal V}^\alpha)^* {\mathcal{F}}^{\Lambda}_{ab} \gamma^{ab} \psi_i - \frac{i}{8} {\bar {\cal N}}_{\Lambda \Sigma} {{\Pi}^{\Lambda}}_{M \alpha}  {L^M}_{\underline{a}} ({\cal V}^\alpha)^* {\mathcal{F}}^{\Sigma}_{ab} \gamma^{ab} \psi_i, \\[2mm]
    \label{rhoel}
    \nonumber \rho_i \supset & - \frac{i}{8} L_{Mij} \mathcal{V}_\alpha {\cal F}^{M \alpha}_{bc} \gamma^{bc} \gamma_a \psi^j \wedge e^a \\[2mm]
    = & \, - \frac{i}{8} {\Pi}_{\Lambda M \alpha} {L^M}_{ij} {\cal V}^{\alpha}  {\mathcal{F}}^{\Lambda}_{bc} \gamma^{bc} \gamma_a \psi^j \wedge e^a \\ & + \frac{i}{8} {\bar{\cal N}}_{\Lambda \Sigma} {{\Pi}^{\Lambda}}_{M \alpha}  {L^M}_{ij} {\cal V}^{\alpha}  {\mathcal{F}}^{\Sigma}_{bc} \gamma^{bc} \gamma_a \psi^j \wedge e^a, \nonumber \\[2mm] \nonumber
    \label{gammaVel}
    \gamma^a V_{ia} \supset & \, \frac{i}{4} \mathcal{V}_{\alpha}^* L_{M \underline{a}} {\cal F}^{M \alpha}_{ab} \gamma^{ab} \lambda^{\underline{a}}_i \\[2mm]
    = & \, \frac{i}{4} {\Pi}_{\Lambda M \alpha} {L^M}_{\underline{a}} ({\cal V}^\alpha)^* {\mathcal{F}}^{\Lambda}_{ab} \gamma^{ab}  \lambda^{\underline{a}}_i -  \frac{i}{4} {\bar{\cal N}}_{\Lambda \Sigma}  {{\Pi}^{\Lambda}}_{M \alpha} {L^M}_{\underline{a}} ({\cal V}^\alpha)^* {\mathcal{F}}^{\Sigma}_{ab} \gamma^{ab} \lambda^{\underline{a}}_i, \\[2mm]  
    \label{gammaLel}
    \nonumber
     \gamma^a \Lambda_{\underline{a} i a} \supset & \, \frac{i}{4} \mathcal{V}_{\alpha}^* L_{M ij}  {\cal F}^{M \alpha}_{ab} \gamma^{ab} \lambda_{\underline{a}}^j + \frac{i}{8} \mathcal{V}_{\alpha} L_{M \underline{a}} {\cal F}^{M \alpha}_{ab} \gamma^{ab} \chi_i \\[2mm]
     = & \,   \frac{i}{4} {\Pi}_{\Lambda M \alpha} {L^M}_{ij} ({\cal V}^{\alpha})^*  {\mathcal{F}}^{\Lambda}_{ab} \gamma^{ab} \lambda_{\underline{a}}^j -  \frac{i}{4} {\cal N}_{\Lambda \Sigma}  {{\Pi}^{\Lambda}}_{M \alpha} {L^M}_{ij} ({\cal V}^{\alpha})^* {\mathcal{F}}^{\Sigma}_{ab} \gamma^{ab} \lambda_{\underline{a}}^j \\[2mm]
     \nonumber & + \frac{i}{8} {\Pi}_{\Lambda M \alpha}  {L^M}_{\underline{a}} {\cal V}^\alpha {\mathcal{F}}^{\Lambda}_{ab} \gamma^{ab} \chi_i -  \frac{i}{8}  {\cal N}_{\Lambda \Sigma} {{\Pi}^{\Lambda}}_{M \alpha}  {L^M}_{\underline{a}} {\cal V}^\alpha  {\mathcal{F}}^{\Sigma}_{ab} \gamma^{ab} {\chi}_i, \\[2mm]
     \label{gammarhoel}
     \gamma^b \rho_{iba} \supset & - \frac{i}{8} {\cal V}_{\alpha} L_{M \underline{a}} {\cal F}^{M \alpha}_{bc}\gamma^{bc} \gamma_a \lambda^{\underline{a}}_i + \frac{i}{8} {\cal V}_{\alpha}^* L_{Mij}  {\cal F}^{M \alpha}_{bc}\gamma^{bc} \gamma_a \chi^j \nonumber \\[2mm]
     = &  - \frac{i}{8} {\Pi}_{\Lambda M \alpha}  {\cal V}^{\alpha} {L^M}_{\underline{a}} {\mathcal{F}}^{\Lambda}_{bc} \gamma^{bc}  \gamma_a \lambda^{\underline{a}}_i + \frac{i}{8} {\cal N}_{\Lambda \Sigma} {{\Pi}^{\Lambda}}_{M \alpha}  {\cal V}^{\alpha} {L^M}_{\underline{a}} {\mathcal{F}}^{\Sigma}_{bc} \gamma^{bc}  \gamma_a \lambda^{\underline{a}}_i \\[2mm]
     & +  \frac{i}{8} {\Pi}_{\Lambda M \alpha}  {L^M}_{ij} ({\cal V}^\alpha)^* {\mathcal{F}}^{\Lambda}_{bc} \gamma^{bc}  \gamma_a  \chi^j - \frac{i}{8} {\cal N}_{\Lambda \Sigma} {{\Pi}^{\Lambda}}_{M \alpha}   {L^M}_{ij}  ({\cal V}^\alpha)^* {\mathcal{F}}^{\Sigma}_{bc} \gamma^{bc}  \gamma_a \chi^j. \nonumber
\end{align} 

{}From the rheonomic parametrizations of the supercurvatures, we can also determine the ${\cal N}=4$ local supersymmetry transformation laws for the spacetime fields of the ungauged theory. 
We recall that, from the superspace point of view, a local supersymmetry transformation parametrized by left-handed Weyl spinors $\epsilon^i$ and their charge conjugates $\epsilon_i$ is a Lie derivative $\ell_{\epsilon}$ along the tangent vector 
\begin{equation}
    \label{tangent}    
   \epsilon = {\bar\epsilon}^i D_i +   {\bar\epsilon}_i D^i , 
\end{equation}
where the basis tangent vectors $D_i, \, D^i$ are dual to the gravitino super-one-forms:
\begin{equation}
    \label{Ddualpsi}
    D_{i\alpha} \left({\bar\psi}^{j \beta}\right) = D^j_{\alpha} \left( {\bar\psi}^{\beta}_i\right) = \delta^j_i \delta^{\beta}_{\alpha}\, ,
\end{equation}
where $\alpha,\beta$ are spinor indices. The above equation implies that $i_{\epsilon} \psi^i = \epsilon^i$ and $i_{\epsilon} \psi_i = \epsilon_i$. 

For the super-one-forms $e^a$, $\psi_i$ and $A^{M \alpha}$ we have 
\begin{align}
\ell_{\epsilon} e^a &=   i_{\epsilon} T^a + {\bar\epsilon}^i \gamma^a \psi_i + {\bar\epsilon}_i \gamma^a \psi^i, \\[2mm]
\ell_{\epsilon} \psi_i &=  D \epsilon_i + i_{\epsilon} {\rho}_i, \\[2mm]
\ell_{\epsilon} A^{M\alpha} & = i_{\epsilon} {\cal F}^{M \alpha} + 2 ({\cal V}^\alpha)^* L^{Mij} {\bar\epsilon}_i \psi_j + 2 {\cal V}^\alpha {L^M}_{ij} {\bar\epsilon}^i \psi^j  , 
\end{align}
where we have used the definitions of the supercurvatures $T^a$, ${\rho}_i$ and ${\cal F}^{M \alpha}$ and 
\begin{equation}
    \label{Depsilon}
   D \epsilon_i \equiv d \epsilon_i + \frac{1}{4} \omega_{ab} \gamma^{ab} \epsilon_i - \frac{i}{2} {\cal A}  \epsilon_i - {\omega_i}^j \epsilon_j.
\end{equation}
For the super-zero-forms, which we denote for short by ${\nu}^I \equiv ({\cal V}_\alpha, {\cal V}_\alpha^*, L_{Mij}, L_{M\underline{a}},\chi^i,\chi_i, \\ \lambda^i_{\underline{a}},\lambda_{\underline{a}i})$, we have the simpler result
\begin{equation}
    \ell_{\epsilon} {\nu}^I = (i_{\epsilon} d + d i_{\epsilon}){\nu}^I = i_{\epsilon} D {\nu}^I.
\end{equation}
Using the parametrizations given for the supercurvatures and identifying the local supersymmetry transformation $\delta_{\epsilon}$ of each spacetime $p$-form with the restriction of the Lie derivative $ \ell_{\epsilon}$ of the corresponding super-$p$-form to spacetime, it is now straightforward to derive the ${\cal N}=4$ local supersymmetry transformations of all the spacetime fields.
The corresponding formulae are
\begin{align}
    \label{dV}
    \delta_{\epsilon} {\cal V}_\alpha = \, & {\cal V}_{\alpha}^* {\bar\epsilon}_i \chi^i\,, \\
    \label{dLij}
     \delta_{\epsilon} L_{Mij} = \, & L_{M \underline{a}} ( 2 {\bar\epsilon}_{[i} \lambda^{\underline{a}}_{j]} + \epsilon_{ijkl} {\bar\epsilon}^k \lambda^{\underline{a}l} )\,, \\
     \label{dLa}
     \delta_{\epsilon} {L_M}^{\underline{a}} = \, & 2 {L_M}^{ij} {\bar\epsilon}_i \lambda^{\underline{a}}_j + c.c.\,, \\
     \label{dchi}
      \delta_{\epsilon} \chi_i =  & - \frac{1}{2} {\cal I}_{\Lambda \Sigma} {{\Pi}^{\Lambda}}_{M \alpha} ({\cal V}^\alpha)^* {L^M}_{ij} {\hat{ \cal F}}^{\Sigma}_{\mu \nu} \gamma^{\mu \nu} \epsilon^j \nonumber \\
            & + \gamma^\mu \epsilon_i ( { P}_{\mu}^* - {\bar\chi}_j \psi^j_\mu  ) - ({\bar\lambda}_{\underline{a}i} \lambda^{\underline{a}}_j) \epsilon^j\,, \\
            \label{dlambda}
      \delta_{\epsilon} \lambda_{\underline{a}i} = & - \frac{1}{4} {\cal I}_{\Lambda \Sigma} {{\Pi}^{\Lambda}}_{M \alpha} ({\cal V}^\alpha)^* {L^M}_{\underline{a}} {\hat{ \cal F}}^{\Sigma}_{\mu \nu} \gamma^{\mu \nu} \epsilon_i \nonumber \\
      & - \gamma^\mu \epsilon^j ( {P}_{\underline{a}ij\mu} + 2 {\bar\lambda}_{\underline{a}[i} \psi_{j]\mu} 
+ \epsilon_{ijkl} {\bar\lambda}^k_{\underline{a}} \psi^l_\mu ) \\
 & + (\bar{\chi}_i \lambda^j_{\underline{a}})\epsilon_j - \frac{1}{2} (\bar{\chi}_j \lambda^j_{\underline{a}}) \epsilon_i\,, \nonumber \\
 \label{dea}
 \delta_{\epsilon} e^a_\mu = \, & {\bar\epsilon}^i \gamma^a \psi_{i \mu} + {\bar\epsilon}_i \gamma^a \psi^i_\mu\,, \\
 \label{dAMa}
  \delta_{\epsilon} A^{M \alpha}_\mu = \, & ({\cal V}^\alpha)^* {L^M}_{ij} {\bar\epsilon}^i \gamma_\mu \chi^j - {\cal V}^\alpha L^{M \underline{a}} {\bar\epsilon}^i \gamma_\mu \lambda_{\underline{a}i} + 2 {\cal V}^\alpha {L^M}_{ij} {\bar\epsilon}^i \psi^j_\mu + c.c.\,, \\
  \label{dpsi}
  \delta_{\epsilon} \psi_{i \mu} = \, & {D}_{\mu} \epsilon_i + \frac{1}{4} {\cal I}_{\Lambda \Sigma} {{\Pi}^{\Lambda}}_{M \alpha} {\cal V}^\alpha {L^M}_{ij}  {\hat{ \cal F}}^{\Sigma}_{\nu \rho} \gamma^{\nu \rho} \gamma_\mu \epsilon^j - \frac{1}{4} \epsilon_{ijkl} ({\bar\lambda}^j_{\underline{a}} \gamma_{\mu \nu} \lambda^{\underline{a}k})\gamma^\nu \epsilon^l \nonumber \\
  & + \frac{1}{4} (\bar\chi_i \gamma_\mu \chi^j)\epsilon_j  -  \frac{1}{4} (\bar\chi_j \gamma_\mu \chi^j)\epsilon_i - \frac{1}{4} (\bar\chi_i \gamma^\nu \chi^j)\gamma_{\mu \nu} \epsilon_j \nonumber \\
  & + \frac{1}{8} (\bar\chi_j \gamma^\nu \chi^j)\gamma_{\mu \nu} \epsilon_i + \frac{1}{2} ({\bar\lambda}^{\underline{a}}_i \gamma_\mu \lambda^j_{\underline{a}}) \epsilon_j - \frac{1}{2} ({\bar\lambda}^{\underline{a}}_i \gamma^\nu \lambda^j_{\underline{a}}) \gamma_{\mu\nu} \epsilon_j \\
  & + \frac{1}{4}  ({\bar\lambda}^{\underline{a}}_j \gamma^\nu \lambda^j_{\underline{a}}) \gamma_{\mu\nu} \epsilon_i - \epsilon_{ijkl} \chi^j {\bar\epsilon}^k \psi^l_\mu  \nonumber \, ,
\end{align}
where ${ P}_\mu$ and ${ P}_{\underline{a} ij \mu}$ are the components of the spacetime one-forms ${P}$ and ${P}_{\underline{a}ij}$ respectively, i.e.  ${P} = { P}_\mu d x^\mu$ and ${P}_{\underline{a}ij} = {P}_{\underline{a} ij \mu} d x^\mu$, ${\hat{ \cal F}}^{\Lambda}_{\mu \nu} = {\Pi^{\Lambda}}_{M \alpha} {\hat{ \cal F}}^{M \alpha}_{\mu \nu}$ and 
\begin{equation}
    \label{Dmepsilon}
    {D}_\mu \epsilon_i \equiv \partial_\mu \epsilon_i + \frac{1}{4} \omega_{\mu a b} (e,\psi) \gamma^{ab} \epsilon_i - \frac{i}{2} {\cal A}_\mu \epsilon_i - {\omega}_{i \hspace{0.1cm} \mu}^{\hspace{0.1cm}j} \epsilon_j, 
\end{equation} 
where
\begin{align}
    {\omega_\mu}^{ab} (e,\psi) = & \, 2 e^{\nu [a} \partial_{[\mu} e^{b]}_{\nu]} - e^{\nu [a} e^{b]\rho} e_{c \mu} \partial_\nu e^c_\rho \nonumber \\[2mm]
    & + {\bar\psi}^i_\mu \gamma^{[a} \psi^{b]}_i +  {\bar\psi}^{i [a} \gamma^{b]} \psi_{i \mu} + {\bar\psi}^{i [a} \gamma_\mu \psi^{b]}_i
\end{align}
is the solution for the spin connection ${\omega_\mu}^{ab}$ of the restriction of the constraint $T^a=0$ to spacetime.

The terms in the local supersymmetry transformations of the fermions that contain ${\hat{ \cal F}}^{\Lambda}_{\mu \nu}$ can also be written in a manifestly SL(2,${\mathbb R}$) $\times$ SO(6,$n$)-covariant form as 
\begin{align}
    \delta_\epsilon \chi_i \supset &  - \frac{1}{2} {\cal I}_{\Lambda \Sigma} {{\Pi}^{\Lambda}}_{M \alpha} ({\cal V}^\alpha)^* {L^M}_{ij} {\hat{ \cal F}}^{\Sigma}_{\mu \nu} \gamma^{\mu \nu} \epsilon^j \nonumber \\[2mm]
    = & - \frac{i}{4} {\cal V}_{\alpha}^* L_{M ij} G^{M \alpha}_{\mu \nu} \gamma^{\mu \nu} \epsilon^j + \gamma_{\mu \nu} \epsilon^j {\bar\chi}_{[i} \gamma^\mu \psi^\nu_{j]} - \frac{1}{2} \epsilon_{ijkl} \gamma^{\mu \nu} \epsilon^j {\bar\psi}^k_\mu \psi^l_\nu ,\\[2mm]
    \delta_\epsilon \lambda_{\underline{a} i} \supset &  - \frac{1}{4} {\cal I}_{\Lambda \Sigma} {{\Pi}^{\Lambda}}_{M \alpha} ({\cal V}^\alpha)^* {L^M}_{\underline{a}} {\hat{ \cal F}}^{\Sigma}_{\mu \nu} \gamma^{\mu \nu} \epsilon_i \nonumber \\[2mm]
    = & \,\frac{i}{8} {\cal V}_{\alpha}^* L_{M \underline{a}} G^{M \alpha}_{\mu \nu} \gamma^{\mu \nu} \epsilon_i + \frac{1}{2} \gamma^{\mu \nu} \epsilon_i {\bar\lambda}_{\underline{a}j} \gamma_\mu \psi^j_\nu, \\[2mm]
     \delta_\epsilon \psi_{i \mu} \supset & \,  \frac{1}{4} {\cal I}_{\Lambda \Sigma} {{\Pi}^{\Lambda}}_{M \alpha} {\cal V}^\alpha {L^M}_{ij}  {\hat{ \cal F}}^{\Sigma}_{\nu \rho} \gamma^{\nu \rho} \gamma_\mu \epsilon^j \nonumber \\[2mm]
     = & - \frac{i}{8} {\cal V}_\alpha L_{M ij} G^{M \alpha}_{\nu \rho} \gamma^{\nu \rho} \gamma_\mu  \epsilon^j + \frac{1}{2} \gamma^{\nu \rho} \gamma_\mu  \epsilon^j {\bar\psi}_{i \nu} \psi_{j \rho} - \frac{1}{4} \epsilon_{ijkl} \gamma^{\nu \rho} \gamma_\mu  \epsilon^j {\bar\chi}^k \gamma_\nu \psi^l_\rho \, ,
\end{align}
where we have introduced the symplectic vector $G^{M\alpha}_{\mu \nu} = (F^{\Lambda}_{\mu \nu}, G_{\Lambda \mu \nu})$.

Using the rheonomic approach, one can also derive the ungauged Lagrangian for the $D=4$, ${\cal N}=4$ Poincar\'{e} supergravity, coupled to $n$ vector multiplets.
In this formalism, the action is obtained by integrating a Lagrangian $\mathcal{L}$ that is a four-form in ${\cal N}=4$ superspace on a four-dimensional bosonic hypersurface ${\cal M}^4$ embedded in superspace, 
\begin{equation}
    \label{Srheon}
    S = \int_{{\cal M}^4 \subset {\cal S} {\cal M}} {\cal L} \, ,  
\end{equation}
where ${\cal S}{\cal M}$ is the ${\cal N} = 4 $ superspace manifold. 
The super-four-form Lagrangian has to be constructed using only differential super-$p$-forms, wedge products among them and their exterior $d$ derivatives, while it must not contain the Hodge duality operator. 
These requirements ensure that ${\cal L}$ is independent of the choice of hypersurface ${\cal M}^4$ and invariant under general coordinate transformations in superspace (superdiffeomorphisms). 
The action \eqref{Srheon} is a functional both of the super-$p$-forms appearing in ${\cal L}$ and of the hypersurface ${\cal M}^4$ on which the integration is performed and one must in principle vary the action with respect to both of them to derive the equations of motion implied by the variational principle $\delta S = 0$. 
However, the variation of ${\cal M}^4$ can be ignored, because any deformation of ${\cal M}^4$ can be compensated by a superdiffeomorphism, which leaves ${\cal L}$ invariant. 
As a result, the hypersurface ${\cal M}^4$ can be chosen arbitrarily and the complete set of variational equations associated with the action \eqref{Srheon} is given by the usual equations of motion obtained by varying $S$ with respect to the various super-$p$-forms on which ${\cal L}$ depends, while keeping the hypersurface ${\cal M}^4$ fixed. These super-$(4-p)$-form equations hold not only on ${\cal M}^4$ but on the whole ${\cal N}=4$ superspace. 

The aforementioned superspace equations can be analyzed along the intrinsic bases of $(4-p)$-forms in superspace, where $p=0,1$, built out of the supervielbein $\{ e^a, \psi^i, \psi_i \}$ by means of the wedge product. 
It turns out that the analysis of these equations of motion along the basis elements that contain only the bosonic vielbein $e^a$ gives dynamical equations for the inner components of the supercurvatures, which must coincide with the corresponding equations implied by the Bianchi identities (equations \eqref{choeom}-\eqref{graveom}). 
The projections of these equations on spacetime are the ordinary spacetime equations of motion of the theory. 
On the other hand, the analysis of the variational equations associated with \eqref{Srheon} along the basis elements featuring at least one of $\psi^i$, $\psi_i$ gives algebraic relations that express the outer components of the supercurvatures in terms of their inner components and the physical superfields (rheonomy principle). 
The outer components of the supercurvatures obtained from the variational principle must be the same as those determined by requiring closure of the Bianchi identities. 

In order to construct the superspace four-form Lagrangian for the ungauged $D=4$, ${\cal N}=4$ matter-coupled supergravity in an arbitrary symplectic frame, we follow the building rules given in volume 2 of \cite{Castellani:1991eu}. 
We first write down an ansatz for the super-four-form Lagrangian in the form of a sum of terms with undetermined coefficients. 
Each of these terms must be invariant under local Lorentz, SO(2), SU(4) and SO($n$) transformations and must have the same scaling behaviour as the Einstein-Hilbert term,
\begin{equation}
    \label{EH}
    {\cal L} \supset \frac{1}{4} \epsilon_{abcd} R^{ab} \wedge e^c \wedge e^d , 
\end{equation}
which scales as $\lambda^2$. Also, from the super-one-forms $A^{M \alpha} = (A^{\Lambda},A_{\Lambda})$, only the electric ones $A^{\Lambda}$ must appear in the superspace Lagrangian. The most general expression for the superspace four-form Lagrangian has the form 
\begin{equation}
\label{Lang}
    \mathcal{L} = {\cal L}_{\text{kin}}  + {\cal L}_{\text{Pauli}}+ {\cal L}_{\text{torsion}} +  {\cal L}_{\text{4fermi}} \, , 
\end{equation}
where 
\begin{align}
    \label{Lkin}
    {\cal L}_{\text{kin}} = & \, \frac{1}{4} \epsilon_{abcd} R^{ab} \wedge e^c \wedge e^d + (k_1 {\bar\psi}_i \wedge \gamma_a {\rho}^i + k_1^* {\bar\psi}^i \wedge \gamma_a {\rho}_i   ) \wedge e^a \nonumber \\[2mm]
    & + \epsilon_{abcd} (k_2 {\bar\chi}_i \gamma^a V^i + k_2^*{\bar\chi}^i \gamma^a V_i + k_3 {\bar\lambda}^{\underline{a}}_i \gamma^a {\Lambda}^i_{\underline{a}} + k_3^*  {\bar\lambda}_{\underline{a}}^i \gamma^a {\Lambda}_i^{\underline{a}}  ) \wedge e^b \wedge e^c \wedge e^d \nonumber \\[2mm]
    & + k_4 \epsilon_{abcd}  S_e^* S^e e^a \wedge e^b \wedge e^c \wedge e^d \nonumber \\[2mm]
    & - 4 k_4 \epsilon_{abcd} \big{[} (S^a)^* (P- {\bar\chi}^i \psi_i) + S^a (P^* -  {\bar\chi}_i \psi^i)\big{]} \wedge e^b \wedge e^c \wedge e^d \nonumber \\[2mm]
    & + k_5 \epsilon_{abcd} R_{\underline{a} ij e}  R^{\underline{a} ij e}  e^a \wedge e^b \wedge e^c \wedge e^d \nonumber \\[2mm]
    & - 8 k_5 \epsilon_{abcd} {R_{\underline{a} i j}}^a ( P^{\underline{a} i j} - 2 {\bar\psi}^i \lambda^{\underline{a}j} - \epsilon^{ijkl} {\bar\psi}_k \lambda^{\underline{a}}_l) \wedge e^b \wedge e^c \wedge e^d  \\[2mm]
    & + \epsilon_{abcd} (k_6 {\bar {\cal N}}_{\Lambda \Sigma} {\cal J}^{\Lambda +}_{ef} {\cal J}^{\Sigma + ef} + k_6^* {\cal N}_{\Lambda \Sigma}  {\cal J}^{\Lambda -}_{ef} {\cal J}^{\Sigma - ef}  )  e^a \wedge e^b \wedge e^c \wedge e^d \nonumber \\[2mm]
    & - 48 i (k_6  {\bar{\cal N}}_{\Lambda \Sigma} {\cal J}^{\Lambda +}_{ab} - k_6^* {\cal N}_{\Lambda \Sigma} {\cal J}^{\Lambda-}_{ab}) \bigg{(} {\cal F}^{\Sigma} + \frac{1}{4} {{\Pi}^\Sigma}_{M \alpha} ({\cal V}^\alpha)^* {L^M}_{ij} {\bar\lambda}^i_{\underline{a}} \gamma_{cd} \lambda^{\underline{a}j} e^c \wedge e^d  \nonumber \\[2mm]
   & + \frac{1}{4}  {{\Pi}^\Sigma}_{M \alpha} {\cal V}^{\alpha} L^{Mij} {\bar \lambda}_{\underline{a} i} \gamma_{cd} \lambda_j^{\underline{a}} e^c \wedge e^d - \frac{1}{4}  {{\Pi}^\Sigma}_{M \alpha} {\cal V}^{\alpha}  L^{M \underline{a}} {\bar\chi}_i \gamma_{cd} \lambda^i_{\underline{a}} e^c \wedge e^d  \nonumber \\[2mm]
   & - \frac{1}{4}  {{\Pi}^\Sigma}_{M \alpha} ({\cal V}^\alpha)^*  L^{M \underline{a}} {\bar\chi}^i \gamma_{cd} \lambda_{\underline{a}i} e^c \wedge e^d - {{\Pi}^\Sigma}_{M \alpha} ({\cal V}^\alpha)^* {L^M}_{ij} {\bar\chi}^i \gamma_c \psi^j \wedge e^c \nonumber \\[2mm] & -  {{\Pi}^\Sigma}_{M \alpha} {\cal V}^{\alpha} L^{Mij}  {\bar\chi}_i \gamma_c \psi_j \wedge e^c - {{\Pi}^\Sigma}_{M \alpha} ({\cal V}^\alpha)^*  L^{M \underline{a}} {\bar\lambda}^i_{\underline{a}} \gamma_c \psi_i \wedge e^c \nonumber \\[2mm]
   & -  {{\Pi}^\Sigma}_{M \alpha} {\cal V}^{\alpha}    L^{M \underline{a}} {\bar\lambda}_{\underline{a}i} \gamma_c \psi^i \wedge e^c \bigg{)} \wedge e^a \wedge e^b, \nonumber \\[2mm]
    \label{LPauli}
    {\cal L}_{\text{Pauli}} = & \, p_1 P^* \wedge {\bar\chi}^i \gamma_{ab} {\psi}_i \wedge e^a \wedge e^b + p_2 {P_{\underline{a}}}^{ij} \wedge {\bar\lambda}^{\underline{a}}_i \gamma_{ab} \psi_j  \wedge e^a \wedge e^b \nonumber \\[2mm]
    & + p_3 {\Pi}_{\Lambda M \alpha} ({\cal V}^\alpha)^* L^{M \underline{a}} {\cal F}^{\Lambda} {\bar\chi}^i \gamma_{ab} {\lambda}_{\underline{a}i} \wedge e^a \wedge e^b \nonumber \\[2mm] 
    & + p_4  {\Pi}_{\Lambda M \alpha} {\cal V}^{\alpha} L^{Mij} {\cal F}^{\Lambda} {\bar\lambda}_{\underline{a}i} \gamma_{ab} {\lambda}^{\underline{a}}_j \wedge e^a \wedge e^b \nonumber \\[2mm]
    & + p_5 {\Pi}_{\Lambda M \alpha} ({\cal V}^\alpha)^* {L^M}_{ij}  {\cal F}^{\Lambda} \wedge {\bar\chi}^i \gamma_a \psi^j \wedge e^a \\[2mm]
    & + p_6   {\Pi}_{\Lambda M \alpha} ({\cal V}^\alpha)^* L^{M \underline{a}}  {\cal F}^{\Lambda} \wedge {\bar\lambda}_{\underline{a}}^i \gamma_a \psi_i \wedge e^a  \nonumber \\[2mm]
    & + p_7 {\Pi}_{\Lambda M \alpha} {\cal V}^{\alpha} {L^M}_{ij} {\cal F}^{\Lambda} \wedge {\bar\psi}^i \wedge \psi^j + c.c., \nonumber \\[2mm]
    \label{Ltorsion}
     {\cal L}_{\text{torsion}} = & \, t_1 {\bar\chi}_i \gamma_a \chi^i T_b \wedge e^a \wedge e^b + t_2 \epsilon_{abcd} {\bar\chi}_i \gamma^a \chi^i T^b \wedge e^c \wedge e^d \nonumber \\[2mm]
     & + t_3  {\bar\lambda}^{\underline{a}}_i \gamma_a {\lambda}_{\underline{a}}^i  T_b \wedge e^a \wedge e^b + t_4 \epsilon_{abcd}  {\bar\lambda}^{\underline{a}}_i \gamma^a {\lambda}_{\underline{a}}^i  T^b \wedge e^c \wedge e^d \\[2mm]
     & + t_5 {\bar\psi}^i \wedge \gamma_a \psi_i \wedge T^a, \nonumber \\[2mm]
     \label{L4fermi}
     {\cal L}_{\text{4fermi}} = & \,  \epsilon_{abcd} ( q_1 {\bar\chi}^i {\chi}^j {\bar\chi}_i {\chi}_j + q_2  {\bar\chi}^i {\lambda}^{\underline{a}}_j {\bar\chi}_i {\lambda}_{\underline{a}}^j + q_3  {\bar\chi}^i {\lambda}^{\underline{a}}_i  {\bar\chi}_j {\lambda}_{\underline{a}}^j  \nonumber \\[2mm]
     &+ q_4 {\bar\lambda}^{\underline{a}}_i   {\lambda}^{\underline{b}}_j  {\bar\lambda}_{\underline{a}}^i   {\lambda}_{\underline{b}}^j + q_5   {\bar\lambda}^{\underline{a}}_i \lambda_{\underline{a}j}  {\bar\lambda}_{\underline{b}}^i \lambda^{\underline{b}j} + q_6   {\bar\lambda}^{\underline{a}}_i {\lambda}^{\underline{b}}_j  {\bar\lambda}_{\underline{b}}^i  {\lambda}_{\underline{a}}^j  ) e^a \wedge e^b \wedge e^c \wedge e^d \nonumber \\[2mm]
     & + (q_7 {\bar\lambda}_{\underline{a}i} \gamma_{ab} {\lambda}^{\underline{a}}_j {\bar\chi}^i \gamma_c \psi^j + q_8  \epsilon_{abcd}  {\bar\lambda}_{\underline{a}i}  {\lambda}^{\underline{a}}_j {\bar\chi}^i \gamma^d \psi^j  + c.c. ) \wedge e^a \wedge e^b  \wedge e^c \nonumber \\[2mm]
     & + (q_9 \epsilon_{ijkl} {\bar\lambda}^{\underline{a}i} \gamma_{ab} {\lambda}_{\underline{a}}^j {\bar\psi}^k \wedge \psi^l + c.c. ) \wedge e^a \wedge e^b \nonumber \\[2mm]
     & + ( r_1 {\bar\chi}_i \gamma_a \chi^j {\bar\psi}^i \wedge \gamma_b \psi_j + r_2  {\bar\chi}_i \gamma_a \chi^i {\bar\psi}^j \wedge \gamma_b \psi_j \nonumber \\[2mm]
     & + r_3 \epsilon_{abcd}  {\bar\chi}_i \gamma^c \chi^j  {\bar\psi}^i \wedge \gamma^d \psi_j + r_4 \epsilon_{abcd}  {\bar\chi}_i \gamma^c \chi^i  {\bar\psi}^j \wedge \gamma^d \psi_j \nonumber \\[2mm]
     & + r_5 {\bar\lambda}^{\underline{a}}_i \gamma_a \lambda^j_{\underline{a}} {\bar\psi}^i \wedge \gamma_b \psi_j  + r_6 {\bar\lambda}^{\underline{a}}_i \gamma_a \lambda^i_{\underline{a}} {\bar\psi}^j \wedge \gamma_b \psi_j \nonumber \\[2mm]
     & + r_7 \epsilon_{abcd}  {\bar\lambda}^{\underline{a}}_i \gamma^c \lambda^j_{\underline{a}}  {\bar\psi}^i \wedge \gamma^d \psi_j + r_8 \epsilon_{abcd}  {\bar\lambda}^{\underline{a}}_i \gamma^c \lambda^i_{\underline{a}}  {\bar\psi}^j \wedge \gamma^d \psi_j ) \wedge e^a \wedge e^b \nonumber \\[2mm]
     & + \Big{[}  \epsilon_{abcd} {\Pi}_{\Lambda M \alpha} {{\Pi}^{\Lambda}}_{N \beta}  \big{(} r_9   ({\cal V}^\alpha)^* {\cal V}^{\beta} {L^M}_{\underline{a}} L^{Njk} {\bar\chi}^i \gamma_{ef} \lambda^{\underline{a}}_i {\bar\lambda}_{\underline{b}j} \gamma^{ef} \lambda^{\underline{b}}_k    \nonumber \\[2mm]
     & + s_1 ({\cal V}^\alpha)^* ({\cal V}^{\beta})^* {L^M}_{\underline{a}} {L^N}_{\underline{b}}   {\bar\chi}^i \gamma_{ef} \lambda^{\underline{a}}_i {\bar\chi}^j \gamma^{ef} \lambda_j^{\underline{b}} \nonumber \\[2mm]
     & + s_2 {\cal V}^{\alpha}  {\cal V}^{\beta} L^{Mij} L^{Nkl} {\bar\lambda}_{\underline{a}i} \gamma_{ef} \lambda^{\underline{a}}_j {\bar\lambda}_{\underline{b}k} \gamma^{ef} \lambda^{\underline{b}}_l \big{)}  e^a \wedge e^b \wedge e^c \wedge e^d  \nonumber \\[2mm]      & +  {\Pi}_{\Lambda M \alpha} {{\Pi}^{\Lambda}}_{N \beta}\big{(} s_3    ({\cal V}^\alpha)^* {\cal V}^{\beta} {L^M}_{\underline{a}} L^{Njk} {\bar\chi}^i \gamma_{ab} \lambda^{\underline{a}}_i {\bar\chi}_j \gamma_c \psi_k \nonumber \\[2mm]
     &+ s_4 ({\cal V}^\alpha)^* ({\cal V}^\beta)^* {L^M}_{\underline{a}} L^{N \underline{b}} {\bar\chi}^i \gamma_{ab} \lambda^{\underline{a}}_i {\bar \lambda}^j_{\underline{b}} \gamma_c \psi_j \nonumber \\[2mm]
     & + s_5  ({\cal V}^\alpha)^* ({\cal V}^\beta)^* {L^M}_{ij} {L^N}_{kl} {\bar\lambda}^k_{\underline{a}} \gamma_{ab} \lambda^{\underline{a}l} {\bar\chi}^i \gamma_c \psi^j \nonumber \\[2mm]
     & + s_6 ({\cal V}^\alpha)^* ({\cal V}^\beta)^* {L^M}_{\underline{a}} {L^N}_{jk} {\bar\chi}^i \gamma_{ab} \lambda^{\underline{a}}_i {\bar\chi}^j \gamma_c \psi^k \nonumber \\[2mm] 
     & + s_7  ({\cal V}^\alpha)^* {\cal V}^\beta {L^M}_{\underline{a}}  {L^N}_{\underline{b}} \epsilon_{abcd} {\bar\lambda}^{\underline{a}}_i \lambda^{\underline{b}}_j   {\bar\chi}^i \gamma^d \psi^j \nonumber \\[2mm]
     & + s_8  ({\cal V}^\alpha)^* {\cal V}^\beta {L^M}_{\underline{a}}  {L^N}_{\underline{b}}  {\bar\lambda}^{\underline{a}}_i \gamma_{ab} \lambda^{\underline{b}}_j {\bar\chi}^i \gamma_c \psi^j \nonumber \\[2mm]
     & + s_9 ({\cal V}^\alpha)^* {\cal V}^\beta {L^M}_{ij} L^{Nkl} {\bar\lambda}_{\underline{a} k } \gamma_{ab} \lambda^{\underline{a}}_l  {\bar\chi}^i \gamma_c \psi^j \nonumber \\[2mm]
     & + w_1  {\cal V}^{\alpha} ({\cal V}^\beta)^* L^{Mij} L^{N \underline{b}} {\bar\lambda}^{\underline{a}}_i \gamma_{ab} \lambda_{\underline{a} j } {\bar\lambda}^k_{\underline{b}} \gamma_c \psi_k \nonumber \\[2mm]
     & + w_2  {\cal V}^{\alpha} {\cal V}^\beta L^{Mij} L^{N \underline{b}} {\bar\lambda}_{\underline{a}i} \gamma_{ab} \lambda^{\underline{a}}_j {\bar\lambda}_{\underline{b}k} \gamma_c \psi^k \big{)} \wedge e^a \wedge e^b \wedge e^c \nonumber \\[2mm]
     & + {\Pi}_{\Lambda M \alpha} {{\Pi}^{\Lambda}}_{N \beta} \big{(} w_3 ({\cal V}^\alpha)^* ({\cal V}^\beta)^* {L^M}_{ij} L^{N \underline{a}}  {\bar\chi}^i \gamma_a \psi^j \wedge {\bar\lambda}^k_{\underline{a}} \gamma_b \psi_k \nonumber \\[2mm]
     & + w_4 ({\cal V}^\alpha)^* {\cal V}^\beta {L^M}_{ij} L^{Nkl} \epsilon_{abcd} {\bar\chi}^i \gamma^c \chi_k {\bar\psi}^j \wedge \gamma^d \psi_l \nonumber \\[2mm]
     & + w_5  ({\cal V}^\alpha)^* ({\cal V}^\beta)^* {L^M}_{ij} {L^N}_{kl}  {\bar\chi}^i \gamma_a \psi^j \wedge {\bar\chi}^k \gamma_b \psi^l \nonumber \\[2mm]
     & + w_6  ({\cal V}^\alpha)^* {\cal V}^\beta {L^M}_{ij} {L^N}_{\underline{a}} {\bar\chi}^i \gamma_a \psi^j \wedge {\bar\lambda}_k^{\underline{a}} \gamma_b \psi^k \nonumber \\[2mm]
     & + w_7 {\cal V}^{\alpha} {\cal V}^\beta  {L^M}_{\underline{a}} {L^N}_{\underline{b}} {\bar\lambda}^{\underline{a}}_i \gamma_a \psi^i \wedge {\bar\lambda}^{\underline{b}}_j \gamma_b \psi^j \nonumber \\[2mm]
     & + w_8  {\cal V}^{\alpha} ({\cal V}^\beta)^* {L^M}_{\underline{a}} L^{N \underline{b}} \epsilon_{abcd} {\bar\lambda}^{\underline{a}}_i \gamma^c \lambda^j_{\underline{b}} {\bar\psi}^i \wedge \gamma^d \psi_j \big{)} \wedge e^a  \wedge e^b \nonumber \\[2mm]
     & +  {\Pi}_{\Lambda M \alpha} {{\Pi}^{\Lambda}}_{N \beta}  \big{(} z_1 {\cal V}^{\alpha} {\cal V}^\beta {L^M}_{ij} {L^N}_{kl} {\bar\psi}^i \wedge \psi^j \wedge  {\bar\psi}^k \wedge \psi^l  \nonumber \\[2mm]
     & + z_2 ({\cal V}^\alpha)^*  {\cal V}^\beta  L^{Mij} {L^N}_{kl} {\bar\psi}_i \wedge \psi_j \wedge  {\bar\psi}^k \wedge \psi^l \big{)} + c.c. \Big{]},
\end{align}
where $S_a , \, R_{\underline{a}ija}  =  (R_{\underline{a}}{}^{ij}{}_a)^* = \frac{1}{2} \epsilon_{ijkl} R_{\underline{a}}{}^{kl}{}_a$ and ${\cal J}^{\Lambda}_{ab} = ({\cal J}^{\Lambda}_{ab})^*$ are auxiliary super-zero-forms which are identified, through their equations of motion, with the inner components $P_a$, $P_{\underline{a} i j a}$ and ${\cal F}^{\Lambda}_{ab}$ of the supercurvatures $P$, $P_{\underline{a}ij}$ and ${\cal F}^{\Lambda}$ respectively. 
They provide a first-order description of the kinetic terms of the bosonic superfields, which avoids the use of the Hodge duality operator, whose presence would imply a dependence of the superspace Lagrangian and the equations of motion associated with the action \eqref{Srheon} on the hypersurface of integration ${\cal M}^4$ and its metric. 

We then fix the coefficients by requiring that the equations of motion that arise from the variation of the action with respect to the super-zero-forms $S_a , \, R_{\underline{a}ija}, \, {\cal J}^{\Lambda}_{ab}$, $\chi^i$, $\lambda^{\underline{a}i}$ and the super-one-forms $\omega^{ab}$ and $\psi^i$ be solved by the constraint \eqref{Ra=0}, the rheonomic equations \eqref{P}-\eqref{rhoi} and the superspace equations of motion \eqref{choeom}-\eqref{graveom} (expressed in terms of ${\cal F}^{\Lambda}_{ab}$ only), which are obtained from the Bianchi identities. 
The results are 
\begin{align*}
    \text{Im} k_1 & = 1, \, \text{Re} k_2 = - \frac{1}{12}, \,  \text{Re} k_3 = - \frac{1}{6}, \, k_4 = \frac{1}{24}, \, k_5 = \frac{1}{48}, \, k_6 = - \frac{i}{96}, \\[2mm]
    p_1 & = - \frac{i}{2}, \, p_2 = i , \, p_3 = -\frac{1}{4} , \, p_4 = \frac{1}{4} , \, p_5 = -1, \, p_6 = - 1, \, p_7=-1, \\[2mm] 
    t_1 & = \frac{i}{4}, \, t_2 = 3i \text{Im} k_2, \, t_3 = - \frac{i}{2}, \, t_4= 3i\text{Im} k_3, \, t_5 = - \text{Re} k_1, \\[2mm]
    q_1 & = \frac{1}{64}, \, q_2=- \frac{1}{48}, \, q_3=-\frac{1}{24}, \, q_4 = - \frac{1}{48} ,  \\[2mm] 
    q_5 &= - \frac{1}{24}, \,
    q_6  = \frac{1}{12}, \, q_7 = 0, \, q_8 = \frac{1}{6}, \, q_9 = \frac{i}{4}, \\[2mm]
    r_1 & = \frac{i}{2}, \, r_2=0, \, r_3 = - i \text{Re} w_4, \, r_4 = i(\text{Re} w_4 + 3 \text{Im} k_2), \\[2mm] 
    r_5 & = i , r_6 = -i, r_7 = -2 i \text{Re} w_8, \, r_8= 3i \text{Im} k_3, \, r_9 = \frac{i}{192}, \\[2mm]
    s_1 & = - \frac{i}{384}, \, s_2  = - \frac{i}{384}, \, s_3 = \frac{1}{4}, \, s_4=\frac{1}{4}, \\[2mm]
    s_5&= - \frac{1}{4},\, s_6 = \frac{1}{4},\, s_7 = - \frac{i}{8}, \, s_8 = -\frac{1}{8}, \, s_9 = - \frac{1}{4}, \\[2mm]
    w_1 & = - \frac{1}{4}, \, w_2 = - \frac{1}{4}, \, w_3=-1 , \, \text{Im} w_4 = \frac{1}{4}, \\[2mm] 
    w_5 & = - \frac{1}{2}, \, w_6 = -1, \, w_7 = - \frac{1}{2}, \, \text{Im} w_8 = \frac{1}{4} , \\[2mm]
    z_1 & = - \frac{1}{2}, \, \text{Re} z_2 = - \frac{1}{2}, \, \text{Im} z_2 = - \text{Re} k_1\,.
    \end{align*}
 The terms that involve the undetermined Re$k_1$, Im$k_2$ and Im$k_3$ combine to a total derivative and thus do not contribute to the action \eqref{Srheon}, while those that contain Re$w_4$ and Re$w_8$ cancel. 

The spacetime Lagrangian then follows from restricting the superspace four-form Lagrangian to spacetime, that is the $\theta^i=d \theta^i=0$ hypersurface. 
In practice, one first goes to the second-order formalism by identifying the auxiliary super-zero-forms $S_a , \, R_{\underline{a}ija}$ and ${\cal J}^{\Lambda}_{ab} $ with $P_a , \, P_{\underline{a}ija}$ and ${\cal F}^{\Lambda}_{ab} $ respectively and setting $T^a=0$. 
Then, one expands all the forms along the $dx^\mu$ differentials and restricts the superfields to their lowest ($\theta^i=0$) components. Using the fact that
\begin{equation}
 \label{volform}
     dx^\mu \wedge dx^\nu \wedge dx^\rho \wedge dx^\sigma = - e \epsilon_{\mu \nu \rho \sigma} d^4 x\,,
\end{equation}
we find that the spacetime Lagrangian for the ungauged theory takes the form
\begin{equation}
 \label{Lsp}
     {\cal L}=   {\cal L}_{\text{kin}} + {\cal L}_{\text{Pauli}}   + {\cal L}_{\text{4fermi}} \, ,
 \end{equation}
 where
 \begin{align}
   \label{Lkinsp}
   e^{-1} {\cal L}_{\text{kin}} = \, &  \frac{1}{2} R + \frac{i}{2} \epsilon^{\mu \nu \rho \sigma} \left({\bar\psi}^i_\mu \gamma_{\nu} {\rho}_{i \rho \sigma} - {\bar\psi}_{i \mu} \gamma_{\nu} {\rho}^i_{\rho\sigma} \right) \nonumber \\[2mm]
 & - \frac{1}{2} \left( {\bar\chi}^i \gamma^\mu {D}_{\mu} \chi_i + {\bar\chi}_i \gamma^\mu {D}_{\mu} \chi^i \right) - \left( {\bar\lambda}^{\underline{a}}_i \gamma^\mu {D}_{\mu} \lambda^i_{\underline{a}} +  {\bar\lambda}_{\underline{a}}^i \gamma^\mu {D}_{\mu} \lambda_i^{\underline{a}} \right) \\[2mm]
   & - {P}_\mu^* {P}^\mu - \frac{1}{2} { P}_{\underline{a} i j \mu} {P}^{\underline{a} i j \mu} + \frac{1}{4} {\cal I}_{\Lambda \Sigma} F^{\Lambda}_{\mu \nu} F^{\Sigma \mu \nu} + \frac{1}{8} \epsilon^{\mu \nu \rho \sigma} {\cal R}_{\Lambda \Sigma} F^{\Lambda}_{\mu \nu} {F}^{\Sigma}_{\rho \sigma}\,, \nonumber \\[2mm]
   \label{LPaulisp}
   e^{-1} {\cal L}_{\text{Pauli}} = \, & {P}_\mu^* \left( {\bar\chi}^i \psi^{\mu}_i - {\bar\chi}^i \gamma^{\mu \nu} \psi_{i\nu} \right) + {P}_\mu \left( {\bar\chi}_i \psi^{i \mu} - {\bar\chi}_i \gamma^{\mu \nu} \psi^i_\nu \right) \nonumber \\[2mm]
   & - 2 {P}_{\underline{a} i j \mu} \left( {\bar\lambda}^{\underline{a}i} \psi^{j \mu} - {\bar\lambda}^{\underline{a}i} \gamma^{\mu \nu} \psi^j_\nu \right) - 2 {P}^{\underline{a} i j \mu} \left( {\bar\lambda}_{\underline{a}i} \psi_{j \mu} - {\bar\lambda}_{\underline{a}i} \gamma_{\mu \nu} \psi_j^\nu\right) \\[2mm]
  & + \frac{1}{2} F^{\Lambda}_{\mu \nu} O_{\Lambda}^{\mu \nu}, \nonumber 
   \end{align}
${\cal L}_{4\text{fermi}}$ is given by \eqref{L4fsp} and we have defined 
 \begin{align}
 {\rho}_{i \mu \nu} \equiv & \, 2 \partial_{[\mu|} \psi_{i|\nu]} + \frac{1}{2} {\omega_{[\mu|}}^{ab}(e,\psi) \gamma_{ab} \psi_{i|\nu]} - i {\cal A}_{[\mu|} \psi_{i|\nu]} - 2 {\omega}_{i \hspace{0.1cm} [\mu|}^{\hspace{0.1cm}j} \psi_{j|\nu]}, \\[2mm]
 {D}_{\mu} \chi_i \equiv & \, \partial_\mu \chi_i + \frac{1}{4} {\omega_{\mu}}^{ab}(e,\psi) \gamma_{ab} \chi_i + \frac{3i}{2} {\cal A}_{\mu} \chi_i -  {\omega}_{i \hspace{0.1cm} \mu}^{\hspace{0.1cm}j} \chi_j \, , \\[2mm]
 {D}_{\mu} \lambda_{\underline{a}i} \equiv & \, \partial_\mu \lambda_{\underline{a}i} + \frac{1}{4} {\omega_{\mu}}^{ab}(e,\psi) \gamma_{ab} \lambda_{\underline{a}i} + \frac{i}{2} {\cal A}_{\mu} \lambda_{\underline{a}i} -  {\omega}_{i \hspace{0.1cm} \mu}^{\hspace{0.1cm}j} \lambda_{\underline{a}j} + {\omega}_{\underline{a} \hspace{0.15cm}\mu}^{\hspace{0.15cm} \underline{b}} \lambda_{\underline{b}i} \, .
 \end{align}
The Lagrangian \eqref{Lsp} is invariant up to a total derivative under the local supersymmetry transformations \eqref{dV}-\eqref{dpsi}.

The introduction of a gauging requires the modification of the supercurvatures by promoting the exterior differentials to gauge covariant differentials and the connections to their gauged counterparts, as described in section \ref{sec:duality_covariant_gauging}, as well as the introduction of new super-two-forms $B^{MN}=B^{[MN]}$ and $B^{\alpha \beta} = B^{(\alpha \beta)}$.

The appropriate definitions for the gauged supercurvatures are the following 
\begin{align}
R^{ab}= \,  & d \omega^{ab} + \omega^{ac} \wedge {\omega_c}^b,  \\[2mm]
T^a = \, &   d e^a + {\omega^a}_b \wedge e^b - {\bar\psi}^i \wedge \gamma^a \psi_i = \hat{D} e^a  - {\bar\psi}^i \wedge \gamma^a \psi_i,  \\[2mm]
\hat{\rho}_i  = \hat{D} \psi_i = \, &  d \psi_i + \frac{1}{4} \omega^{ab} \wedge \gamma_{ab} \psi_i - \frac{i}{2} \hat{\mathcal{A}} \wedge \psi_i - {\hat{\omega}_i}{}^{j} \wedge \psi_j, \\[2mm]
{\hat V}_i = \hat{D} \chi_i = \,  & d \chi_i + \frac{1}{4} \omega^{ab} \gamma_{ab} \chi_i + \frac{3 i}{2} \hat{\mathcal{A}} \chi_i -  {\hat{\omega}_i}{}^{j} \chi_j , \\[2mm]
 {\hat \Lambda}_{\underline{a}i} = \hat{D} \lambda_{\underline{a}i}= \, & d \lambda_{\underline{a}i} + \frac{1}{4} \omega^{ab}  \gamma_{ab}  \lambda_{\underline{a}i}  + \frac{i}{2} \hat{\mathcal{A}} \lambda_{\underline{a}i} - {\hat{\omega}_i}{}^{j} \lambda_{\underline{a}j} + {\hat{\omega}_{\underline{a}}}{}^{\underline{b}} \lambda_{\underline{b}i} , \\[2mm]
 \label{Hdef}
{\cal H}^{M \alpha}  = \, & d A^{M \alpha} - \frac{g}{2} {{\hat{f}}_{\beta N P}}{}^{ M} A^{N \beta}  \wedge A^{P \alpha} - \frac{g}{2} {{\Theta}^{\alpha M}}_{NP} B^{NP} + \frac{g}{2} \xi^M_\beta B^{\alpha \beta} \nonumber \\[2mm] 
&- (\mathcal{V}^\alpha)^* L^{Mij} {\bar\psi}_i \wedge \psi_j - \mathcal{V}^\alpha  {L^M}_{ij} {\bar \psi}^i \wedge \psi^j, \\[2mm]
{\cal H}^{(3) MN} = \, & {\hat d}B^{MN}   + \epsilon_{\alpha \beta} A^{[M|\alpha} \wedge \left( dA^{|N]\beta} + \frac{g}{3} {X_{P \gamma Q \delta}}^{|N]\beta} A^{P \gamma} \wedge A^{Q \delta} \right), \\[2mm]
{\cal H}^{(3) \alpha \beta} = \, & {\hat d}B^{\alpha \beta} - \eta_{MN} A^{M(\alpha|} \wedge \left(  dA^{N|\beta)} + \frac{g}{3} {X_{P \gamma Q \delta}}^{N|\beta)} A^{P \gamma} \wedge A^{Q \delta} \right) \,, \\[2mm]
{\hat P} = \, & \frac{i}{2} \epsilon^{\alpha \beta} {\cal V}_{\alpha} {\hat d} {\cal V}_\beta, \\[2mm]
{\hat P}_{\underline{
a}ij} = \, & {L_{\underline{a}}}^M {\hat d} L_{Mij}  ,
\end{align} 
where $\hat{\mathcal{A}}$, ${\hat{\omega}_i}{}^{j}$ and ${\hat{\omega}_{\underline{a}}}{}^{\underline{b}}$ are the extensions of the gauged SO(2), SU(4) and SO($n$) connections to ${\cal N}=4$ superspace respectively and $\hat{D}$ is the exterior derivative that is covariant with respect to local Lorentz, SO(2), SU(4), SO($n$) and gauge transformations.  The definitions of the super-field strengths ${\cal H}^{(3)MN}$ and ${\cal H}^{(3)\alpha \beta}$ of the super-two-forms $B^{MN}$ and $B^{\alpha \beta}$ respectively are constructed according to the rules in \cite{deWit:2005hv}. 

By acting on the gauged supercurvatures with the exterior derivative $d$ and using the fact that $d^2=0$, we obtain the following Bianchi identities
\begin{align}
\label{DhatRab}
    {\hat D} R^{ab} =& \, 0 ,\\[2mm]
    \label{DhatRa}
{\hat D} T^a =& \,  {R^a}_b \wedge e^b +  {\bar\psi}_i \wedge \gamma^a {\hat{\rho}}^i + {\bar\psi}^i \wedge \gamma^a {\hat\rho}_i, \\[2mm]
\label{Dhatrho}
\hat{D} {\hat\rho}_i   =& \, \frac{1}{4} R^{ab} \wedge \gamma_{ab} \psi_i - \frac{i}{2} \hat{F} \wedge \psi_i - {\hat{R}_i}{}^{ j} \wedge \psi_j, \\[2mm] 
\label{DhatX}
\hat{D} \hat{V}_i   =& \, \frac{1}{4} R^{ab}  \gamma_{ab} \chi_i + \frac{3i}{2}  \hat{F} \chi_i - {\hat{R}_i}{}^{j}   \chi_j, \\[2mm] 
\label{DhatL}
\hat{D}  \hat{\Lambda}_{\underline{a}i} = & \, \frac{1}{4} R^{ab}  \gamma_{ab} \lambda_{\underline{a}i} + \frac{i}{2}  \hat{F} \lambda_{\underline{a}i} -  {\hat{R}_i}{}^{j}\lambda_{\underline{a}j} +   {{\hat{R}}_{\underline{a}}}{}^{\underline{b}} \lambda_{\underline{b}i}, \\[2mm]
\label{DhatH}
\hat{D} {\cal H}^{M \alpha} =& - \mathcal{V}^\alpha L^{Mij}  {\hat{P}}^* \wedge \bar\psi_i \wedge \psi_j - (\mathcal{V}^\alpha )^* L^{M \underline{a}} {{\hat{P}}_{\underline{a}}}^{\hspace{0.15cm} ij} \wedge \bar\psi_i \wedge \psi_j + 2 (\mathcal{V}^\alpha)^* L^{Mij}  \bar\psi_i \wedge {\hat\rho}_j \nonumber  \\[2mm] 
 & - (\mathcal{V}^\alpha)^* {L^M}_{ij}  \hat{P} \wedge \bar\psi^i \wedge \psi^j - \mathcal{V}^\alpha L^{M \underline{a}} \hat{P}_{\underline{a}ij} \wedge \bar\psi^i \wedge \psi^j + 2 \mathcal{V}^\alpha  {L^M}_{ij}  \bar\psi^i \wedge {\hat \rho}^j \\[2mm]
 & - \frac{g}{2} {{\Theta}^{\alpha M}}_{NP} {\cal H}^{(3)NP} + \frac{g}{2} \xi^M_{\beta} {\cal H}^{(3) \alpha \beta} ,  \nonumber \\[2mm] 
 \label{2formBianchi}
 -\frac{1}{2} {{\Theta}^{\alpha M}}_{NP} & \hat{D} {\cal H}^{(3)NP} +  \frac{1}{2} \xi^M_\beta \hat{D} {\cal H}^{(3) \alpha \beta} =  \nonumber {X_{N \beta P \gamma}}^{M \alpha} \Big{[} {\cal H}^{N \beta} + ({\cal V}^\beta)^* L^{Nij} {\bar \psi}_i \wedge \psi_j \\[2mm]
 & + {\cal V}^\beta {L^N}_{ij} {\bar \psi}^i \wedge \psi^j \Big{]} \wedge \Big{[} {\cal H}^{P \gamma}  + ({\cal V}^\gamma)^* L^{Pkl} {\bar \psi}_k \wedge \psi_l + {\cal V}^\gamma {L^P}_{kl} {\bar \psi}^k \wedge \psi^l \Big{]} ,\\[2mm]
 \label{DhatP}
 \hat{D} \hat{P} = & \, \frac{i}{2} g \xi_{\alpha M} {\cal V}^\alpha  {\cal V}_\beta {\cal H}^{M \beta} - g \xi_{\alpha M} {\cal V}^\alpha L^{Mij} {\bar \psi}_i \wedge \psi_j , \\[2mm]
 \label{DhatPaij}
  \hat{D} {{\hat{P}}_{\underline{a} ij}}  = \, & g {{\Theta}_{\alpha M }}^{N P} L_{N \underline{a}} L_{P ij} \left[{\cal H}^{M \alpha} + ({\cal V}^\alpha)^* L^{Mkl} {\bar\psi}_k \wedge \psi_l + {\cal V}^\alpha {L^M}_{kl} {\bar\psi}^k \wedge \psi^l\right]  ,
 \end{align}
 where $\hat{F}$, ${\hat{R}_i}{}^{ j}$ and ${{\hat{R}}_{\underline{a}}}{}^{ \underline{b}}$ are the superspace gauged SO(2), SU(4) and SO($n$) curvatures respectively, given by equations \eqref{gSO(2)curv}, \eqref{gSU(4)curv} and \eqref{gSO(n)curv}, which are now to be viewed as superspace equations. 

In the same way as in the ungauged theory, the Bianchi identities \eqref{DhatRab}-\eqref{DhatPaij} can be solved by providing suitable rheonomic parametrizations of the supercurvatures.
These can be found by starting from the corresponding results for the ungauged theory and focusing on the terms proportional to the gauge coupling $g$.
The result is the following:
 \begin{align}
 \label{hatP}
     \hat{P}  = & {\hat{P}}_a e^a + {\bar\psi}_i \chi^i, \\[2mm]
 \label{hatPaij}    
{\hat{P}}_{\underline{a}ij} = & {\hat{P}}_{\underline{a}ij a} e^a + 2 {\bar{\psi}}_{[i|} \lambda_{\underline{a} |j]} + \epsilon_{ijkl} {\bar{\psi}}^k \lambda^l_{\underline{a}}, \\[2mm] \label{hatVi}
 {\hat{V}}_i  = & {\hat{V}}_{ia} e^a  - \frac{i}{4} L_{Mij} \mathcal{V}_\alpha^* {\cal H}^{M \alpha}_{ab}\gamma^{ab} \psi^j - ({\bar\lambda}_{\underline{a} i} \lambda^{\underline{a}}_j) \psi^j + \gamma^a {\hat{P}}_a^* \psi_i + \frac{2}{3} g {\bar A}_{2ij}\psi^j,
\\[2mm] \label{hatLai}
 {\hat{\Lambda}}_{\underline{a}i} = & {\hat{\Lambda}}_{\underline{a}i a}  e^a - {\hat{P}}_{\underline{a}ij a} \gamma^a \psi^j + \frac{i}{8} L_{M \underline{a}} \mathcal{V}_\alpha^*  {\cal H}^{M \alpha}_{ab}\gamma^{ab} \psi_i  +  ({\bar\chi}_i \lambda^j_{\underline{a}}) \psi_j -\frac{1}{2} ({\bar\chi}_j \lambda^j_{\underline{a}}) \psi_i \nonumber \\ 
 & + g {\bar A}_{2 \underline{a}}{}^j{}_i \psi_j ,\\[2mm]
{\cal H}^{M \alpha}  =&  \frac{1}{2} {\cal H}^{M \alpha}_{ab} e^a \wedge e^b + \bigg{(}-\frac{1}{4} \mathcal{V}^\alpha L^{Mij} {\bar\lambda}_{\underline{a}i} \gamma_{ab} \lambda^{\underline{a}}_j \, e^a \wedge e^b + \frac{1}{4} \mathcal{V}^\alpha L^{M \underline{a}} {\bar\chi}_i \gamma_{ab} \lambda^i_{\underline{a}} \, e^a \wedge e^b \nonumber  \\[2mm]
\label{Hrheon}
&+ (\mathcal{V}^\alpha)^* {L^M}_{ij} {\bar \chi}^i \gamma_a \psi^j \wedge e^a +  (\mathcal{V}^\alpha)^* L^{M \underline{a}} {\bar\lambda}^i_{\underline{a}} \gamma_a \psi_i \wedge e^a  + c.c. \bigg{)},\\[2mm]
\label{hatrhoi}
{\hat{\rho}}_i =& \frac{1}{2} {\hat{\rho}}_{iab} e^a \wedge e^b - \frac{i}{8} L_{Mij} \mathcal{V}_\alpha {\cal H}^{M \alpha}_{bc} \gamma^{bc} \gamma_a \psi^j \wedge e^a  \nonumber \\[2mm]
&+ \frac{1}{4} \epsilon_{ijkl} ({\bar{\lambda}}^j_{\underline{a}} \gamma_{ab} \lambda^{\underline{a} k }) \gamma^a \psi^l \wedge e^b
 + \frac{1}{4} (\bar{\chi}_i \gamma_a \chi^j) \psi_j \wedge e^a - \frac{1}{4}  (\bar{\chi}_j \gamma_a \chi^j) \psi_i \wedge e^a  \nonumber \\[2mm]
& +\frac{1}{4} (\bar{\chi}_i \gamma^a \chi^j) \gamma_{ab} \psi_j \wedge e^b - \frac{1}{8}(\bar{\chi}_j \gamma^a \chi^j) \gamma_{ab} \psi_i \wedge e^b \\[2mm]
& +\frac{1}{2} ({\bar\lambda}^{\underline{a}}_i \gamma_a \lambda^j_{\underline{a}} ) \psi_j \wedge e^a + \frac{1}{2}  ({\bar\lambda}^{\underline{a}}_i \gamma^a \lambda^j_{\underline{a}} ) \gamma_{ab} \psi_j \wedge e^b
\nonumber \\[2mm]
& - \frac{1}{4} ({\bar\lambda}^{\underline{a}}_j \gamma^a \lambda^j_{\underline{a}} ) \gamma_{ab} \psi_i \wedge e^b - \frac{1}{2} \epsilon_{ijkl} \chi^j ({\bar\psi}^k \wedge \psi^l)  - \frac{1}{3} g {\bar A}_{1ij}  \gamma_a \psi^j \wedge e^a, \nonumber \\[2mm]
\label{Rabg}
R_{ab} = & \frac{1}{2} R_{cdab} e^c \wedge e^d + {\bar{\hat{\theta}}}^i_{abc} \psi_i \wedge e^c + {\bar{\hat{\theta}}}_{iabc} \psi^i \wedge e^c \nonumber \\[2mm]
& + \frac{i}{4} \mathcal{V}_\alpha L_{Mij} {\cal H}^{M \alpha}_{ab} {\bar\psi}^i \wedge \psi^j + \frac{1}{8} \mathcal{V}_\alpha L_{Mij} \epsilon_{abcd} {\cal H}^{M \alpha cd} {\bar\psi}^i \wedge \psi^j \nonumber \\[2mm]
& - \frac{i}{4} \mathcal{V}_\alpha^* {L_M}^{ij} {\cal H}^{M \alpha}_{ab} {\bar\psi}_i \wedge \psi_j + \frac{1}{8}  \mathcal{V}_\alpha^* {L_M}^{ij}  \epsilon_{abcd} {\cal H}^{M \alpha cd} {\bar\psi}_i \wedge \psi_j \nonumber \\[2mm] 
& - \frac{1}{4} \epsilon_{ijkl} ({\bar{\lambda}}^i_{\underline{a}} \gamma_{ab} \lambda^{\underline{a} j }) {\bar\psi}^k \wedge \psi^l - \frac{1}{4} \epsilon^{ijkl} ({\bar{\lambda}}_i^{\underline{a}} \gamma_{ab} \lambda_{\underline{a} j }) {\bar\psi}_k \wedge \psi_l 
\\[2mm]
& + \frac{1}{2} ({\bar\chi}_i \gamma^c \chi^j) {\bar\psi}^i \wedge \gamma_{abc} \psi_j - \frac{1}{4} ({\bar\chi}_j \gamma^c \chi^j) {\bar\psi}^i \wedge \gamma_{abc} \psi_i \nonumber \\[2mm]  \nonumber
& + ({\bar\lambda}^{\underline{a}}_i \gamma^c \lambda^j_{\underline{a}}) {\bar\psi}^i \wedge \gamma_{abc} \psi_j - \frac{1}{2} ({\bar\lambda}^{\underline{a}}_j \gamma^c \lambda^j_{\underline{a}}) {\bar\psi}^i \wedge \gamma_{abc} \psi_i \, \\[2mm]
& + \frac{1}{3} g {\bar A}_{1ij} {\bar\psi}^i \wedge \gamma_{ab} \psi^j + \frac{1}{3} g A_1^{ij} {\bar\psi}_i \wedge \gamma_{ab} \psi_j,    \nonumber \\[2mm]
\label{H3Ma}
{\cal H}^{(3) M \alpha} \equiv & - \frac{1}{2} {{\Theta}^{\alpha M}}_{NP} {\cal H}^{(3) NP} + \frac{1}{2} \xi^M_\beta {\cal H}^{(3) \alpha \beta}
=  \frac{1}{6} {\cal H}^{(3) M \alpha}_{abc} e^a \wedge e^b \wedge e^c \nonumber\\[2mm]
& + i {\Theta}^{\alpha M N P} {L_N}^{\underline{a}} {L_P}^{ij} {\bar{\lambda}}_{\underline{a}i} \gamma_{ab} \psi_j \wedge e^a \wedge e^b  \nonumber \\[2mm]
& - \frac{1}{4} \xi^M_\beta ({\cal V}^\alpha)^*  ({\cal V}^\beta)^* {\bar\chi}^i \gamma_{ab} \psi_i \wedge e^a \wedge e^b \nonumber \\[2mm]
&- i  {\Theta}^{\alpha M N P} {L_N}^{\underline{a}} L_{Pij} {\bar\lambda}^i_{\underline{a}} \gamma_{ab} \psi^j \wedge e^a \wedge e^b  \\[2mm]
& - \frac{1}{4} \xi^M_\beta {\cal V}^\alpha {\cal V}^\beta {\bar\chi}_i \gamma_{ab} \psi^i \wedge e^a \wedge e^b \nonumber
\\[2mm] & + 2i {\Theta}^{\alpha MNP} {{L_N}}^{ik} L_{Pjk} {\bar\psi}^j \wedge \gamma_a \psi_i \wedge e^a \nonumber \\[2mm]
& - \frac{1}{2} \xi^M_\beta M^{\alpha \beta}  {\bar\psi}^i \wedge \gamma_a \psi_i \wedge e^a, \nonumber
 \end{align}
supplemented with the constraint $T^a=0$.
Here ${\cal H}^{M \alpha}_{ab}$ satisfy 
\begin{equation}
\label{moddualg}
\epsilon_{abcd} {\cal H}^{M \alpha cd}= -2   {M^M}_N {{M}^{\alpha}}_{\beta} { \cal H}^{N \beta}_{ab} ,   
\end{equation}
${\hat{\theta}}^i_{abc}$ equals
\begin{equation}
    {\hat{\theta}}^i_{abc}  =  \gamma_{[a} {\hat{\rho}}^i_{b]c} - \frac{1}{2} \gamma_c {\hat{\rho}}^i_{ab} 
\end{equation}
and the fermion shift matrices are given by \cite{Schon:2006kz}     
\begin{align}     
     \label{dilshiftapp}
   A_2^{ij} & =  f_{\alpha MNP} {\cal V}^\alpha {L^M}_{kl} L^{Nik} L^{P jl} + \frac{3}{2} \xi_{\alpha M} {\cal V}^\alpha L^{M ij}, \\[2mm]
    \label{gaushiftapp}
    A_{2 \underline{a} i}{}^j & =  f_{\alpha MNP} {\cal V}^\alpha {L_{\underline{a}}}^M {L^N}_{ik} L^{Pjk} - \frac{1}{4} \delta^j_i \xi_{\alpha M}  {\cal V}^\alpha {L_{\underline{a}}}^M, \\[2mm] 
    \label{gravshiftapp}
   A_1^{ij} & =   f_{\alpha MNP} ({\cal V}^\alpha)^* L^M{}_{kl} L^{Nik} L^{Pjl}.
\end{align}

Furthermore, the ${\bar\psi}^i \wedge \gamma^a \psi_i$ sector of the Bianchi identity \eqref{DhatX} implies the following superspace equations of motion for the dilatini
\begin{align}
    \label{choeomg}
    \gamma^a {\hat{V}}_{i a} = & \, \frac{i}{4} \mathcal{V}_{\alpha}^* L_{M \underline{a}} {\cal H}^{M \alpha}_{ab} \gamma^{ab} \lambda^{\underline{a}}_i + \frac{3}{4} \chi^j \bar{\chi}_i \chi_j - \frac{1}{2} \lambda^{\underline{a}}_j {\bar\lambda}^j_{\underline{a}} \chi_i - \lambda^{\underline{a}}_i {\bar\lambda}^j_{\underline{a}} \chi_j \nonumber \\[2mm]  
    & -2 g {\bar A}_2{}^{\underline{a} j}{}_i \lambda_{\underline{a}j} + 2 g {\bar A}_2{}^{\underline{a} j}{}_j  \lambda_{\underline{a}i}  , 
\end{align}
while the corresponding sector of the Bianchi identity \eqref{DhatL} gives the following superspace equations of motion for the gaugini
\begin{align}
    \label{lameomg}
     \gamma^a {\hat \Lambda}_{\underline{a} i a} = & \, \frac{i}{4} \mathcal{V}_{\alpha}^* L_{Mij}  {\cal H}^{M \alpha}_{ab} \gamma^{ab} \lambda^j_{\underline{a}} + \frac{i}{8} \mathcal{V}_{\alpha} L_{M \underline{a}} {\cal H}^{M \alpha}_{ab} \gamma^{ab} \chi_i  \nonumber \\[2mm]
    & - \frac{1}{2} \lambda_{\underline{b}}^j {\bar\lambda}_j^{\underline{b}} \lambda_{\underline{a}i} - \lambda_{\underline{a}}^j  {\bar\lambda}_{\underline{b} i } \lambda^{\underline{b}}_j + 2 \lambda_{\underline{b}}^j {\bar\lambda}_i^{\underline{b}} \lambda_{\underline{a}j} - \frac{1}{4} \chi_j \bar{\chi}^j \lambda_{\underline{a}i} - \frac{1}{2} \chi_i {\bar\chi}^j \lambda_{\underline{a}j}  \\[2mm]
    & - g A_{2 \underline{a} i}{}^j \chi_j + g A_{2 \underline{a} j}{}^j \chi_i + 2 g {\bar A}_{\underline{a} \underline{b} ij} \lambda^{\underline{b}j} + \frac{2}{3} g {\bar A}_{2 (ij)} \lambda^j_{\underline{a}} \nonumber , 
\end{align}
where
\begin{equation}
    \label{Aabijapp}
    A_{\underline{a} \underline{b}}{}^{ij} \equiv f_{\alpha MNP} {\cal V}^{\alpha} {L^M}{}_{\underline{a}} {L^N}{}_{\underline{b}} L^{Pij}.
\end{equation}
Moreover, by considering the ${\bar\psi}^i \wedge \gamma^a \psi_i \wedge e^b$ sector of the Bianchi identity \eqref{Dhatrho}, one can specify the superspace equations of motion for the gravitini in the gauged theory
\begin{align}
 \label{psieomg}
 \gamma^b \hat{\rho}_{iba} = & \frac{i}{2}  \mathcal{V}_\alpha L_{M \underline{a}} {\cal H}^{M \alpha}_{ab} \gamma^b \lambda^{\underline{a}}_i - \frac{i}{2}  \mathcal{V}_{\alpha}^*  L_{Mij}  {\cal H}^{M \alpha}_{ab} \gamma^b \chi^j \nonumber \\[2mm] &  + {\hat{P}}_a \chi_i + 2 {\hat{P}}_{\underline{a} ija} \lambda^{\underline{a}j} - \frac{1}{2} \gamma_a \lambda_{\underline{a} j} {\bar\lambda}^{\underline{a}}_i \chi^j \\[2mm] & + \frac{1}{3} g {\bar A}_{2ji} \gamma_a \chi^j + g A_{2 \underline{a} i}{}^j \gamma_a \lambda^{\underline{a}}_j. \nonumber
 \end{align}
By taking the covariant derivative $\hat{D}$ of the above equation and considering the $\psi_i$ sector of the resulting one-form equation in ${\cal N}=4$ superspace we obtain the superspace Einstein equation
\begin{align}
    \label{Einsteing}
    {\cal R}_{ab} - \, &\frac{1}{2} {\bar\chi}_i \gamma_{(a} {\hat V}^i_{b)} - \frac{1}{2} {\bar\chi}^i \gamma_{(a|} {\hat V}_{i|b)} - {\bar\lambda}^i_{\underline{a}}  \gamma_{(a|} {\hat\Lambda}^{\underline{a}}_{i|b)}  - {\bar\lambda}^{\underline{a}}_i \gamma_{(a|} {\hat\Lambda}^i_{\underline{a}|b)} = \nonumber \\[2mm]
    & {\hat P}^*_a {\hat P}_b + {\hat P}_a {\hat P}^*_b + {{\hat P}^{\underline{a} i j}}_{\hspace{0.4cm}a} {\hat P}_{\underline{a} i j b} \, + \frac{1}{2} M_{MN} M_{\alpha \beta} {\cal H}^{M \alpha}_{ac} {{{\cal H}^{N \beta}}_b}^c \nonumber \\[2mm] 
    & - \frac{1}{2} {\bar\chi}^i \gamma_{ac} \lambda^{\underline{a}}_i {\bar\chi}_j {\gamma_b}^c \lambda^j_{\underline{a}} - \frac{1}{2} {\bar\lambda}_{\underline{a}i} \gamma_{ac} \lambda^{\underline{a}}_j {\bar\lambda}_{\underline{b}}^i {\gamma_b}^c \lambda^{\underline{b}j}  \\[2mm]
    & - g  \eta_{ab} \left( - {\bar A}_2{}^{\underline{a}j}{}_i {\bar\chi}^i \lambda_{\underline{a}j} +  {\bar A}_2{}^{\underline{a}i}{}_i {\bar\chi}^j \lambda_{\underline{a}j} + A_{\underline{a} \underline{b}}{}^{ij} {\bar\lambda}^{\underline{a}}_i \lambda^{\underline{b}}_j + \frac{1}{3} A_2^{ij} {\bar\lambda}^{\underline{a}}_i \lambda_{\underline{a}j} + c.c. \right) \nonumber \\[2mm]
    & - g^2 \eta_{ab} \left( \frac{1}{3} A_1^{ij} {\bar A}_{1ij} - \frac{1}{9} A_2^{ij} {\bar A}_{2 ij} - \frac{1}{2} A_{2 \underline{a} i}{}^j {\bar A}_2{}^{\underline{a}i}{}_j \right), \nonumber
\end{align}
where ${\cal R}_{ab} \equiv {R_{acb}}^c = {\cal R}_{ba}$ and we have used \eqref{Wardmain}.

Also, the Bianchi identity \eqref{2formBianchi} constrains the inner components of ${\cal H}^{(3)M \alpha}$ to be equal to 
\begin{align}
  \label{H3abc}
  {\cal H}^{(3)M \alpha}_{abc} = \, & \epsilon_{abcd} {\Theta}^{\alpha MNP} \Big{(} L_{N \underline{a}} L_{Pij} {\hat P}^{\underline{a}ijd}  - {L_N}^{ik} L_{Pjk} {\bar\chi}_i \gamma^d \chi^j 
 \nonumber 
  \\[2mm] & -2 {L_N}^{ik} L_{Pjk} {\bar\lambda}^{\underline{a}}_i \gamma^d \lambda^j_{\underline{a}} + 2 L_{N \underline{a}} L_{P \underline{b}} {\bar\lambda}^{\underline{a}}_i \gamma^d \lambda^{\underline{b}i} \Big{)} \nonumber \\[2mm]
  & + \epsilon_{abcd} \xi^M_{\beta} \bigg{[} \frac{i}{2} {\cal V}^\alpha {\cal V}^{\beta} ({\hat P}^d)^* - \frac{i}{2} ({\cal V}^\alpha)^* ({\cal V}^{\beta})^* {\hat P}^d  \\[2mm]
  & +  2 M^{\alpha \beta} \left( \frac{3i}{8} {\bar\chi}_i \gamma^d \chi^i + \frac{i}{4} {\bar\lambda}^{\underline{a}}_i \gamma^d \lambda^i_{\underline{a}}  \right) \bigg{]}.\nonumber
\end{align}

In addition, equations \eqref{moddualg} and \eqref{Mmatrixapp} imply the following expression for  the inner components ${\cal H}_{\Lambda ab} = \Pi_{\Lambda M \alpha} {\cal H}^{M \alpha}_{ab}$ of the super-field strengths ${\cal H}_{\Lambda} = \Pi_{\Lambda M \alpha}{\cal H}^{M \alpha}$ of the magnetic super-one-forms $A_{\Lambda} = \Pi_{\Lambda M \alpha} A^{M \alpha}$ in terms of the inner components ${\cal H}^{\Lambda}_{ab} = {{\Pi}^{\Lambda}}_{M \alpha} {\cal H}^{M \alpha}_{ab} $ of the super-field strengths ${\cal H}^{\Lambda} = {{\Pi}^{\Lambda}}_{M \alpha} {\cal H}^{M \alpha}$ of the electric super-one-forms $A^{\Lambda} = {{\Pi}^{\Lambda}}_{M \alpha} A^{M \alpha}$
\begin{equation}
 \label{Hmag}
    {\cal H}_{\Lambda ab} = - \frac{1}{2} \epsilon_{abcd} {\cal I}_{\Lambda \Sigma} {\cal H}^{\Sigma cd} + {\cal R}_{\Lambda \Sigma} {\cal H}^{\Sigma}_{ab}.
\end{equation}    
Using the above equation and \eqref{PP=C}, we can express the terms in the rheonomic parametrizations of the fermionic gauged supercurvatures and the superspace equations of motion \eqref{choeomg}, \eqref{lameomg} and \eqref{psieomg} that involve ${\cal H}^{M \alpha}_{ab}$ solely in terms of ${\cal H}^{\Lambda}_{ab}$. Those expressions are similar to the corresponding ones in the ungauged theory and are given by equations \eqref{Vel}-\eqref{gammarhoel} with $V_i$, ${\Lambda}_{\underline{a}i}$, $\rho_i$, $V_{ia}$, $\Lambda_{\underline{a}ia}$, $\rho_{iba}$ and ${\cal F}^{\Lambda}_{ab}$ replaced by ${\hat V}_i$, ${\hat \Lambda}_{\underline{a} i}$, ${\hat \rho}_i$, ${\hat V}_{ia}$, ${\hat \Lambda}_{\underline{a}ia}$, ${\hat \rho}_{iba}$ and ${\cal H}^{\Lambda}_{ab}$ respectively. Furthermore, using equations \eqref{PP=C}, \eqref{Iinv}-\eqref{I + RIinvR} and \eqref{Hmag} we can write the fourth term on the right-hand side of \eqref{Einsteing} as 
\begin{equation}
     \frac{1}{2} M_{MN} M_{\alpha \beta} {\cal H}^{M \alpha}_{ac} {{{\cal H}^{N \beta}}_b}^c = - 2 {\cal I}_{\Lambda \Sigma} {\cal H}^{\Lambda +}_{ac} {{{\cal H}^{\Sigma -}}_b}^c\,.
 \end{equation}

 From the rheonomic parametrizations of the gauged supercurvatures, we can derive the local supersymmetry transformations of the spacetime fields in the gauged $D=4$, ${\cal N}=4$ Poincar\'{e} supergravity, as we specified the corresponding transformations in the ungauged theory. The Lie derivatives of the super-one-forms $e^a$, $\psi_i$ and $A^{M \alpha}$ along the tangent vector \eqref{tangent} are given by  
\begin{align}
\ell_{\epsilon} e^a = \,& (i_{\epsilon} d + d i_{\epsilon}) e^a =  i_{\epsilon} T^a + {\bar\epsilon}^i \gamma^a \psi_i + {\bar\epsilon}_i \gamma^a \psi^i\,, \\[2mm]
\ell_{\epsilon} \psi_i =  \,& (i_{\epsilon} d + d i_{\epsilon}) \psi_i = {\hat D} \epsilon_i + i_{\epsilon} {\hat\rho}_i\,, \\[2mm]
\ell_{\epsilon} A^{M\alpha} = \, & (i_{\epsilon} d + d i_{\epsilon})A^{M\alpha} = i_{\epsilon} {\cal H}^{M \alpha} + 2 ({\cal V}^\alpha)^* L^{Mij} {\bar\epsilon}_i \psi_j + 2 {\cal V}^\alpha {L^M}_{ij} {\bar\epsilon}^i \psi^j\,  , 
\end{align}
where we have used the definitions of the superspace curvatures $T^a$, ${\hat\rho}_i$ and ${\cal H}^{M \alpha}$ and 
\begin{equation}
    \label{Dhatepsilon}
    {\hat D} \epsilon_i \equiv d \epsilon_i + \frac{1}{4} \omega_{ab} \gamma^{ab} \epsilon_i - \frac{i}{2} \hat{\cal A}  \epsilon_i - {{\hat\omega}_i}^{\hspace{0.1cm}j} \epsilon_j.
\end{equation} 
For the super-zero-forms ${\nu}^I \equiv ({\cal V}_\alpha, {\cal V}_\alpha^*, L_{Mij}, L_{M\underline{a}},\chi^i,\chi_i, \lambda^i_{\underline{a}},\lambda_{\underline{a}i})$ we have the simpler result
\begin{equation}
    \ell_{\epsilon} {\nu}^I = (i_{\epsilon} d + d i_{\epsilon}){\nu}^I = i_{\epsilon} {\hat D} {\nu}^I.
\end{equation}
Furthermore, for the super-two-forms $B^{M \alpha} \equiv  - \frac{1}{2} {{\Theta}^{\alpha M}}_{NP} B^{NP} + \frac{1}{2} \xi^M_\beta B^{\alpha \beta}$ we find
\begin{align}
    \ell_{\epsilon} B^{M \alpha} = (i_{\epsilon} d + d i_{\epsilon})B^{M \alpha} = & \, i_{\epsilon} {\cal H}^{(3)M\alpha} - \frac{1}{2} {{\Theta}^{\alpha M}}_{NP} \epsilon_{\beta \gamma} A^{N \beta} \wedge \ell_{\epsilon} A^{P \gamma} \nonumber \\
    &  - \frac{1}{2} \xi^M_\beta \eta_{NP} A^{N(\alpha|} \wedge \ell_{\epsilon} A^{P|\beta)}.
\end{align}
Using the parametrizations given for the gauged supercurvatures and identifying the local supersymmetry transformation $\delta_{\epsilon}$ of each spacetime $p$-form with the projection of the Lie derivative $ \ell_{\epsilon}$ of the corresponding super-$p$-form on spacetime it is straightforward to determine the ${\cal N}=4$ local supersymmetry transformations of all the spacetime fields in the gauged theory. 
The results have been presented in section \ref{sec:lagrangian_and_supersymmetry_rules}.

Using the rheonomic approach, one can also construct the spacetime Lagrangian for the gauged $D=4$, ${\cal N}=4$ matter-coupled Poincar\'{e} supergravity in an arbitrary symplectic frame. As we have already mentioned, in this approach the gauged action is given by the integral of a superspace four-form Lagrangian ${\cal L}$ on a four-dimensional bosonic hypersurface ${\cal M}^4$ immersed in ${\cal N}=4$ superspace, 
 \begin{equation}
     \label{Srheong}
      S = \int_{{\cal M}^4 \subset {\cal S} {\cal M}} {\cal L} \, . 
 \end{equation}
The superspace Lagrangian ${\cal L}$ for the gauged theory contains the corresponding Lagrangian for the ungauged theory, which is given by equations \eqref{Lang}-\eqref{L4fermi} (with the coefficients replaced by their specified values), with the supercurvatures $\rho_i$, $V_i$, $\Lambda_{\underline{a}i}$, $P$, $P_{\underline{a} i j }$ and ${\cal F}^{\Lambda}$ replaced by their gauged counterparts ${\hat\rho}_i$, ${\hat V}_i$, ${\hat\Lambda}_{\underline{a}i}$, ${\hat P}$, ${\hat P}_{\underline{a} ij}$ and ${\cal H}^{\Lambda}$ respectively, i.e.
\begin{equation}
    {\cal L} \supset  {\cal L}_{\text{kin}} + {\cal L}_{\text{Pauli}} + {\cal L}_{\text{torsion}} + {\cal L}_{\text{4fermi}} \, , 
\end{equation}
where
\begin{align}
    \label{Lking}
    {\cal L}_{\text{kin}} = & \, \frac{1}{4} \epsilon_{abcd} R^{ab} \wedge e^c \wedge e^d + i( {\bar\psi}_i \wedge \gamma_a {\hat{\rho}}^i - {\bar\psi}^i \wedge \gamma_a {\hat{\rho}}_i   ) \wedge e^a \nonumber \\[2mm]
    & - \frac{1}{12} \epsilon_{abcd} ( {\bar\chi}_i \gamma^a {\hat{V}}^i + {\bar\chi}^i \gamma^a {\hat{V}}_i + 2 {\bar\lambda}^{\underline{a}}_i \gamma^a {\hat{\Lambda}}^i_{\underline{a}} + 2  {\bar\lambda}_{\underline{a}}^i \gamma^a {\hat{\Lambda}}_i^{\underline{a}}  ) \wedge e^b \wedge e^c \wedge e^d \nonumber \\[2mm]
    & +\frac{1}{24} \epsilon_{abcd}  {\hat{S}}_e^* \hat{S}^e e^a \wedge e^b \wedge e^c \wedge e^d \nonumber \\[2mm]
    & - \frac{1}{6} \epsilon_{abcd} \big{[} (\hat{S}^a)^* (\hat{P}- {\bar\chi}^i \psi_i) + \hat{S}^a (\hat{P}^* -  {\bar\chi}_i \psi^i)\big{]} \wedge e^b \wedge e^c \wedge e^d \nonumber \\[2mm]
    & + \frac{1}{48} \epsilon_{abcd} \hat{R}_{\underline{a} ij e}  \hat{R}^{\underline{a} ij e}  e^a \wedge e^b \wedge e^c \wedge e^d \nonumber \\[2mm]
    & -\frac{1}{6} \epsilon_{abcd} {\hat{R}_{\underline{a} i j}}{}^{ a} ( \hat{P}^{\underline{a} i j} - 2 {\bar\psi}^i \lambda^{\underline{a}j} - \epsilon^{ijkl} {\bar\psi}_k \lambda^{\underline{a}}_l) \wedge e^b \wedge e^c \wedge e^d  \\[2mm]
    & - \frac{i}{96} \epsilon_{abcd} ({\bar{\cal N}}_{\Lambda \Sigma} {\cal K}^{\Lambda +}_{ef} {\cal K}^{\Sigma + ef} - {\cal N}_{\Lambda \Sigma}  {\cal K}^{\Lambda -}_{ef} {\cal K}^{\Sigma - ef}  )  e^a \wedge e^b \wedge e^c \wedge e^d \nonumber \\[2mm]
    & - \frac{1}{2}(  {\bar{\cal N}}_{\Lambda \Sigma} {\cal K}^{\Lambda +}_{ab} +  {\cal{N}}_{\Lambda \Sigma} {\cal K}^{\Lambda-}_{ab}) \bigg{(} {\cal H}^{\Sigma} + \frac{1}{4} {{\Pi}^\Sigma}_{M \alpha} ({\cal V}^\alpha)^* {L^M}_{ij} {\bar\lambda}^i_{\underline{a}} \gamma_{cd} \lambda^{\underline{a}j} e^c \wedge e^d  \nonumber \\[2mm]
   & + \frac{1}{4}  {{\Pi}^\Sigma}_{M \alpha} {\cal V}^{\alpha} L^{Mij} {\bar \lambda}_{\underline{a} i} \gamma_{cd} \lambda_j^{\underline{a}} e^c \wedge e^d - \frac{1}{4}  {{\Pi}^\Sigma}_{M \alpha} {\cal V}^{\alpha}  L^{M \underline{a}} {\bar\chi}_i \gamma_{cd} \lambda^i_{\underline{a}} e^c \wedge e^d  \nonumber \\[2mm]
   & - \frac{1}{4}  {{\Pi}^\Sigma}_{M \alpha} ({\cal V}^\alpha)^*  L^{M \underline{a}} {\bar\chi}^i \gamma_{cd} \lambda_{\underline{a}i} e^c \wedge e^d - {{\Pi}^\Sigma}_{M \alpha} ({\cal V}^\alpha)^* {L^M}_{ij} {\bar\chi}^i \gamma_c \psi^j \wedge e^c \nonumber \\[2mm] & -  {{\Pi}^\Sigma}_{M \alpha} {\cal V}^{\alpha} L^{Mij}  {\bar\chi}_i \gamma_c \psi_j \wedge e^c - {{\Pi}^\Sigma}_{M \alpha} ({\cal V}^\alpha)^*  L^{M \underline{a}} {\bar\lambda}^i_{\underline{a}} \gamma_c \psi_i \wedge e^c \nonumber \\[2mm]
   & -  {{\Pi}^\Sigma}_{M \alpha} {\cal V}^{\alpha}    L^{M \underline{a}} {\bar\lambda}_{\underline{a}i} \gamma_c \psi^i \wedge e^c \bigg{)} \wedge e^a \wedge e^b, \nonumber \\[2mm]
    \label{LPaulig}
    {\cal L}_{\text{Pauli}} = & - \frac{i}{2} \hat{P}^* \wedge {\bar\chi}^i \gamma_{ab} {\psi}_i \wedge e^a \wedge e^b +i {\hat{P}_{\underline{a}}}{}^{ ij} \wedge {\bar\lambda}^{\underline{a}}_i \gamma_{ab} \psi_j  \wedge e^a \wedge e^b \nonumber \\[2mm]
    & - \frac{1}{4} {\Pi}_{\Lambda M \alpha} ({\cal V}^\alpha)^* L^{M \underline{a}} {\cal H}^{\Lambda} {\bar\chi}^i \gamma_{ab} {\lambda}_{\underline{a}i} \wedge e^a \wedge e^b \nonumber \\[2mm] 
    & + \frac{1}{4}  {\Pi}_{\Lambda M \alpha} {\cal V}^{\alpha} L^{Mij} {\cal H}^{\Lambda} {\bar\lambda}_{\underline{a}i} \gamma_{ab} {\lambda}^{\underline{a}}_j \wedge e^a \wedge e^b \nonumber \\[2mm]
    & - {\Pi}_{\Lambda M \alpha} ({\cal V}^\alpha)^* {L^M}_{ij}  {\cal H}^{\Lambda} \wedge {\bar\chi}^i \gamma_a \psi^j \wedge e^a \\[2mm]
    & -  {\Pi}_{\Lambda M \alpha} ({\cal V}^\alpha)^* L^{M \underline{a}}  {\cal H}^{\Lambda} \wedge {\bar\lambda}_{\underline{a}}^i \gamma_a \psi_i \wedge e^a  \nonumber \\[2mm]
    & - {\Pi}_{\Lambda M \alpha} {\cal V}^{\alpha} {L^M}_{ij} {\cal H}^{\Lambda} \wedge {\bar\psi}^i \wedge \psi^j + c.c., \nonumber \\[2mm]
    \label{Ltorsiong}
     {\cal L}_{\text{torsion}} = & \, \frac{i}{4} {\bar\chi}_i \gamma_a \chi^i R_b \wedge e^a \wedge e^b - \frac{i}{2}  {\bar\lambda}^{\underline{a}}_i \gamma_a {\lambda}_{\underline{a}}^i  R_b \wedge e^a \wedge e^b, \\[2mm]
    \label{L4f}
    {\cal L}_{\text{4fermi}} = & \,  \epsilon_{abcd} \bigg{(} \frac{1}{64}  {\bar\chi}^i {\chi}^j {\bar\chi}_i {\chi}_j - \frac{1}{48}  {\bar\chi}^i {\lambda}^{\underline{a}}_j {\bar\chi}_i {\lambda}_{\underline{a}}^j - \frac{1}{24}  {\bar\chi}^i {\lambda}^{\underline{a}}_i  {\bar\chi}_j {\lambda}_{\underline{a}}^j  \nonumber \\[2mm]
     &- \frac{1}{48}{\bar\lambda}^{\underline{a}}_i   {\lambda}^{\underline{b}}_j  {\bar\lambda}_{\underline{a}}^i   {\lambda}_{\underline{b}}^j - \frac{1}{24}   {\bar\lambda}^{\underline{a}}_i \lambda_{\underline{a}j}  {\bar\lambda}_{\underline{b}}^i \lambda^{\underline{b}j} + \frac{1}{12}  {\bar\lambda}^{\underline{a}}_i {\lambda}^{\underline{b}}_j  {\bar\lambda}_{\underline{b}}^i  {\lambda}_{\underline{a}}^j  \bigg{)} e^a \wedge e^b \wedge e^c \wedge e^d \nonumber \\[2mm]
     & + \left( \frac{1}{6}  \epsilon_{abcd}  {\bar\lambda}_{\underline{a}i}  {\lambda}^{\underline{a}}_j {\bar\chi}^i \gamma^d \psi^j  + c.c.  \right) \wedge e^a \wedge e^b  \wedge e^c \nonumber \\[2mm]
     & + \left(\frac{i}{4} \epsilon_{ijkl} {\bar\lambda}^{\underline{a}i} \gamma_{ab} {\lambda}_{\underline{a}}^j {\bar\psi}^k \wedge \psi^l + c.c. \right) \wedge e^a \wedge e^b \nonumber \\[2mm]
     & + \left(\frac{i}{2} {\bar\chi}_i \gamma_a \chi^j {\bar\psi}^i \wedge \gamma_b \psi_j  + i {\bar\lambda}^{\underline{a}}_i \gamma_a \lambda^j_{\underline{a}} {\bar\psi}^i \wedge \gamma_b \psi_j - i  {\bar\lambda}^{\underline{a}}_i \gamma_a \lambda^i_{\underline{a}} {\bar\psi}^j \wedge \gamma_b \psi_j \right) \wedge e^a \wedge e^b  \nonumber \\[2mm]
     & + \Bigg{[}  \epsilon_{abcd} {\Pi}_{\Lambda M \alpha} {{\Pi}^{\Lambda}}_{N \beta}  \bigg{(} \frac{i}{192}   ({\cal V}^\alpha)^* {\cal V}^{\beta} {L^M}_{\underline{a}} L^{Njk} {\bar\chi}^i \gamma_{ef} \lambda^{\underline{a}}_i {\bar\lambda}_{\underline{b}j} \gamma^{ef} \lambda^{\underline{b}}_k    \nonumber \\[2mm]
     & - \frac{i}{384} ({\cal V}^\alpha)^* ({\cal V}^{\beta})^* {L^M}_{\underline{a}} {L^N}_{\underline{b}}   {\bar\chi}^i \gamma_{ef} \lambda^{\underline{a}}_i {\bar\chi}^j \gamma^{ef} \lambda_j^{\underline{b}} \nonumber \\[2mm]
     & - \frac{i}{384} {\cal V}^{\alpha}  {\cal V}^{\beta} L^{Mij} L^{Nkl} {\bar\lambda}_{\underline{a}i} \gamma_{ef} \lambda^{\underline{a}}_j {\bar\lambda}_{\underline{b}k} \gamma^{ef} \lambda^{\underline{b}}_l \bigg{)}  e^a \wedge e^b \wedge e^c \wedge e^d  \nonumber \\[2mm]      & +  {\Pi}_{\Lambda M \alpha} {{\Pi}^{\Lambda}}_{N \beta}\bigg{(} \frac{1}{4}   ({\cal V}^\alpha)^* {\cal V}^{\beta} {L^M}_{\underline{a}} L^{Njk} {\bar\chi}^i \gamma_{ab} \lambda^{\underline{a}}_i {\bar\chi}_j \gamma_c \psi_k \nonumber \\[2mm]
     &+ \frac{1}{4} ({\cal V}^\alpha)^* ({\cal V}^\beta)^* {L^M}_{\underline{a}} L^{N \underline{b}} {\bar\chi}^i \gamma_{ab} \lambda^{\underline{a}}_i {\bar \lambda}^j_{\underline{b}} \gamma_c \psi_j \nonumber \\[2mm]
     & - \frac{1}{4} ({\cal V}^\alpha)^* ({\cal V}^\beta)^* {L^M}_{ij} {L^N}_{kl} {\bar\lambda}^k_{\underline{a}} \gamma_{ab} \lambda^{\underline{a}l} {\bar\chi}^i \gamma_c \psi^j \nonumber \\[2mm]
     & + \frac{1}{4} ({\cal V}^\alpha)^* ({\cal V}^\beta)^* {L^M}_{\underline{a}} {L^N}_{jk} {\bar\chi}^i \gamma_{ab} \lambda^{\underline{a}}_i {\bar\chi}^j \gamma_c \psi^k \nonumber \\[2mm] 
     & - \frac{i}{8}  ({\cal V}^\alpha)^* {\cal V}^\beta {L^M}_{\underline{a}}  {L^N}_{\underline{b}} \epsilon_{abcd} {\bar\lambda}^{\underline{a}}_i \lambda^{\underline{b}}_j   {\bar\chi}^i \gamma^d \psi^j \nonumber \\[2mm]
     &- \frac{1}{8} ({\cal V}^\alpha)^* {\cal V}^\beta {L^M}_{\underline{a}}  {L^N}_{\underline{b}}  {\bar\lambda}^{\underline{a}}_i \gamma_{ab} \lambda^{\underline{b}}_j {\bar\chi}^i \gamma_c \psi^j \nonumber \\[2mm]
     & - \frac{1}{4} ({\cal V}^\alpha)^* {\cal V}^\beta {L^M}_{ij} L^{Nkl} {\bar\lambda}_{\underline{a} k } \gamma_{ab} \lambda^{\underline{a}}_l  {\bar\chi}^i \gamma_c \psi^j \nonumber \\[2mm]
     & - \frac{1}{4}  {\cal V}^{\alpha} ({\cal V}^\beta)^* L^{Mij} L^{N \underline{b}} {\bar\lambda}^{\underline{a}}_i \gamma_{ab} \lambda_{\underline{a} j } {\bar\lambda}^k_{\underline{b}} \gamma_c \psi_k \nonumber \\[2mm]
     & - \frac{1}{4}  {\cal V}^{\alpha} {\cal V}^\beta L^{Mij} L^{N \underline{b}} {\bar\lambda}_{\underline{a}i} \gamma_{ab} \lambda^{\underline{a}}_j {\bar\lambda}_{\underline{b}k} \gamma_c \psi^k \bigg{)} \wedge e^a \wedge e^b \wedge e^c \nonumber \\[2mm]
     & + {\Pi}_{\Lambda M \alpha} {{\Pi}^{\Lambda}}_{N \beta} \bigg{(} - ({\cal V}^\alpha)^* ({\cal V}^\beta)^* {L^M}_{ij} L^{N \underline{a}}  {\bar\chi}^i \gamma_a \psi^j \wedge {\bar\lambda}^k_{\underline{a}} \gamma_b \psi_k \nonumber \\[2mm]
     & +\frac{i}{4} ({\cal V}^\alpha)^* {\cal V}^\beta {L^M}_{ij} L^{Nkl} \epsilon_{abcd} {\bar\chi}^i \gamma^c \chi_k {\bar\psi}^j \wedge \gamma^d \psi_l \nonumber \\[2mm]
     & - \frac{1}{2} ({\cal V}^\alpha)^* ({\cal V}^\beta)^* {L^M}_{ij} {L^N}_{kl}  {\bar\chi}^i \gamma_a \psi^j \wedge {\bar\chi}^k \gamma_b \psi^l \nonumber \\[2mm]
     & -  ({\cal V}^\alpha)^* {\cal V}^\beta {L^M}_{ij} {L^N}_{\underline{a}} {\bar\chi}^i \gamma_a \psi^j \wedge {\bar\lambda}_k^{\underline{a}} \gamma_b \psi^k \nonumber \\[2mm]
     & - \frac{1}{2} {\cal V}^{\alpha} {\cal V}^\beta  {L^M}_{\underline{a}} {L^N}_{\underline{b}} {\bar\lambda}^{\underline{a}}_i \gamma_a \psi^i \wedge {\bar\lambda}^{\underline{b}}_j \gamma_b \psi^j \nonumber \\[2mm]
     & + \frac{i}{4} {\cal V}^{\alpha} ({\cal V}^\beta)^* {L^M}_{\underline{a}} L^{N \underline{b}} \epsilon_{abcd} {\bar\lambda}^{\underline{a}}_i \gamma^c \lambda^j_{\underline{b}} {\bar\psi}^i \wedge \gamma^d \psi_j \bigg{)} \wedge e^a  \wedge e^b \nonumber \\[2mm]
     & +  {\Pi}_{\Lambda M \alpha} {{\Pi}^{\Lambda}}_{N \beta}  \bigg{(} - \frac{1}{2} {\cal V}^{\alpha} {\cal V}^\beta {L^M}_{ij} {L^N}_{kl} {\bar\psi}^i \wedge \psi^j \wedge  {\bar\psi}^k \wedge \psi^l  \nonumber \\[2mm]
     & - \frac{1}{2} ({\cal V}^\alpha)^*  {\cal V}^\beta  L^{Mij} {L^N}_{kl} {\bar\psi}_i \wedge \psi_j \wedge  {\bar\psi}^k \wedge \psi^l \bigg{)} + c.c. \Bigg{]} \, , 
 \end{align}
where we have dropped a total derivative and $\hat{S}_a , \, {\hat R}_{\underline{a}ija}= ({\hat R}_{\underline{a}}{}^{ij}{}_a)^* = \frac{1}{2} \epsilon_{ijkl} {\hat R}_{\underline{a}}{}^{kl}{}_a $ and $ {\cal K}^{\Lambda}_{ab} =({\cal K}^{\Lambda}_{ab})^*$   are auxiliary super-zero-forms that are identified, through their equations of motions, with the inner components ${\hat P}_{a}$, ${\hat P}_{\underline{a} ija}$ and ${\cal H}^{\Lambda}_{ab}$ of the supercurvatures $\hat{P}$, ${\hat P}_{\underline{a} ij }$ and ${\cal H}^{\Lambda}$ respectively. They provide a first-order description of the kinetic terms of the bosonic superfields which avoids the use of the Hodge duality operator. 

The equations of motion that arise from the variation of the gauged action with respect to the superforms $\chi^i$, $\lambda^{\underline{a} i}$ and $\psi^i$ must be solved by the constraint $T^a=0$, the rheonomic equations \eqref{hatP}-\eqref{hatrhoi} and the equations of motion \eqref{choeomg}, \eqref{lameomg} and \eqref{psieomg} (expressed in terms of ${\cal H}^{\Lambda}_{ab}$ only), which are obtained by requiring closure of the Bianchi identities. In order for this condition to be satisfied, the following fermionic mass terms have to be added to the superspace Lagrangian for the gauged theory
\begin{align}
    \label{Lfmass}
     {\cal L}_{\text{fermion mass}} = & \, \frac{1}{12} g \epsilon_{abcd} \left(  - {\bar A}_2{}^{\underline{a} j}{}_i {\bar\chi}^i \lambda_{\underline{a}j} + {\bar A}_2{}^{\underline{a} i}{}_i {\bar\chi}^j \lambda_{\underline{a}j}  + A_{\underline{a} \underline{b}}{}^{ij} {\bar\lambda}^{\underline{a}}_i \lambda^{\underline{b}}_j + \frac{1}{3} A_2^{ij} {\bar\lambda}_{\underline{a}i} \lambda^{\underline{a}}_j \right) \nonumber \\[2mm]
      & e^a \wedge e^b \wedge e^c \wedge e^d  + \frac{1}{3} g \epsilon_{abcd} \left( \frac{1}{3} {\bar A}_{2ij} {\bar\chi}^i \gamma^a \psi^j + A_{2 \underline{a} j}{}^i  {\bar\lambda}^{\underline{a}}_i \gamma^a \psi^j  \right) \wedge e^b \wedge e^c \wedge e^d \nonumber\\[2mm]
      & + \frac{i}{3} g {\bar A}_{1ij} {\bar\psi}^i \wedge \gamma_{ab} \psi^j \wedge e^a \wedge e^b + c.c. \,.
\end{align}

We also require that the superspace Einstein equation obtained from the analysis of the super-three-form equation of motion for the bosonic vielbein $e^a$ following from the variational principle along the elements $e^a \wedge e^b \wedge e^c$ of the intrinsic basis of three-forms in ${\cal N}=4$ superspace be the same as \eqref{Einsteing}, which follows from the Bianchi identities. This is achieved if we add the following scalar potential term to the superspace Lagrangian
\begin{equation}
    \label{Lpot}
      {\cal L}_{\text{potential}} = \frac{1}{72} g^2   \left( A_1^{ij} {\bar A}_{1ij} - \frac{1}{3} A_2^{ij} {\bar A}_{2 ij} - \frac{3}{2} A_{2 \underline{a} i}{}^j {\bar A}_2{}^{\underline{a}i}{}_j \right) \epsilon_{abcd} e^a \wedge e^b \wedge e^c \wedge e^d.
\end{equation}

Finally, the superspace four-form Lagrangian for the gauged $D=4$, ${\cal N}=4$ Poincar\'{e} supergravity must contain the topological term \cite{deWit:2005ub}
\begin{align}
     \label{Ltop} 
     {\cal L}_{\text{top}} = & - \frac{1}{2} g {{\Pi}^\Lambda}_{M \alpha} {\Pi}_{\Lambda N \beta} \left( {{\Theta}^{\alpha M}}_{PQ} B^{PQ} - \xi^M_\gamma B^{\alpha \gamma}\right) \wedge \nonumber \\
    & \left( {\cal H}^{N \beta} + \frac{g}{4} {{\Theta}^{\beta N}}_{RS} B^{RS} - \frac{g}{4} \xi^N_\delta B^{\beta \delta} + ({\cal V}^\beta)^* L^{Nij} {\bar\psi}_i \wedge \psi_j + {\cal V}^\beta {L^N}_{ij} {\bar \psi}^i \wedge \psi^j \right) \nonumber \\
    & + \frac{1}{6} g \left( {{\Pi}^\Lambda}_{R \epsilon}  {\Pi}_{\Lambda S \zeta}  + 2 {\Pi}_{\Lambda R \epsilon} {{\Pi}^\Lambda}_{S \zeta}  \right) {X_{M \alpha N \beta}}^{R \epsilon} A^{M \alpha} \wedge A^{N \beta} \wedge \\
    & \left( d A^{S \zeta} + \frac{1}{4} g {X_{P \gamma Q \delta}}^{S \zeta} A^{P \gamma} \wedge A^{Q \delta} \right). \nonumber
\end{align}
This term ensures that the superspace equations of motion arising from the variation of the gauged action with respect to $B^{MN}$, $B^{\alpha \beta}$ and $A_{\Lambda}$ are solved by the rheonomic equations \eqref{Hrheon} and \eqref{H3Ma} and the constraints \eqref{H3abc} and \eqref{Hmag}. 

In summary, the superspace Lagrangian for the the gauged $D=4$, ${\cal N}=4$ matter-coupled supergravity in an arbitrary symplectic frame is given by
\begin{align}
\label{L4formgapp}
    {\cal L}= &  {\cal L}_{\text{kin}} + {\cal L}_{\text{Pauli}} + {\cal L}_{\text{torsion}} +  {\cal L}_{\text{fermion mass}} \nonumber \\
   & +  {\cal L}_{\text{potential}}  +  {\cal L}_{\text{top}} + {\cal L}_{\text{4fermi}} \, , 
\end{align}
 where the various terms on the right-hand side are given by equations \eqref{Lking}-\eqref{Ltop}. 

 In order to obtain the gauged spacetime Lagrangian, we restrict the superspace four-form Lagrangian \eqref{L4formgapp} to spacetime ($\theta^i=d \theta^i = 0$ hypersurface). As we did for the ungauged theory, we first go to the second-order formalism by identifying the auxiliary super-zero-forms $\hat{S}_a , \, {\hat R}_{\underline{a}ija}$ and ${\cal K}^{\Lambda}_{ab} $ with $\hat{P}_a , \, {\hat P}_{\underline{a}ija}$ and ${\cal H}^{\Lambda}_{ab} $ respectively and setting $T^a=0$. Then, we expand all the forms along the $dx^\mu$ differentials and restrict the superfields to their lowest ($\theta^i=0$) components. The result is given in section \ref{sec:lagrangian_and_supersymmetry_rules}.

\section{T-tensor Identities} 
\label{sec:t_tensor_identities}

In this appendix we derive the quadratic constraints satisfied by the T-tensor by appropriately dressing the quadratric constraints on the embedding tensor \eqref{xixi}-\eqref{ff + 3 xi f}  with the representatives of the coset spaces SL(2,${\mathbb R}$)/SO(2) and SO(6,$n$)/SO(6) $ \times$ SO($n$). Many of these constraints have been used for the derivation of the results of section \ref{sec:vacua_masses_gradient_flow_and_supertrace_relations} and their form and structure can be analyzed by classifying them according to their $H =$ SO(2) $\times$ SO(6) $\times$ SO($n$) representation.

\subsection{The T-tensor} 
\label{sub:the_t_tensor}

Let us join the coset representatives of SL(2,$\mathbb R$) and SO(6,$n$) into a single object
\begin{align}
    {\mathbb L}_{(R)}=\mathcal S\otimes L\implies ({\mathbb L}_{(R)})_\mathcal M{}^{\underline{\cal M}}= ({\mathbb L}_{(R)})_{M \alpha}{}^{\underline{M} \underline{\alpha} }=\mathcal S_\alpha{}^{\underline{\alpha}} L_M{}^{\underline{M}}.
\end{align}
We introduce a complex representative $\mathbb{L}$ of the coset space $\frac{\text{SL(2,${\mathbb R}$)}}{\text{SO(2)}} \times \frac{\text{SO(6,$n$)}}{\text{SO(6)} \times \text{SO($n$)}}$ defined by 
\begin{equation}
\mathbb{L} = {\mathbb L}_{(R)} {\bf A}^\dagger \, , 
\end{equation}
where 
\begin{equation}
{\bf A}^\dagger = {\cal A}^{\dagger} \otimes \mathbb{1}_{n+6} \, ,
\end{equation}
where $\mathbb{1}_{n+6}$ is the $(n+6)\times(n+6)$ identity matrix and ${\cal A}$ is the unitary $2\times2$ matrix with entries 
\begin{equation}
\label{Cayley}
{\cal A} = \frac{1}{\sqrt{2}} \begin{pmatrix}
1 & i \\ 1 & -i  
\end{pmatrix}\,.
\end{equation}
The elements of the complex matrix $\mathbb{L}$ are given by 
\begin{align}
  \mathbb{L}_{\cal M}{}^{\underline{\mathcal M}}&={\mathbb L}_{M\alpha}{}^{\underline{M} \underline{\alpha}}  =L_M{}^{\underline{M}} S_{\alpha}{}^{\underline{\beta}} ({\cal A})^{\dagger}_{\underline{\beta}}{}^{\underline{\alpha}}=  \left( {\mathbb{L}}_{M \alpha}{}^{\underline{M} 1} ,  {\mathbb{L}}_{M \alpha}{}^{\underline{M} 2} \right)\\
  &=\left({\frac{1}{\sqrt{2}}} {\mathcal V}_\alpha^* L_M{}^{\underline M},{\frac{1}{\sqrt{2}}}\mathcal V_\alpha L_M{}^{\underline M} \right).
\end{align}
The inverse matrix is obtained from the relation $\mathbb L^T_{(R)}\mathbb C\mathbb L_{(R)}=\mathbb C$, where $\mathbb C_{\mathcal M\mathcal N}=\epsilon_{\alpha\beta}\eta_{MN}$ (the subscript $(R)$ indicates that we are referring to the matrix with real entries). The inverse of the real coset representative is $\mathbb L^{-1}_{(R)}=-\mathbb C\mathbb L^T_{(R)}\mathbb C$, while the inverse of the complex one is
\begin{align}\nonumber
    \mathbb L^{-1}&=(\mathbb L_{(R)}{\bf A}^\dagger)^{-1}=({\bf A}^\dagger)^{-1}\mathbb L^{-1}_{(R)}=-({\bf A}^\dagger)^{-1}\mathbb C\mathbb L^T_{(R)}\mathbb C\\
    &=-{\bf A}\mathbb C\mathbb L^T_{(R)}\mathbb C=-{\bf A}\mathbb C{\bf A}^\dagger{\bf A}\mathbb L^T_{(R)}\mathbb C=-\varpi\mathbb L^\dagger\mathbb C,
\end{align}
where we have defined $\varpi\equiv {\bf A}\mathbb C{\bf A}^\dagger = -i\sigma^3\otimes\eta$. With indices, we have
\begin{align}
    (\mathbb L^{-1})_{\underline{\mathcal M}}{}^{\mathcal N}={\varpi}_{\underline{\mathcal M}\underline{\mathcal L}}\overline{\mathbb L}_{\mathcal M}{}^{\underline{\mathcal L}}\mathbb C^{\mathcal N\mathcal M} \, , 
\end{align}
or equivalently
\begin{equation}
 (\mathbb L^{-1})_{\underline{M} \underline{\alpha}}{}^{M \alpha} = i (\sigma^3)_{\underline{\alpha} \underline{\beta}} \eta_{\underline{M} \underline{N}} \overline{\mathbb L}_{N \beta}{}^{\underline{N} \underline{\beta}} \eta^{MN} \epsilon^{\beta \alpha}
\end{equation}
Therefore, the various elements of $\mathbb{L}^{-1}$ are
\begin{align}
   (\mathbb L^{-1})_{ij 1}{}^{M\alpha} & =\frac{i}{\sqrt2}\mathcal V^\alpha L^M{}_{ij}\\
    (\mathbb L^{-1})_{\underline a 1}{}^{M\alpha}&=\frac {i}{\sqrt2}\mathcal V^\alpha L^M{}_{\underline a}\\ 
    (\mathbb L^{-1})_{ij 2}{}^{M \alpha}&=-\frac {i}{\sqrt2}({\mathcal V}^\alpha)^* L^M{}_{ij}\\
    (\mathbb L^{-1})_{\underline a 2}{}^{M\alpha}&=-\frac {i}{\sqrt2} ({\mathcal V}^\alpha)^* L^M{}_{\underline a}. 
\end{align}
The $T$-tensor is obtained from the ``dressing'' of the structure constants \eqref{XMNP} with the above defined coset representatives and its explicit expression is
\begin{align}
    T_{\underline{\mathcal M}\underline{\mathcal N}}{}^{\underline{\mathcal P}}=(\mathbb L^{-1})_{\underline{\mathcal M}}{}^{\mathcal M}(\mathbb L^{-1})_{\underline{\mathcal N}}{}^{\mathcal N}X_{{\mathcal M}{\mathcal N}}{}^{{\mathcal P}}\mathbb L_{\mathcal P}{}^{\underline{\mathcal P}}.
\end{align}
The $T$-tensor contains the expressions for all the fermion shifts that have to be added to the rheonomic parametrizations during the gauging procedure, that is \eqref{dilshiftapp}-\eqref{gravshiftapp}. To help ourselves in the quest of extracting these expressions out of all the components of the $T$-tensor, let us recall the $U(1)$ charges of the fermion shifts:
\begin{center}
\begin{tabular}{ |c|c| } 
 \hline
 Field & U(1) charge\\ 
 \hline
 ${\bar A}_{1ij}$ & $1$\\ 
 \hline
 ${\bar A}_{2 \underline{a}}{}^i{}_j$ & $-1$\\ 
 \hline
 ${\bar A}_{2ij}$ & $-1$\\ 
 \hline
\end{tabular}
\end{center}
Let us then consider the following component of the $T$-tensor with charge $+1$:
\begin{align}\nonumber
    T_{ij1kl1}{}^{mn1}&=-\frac{1}{2\sqrt2}\mathcal  V^\alpha\bigg(2i f_{\alpha MN}{}^PL^M{}_{ij}L^N{}_{kl}L_P{}^{mn}-iL^M{}_{ij}\delta^{[m}_k\delta^{n]}_l\xi_{\alpha M}\nonumber\\
    &\qquad\qquad\quad +iL^M{}_{kl}\delta^{[m}_i\delta^{n]}_j\xi_{\alpha M}-\frac{i}{2}\epsilon_{ijkl}L_M{}^{mn}\xi^M_\alpha\bigg).
\end{align}
This component is an element of the SU$(4)$ algebra and, as such, can be expressed as 
\begin{align}
    T_{ij 1kl 1}{}^{mn 1}=4T_{ij[k}{}^{[m}\delta^{n]}_{l]}.
\end{align}
By contracting the above equation first with $\delta^l_n$ and then with $\delta^k_m$, one can find the expression for $T_{ijk}{}^m$:
\begin{align}\label{smallT}
    T_{ijk}{}^{m}&=\frac12\left(T_{ij1kl1}{}^{ml1}-\frac16\delta^m_k T_{ij1sl1}{}^{sl1}\right)\nonumber\\
    &=-\frac{i}{2\sqrt2}\mathcal V^\alpha\left(f_{\alpha MN}{}^PL^M{}_{ij}L^{N}{}_{kl}L_P{}^{ml}+\delta^m_{[i}L^M{}_{k|j]}\xi_{\alpha M}-\frac12\delta^m_kL^M{}_{ij}\xi_{\alpha M}\right).
\end{align}
By contracting \eqref{smallT} with $\epsilon^{ijkp}$, we precisely get
\begin{align}
     A_2^{pm} = i \sqrt2 T_{ijk}{}^{m}\epsilon^{ijkp}.
\end{align}
Moreover, by contracting \eqref{smallT} with $\delta^i_m$, we obtain
\begin{align}
    T_{ijk}{}^i=-\frac{i}{2\sqrt2}\mathcal V^\alpha\left(f_{\alpha MN}{}^PL^M{}_{ij}L^N{}_{kl}L_P{}^{il}+\delta^i_{[i}L^M{}_{k|j]}\xi_{\alpha M}-\frac12L^M{}_{kj}\xi_{\alpha M}\right)
\end{align}
and by further symmetrising in $(jk)$, we get the following relation 
\begin{align}
    {\bar A}_{1jk}= - 2 i \sqrt{2} T_{i(jk)}{}^i . 
\end{align}
 Therefore, the tensor $T_{ijk}{}^m$ can be written as
\begin{align}
\label{Tijkm}
    T_{ijk}{}^m=-\frac{i}{6\sqrt2}\epsilon_{ijkp} A_2^{pm} +\frac{i}{3 \sqrt2}\delta^m_{[i} {\bar A}_{1j]k}.
\end{align}
To derive the expression for ${\bar A}_{2 \underline{a}}{}^i{}_j$ in terms of the $T$-tensor, we instead need to consider the following component of the $T$-tensor, with U(1) charge $-1$:
\begin{align}
    T_{\underline a 2ij 1}{}^{kl 1}&=\frac{i}{\sqrt2}({\mathcal V}^\alpha)^* L^M{}_{\underline a}\left(L^N{}_{ij}L_P{}^{kl}f_{\alpha MN}{}^P+\frac12\delta^{[k}_i\delta^{l]}_j\xi_{M\alpha}\right)\nonumber\\
    &\equiv4T_{\underline a [i}{}^{[k}\delta^{l]}_{j]},
\end{align}
where $T_{\underline{a}i}{}^k$ is given by
\begin{align}
    T_{\underline ai}{}^k=\frac12\left(T_{\underline a 2ij 1}{}^{kj  1}-\frac16\delta^k_iT_{\underline a 2jl  1}{}^{jl 1}\right).
\end{align}
The explicit expression of the above tensor leads to a relation with ${\bar A}_{2 \underline{a}}{}^k{}_i$ given by \eqref{gaushift}: 
\begin{align}
    {\bar A}_{2 \underline{a}}{}^k{}_i =2\sqrt 2i T_{\underline ai}{}^k.
\end{align}

\subsection{Quadratic identities} 
\label{sub:quadratic_identities}

The quadratic constraints (\ref{xixi})--(\ref{ff + 3 xi f}) sit in definite irreducible representations of SL(2,${\mathbb R}$) $\times$ SO($6,n$) and their contraction with the coset representatives leads to tensorial structures in definite irreducible representations of the isotropy group $H$.
The resulting expressions are quadratic constraints in terms of the scalar tensors $A$ used for the fermion shifts and the fermion mass matrices.
We list them here according to their origin and their representations, using the notation $({\cal R}_{\text{SU}(4)},{\cal R}_{\text{SO}(n)})_{q_{U(1)}}$, where ${\cal R}_{\text{SU}(4)}$ and ${\cal R}_{\text{SO}(n)}$ denote the SU(4) and SO($n$) representations respectively and $q_{U(1)}$ the U(1) charge.
\subsubsection{From (\ref{xixi})}

Irreps $\left(\mathbf{1},\mathbf{1}\right)_{+2}$:
\begin{equation}
\label{T1}
    \frac{2}{9} \epsilon_{ijkl} A_2^{ij} A_2^{kl} = A_{2 \underline{a} i}{}^i A_2{}^{\underline{a}}{}_j{}^j.
\end{equation}
Irreps $\left(\mathbf{1},\mathbf{1}\right)_{0}$:
\begin{equation}
    \label{T2}
     \frac{4}{9} A_2^{[ij]} {\bar A}_{2ij} = A_{2 \underline{a} i}{}^i {\bar A}_2{}^{\underline{a} j}{}_j\,.
\end{equation}

\subsubsection{From (\ref{xif})}

Irreps $\left(\mathbf{15},\mathbf{1}\right)_{+2}$:
\begin{equation}
\label{T3}
     \frac{2}{9} \epsilon_{iklm} A_2^{(jk)} A_2^{lm} - \frac{4}{9}  A_2^{[jk]} {\bar A}_{1ik} = - A_{2 \underline{a} i}{}^j A_2{}^{\underline{a}}{}_k{}^k + \frac{1}{4} \delta^j_i A_{2 \underline{a} k}{}^k A_2{}^{\underline{a}}{}_l{}^l \, .
\end{equation}

\noindent
Irreps $\left(\mathbf{15},\mathbf{1}\right)_{0}$:
\begin{equation}
\label{T4}
    \frac{2}{9} \epsilon_{iklm} A_1^{jk} A_2^{lm} - \frac{2}{9} \epsilon^{jklm} {\bar A}_{1ik} {\bar A}_{2lm} + \frac{4}{9} A_2^{(jk)} {\bar A}_{2[ik]} - \frac{4}{9} A_2^{[jk]} {\bar A}_{2(ik)} = A_{2 \underline{a} k}{}^k {\bar A}_2{}^{\underline{a} j}{}_i - A_{2 \underline{a} i}{}^j {\bar A}_2{}^{\underline{a} k}{}_k.
\end{equation}

\noindent
Irreps $\left(\mathbf{6},\mathbf{n}\right)_{+2}$:
\begin{equation}
\label{T5}
     \epsilon_{ijlm} A_{\underline{a} \underline{b}}{}^{lm} A_2{}^{\underline{b}}{}_k{}^k = \frac{4}{3} A_2^{lm} A_{2 \underline{a} [j}{}^k \epsilon_{i]klm} - \frac{1}{3} \epsilon_{ijlm} A_2^{lm}  A_{2 \underline{a} k}{}^k  \, .
\end{equation}
Irreps $\left( \mathbf{6},\mathbf{n}\right)_{-2}$
\begin{equation}
 \label{T6}
     {\bar A}_{\underline{a} \underline{b} i j} {\bar A}_2{}^{\underline{b} k}{}_k = - \frac{2}{3} {\bar A}_{2[ik]} {\bar A}_{2 \underline{a}}{}^k{}_j + \frac{2}{3} {\bar A}_{2[jk]} {\bar A}_{2 \underline{a}}{}^k{}_i  + \frac{1}{3} {\bar A}_{2[ij]} {\bar A}_{2 \underline{a}}{}^k{}_k \, .
 \end{equation}

\noindent
Irreps $\left(\mathbf{6},\mathbf{n}\right)_{0}$:
\begin{align}
\label{T7}
    {\bar A}_{\underline{a} \underline{b} ij} A_2{}^{\underline{b}}{}_k{}^k + \frac{1}{2} \epsilon_{ijkl} A_{\underline{a} \underline{b}}{}^{kl} {\bar A}_2{}^{\underline{b}m}{}_m = & \,\frac{2}{3} A_2^{lm} {\bar A}_{2 \underline{a}}{}^k{}_{[i} \epsilon_{j]klm}  + \frac{2}{3} {\bar A}_{2[ik]} A_{2 \underline{a} j}{}^k -  \frac{2}{3} {\bar A}_{2[jk]} A_{2 \underline{a} i}{}^k\nonumber \\
    & + \frac{1}{6} \epsilon_{ijkl} A_2^{kl} {\bar A}_{2 \underline{a}}{}^m{}_m - \frac{1}{3} {\bar A}_{2[ij]} A_{2 \underline{a} k}{}^k \, . 
\end{align}

\noindent
Irreps $\left(\mathbf{1},\mathbf{n(n-1)/2}\right)_{+2}$:
\begin{equation}
\label{T8}
    A_{\underline{a} \underline{b} \underline{c}} A_2{}^{\underline{c}}{}_i{}^i = - \frac{1}{3} \epsilon_{ijkl} A_2^{ij} A_{\underline{a} \underline{b}}{}^{kl} \, . 
\end{equation}

\noindent
Irreps $\left(\mathbf{1},\mathbf{n(n-1)/2}\right)_{0}$:
\begin{equation}
\label{T9}
    {\bar A}_{\underline{a} \underline{b} \underline{c}} A_2{}^{\underline{c}}{}_i{}^i + A_{\underline{a} \underline{b} \underline{c}}  {\bar A}_2{}^{\underline{c} i}{}_i = - \frac{2}{3} {\bar A}_{\underline{a} \underline{b} ij} A_2^{ij} - \frac{2}{3} A_{\underline{a} \underline{b}}{}^{ij} {\bar A}_{2ij}\,.
\end{equation}

\subsubsection{From (\eqref{ff + xif})}

\noindent
Irreps $\left(\mathbf{15},\mathbf{1}\right)_{-2}$:
\begin{align}
\label{T10}
   & \frac{4}{3}  A_1^{jk} {\bar A}_{2(ik)} + 3 \left( {\bar A}_2{}^{\underline{a}j}{}_k {\bar A}_{2 \underline{a}}{}^k{}_i - \frac{1}{2}  {\bar A}_2{}^{\underline{a}j}{}_i {\bar A}_{2 \underline{a}}{}^k{}_k \right) + \frac{2}{3}  A_1^{jk} {\bar A}_{2[ik]} + \frac{1}{3} \epsilon^{jklm}  {\bar A}_{2(ik)} {\bar A}_{2lm} = \nonumber \\
   & = \frac{1}{3} \delta^j_i  A_1^{kl} {\bar A}_{2kl} + \frac{3}{4} \delta^j_i \left( {\bar A}_2{}^{\underline{a} k}{}_l {\bar A}_{2 \underline{a}}{}^l{}_k - \frac{1}{2} {\bar A}_2{}^{\underline{a} k}{}_k {\bar A}_{2 \underline{a}}{}^l{}_l \right) \, . 
\end{align}

\noindent
Irreps $\left(\mathbf{15},\mathbf{1}\right)_{0}$:
\begin{align}
\label{T11}
   & \frac{2}{3}  A_1^{jk} {\bar A}_{1ik} + \frac{2}{3}  A_2^{(jk)} {\bar A}_{2(ik)} + \frac{1}{3}  A_2^{[jk]} {\bar A}_{2(ik)} +  \frac{1}{3}  A_2^{(jk)} {\bar A}_{2[ik]} + \frac{1}{6} \epsilon^{jklm} {\bar A}_{1ik} {\bar A}_{2lm} \nonumber\\ 
   &  + \frac{1}{6} \epsilon_{iklm} A_1^{jk} A_2^{lm} - \frac{3}{2} A_{2 \underline{a} k}{}^j {\bar A}_{2 }{}^{\underline{a} k}{}_i - \frac{3}{2} A_{2 \underline{a} i}{}^k {\bar A}_{2 }{}^{\underline{a} j}{}_k + \frac{3}{4} A_{2 \underline{a} i}{}^j {\bar A}_{2 }{}^{\underline{a} k}{}_k +  \frac{3}{4} A_{2 \underline{a} k}{}^k {\bar A}_{2 }{}^{\underline{a} j}{}_i = \\
   & = \frac{1}{6} \delta^j_i A_1^{kl} {\bar A}_{1kl} + \frac{1}{6} \delta^j_i  A_2^{(kl)} {\bar A}_{2kl} - \frac{3}{4} \delta^j_i \left( A_{2 \underline{a} k}{}^l {\bar A}_{2 }{}^{\underline{a} k}{}_l - \frac{1}{2} A_{2 \underline{a} k}{}^k {\bar A}_{2 }{}^{\underline{a} l}{}_l \right)\,. \nonumber
\end{align}

\noindent
Irreps $\left(\mathbf{10},\mathbf{n}\right)_{+2}$:
\begin{equation}
\label{T12}
    \frac{1}{3} A_2^{lm} A_{2 \underline{a} (i}{}^k \epsilon_{j)klm} + \frac{4}{3} A_{2 \underline{a} (i}{}^k {\bar A}_{1 j) k} + A_{\underline{a} \underline{b}}{}^{lm} A_2{}^{\underline{b}}{}_{(i}{}^k \epsilon_{j)klm} = 0 \, .
\end{equation}

\noindent
Irreps $\left(\mathbf{10},\mathbf{n}\right)_{-2}$:
\begin{equation}
\label{T13}
    2 {\bar A}_{2(ij)} {\bar A}_{2 \underline{a}}{}^k{}_k =  {\bar A}_{2[ik]} {\bar A}_{2 \underline{a}}{}^k{}_j +  {\bar A}_{2[jk]} {\bar A}_{2 \underline{a}}{}^k{}_i  + 2 {\bar A}_{2(ik)} {\bar A}_{2 \underline{a}}{}^k{}_j + 2 {\bar A}_{2(jk)} {\bar A}_{2 \underline{a}}{}^k{}_i + 6 {\bar A}_{\underline{a} \underline{b} (i|k}  {\bar A}_2{}^{\underline{b}k}{}_{|j)} .
\end{equation}

\noindent
Irreps $\left(\mathbf{10},\mathbf{n}\right)_{0}$:
\begin{align}
\label{T14}
    -2 {\bar A}_{1ij} {\bar A}_{2 \underline{a}}{}^k{}_k = & \, 2 {\bar A}_{2(ik)} A_{2 \underline{a} j}{}^k + 2 {\bar A}_{2(jk)} A_{2 \underline{a} i}{}^k+ {\bar A}_{2[ik]} A_{2 \underline{a} j}{}^k + {\bar A}_{2[jk]} A_{2 \underline{a} i}{}^k   \\
    & -4 {\bar A}_{2 \underline{a}}{}^k{}_{(i} {\bar A}_{1j)k} + 6 {\bar A}_{\underline{a} \underline{b} (i|k} A_2{}^{\underline{b}}{}_{|j)}{}^k  - A_2^{lm} {\bar A}_{2 \underline{a}}{}^k{}_{(i} \epsilon_{j)klm} - 3 A_{\underline{a} \underline{b}}{}^{lm} {\bar A}_2{}^{\underline{b}k}{}_{(i} \epsilon_{j)klm}\,. \nonumber
\end{align}

\noindent
Irreps $\left(\mathbf{15},\mathbf{n(n-1)/2}\right)_{-2}$:
\begin{align}
\label{T15}
   & -4 {\bar A}_{2 [\underline{a}|}{}^k{}_i {\bar A}_{2 |\underline{b}]}{}^j{}_k  - 2 {\bar A}_{2 [\underline{a}|}{}^j{}_i {\bar A}_{2 |\underline{b}]}{}^k{}_k   - \frac{4}{3} A_1^{jk} {\bar A}_{\underline{a} \underline{b} ik} \nonumber \\ 
   & + \frac{1}{3} \epsilon^{jklm} \left( 2 {\bar A }_{2(ik)} {\bar A}_{\underline{a} \underline{b} lm} +   {\bar A }_{2[ik]} {\bar A}_{\underline{a} \underline{b} lm}- {\bar A }_{2lm} {\bar A}_{\underline{a} \underline{b} ik}  \right)   \\  & - 2 \epsilon^{jklm} {\bar A}_{[\underline{a}| \underline{c} ik} {\bar A}_{|\underline{b}]}{}^{\underline{c}}{}_{lm} - 2 {\bar A}_{\underline{a} \underline{b} \underline{c} } \left( {\bar A}_2{}^{\underline{c}j}{}_i - \frac{1}{4} \delta^j_i  {\bar A}_2{}^{\underline{c}k}{}_k \right)  = 0 \, . \nonumber
\end{align}

\noindent
Irreps $\left(\mathbf{15},\mathbf{n(n-1)/2}\right)_{0}$:
\begin{align}
\label{T16}
    &  - \frac{1}{3} \epsilon_{iklm} A_1^{jk} A_{\underline{a} \underline{b} }{}^{lm} + \frac{1}{3} \epsilon^{jklm} {\bar A}_{1ik}  {\bar A}_{\underline{a} \underline{b} lm} \nonumber \\
    & + \frac{2}{3} {\bar A}_{2ik} A_{\underline{a} \underline{b}}{}^{jk} - \frac{2}{3} A_2^{jk} {\bar A}_{\underline{a} \underline{b} ik} -\frac{1}{6} \delta^j_i \left( {\bar A}_{2kl} A_{\underline{a} \underline{b} }{}^{kl} - A_2^{kl} {\bar A}_{\underline{a} \underline{b} kl} \right) \nonumber \\
    & + 2 A_{2 [\underline{a}|i}{}^k {\bar A}_{2 |\underline{b}]}{}^j{}_k - 2 A_{2 [\underline{a}|k}{}^j {\bar A}_{2 |\underline{b}]}{}^k{}_i + A_{2 [\underline{a}|k}{}^k {\bar A}_{2 |\underline{b}]}{}^j{}_i + A_{2 [\underline{a}|i}{}^j {\bar A}_{2 |\underline{b}]}{}^k{}_k - \frac{1}{2} \delta^j_i A_{2 [\underline{a}|k}{}^k {\bar A}_{2 |\underline{b}]}{}^l{}_l \nonumber \\
    & - A_{\underline{a} \underline{b} \underline{c}} \left( {\bar A}_2{}^{\underline{c}j}{}_i - 
\frac{1}{4} \delta^j_i {\bar A}_2{}^{\underline{c}k}{}_k \right) + {\bar A}_{\underline{a} \underline{b} \underline{c}} \left( A_2{}^{\underline{c}}{}_i{}^j - \frac{1}{4} \delta^j_i  A_2{}^{\underline{c}}{}_k{}^k  \right)  \\
& + 4 A_{[\underline{a}}{}^{\underline{c}jk} {\bar A}_{\underline{b}] \underline{c} ik}  - \delta^j_i A_{[\underline{a}}{}^{\underline{c} kl} {\bar A}_{\underline{b}] \underline{c} kl} = 0 .\nonumber
\end{align}

\noindent
Irreps $\left(\mathbf{6},\mathbf{n(n-1)(n-2)/6}\right)_{+2}$:
\begin{equation}
\label{T17}
    \epsilon_{ijkl} \left( \frac{1}{3} A_2^{kl} A_{\underline{a} \underline{b} \underline{c}} + 3 A_{[\underline{a} \underline{b}| \underline{d}} A_{|\underline{c}]}{}^{\underline{d}kl} \right) = -6 A_{[\underline{a} \underline{b}}{}^{lm} A_{2 \underline{c}][j}{}^k \epsilon_{i]klm} \, .
\end{equation}

\noindent
Irreps $\left(\mathbf{6},\mathbf{n(n-1)(n-2)/6}\right)_{-2}$:
\begin{equation}
\label{T18}
    \frac{1}{3} {\bar A}_{2[ij]} {\bar A}_{\underline{a} \underline{b} \underline{c}} + 3 {\bar A}_{[\underline{a} \underline{b}| \underline{d}} {\bar A}_{|\underline{c}]}{}^{\underline{d}}{}_{ij} =  -3 {\bar A}_{2 [\underline{a}|}{}^k{}_k {\bar A}_{|\underline{b} \underline{c}]ij} - 6 {\bar A}_{2 [\underline{a}|}{}^k{}_{[i|} {\bar A}_{|\underline{b} \underline{c}]j]k} \, .
\end{equation}

\noindent
Irreps $\left(\mathbf{6},\mathbf{n(n-1)(n-2)/6}\right)_{0}$:
\begin{align}
\label{T19}
 &    \frac{1}{3} \epsilon_{ijkl} A_2^{kl} {\bar A}_{\underline{a} \underline{b} \underline{c}} + \frac{2}{3} {\bar A}_{2[ij]} A_{\underline{a} \underline{b} \underline{c}} + 6 A_{[\underline{a} \underline{b}|\underline{d} } {\bar A}_{|\underline{c}]}{}^{\underline{d} }{}_{ij} + 3 \epsilon_{ijkl} {\bar A}_{[\underline{a} \underline{b}|\underline{d} }  A_{|\underline{c}]}{}^{\underline{d} kl } \nonumber = \\ 
   & = - 3 \epsilon_{ijkl} A_{[\underline{a} \underline{b}}{}^{kl} {\bar A}_{2 \underline{c}]}{}^m{}_m  + 12 A_{2 [\underline{a}| [i| }{}^k {\bar A}_{|\underline{b} \underline{c}]j]k} + 6 A_{[\underline{b} \underline{c}}{}^{lm} 
   {\bar A}_{2 \underline{a}]}{}^k{}_{[j}  \epsilon_{i]klm} \, . 
\end{align}

\noindent
Irreps $\left(\mathbf{1},\mathbf{n(n-1)(n-2)(n-3)/24}\right)_{+2}$:
\begin{equation}
\label{T20}
    3 A_{\underline{e} [\underline{a} \underline{b}} A_{\underline{c} \underline{d}]}{}^{\underline{e}} + 2 A_{[\underline{a} \underline{b} \underline{c} } A_{2 \underline{d}] i}{}^i = \frac{3}{2} \epsilon_{ijkl} A_{[\underline{a} \underline{b}}{}^{ij} A_{\underline{c} \underline{d}]}{}^{kl}.
\end{equation}

\noindent
Irreps $\left(\mathbf{1},\mathbf{n(n-1)(n-2)(n-3)/24}\right)_{0}$:
\begin{equation}
\label{T21}
    3 A_{\underline{e} [\underline{a} \underline{b}} {\bar A}_{\underline{c} \underline{d}]}{}^{\underline{e}} + {\bar A}_{[\underline{a} \underline{b} \underline{c} }  A_{2 \underline{d}]i}{}^i + A_{[\underline{a} \underline{b} \underline{c} } {\bar A}_{2 \underline{d}]}{}^i{}_i = 3 A_{[\underline{a} \underline{b}}{}^{ij} {\bar A}_{\underline{c} \underline{d}]ij} \, .
\end{equation}

\subsubsection{From \eqref{xif + xixi}.}

\noindent
Irreps $\left(\mathbf{15},\mathbf{1}\right)_{0}$:
\begin{align}
\label{T22}
    \frac{2}{9} \epsilon_{iklm} A_1^{jk} A_2^{lm}  + \frac{2}{9} \epsilon^{jklm} {\bar A}_{1ik} {\bar A}_{2lm} = & \, A_{2 \underline{a} k}{}^k {\bar A}_2{}^{\underline{a}j}{}_i + A_{2 \underline{a} i}{}^j  {\bar A}_2{}^{\underline{a}k}{}_k- \frac{1}{2} \delta^j_i  A_{2 \underline{a} k}{}^k  {\bar A}_2{}^{\underline{a}l}{}_l \nonumber  \\ & + \frac{4}{9} A_2^{(jk)} {\bar A}_{2 [ik]}  + \frac{4}{9} A_2^{[jk]} {\bar A}_{2 (ik)} -  \frac{8}{9} A_2^{[jk]} {\bar A}_{2 [ik]}  \\ &+ \frac{2}{9} \delta^j_i A_2^{[kl]} {\bar A}_{2kl}  \, . \nonumber
\end{align}

\noindent
Irreps $\left(\mathbf{6},\mathbf{n}\right)_{0}$:
\begin{align}
\label{T23}
    {\bar A}_{\underline{a} \underline{b} ij} A_2{}^{\underline{b}}{}_k{}^k - \frac{1}{2} \epsilon_{ijkl} A_{\underline{a} \underline{b}}{}^{kl} {\bar A}_2{}^{\underline{b} m}{}_m = & - \frac{2}{3} {\bar A}_{2[ik]} A_{2 \underline{a} j
    }{}^k + \frac{2}{3}{\bar  A}_{2[jk]} A_{2 \underline{a} i
    }{}^k +  {\bar  A}_{2[ij]} A_{2 \underline{a} k
    }{}^k \nonumber \\
    & - \frac{2}{3} A_2^{lm} {\bar A}_{2 \underline{a}}{}^k{}_{[j} \epsilon_{i]klm} - \frac{1}{6} \epsilon_{ijlm} A_2^{lm} {\bar A}_{2 \underline{a}}{}^k{}_k \, .
\end{align}

\noindent
Irreps $\left(\mathbf{1},\mathbf{n(n-1)/2}\right)_{0}$:
\begin{equation}
\label{T24}
    A_{\underline{a} \underline{b} \underline{c}} {\bar A}_2{}^{\underline{c} i}{}_i - {\bar A}_{\underline{a} \underline{b} \underline{c}} A_2{}^{\underline{c}}{}_i{}^i + 2 A_{2 [\underline{a}|i}{}^i {\bar A}_{2 |\underline{b}]}{}^j{}_j = \frac{2}{3} {\bar A}_{\underline{a} \underline{b} ij} A_2^{ij} - \frac{2}{3} A_{\underline{a} \underline{b}}{}^{ij} {\bar A}_{2ij} .
\end{equation}

\subsubsection{From \eqref{ff + 3 xi f}}

\noindent
Irreps $\left((\mathbf{15}\times \mathbf{15})_A,\mathbf{1}\right)_{0}$:
\begin{align}
\label{T25}
  &  - \frac{2}{9} \delta^j_i A_1^{lm} {\bar A}_{1km} + \frac{2}{9} \delta^l_k A_1^{jm} {\bar A}_{1im} +\frac{2}{9} \delta^j_i A_2^{(lm)} {\bar A}_{2(km)} - \frac{2}{9} \delta^l_k A_2^{(jm)} {\bar A}_{2(im)} \nonumber \\
  & - \frac{2}{9} \epsilon_{ikmn} \left( A_1^{jm} A_2^{(ln)} - A_2^{(jm)} A_1^{ln}\right) - \frac{2}{9} \epsilon^{jlmn} \left( {\bar A}_{1im} {\bar A}_{2 (kn)} - {\bar A}_{2(im)} {\bar A}_{1kn} \right) \nonumber \\
  & - \frac{4}{9} A_2^{(jl)} {\bar A}_{2 [ik]} - \frac{4}{9} A_2^{[jl]} {\bar A}_{2 (ik)} - \frac{1}{9} \delta^j_i \left( A_2^{(lm)} {\bar A}_{2 [km]} + A_2^{[lm]} {\bar A}_{2 (km)} \right)  \nonumber \\ &  + \frac{1}{9} \delta^j_k \left( A_2^{(lm)} {\bar A}_{2 [im]} -  A_2^{[lm]} {\bar A}_{2 (im)} \right)  + \frac{1}{9} \delta^l_k \left( A_2^{(jm)} {\bar A}_{2 [im]} +  A_2^{[jm]} {\bar A}_{2 (im)} \right) \nonumber \\ & - \frac{1}{9} \delta^l_i \left(  A_2^{(jm)} {\bar A}_{2 [km]} - A_2^{[jm]} {\bar A}_{2 (km)} \right)  + \frac{1}{9} \epsilon_{ikmn} \left( A_1^{jl} A_2^{mn} +  A_1^{jm} A_2^{[ln]} - A_2^{[jm]} A_1^{ln} \right) \\ 
  &  + \frac{1}{9} \epsilon^{jlmn} \left( {\bar A}_{1ik} {\bar A}_{2mn} + {\bar A}_{1im} {\bar A}_{2 [kn]} - {\bar A}_{2 [im]} {\bar A}_{1kn}  \right) + A_{2 \underline{a} k }{}^j {\bar A}_2{}^{\underline{a} l}{}_i -  A_{2 \underline{a} i }{}^l {\bar A}_2{}^{\underline{a} j}{}_k \nonumber \\
  &  + \frac{1}{4} \delta^j_i \left(   A_{2 \underline{a} m }{}^m {\bar A}_2{}^{\underline{a} l}{}_k + A_{2 \underline{a} k }{}^l {\bar A}_2{}^{\underline{a} m}{}_m \right) - \frac{1}{4} \delta^j_k \left( A_{2 \underline{a} m }{}^m {\bar A}_2{}^{\underline{a} l}{}_i - A_{2 \underline{a} i }{}^l {\bar A}_2{}^{\underline{a} m}{}_m \right) \nonumber \\
  & - \frac{1}{4} \delta^l_k \left( A_{2 \underline{a} m }{}^m {\bar A}_2{}^{\underline{a} j}{}_i  + A_{2 \underline{a} i }{}^j {\bar A}_2{}^{\underline{a} m}{}_m
       \right)  + \frac{1}{4} \delta^l_i \left(   A_{2 \underline{a} m }{}^m {\bar A}_2{}^{\underline{a} j}{}_k - A_{2 \underline{a} k }{}^j {\bar A}_2{}^{\underline{a} m}{}_m  \right) = 0 \,. , \nonumber 
\end{align}
The tensor product $(\mathbf{15} \times \mathbf{15})_A$ of SU(4) decomposes as 
\begin{equation}
    (\mathbf{15} \times \mathbf{15})_A = \mathbf{15} + \mathbf{45} + \overline{\mathbf{45}}\,.
\end{equation}
The component of the quadratic constraint \eqref{T25} that transforms in the $\mathbf{15}$ of SU(4) follows from contracting \eqref{T25} with $\delta^l_k$, which yields
\begin{align}
    \label{T2515}
  (\mathbf{15},\mathbf{1})_0: \hspace{0.5cm} & \frac{8}{9} \left( A_1^{jk} {\bar A}_{1ik} - A_2^{(jk)} {\bar A}_{2 (ik)} \right) - \frac{2}{9} \delta^j_i \left( A_1^{kl} {\bar A}_{1kl} - A_2^{(kl)} {\bar A}_{2 kl} \right) \nonumber \\
   & + A_{2 \underline{a} k}{}^j {\bar A}_{2}{}^{\underline{a}k}{}_i - A_{2 \underline{a} i}{}^k {\bar A}_2{}^{\underline{a}j}{}_k  -  A_{2 \underline{a} i}{}^j {\bar A}_{2}{}^{\underline{a}k}{}_k - A_{2 \underline{a} k}{}^k {\bar A}_{2}{}^{\underline{a}j}{}_i  \\ 
   & + \frac{1}{2} \delta_i^j A_{2 \underline{a} k}{}^k {\bar A}_{2}{}^{\underline{a}l}{}_l = 0 \,.\nonumber
\end{align}

\noindent
Irreps $\left(\mathbf{15}\times \mathbf{6},\mathbf{n}\right)_{0}$:
\begin{align}
\label{T26}
    & \frac{2}{3} {\bar A}_{2 \underline{a} }{}^l{}_{[i} {\bar A}_{1j]k} - \frac{2}{3} {\bar A}_{1km} {\bar A}_{2 \underline{a}}{}^m{}_{[i} \delta^l_{j]}  + \frac{2}{3} A_1^{lm} A_{2 \underline{a} [i}{}^n \epsilon_{j]kmn}  + \frac{1}{3} \epsilon_{ijkm} A_1^{lm} A_{2 \underline{a}n}{}^n \nonumber \\
    & - \frac{1}{3} {\bar A}_{2 (ik)} A_{2 \underline{a} j}{}^l  + \frac{1}{3} {\bar A}_{2 (jk)} A_{2 \underline{a} i}{}^l - \frac{2}{3} {\bar A}_{2 (km)} A_{2 \underline{a} [i}{}^m \delta^l_{j]} - \frac{1}{3} \delta^l_i {\bar A}_{2 (jk)} A_{2  \underline{a} m}{}^m  +  \frac{1}{3} \delta^l_j {\bar A}_{2 (ik)} A_{2  \underline{a} m}{}^m\nonumber \\
    &   + \frac{2}{3} A_2^{(lm)} {\bar A}_{2 \underline{a}}{}^n{}_{[i} \epsilon_{j]kmn} - \frac{1}{3} {\bar A}_{2 [ij]} A_{2 \underline{a} k}{}^l - \frac{1}{3} {\bar A}_{2 [ik]} A_{2 \underline{a} j}{}^l  + \frac{1}{3} {\bar A}_{2 [jk]} A_{2 \underline{a} i}{}^l - \frac{1}{4} \delta^l_i {\bar A}_{2 [jm]} A_{2 \underline{a} k}{}^m \nonumber \\ 
    &  + \frac{1}{4} \delta^l_j {\bar A}_{2 [im]} A_{2 \underline{a} k}{}^m  + \frac{1}{3} {\bar A}_{2 [km]} A_{2 \underline{a} [i}{}^m \delta^l_{j]}  + \frac{1}{6} {\bar A}_{2 [jm]} A_{2 \underline{a} [k}{}^m \delta^l_{i]}   + \frac{1}{6} {\bar A}_{2 [im]} A_{2 \underline{a} [j}{}^m \delta^l_{k]}\nonumber \\ 
    & - \frac{1}{12} \delta^l_i {\bar A}_{2 [jk]} A_{2 \underline{a} m}{}^m  + \frac{1}{12} \delta^l_j {\bar A}_{2 [ik]} A_{2 \underline{a} m}{}^m + \frac{1}{24}  \delta^l_k {\bar A}_{2 [ij]} A_{2 \underline{a} m}{}^m - \frac{1}{12} \epsilon_{ijkm} \left( A_2^{[ln]} {\bar A}_{2 \underline{a}}{}^m{}_n + A_2^{[mn]} {\bar A}_{2 \underline{a}}{}^l{}_n  \right)  \nonumber \\ 
    & + \frac{1}{6} {\bar A}_{2 \underline{a}}{}^l{}_{[i} \epsilon_{j]kmn} A_2^{mn} + \frac{1}{6} \epsilon_{ijmn} \left( A_2^{[lm]} {\bar A}_{2 \underline{a}}{}^n{}_k -A_2^{mn} {\bar A}_{2 \underline{a}}{}^l{}_k  \right)  + \frac{1}{24} \epsilon_{ijkm} A_2^{[lm]} {\bar A}_{2 \underline{a}}{}^n{}_n \\
    &  - \frac{1}{12} \delta^l_{[i} \epsilon_{j]mnp} A_2^{np} {\bar A}_{2 \underline{a}}{}^m{}_k   + \frac{1}{12} \epsilon_{kmnp} A_2^{np} {\bar A}_{2 \underline{a}}{}^m{}_{[i} \delta^l_{j]} + \frac{1}{24} \delta^l_k \epsilon_{ijmn} A_2^{mn} {\bar A}_{2 \underline{a}}{}^p{}_p \nonumber \\
    &  - \frac{1}{24} \delta^l_{[i} \epsilon_{j]kmn} A_2^{mn}  {\bar A}_{2 \underline{a}}{}^p{}_p + {\bar A}_{\underline{a} \underline{b} ij} A_2{}^{\underline{b}}{}_k{}^l - \frac{3}{8} \delta^l_k {\bar A}_{\underline{a} \underline{b} ij}  A_2{}^{\underline{b}}{}_m{}^m - \frac{1}{2} \delta^l_{[i|} {\bar A}_{\underline{a} \underline{b} |j] k}  A_2{}^{\underline{b}}{}_m{}^m   \nonumber \\
    &  + \frac{1}{2} \epsilon_{ijnp} A_{\underline{a} \underline{b}}{}^{np} {\bar A}_2{}^{\underline{b}l}{}_k  + \frac{1}{8} \epsilon_{ijkm} A_{\underline{a} \underline{b}}{}^{lm} {\bar A}_2{}^{\underline{b}n}{}_n - \frac{1}{8} \delta^l_k \epsilon_{ijmn} A_{\underline{a} \underline{b}}{}^{mn} {\bar A}_2{}^{\underline{b}p}{}_p \nonumber \\
    &  + \frac{1}{8} \delta^l_{[i} \epsilon_{j]kmn} A_{\underline{a} \underline{b}}{}^{mn} {\bar A}_2{}^{\underline{b}p}{}_p = 0 \nonumber \, . 
    \end{align}
We have the decomposition 
\begin{equation}
    \mathbf{15} \times \mathbf{6} = \mathbf{6} + \mathbf{10} + \overline{\mathbf{10}} + \mathbf{64} \,.
\end{equation}
In order to specify the components of \eqref{T26} in the 
$\mathbf{10}$ and $\mathbf{6}$ representations of SU(4), we first contract \eqref{T26} with $\delta^j_l$. To obtain the $\mathbf{10}$ component, we symmetrize the resulting identity in $i$ and $k$, whereas to get the $\mathbf{6}$ component, we antisymmetrize in $i$ and $k$. The results are 
\begin{align}
\label{T2610}
    (\mathbf{10},\mathbf{n})_0: \hspace{0.5cm} & \frac{2}{3} {\bar A}_{2 \underline{a}}{}^j{}_{(i} {\bar A}_{1k)j} + \frac{1}{3} {\bar A}_{1ik} {\bar A}_{2 \underline{a}}{}^j{}_j - \frac{2}{3} {\bar A}_{2(ik)} A_{2 \underline{a} j}{}^j + \frac{1}{3} {\bar A}_{2(jk)} A_{2 \underline{a} i}{}^j \nonumber \\
    & + \frac{1}{3} {\bar A}_{2 (ij)} A_{2 \underline{a}k}{}^j - \frac{1}{6} {\bar A}_{2[ij]} A_{2 \underline{a}k}{}^j + \frac{1}{6} {\bar A}_{2[jk]} A_{2 \underline{a}i}{}^j \\
    & - {\bar A}_{\underline{a} \underline{b} (i| j} A_2{}^{\underline{b}}{}_{|k)}{}^j - \frac{1}{6} {\bar A}_{2 \underline{a}}{}^l{}_{(k} \epsilon_{i)lmn} A_2^{mn} - \frac{1}{2} {\bar A}_2{}^{\underline{b} l}{}_{(k} \epsilon_{i)lmn} A_{\underline{a} \underline{b}}{}^{mn} = 0\,. \nonumber  
\end{align}
\begin{align}
\label{T266}
     (\mathbf{6},\mathbf{n})_0: \hspace{0.5cm} & -  \frac{2}{3} {\bar A}_{2 \underline{a}}{}^j{}_{[i} {\bar A}_{1k]j} + \frac{1}{3} {\bar A}_{2(ij)} A_{2 \underline{a} k}{}^j - \frac{1}{3} {\bar A}_{2(jk)} A_{2 \underline{a} i}{}^j \nonumber \\
     & + \frac{1}{12} {\bar A}_{2 [ij]} A_{2 \underline{a} k}{}^j + \frac{1}{12} {\bar A}_{2 [jk]} A_{2 \underline{a} i}{}^j - \frac{1}{24} {\bar A}_{2[ik]} A_{2 \underline{a} j}{}^j \nonumber \\
     & + {\bar A}_{\underline{a} \underline{b} [i|j} A_2{}^{\underline{b}}{}_{|k]}{}^j + \frac{3}{8} {\bar A}_{\underline{a} \underline{b} ik} A_2{}^{\underline{b}}{}_m{}^m - \frac{1}{3} \epsilon_{iklm} A_1^{ln} A_{2 \underline{a}n}{}^m  \\
     & + \frac{1}{12} {\bar A}_{2 \underline{a}}{}^l{}_{[k} \epsilon_{i]lmn} A_2^{mn} - \frac{1}{3} \epsilon_{iklm} A_2^{(ln)} {\bar A}_{2 \underline{a}}{}^m{}_n - \frac{1}{48} \epsilon_{iklm} A_2^{lm}  {\bar A}_{2 \underline{a}}{}^n{}_n \nonumber \\
     & + \frac{1}{2} \epsilon_{iklm} A_{\underline{a} \underline{b}}{}^{mn} {\bar A}_2{}^{\underline{b}l}{}_n  - \frac{3}{16} \epsilon_{iklm} A_{\underline{a} \underline{b}}{}^{lm} {\bar A}_2{}^{\underline{b} n}{}_n = 0\,. \nonumber
    \end{align}
\noindent
Irreps $(((\mathbf{6},\mathbf{n})\times(\mathbf{6},\mathbf{n}))_A)_0$:
\begin{align}
\label{T27}
    & - 4 A_{2 [\underline{a} [i|}{}^{[k} {\bar A}_{2 |\underline{b}]}{}^{l]}{}_{|j]} - 2  A_{2 \underline{a} [i|}{}^m {\bar A}_{2 \underline{b}}{}^{[k}{}_m \delta^{l]}_{|j]}  + 2 {\bar A}_{2 \underline{a}}{}^m{}_{[i|} A_{2 \underline{b} m}{}^{[k}  \delta^{l]}_{|j]} \nonumber \\
    &  - 2 {\bar A}_{2 [\underline{a}|}{}^m{}_m A_{2 |\underline{b}][i}{}^{[k} \delta^{l]}_{j]} 
 + 2 A_{2 [\underline{a}|m}{}^m {\bar A}_{2 |\underline{b}]}{}^{[k}{}_{[i} \delta^{l]}_{j]} - \frac{1}{2} \delta^{[k}_i \delta^{l]}_j A_{2 [\underline{a}|m}{}^m {\bar A}_{2 |\underline{b}]}{}^n{}_n \nonumber \\
 &  + 2 {\bar A}_{2 (\underline{a}|}{}^m{}_m A_{2 |\underline{b})[i}{}^{[k} \delta^{l]}_{j]} 
 + 2 A_{2 (\underline{a}|m}{}^m {\bar A}_{2 |\underline{b})}{}^{[k}{}_{[i} \delta^{l]}_{j]} -  \delta^{[k}_i \delta^{l]}_j A_{2 (\underline{a}|m}{}^m {\bar A}_{2 |\underline{b})}{}^n{}_n \nonumber \\
 & - \frac{1}{2} \delta_{\underline{a} \underline{b}} A_{2 \underline{c} m}{}^m {\bar A}_2{}^{\underline{c}[k}{}_{[i} \delta^{l]}_{j]} - \frac{1}{2} \delta_{\underline{a} \underline{b}} {\bar A}_2{}^{\underline{c} m}{}_m A_{2 \underline{c} [i}{}^{[k} \delta^{l]}_{j]}  + \frac{1}{4} \delta_{\underline{a} \underline{b}} \delta^{[k}_i \delta^{l]}_j  A_{2 \underline{c} m}{}^m {\bar A}_2{}^{\underline{c} n}{}_n  \nonumber \\
 & + 2 A_{[\underline{a}}{}^{\underline{c} kl} {\bar A}_{\underline{b}] \underline{c} ij } - 4 \delta^{[k}_{[i|} A_{\underline{a} \underline{c}}{}^{l] m} {\bar A}_{\underline{b}}{}^{\underline{c}}{}_{|j]m} + \delta^{[k}_i \delta^{l]}_j A_{\underline{a} \underline{c}}{}^{mn} {\bar A}_{\underline{b}}{}^{\underline{c}}{}_{mn} \nonumber \\ & + \frac{1}{9} \delta_{\underline{a} \underline{b}} \delta^{[k}_{[i} \epsilon_{j]mnp} A_1^{l]m} A_2^{np} + \frac{1}{9} \delta_{\underline{a} \underline{b}} \delta^{[k}_{[i} {\bar A}_{1j]m} \epsilon^{l]mnp} {\bar A}_{2np} \nonumber \\
 & - \frac{1}{18} \delta_{\underline{a} \underline{b}} \delta^k_i \left( A_2^{(lm)} {\bar A}_{2 [jm]} +  A_2^{[lm]} {\bar A}_{2 (jm)} \right)  + \frac{1}{18} \delta_{\underline{a} \underline{b}} \delta^k_j \left( A_2^{(lm)} {\bar A}_{2 [im]} +  A_2^{[lm]} {\bar A}_{2 (im)} \right) \nonumber \\
 & - \frac{1}{18} \delta_{\underline{a} \underline{b}} \delta^l_j \left( A_2^{(km)} {\bar A}_{2 [im]} +  A_2^{[km]}{\bar A}_{2 (im)} \right) + \frac{1}{18} \delta_{\underline{a} \underline{b}} \delta^l_i \left( A_2^{(km)} {\bar A}_{2 [jm]} +  A_2^{[km]}{\bar A}_{2 (jm)} \right) \nonumber \\
 & + \frac{2}{3} {A}_2^{[kl]} {\bar A}_{\underline{a} \underline{b} ij} - \frac{2}{3} \delta^k_{[i|} A_2^{[lm]} {\bar A}_{\underline{a} \underline{b} |j] m} + \frac{2}{3} \delta^l_{[i|} A_2^{[km]} {\bar A}_{\underline{a} \underline{b} |j] m}+ \frac{1}{6} \delta^{[k}_i \delta^{l]}_j A_2^{mn} {\bar A}_{\underline{a} \underline{b} mn} \nonumber \\
 & - \frac{2}{3} {\bar A }_{2[ij]} A_{\underline{a} \underline{b}}{}^{kl} + \frac{2}{3} \delta^{[k}_i {\bar A}_{2[jm]} A_{\underline{a} \underline{b}}{}^{l]m} -  \frac{2}{3} \delta^{[k}_j {\bar A}_{2[im]} A_{\underline{a} \underline{b}}{}^{l]m} - \frac{1}{6} \delta^{[k}_i \delta^{l]}_j {\bar A}_{2mn} A_{\underline{a} \underline{b}}{}^{mn} \nonumber \\
 & + \frac{1}{4} \delta^{[k}_i \delta^{l]}_j A_{\underline{a} \underline{b} \underline{c}} {\bar A}_2{}^{\underline{c} m}{}_m - \frac{1}{4} \delta^{[k}_i \delta^{l]}_j {\bar A}_{\underline{a} \underline{b} \underline{c}}  A_2{}^{\underline{c}}{}_m{}^m = 0 .
\end{align}
The tensor product $((\mathbf{6},\mathbf{n})\times(\mathbf{6},\mathbf{n}))_A$ of SU(4)$\times$SO($n$) decomposes as
\begin{align}
((\mathbf{6},\mathbf{n})\times(\mathbf{6},\mathbf{n}))_A = & \, \left(\mathbf{1},\mathbf{ n (n-1)/2} \right) + \left({\mathbf{20}}^{'},\mathbf{ n (n-1)/2}\right) \nonumber \\
& + \left( \mathbf{15}, \mathbf{ n (n+1)/2 -1}  \right) + \left( \mathbf{15},\mathbf{1}  \right)\,.
\end{align} 
In order to specify the component of \eqref{T27} transforming in the (reducible) $\left( \mathbf{15}, \mathbf{ n (n+1) /2 }\right)$ representation of SU(4)$\times$SO($n$), we contract \eqref{T27} with $\delta^j_l$ and we then symmetrize the resulting equation in $\underline{a}$ and $\underline{b}$. We find
\begin{align}
    (\mathbf{15},\mathbf{n(n+1)/2})_0 :\hspace{0.5cm} & - A_{2 (\underline{a}|i}{}^j {\bar A}_{2|\underline{b})}{}^k{}_j + A_{2 (\underline{a}|j}{}^k {\bar A}_{2|\underline{b})}{}^j{}_i + A_{2 (\underline{a}|i}{}^k {\bar A}_{2|\underline{b})}{}^j{}_j \nonumber \\ 
     &+ A_{2 (\underline{a}|j}{}^j {\bar A}_{2|\underline{b})}{}^k{}_i - \frac{1}{2} \delta^k_i  A_{2 (\underline{a}|j}{}^j {\bar A}_{2|\underline{b})}{}^l{}_l - \frac{1}{4} \delta_{\underline{a} \underline{b}} A_{2 \underline{c} j}{}^j {\bar A}_2{}^{\underline{c} k}{}_i \nonumber \\
     & - \frac{1}{4} \delta_{\underline{a} \underline{b}} A_{2 \underline{c} i}{}^k {\bar A}_2{}^{\underline{c} j}{}_j + \frac{1}{8} \delta_{\underline{a} \underline{b}} \delta^k_i A_{2 \underline{c} j}{}^j {\bar A}_2{}^{\underline{c} l}{}_l - 2 A_{(\underline{a}|\underline{c}}{}^{kj} {\bar A}_{|\underline{b})}{}^{\underline{c}}{}_{ij} \\
     &  + \frac{1}{2} \delta^k_i A_{(\underline{a}|\underline{c}}{}^{jl} {\bar A}_{|\underline{b})}{}^{\underline{c}}{}_{jl} + \frac{1}{18} \delta_{\underline{a} \underline{b}} \epsilon_{ilmn} A_1^{kl} A_2^{mn} \nonumber \\
     & + \frac{1}{18} \delta_{\underline{a} \underline{b}} \epsilon^{klmn} {\bar A}_{1il} {\bar A}_{2mn}  - \frac{1}{9} \delta_{\underline{a} \underline{b}} \left( A_2^{(kj)} {\bar A}_{2 [ij]} + A_2^{[kj]} {\bar A}_{2(ij)} \right) = 0\,.  \nonumber
 \end{align}
On the other hand, the $(\mathbf{1},\mathbf{n(n-1)/2})_0$ component of the quadratic constraint \eqref{T27} follows from contracting \eqref{T27} with $\delta^i_k \delta_l^j$ and then antisymmetrizing the resulting identity in $\underline{a}$ and $\underline{b}$, which gives
\begin{align}
(\mathbf{1},\mathbf{n(n-1)/2})_0: \hspace{0.5cm} & - 4 A_{2 [\underline{a}|i}{}^j {\bar A}_{2 |\underline{b}]}{}^i{}_j + A_{2 [\underline{a}|i}{}^i {\bar A}_{2 |\underline{b}]}{}^j{}_j + 2 A_{[\underline{a}}{}^{\underline{c}ij} {\bar A}_{\underline{b}] \underline{c} ij} \nonumber \\
& - \frac{1}{3} \left( A_2^{ij} {\bar A}_{\underline{a} \underline{b} ij} - {\bar A}_{2ij} A_{\underline{a} \underline{b}}{}^{ij} \right) + \frac{3}{2} \left( A_{\underline{a} \underline{b} \underline{c}} {\bar A}_{2}{}^{\underline{c}i}{}_i - {\bar A}_{\underline{a} \underline{b} \underline{c}} A_2{}^{\underline{c}}{}_i{}^i\right) = 0\,. 
\end{align}
\noindent
Irreps $\left(\mathbf{15},\mathbf{n(n-1)/2}\right)_{0}$:
\begin{align}
\label{T28}
    & \frac{2}{3} A_2^{(jk)} {
    \bar A}_{\underline{a} \underline{b} ik} + \frac{2}{3} {\bar A}_{2(ik)} A_{\underline{a} \underline{b}}{}^{jk} - \frac{1}{3} \epsilon_{iklm} A_1^{jk} A_{\underline{a} \underline{b}}{}^{lm} - \frac{1}{3} \epsilon^{jklm} {\bar A}_{1ik} {\bar A}_{\underline{a} \underline{b} lm} \nonumber \\ 
    & - \frac{2}{3} A_2^{[jk]} {\bar A}_{\underline{a} \underline{b} ik} + \frac{1}{6} \delta^j_i A_2^{kl} {\bar A}_{\underline{a} \underline{b} kl} - \frac{2}{3} {\bar  A}_{2[ik]} A_{\underline{a} \underline{b}}{}^{jk} + \frac{1}{6} \delta^j_i {\bar A}_{2kl} A_{\underline{a} \underline{b}}{}^{kl} \nonumber \\
    & - A_{\underline{a} \underline{b} \underline{c}} \left( {\bar A}_2{}^{\underline{c}j}{}_i - \frac{1}{4} \delta^j_i {\bar A}_2{}^{\underline{c}k}{}_k  \right) - {\bar A}_{\underline{a} \underline{b} \underline{c}} \left( A_2{}^{\underline{c}}{}_i{}^j - \frac{1}{4} \delta^j_i A_2{}^{\underline{c}}{}_k{}^k \right) \\
    &  - A_{2 [\underline{a}| k}{}^k {\bar A}_{2 |\underline{b}]}{}^j{}_i  +  A_{2 [\underline{a}| i}{}^j {\bar A}_{2 |\underline{b}]}{}^k{}_k=  0 \, .\nonumber
\end{align}

\noindent
Irreps $\left(\mathbf{6},\mathbf{n(n-1)/2} \times \mathbf{n}\right)_{0}$:
\begin{align}
\label{T29}
    & \frac{1}{3} \delta_{\underline{c} [\underline{a} } \left( A_{2 \underline{b}]i}{}^k {\bar A}_{2 [jk]} - A_{2 \underline{b}]j}{}^k {\bar A}_{2 [ik]} + \frac{1}{2} A_{2 \underline{b}]k}{}^k {\bar A}_{2 [ij]} \right) + \frac{1}{3} \delta_{\underline{c} [\underline{a} } {\bar A}_{2 \underline{b}]}{}^k{}_{[i} \epsilon_{j]klm} A_2^{lm} + \frac{1}{12} \epsilon_{ijkl} A_2^{kl} \delta_{\underline{c} [\underline{a} } {\bar A}_{2 \underline{b}]}{}^m{}_m \nonumber \\
    & - 2 A_{2 \underline{c} [i|}{}^k {\bar A}_{\underline{a} \underline{b} |j]k} + A_{2 [\underline{a}| k}{}^k {\bar A}_{|\underline{b}] \underline{c} ij} - A_{2 \underline{c} k}{}^k {\bar A}_{\underline{a} \underline{b} ij} - {\bar A}_{2 \underline{c}}{}^k{}_{[i}
\epsilon_{j]klm} A_{\underline{a} \underline{b}}{}^{lm} - \frac{1}{2} \epsilon_{ijlm} {\bar A}_{2 [\underline{a}|}{}^k{}_k
   A_{|\underline{b}] \underline{c}}{}^{lm} \nonumber \\ & - \frac{1}{2} \delta_{\underline{c} [\underline{a}} \left( {\bar A}_{\underline{b}] \underline{d} ij} A_2{}^{\underline{d}}{}_k{}^k - \frac{1}{2} \epsilon_{ijkl} A_{\underline{b}] \underline{d}}{}^{kl} {\bar A}_{2}{}^{\underline{d}m}{}_m  \right)  + \frac{1}{3} {\bar A}_{2[ij]} A_{\underline{a} \underline{b} \underline{c}} - \frac{1}{6} \epsilon_{ijkl} A_2^{kl} {\bar A}_{\underline{a} \underline{b} \underline{c}} \\ 
    & - A_{\underline{a} \underline{b} }{}^{\underline{d}} {\bar A}_{\underline{c} \underline{d} ij} + \frac{1}{2} \epsilon_{ijkl}  {\bar A}_{\underline{a} \underline{b} }{}^{\underline{d}} A_{\underline{c} \underline{d}}{}^{kl} =  0 \, . \nonumber
\end{align}

\noindent
Irreps $\left(\mathbf{1},(\mathbf{n(n-1)/2} \times \mathbf{n(n-1)/2})_A\right)_{0}$:
\begin{align}
\label{T30}
   & - A_{\underline{a} \underline{b}}{}^{ij} {\bar A}_{\underline{c} \underline{d} ij}+ {\bar A}_{\underline{a} \underline{b} ij} A_{\underline{c} \underline{d}}{}^{ij} + A_{\underline{a} \underline{b} \underline{e}} {\bar A}_{\underline{c} \underline{d}}{}^{\underline{e}} - {\bar A}_{\underline{a} \underline{b} \underline{e}} A_{\underline{c} \underline{d}}{}^{\underline{e}} \nonumber \\
   & - \delta_{[\underline{a}[\underline{c}} \left( 
 \frac{2}{3} {\bar A}_{\underline{d}] \underline{b}]ij} A_2^{ij} - \frac{2}{3} A_{\underline{d}] \underline{b}]}{}^{ij} {\bar A}_{2ij} + {\bar A}_{\underline{d}] \underline{b}] \underline{e}} A_2{}^{\underline{e}}{}_i{}^i - A_{\underline{d}] \underline{b}] \underline{e}} {\bar A}_2{}^{\underline{e}i}{}_i \right) \\ 
   & + A_{2 [\underline{a}|i}{}^i {\bar A}_{|\underline{b}] \underline{c} \underline{d}} - A_{2 [\underline{c}|i}{}^i {\bar A}_{|\underline{d}] \underline{a} \underline{b}} - {\bar A}_{2 [\underline{a}|}{}^i{}_i A_{|\underline{b}] \underline{c} \underline{d}} + {\bar A}_{2 [\underline{c}|}{}^i{}_i A_{|\underline{d}] \underline{a} \underline{b}} = 0\,. \nonumber \end{align} 




\begin{thebibliography}{100}

\bibitem{Sen:1994fa} A.~Sen, ``Strong - weak coupling duality in four-dimensional string theory,'' Int. J. Mod. Phys. A \textbf{9} (1994), 3707-3750 [arXiv:hep-th/9402002 [hep-th]].

\bibitem{Maldacena:1996gb} J.~M.~Maldacena and A.~Strominger,  ``Statistical entropy of four-dimensional extremal black holes,'' Phys. Rev. Lett. \textbf{77} (1996), 428-429 [arXiv:hep-th/9603060 [hep-th]].

\bibitem{Dijkgraaf:1996it} R.~Dijkgraaf, E.~P.~Verlinde and H.~L.~Verlinde, ``Counting dyons in N=4 string theory,'' Nucl. Phys. B \textbf{484} (1997), 543-561 [arXiv:hep-th/9607026 [hep-th]].

\bibitem{Dibitetto:2012rk} G.~Dibitetto, J.~J.~Fernandez-Melgarejo, D.~Marques and D.~Roest, ``Duality orbits of non-geometric fluxes,'' Fortsch. Phys. \textbf{60} (2012), 1123-1149 [arXiv:1203.6562 [hep-th]].

\bibitem{Das:1977uy}
A.~K.~Das,
``SO(4) Invariant Extended Supergravity,''
Phys. Rev. D \textbf{15} (1977), 2805

\bibitem{Cremmer:1977tc}
E.~Cremmer and J.~Scherk, ``Algebraic Simplifications in Supergravity Theories,''
Nucl. Phys. B \textbf{127} (1977), 259-268

\bibitem{Cremmer:1977tt}
E.~Cremmer, J.~Scherk and S.~Ferrara,
``SU(4) Invariant Supergravity Theory,''
Phys. Lett. B \textbf{74} (1978), 61-64

\bibitem{Freedman:1978ra}
D.~Z.~Freedman and J.~H.~Schwarz,
``N=4 Supergravity Theory with Local SU(2) x SU(2) Invariance,''
Nucl. Phys. B \textbf{137} (1978), 333-339

\bibitem{deRoo:1984zyh} M.~de Roo,
``Matter Coupling in N=4 Supergravity,''
Nucl. Phys. B \textbf{255} (1985), 515-531

\bibitem{Bergshoeff:1985ms} 
E.~Bergshoeff, I.~G.~Koh and E.~Sezgin,
``Coupling of Yang-Mills to N=4, D=4 Supergravity,''
Phys. Lett. B \textbf{155} (1985), 71

\bibitem{deRoo:1985np}
M.~de Roo,
``GAUGED N=4 MATTER COUPLINGS,''
Phys. Lett. B \textbf{156} (1985), 331-334


\bibitem{deRoo:1985jh} M.~de Roo and P.~Wagemans,
``Gauged Matter Coupling in $N=4$ Supergravity,''
Nucl. Phys. B \textbf{262} (1985), 644



\bibitem{Perret:1987nk} R.~E.~C.~Perret, ``GEOMETRIC N=4 SUPERGRAVITY,''
Class. Quant. Grav. \textbf{5} (1988), 1109

\bibitem{Perret:1988jq} R.~E.~C.~Perret,
``GEOMETRIC STRUCTURE OF N=4 MATTER COUPLED SUPERGRAVITY,''
Class. Quant. Grav. \textbf{5} (1988), 1115

\bibitem{Frey:2002hf} A.~R.~Frey and J.~Polchinski,
``N=3 warped compactifications,''
Phys. Rev. D \textbf{65} (2002), 126009
[arXiv:hep-th/0201029 [hep-th]].

\bibitem{Kachru:2002he} S.~Kachru, M.~B.~Schulz and S.~Trivedi,
``Moduli stabilization from fluxes in a simple IIB orientifold,''
JHEP \textbf{10} (2003), 007 [arXiv:hep-th/0201028 [hep-th]].

\bibitem{DAuria:2002qje} 
R.~D'Auria, S.~Ferrara and S.~Vaula,
``N=4 gauged supergravity and a IIB orientifold with fluxes,''
New J. Phys. \textbf{4} (2002), 71
[arXiv:hep-th/0206241 [hep-th]].

\bibitem{DAuria:2003nhg} R.~D'Auria, S.~Ferrara, F.~Gargiulo, M.~Trigiante and S.~Vaula,
``$N=4$ supergravity Lagrangian for type IIB on $T^6 / \mathbb{Z}_2$ in presence of fluxes and $D3$-branes,''
JHEP \textbf{06} (2003), 045
[arXiv:hep-th/0303049 [hep-th]].

\bibitem{Berg:2003ri}
M.~Berg, M.~Haack and B.~Kors,
``An Orientifold with fluxes and branes via T duality'',
Nucl. Phys. B \textbf{669} (2003), 3-56
[arXiv:hep-th/0305183 [hep-th]].

\bibitem{Angelantonj:2003rq}
C.~Angelantonj, S.~Ferrara and M.~Trigiante,
``New D = 4 gauged supergravities from N=4 orientifolds with fluxes'',
JHEP \textbf{10} (2003), 015
[arXiv:hep-th/0306185 [hep-th]].

\bibitem{Angelantonj:2003up}
C.~Angelantonj, S.~Ferrara and M.~Trigiante,
``Unusual gauged supergravities from type IIA and type IIB orientifolds,''
Phys. Lett. B \textbf{582} (2004), 263-269 [arXiv:hep-th/0310136 [hep-th]].

\bibitem{Villadoro:2004ci} G.~Villadoro and F.~Zwirner, ``The Minimal N=4 no-scale model from generalized dimensional reduction,'' JHEP \textbf{07} (2004), 055 [arXiv:hep-th/0406185 [hep-th]].

\bibitem{Derendinger:2004jn} J.~P.~Derendinger, C.~Kounnas, P.~M.~Petropoulos and F.~Zwirner,
``Superpotentials in IIA compactifications with general fluxes,'' Nucl. Phys. B \textbf{715} (2005), 211-233 [arXiv:hep-th/0411276 [hep-th]].

\bibitem{Villadoro:2005cu} G.~Villadoro and F.~Zwirner, ``N=1 effective potential from dual type-IIA D6/O6 orientifolds with general fluxes,'' JHEP \textbf{06} (2005), 047 
[arXiv:hep-th/0503169 [hep-th]].

\bibitem{DallAgata:2009wsi} G.~Dall'Agata, G.~Villadoro and F.~Zwirner, ``Type-IIA flux compactifications and N=4 gauged supergravities,'' JHEP \textbf{08} (2009), 018 [arXiv:0906.0370 [hep-th]].

\bibitem{Schon:2006kz} J.~Schon and M.~Weidner,
``Gauged N=4 supergravities,''
JHEP \textbf{05} (2006), 034 [arXiv:hep-th/0602024 [hep-th]].

\bibitem{Gaillard:1981rj} M.~K.~Gaillard and B.~Zumino, ``Duality Rotations for Interacting Fields'', Nucl. Phys. B \textbf{193} (1981), 221-244

\bibitem{Samtleben:2008pe} H.~Samtleben,
``Lectures on Gauged Supergravity and Flux Compactifications,''
Class. Quant. Grav. \textbf{25} (2008), 214002
[arXiv:0808.4076 [hep-th]].

\bibitem{Trigiante:2016mnt} 
M.~Trigiante, ``Gauged Supergravities,''
Phys. Rept. \textbf{680} (2017), 1-175
[arXiv:1609.09745 [hep-th]].

\bibitem{DallAgata:2021uvl} 
G.~Dall\textquoteright{}Agata and M.~Zagermann,
``Supergravity: From First Principles to Modern Applications,''
Lect. Notes Phys. \textbf{991} (2021), 1-263

\bibitem{Cordaro:1998tx}
F.~Cordaro, P.~Fre, L.~Gualtieri, P.~Termonia and M.~Trigiante,
``N=8 gaugings revisited: An Exhaustive classification,''
Nucl. Phys. B \textbf{532} (1998), 245-279 [arXiv:hep-th/9804056 [hep-th]].

\bibitem{Nicolai:2000sc} H.~Nicolai and H.~Samtleben,
``Maximal gauged supergravity in three-dimensions,''
Phys. Rev. Lett. \textbf{86} (2001), 1686-1689
[arXiv:hep-th/0010076 [hep-th]].

\bibitem{Nicolai:2001sv} H.~Nicolai and H.~Samtleben,
``Compact and noncompact gauged maximal supergravities in three-dimensions,''
JHEP \textbf{04} (2001), 022
[arXiv:hep-th/0103032 [hep-th]].

\bibitem{deWit:2002vt} B.~de Wit, H.~Samtleben and M.~Trigiante,
``On Lagrangians and gaugings of maximal supergravities,''
Nucl. Phys. B \textbf{655} (2003), 93-126
[arXiv:hep-th/0212239 [hep-th]].

\bibitem{deWit:2004nw} 
B.~de Wit, H.~Samtleben and M.~Trigiante,
``The Maximal D=5 supergravities,''
Nucl. Phys. B \textbf{716} (2005), 215-247 [arXiv:hep-th/0412173 [hep-th]].

\bibitem{deWit:2005ub} B.~de Wit, H.~Samtleben and M.~Trigiante,
``Magnetic charges in local field theory,''
JHEP \textbf{09} (2005), 016
[arXiv:hep-th/0507289 [hep-th]].

\bibitem{deWit:2007kvg} B.~de Wit, H.~Samtleben and M.~Trigiante,
``The Maximal D=4 supergravities,''
JHEP \textbf{06} (2007), 049
[arXiv:0705.2101 [hep-th]].

\bibitem{Louis:2014gxa} J.~Louis and H.~Triendl, ``Maximally supersymmetric AdS$_{4}$ vacua in N = 4 supergravity'', JHEP \textbf{10} (2014), 007 [arXiv:1406.3363 [hep-th]].

\bibitem{DallAgata:2012tne} G.~Dall'Agata and F.~Zwirner, ``Quantum corrections to broken N = 8 supergravity'', JHEP \textbf{09} (2012), 078
[arXiv:1205.4711 [hep-th]].  

\bibitem{Polchinski:1998rr}
J.~Polchinski,
``String theory. Vol. 2: Superstring theory and beyond,''
Cambridge University Press, 2007

\bibitem{DallAgata:2005fjb}
G.~Dall'Agata and S.~Ferrara,
``Gauged supergravity algebras from twisted tori compactifications with fluxes'',
Nucl. Phys. B \textbf{717} (2005), 223-245
[arXiv:hep-th/0502066 [hep-th]].

\bibitem{Louis:2002ny} J.~Louis and A.~Micu, ``Type 2 theories compactified on Calabi-Yau threefolds in the presence of background fluxes'',
Nucl. Phys. B \textbf{635} (2002), 395-431
[arXiv:hep-th/0202168 [hep-th]].

\bibitem{DallAgata:2003sjo} G.~Dall'Agata, R.~D'Auria, L.~Sommovigo and S.~Vaula, ``D = 4, N=2 gauged supergravity in the presence of tensor multiplets'',
Nucl. Phys. B \textbf{682} (2004), 243-264
[arXiv:hep-th/0312210 [hep-th]].

\bibitem{DallAgata:2005xvr} G.~Dall'Agata, R.~D'Auria and S.~Ferrara, ``Compactifications on twisted tori with fluxes and free differential algebras'',
Phys. Lett. B \textbf{619} (2005), 149-154
[arXiv:hep-th/0503122 [hep-th]].

\bibitem{deWit:2005hv}  
B.~de Wit and H.~Samtleben,
``Gauged maximal supergravities and hierarchies of nonabelian vector-tensor systems,''
Fortsch. Phys. \textbf{53} (2005), 442-449 [arXiv:hep-th/0501243 [hep-th]].

\bibitem{deVroome:2007unr} M.~de Vroome and B.~de Wit,
``Lagrangians with electric and magnetic charges of N=2 supersymmetric gauge theories,''
JHEP \textbf{08} (2007), 064 [arXiv:0707.2717 [hep-th]].

\bibitem{deWit:1975veh} B.~de Wit and D.~Z.~Freedman,
``On Combined Supersymmetric and Gauge Invariant Field Theories,''
Phys. Rev. D \textbf{12} (1975), 2286

\bibitem{Jackiw:1978ar} R.~Jackiw,
``GAUGE COVARIANT CONFORMAL TRANSFORMATIONS,''
Phys. Rev. Lett. \textbf{41} (1978), 1635

\bibitem{Freedman:2012zz} D.~Z.~Freedman and A.~Van Proeyen,
``Supergravity,''
Cambridge Univ. Press, 2012

\bibitem{DAuria:2001rlt} R.~D'Auria and S.~Ferrara, ``On fermion masses, gradient flows and potential in supersymmetric theories'', JHEP \textbf{05} (2001), 034 [arXiv:hep-th/0103153 [hep-th]].

\bibitem{deWit:1983gs} B.~de Wit and H.~Nicolai,
``The Parallelizing S(7) Torsion in Gauged $N=8$ Supergravity,''
Nucl. Phys. B \textbf{231} (1984), 506-532

\bibitem{Ferrara:2016ntj} S.~Ferrara and A.~Van Proeyen,
``Mass Formulae for Broken Supersymmetry in Curved Space-Time,''
Fortsch. Phys. \textbf{64} (2016) no.11-12, 896-902 [arXiv:1609.08480 [hep-th]].

 \bibitem{Coleman:1973jx} S.~R.~Coleman and E.~J.~Weinberg,
``Radiative Corrections as the Origin of Spontaneous Symmetry Breaking,''
Phys. Rev. D \textbf{7} (1973), 1888-1910

 \bibitem{Weinberg:1973ua} S.~Weinberg,
``Perturbative Calculations of Symmetry Breaking,''
Phys. Rev. D \textbf{7} (1973), 2887-2910



\bibitem{Dibitetto:2011eu} G.~Dibitetto, A.~Guarino and D.~Roest, ``How to halve maximal supergravity,'' JHEP \textbf{06} (2011), 030 [arXiv:1104.3587 [hep-th]]. 

\bibitem{Castellani:1991eu} L.~Castellani, R.~D'~Auria and P.~Fr\'{e}, ``Supergravity and Superstrings: A Geometric Perspective," World Scientific (1991), Vol. 1,2,3


\end{thebibliography}
\end{document}